\PassOptionsToPackage{unicode}{hyperref}
\PassOptionsToPackage{hyphens}{url}
\PassOptionsToPackage{dvipsnames,svgnames,x11names}{xcolor}
\documentclass[12pt]{article}

\usepackage{amsmath,amssymb}
\usepackage{iftex}
\ifPDFTeX
  \usepackage[T1]{fontenc}
  \usepackage[utf8]{inputenc}
  \usepackage{textcomp} %
\else %
  \usepackage{unicode-math}
  \defaultfontfeatures{Scale=MatchLowercase}
  \defaultfontfeatures[\rmfamily]{Ligatures=TeX,Scale=1}
\fi
\usepackage{lmodern}
\ifPDFTeX\else  
\fi
\IfFileExists{upquote.sty}{\usepackage{upquote}}{}
\IfFileExists{microtype.sty}{%
  \usepackage[]{microtype}
  \UseMicrotypeSet[protrusion]{basicmath} %
}{}

\usepackage{xcolor}
\makeatletter
\ifx\paragraph\undefined\else
  \let\oldparagraph\paragraph
  \renewcommand{\paragraph}{
    \@ifstar
      \xxxParagraphStar
      \xxxParagraphNoStar
  }
  \newcommand{\xxxParagraphStar}[1]{\oldparagraph*{#1}\mbox{}}
  \newcommand{\xxxParagraphNoStar}[1]{\oldparagraph{#1}\mbox{}}
\fi
\ifx\subparagraph\undefined\else
  \let\oldsubparagraph\subparagraph
  \renewcommand{\subparagraph}{
    \@ifstar
      \xxxSubParagraphStar
      \xxxSubParagraphNoStar
  }
  \newcommand{\xxxSubParagraphStar}[1]{\oldsubparagraph*{#1}\mbox{}}
  \newcommand{\xxxSubParagraphNoStar}[1]{\oldsubparagraph{#1}\mbox{}}
\fi
\makeatother

\usepackage{longtable,booktabs,array}
\usepackage{calc} %
\usepackage{etoolbox}
\makeatletter
\patchcmd\longtable{\par}{\if@noskipsec\mbox{}\fi\par}{}{}
\makeatother
\IfFileExists{footnotehyper.sty}{\usepackage{footnotehyper}}{\usepackage{footnote}}
\makesavenoteenv{longtable}
\usepackage{graphicx}
\makeatletter
\def\maxwidth{\ifdim\Gin@nat@width>\linewidth\linewidth\else\Gin@nat@width\fi}
\def\maxheight{\ifdim\Gin@nat@height>\textheight\textheight\else\Gin@nat@height\fi}
\makeatother
\setkeys{Gin}{width=\maxwidth,height=\maxheight,keepaspectratio}
\makeatletter
\def\fps@figure{htbp}
\makeatother

\makeatletter
\@ifpackageloaded{caption}{}{\usepackage{caption}}
\AtBeginDocument{%
\ifdefined\contentsname
  \renewcommand*\contentsname{Table of contents}
\else
  \newcommand\contentsname{Table of contents}
\fi
\ifdefined\listfigurename
  \renewcommand*\listfigurename{List of Figures}
\else
  \newcommand\listfigurename{List of Figures}
\fi
\ifdefined\listtablename
  \renewcommand*\listtablename{List of Tables}
\else
  \newcommand\listtablename{List of Tables}
\fi
\ifdefined\figurename
  \renewcommand*\figurename{Figure}
\else
  \newcommand\figurename{Figure}
\fi
\ifdefined\tablename
  \renewcommand*\tablename{Table}
\else
  \newcommand\tablename{Table}
\fi
}
\@ifpackageloaded{float}{}{\usepackage{float}}
\floatstyle{ruled}
\@ifundefined{c@chapter}{\newfloat{codelisting}{h}{lop}}{\newfloat{codelisting}{h}{lop}[chapter]}
\floatname{codelisting}{Listing}

\makeatother
\makeatletter
\@ifpackageloaded{caption}{}{\usepackage{caption}}
\@ifpackageloaded{subcaption}{}{\usepackage{subcaption}}
\makeatother

\ifLuaTeX
  \usepackage{selnolig}  %
\fi
\usepackage[]{natbib}
\usepackage{bookmark}

\IfFileExists{xurl.sty}{\usepackage{xurl}}{} %
\hypersetup{
  pdftitle={Unified Framework for Multiple Exposure Distributed Lag Non-Linear Models for Air Pollution Epidemiology},
  pdfauthor={Tianyi Pan; Hwashin Hyun Shin; Alex Stringer; Glen McGee},
  pdfkeywords={Air pollution; Environmental mixtures; Distributed lag non-linear models; Single-index models; Stacking},
  colorlinks=true,
  linkcolor={blue},
  filecolor={Maroon},
  citecolor={Blue},
  urlcolor={Blue},
  pdfcreator={LaTeX via pandoc}}

\usepackage{moreverb}
\usepackage{booktabs}
\usepackage{eqparbox} %

\usepackage{lineno}

\let\oldequation\equation
\let\oldendequation\endequation
\renewenvironment{equation}
  {\linenomathNonumbers\oldequation}
  {\oldendequation\endlinenomath}

\usepackage{xr-hyper}
\usepackage{hyperref}

\usepackage{caption}
\usepackage{subcaption}

\usepackage{float}

\usepackage{makecell}
\usepackage{multirow}
\usepackage{setspace}%

\usepackage{amsthm}

\usepackage{algorithm}
\usepackage{algpseudocode}

\usepackage{arydshln} %

\usepackage[dvipsnames]{xcolor}

\usepackage{enumitem}

\newcommand{\anon}{1}

\usepackage{minitoc}

\begin{document}

\def\spacingset#1{\renewcommand{\baselinestretch}%
{#1}\small\normalsize} \spacingset{1}

\if1\anon
{
  \title{\bf A Unified Framework for Multiple Exposure Distributed Lag Non-Linear Models for Air Pollution Epidemiology}
  \author{Tianyi Pan\\
    Department of Statistics and Actuarial Science, University of Waterloo\\
    \\
    Hwashin Hyun Shin \\ 
    Environmental Health Science and Research Bureau, Health Canada \\ 
    Department of Mathematics and Statistics, Queen's University \\ 
    \\
    Alex Stringer \\ 
    Department of Statistics and Actuarial Science, University of Waterloo\\
    \\
    Glen McGee \\ 
    Department of Statistics and Actuarial Science, University of Waterloo}
  \date{}
  \maketitle
} \fi

\if0\anon
{
  \title{\bf A Unified Framework for Multiple Exposure Distributed Lag Non-Linear Models for Air Pollution Epidemiology}
  \date{}
  \maketitle
} \fi


\begin{abstract}

This study quantifies the association between air pollution and mortality in Ontario, Canada. Exposure-response relationships in air pollution epidemiology are complex due to three features: time-lagged associations, non-linear associations, and multiple pollutants. 
To address the first two features, two distinct classes of distributed lag non-linear model (DLNM) have been proposed, but extending them to multiple exposures and selecting an appropriate model remain challenging. 
We propose a unified framework for multiple exposure DLNMs, integrating model specification, estimation, selection and stacking. The framework applies to four different model structures: two additive and two proposed single-index DLNMs, all applicable to general outcome types, including the mortality counts in the motivating application. We develop an estimation approach that applies to all four models. Choosing among the candidate DLNMs is challenging \textit{a priori}, and we derive an AIC to select among them. As an alternative to selecting a single model, we also extend a model stacking approach to combine inferences across the four DLNMs and propose an implementation scalable to our dataset with 106,346 observations. In the motivating analysis, the four DLNMs yield different estimates, and the proposed stacking approach identifies significant associations between respiratory mortality and a mixture of PM$_{2.5}$, O$_3$ and NO$_2$. 

\end{abstract}

\noindent
{\it Keywords:} Air pollution; Environmental mixtures; Distributed lag non-linear models; Single-index models; Stacking
\vfill

\spacingset{1.8} %

\doparttoc 
\faketableofcontents 
\part{}  

\vspace{-2cm}
\section{Introduction}

Air pollution has been recognized as a major public health concern, with adverse effects confirmed by evidence from Canada \citep{huang2023multi} and worldwide \citep{liu2019ambient}. %
Despite improvements in air quality, the World Health Organization reported that the global health burden attributable to air pollution has shown little reduction since the 1990s \citep{who2021guidelines}. 
The Canadian government reported that more than 15,000 premature deaths per year are attributable to PM$_{2.5}$, O$_3$ and NO$_2$, 
with more than 40\% occurring in Ontario \citep{health_canada_2021_air_pollution_report}. 
This study assesses the associations between a mixture of PM$_{2.5}$, O$_3$ and NO$_2$ and mortality counts in Ontario, Canada, between 2001 and 2015. 
However, air pollution analysis is challenged by three %
features: time-lagged associations \citep{farhat2013short}, non-linear associations \citep{liu2019ambient}, and multiple pollutants \citep{stieb2008new}. 
This paper develops a framework that jointly addresses these challenges.

Ignoring lagged exposures can lead to biased estimates of health effects \citep{bell2004ozone}.  %
A common practical approach %
is to take a simple average of concentrations over several days \citep{chen2017fine, liu2019ambient}, but this assumes constant contributions across lags, which is unlikely to hold in practice. 
Distributed lag models (DLMs; \citealp{schwartz2000distributed, zanobetti2000generalized}) instead estimate lag weights from the data, but assume linear associations. 
To extend DLMs to allow for non-linear exposure-response relationships, two distinct classes of distributed lag non-linear models (DLNMs) have been proposed. 
One class, called adaptive cumulative exposure DLNMs (ACE-DLNMs), specifies a univariate non-linear function of the cumulative exposure over lags \citep{pan2025estimating}. The other, referred to as distributed response function DLNMs (DRF-DLNMs), models non-linear functions of exposure at each lag through a bivariate surface \citep{gasparrini2017penalized}. %

Humans are simultaneously exposed to many air pollutants, which may be highly correlated, and recent studies have considered extensions of DLNMs to mixtures of multiple pollutants. 
Within an ACE-DLNM structure, \citet{wilson2022kernel} proposed a Bayesian kernel machine regression for continuous outcomes, allowing for non-linear high-order interactions. 
However, estimating such a flexible surface is challenging when effect sizes are small, as is common in air pollution studies such as our own. 
In this paper, we therefore focus on the alternative class of approaches that impose parsimonious structures to improve efficiency, including additive and single-index structures.
Additive versions of both ACE- and DRF-DLNMs have been proposed by \citet{pan2025estimating} and \citet{gasparrini2017penalized}, respectively, but they require estimating separate non-linear functions for each exposure, which is challenging in analyses with low signal and multiple highly correlated exposures. 
An alternative approach is to impose a single-index structure, reducing multiple exposures to a linear index and capturing its non-linear effect through a univariate function, which can improve efficiency in studies with small effect sizes, like our analysis of air pollution in Ontario. 
\citet{wang2023semiparametric} and \citet{li_dynamic_2025} introduced single-index structures into ACE-DLNM and DRF-DLNM respectively; 
however, these models are restricted to continuous outcomes, which precludes them from being applied to our analysis of the mortality counts. 
This therefore motivates the development of single-index DLNMs that accommodate general types of outcomes.

These different DLNMs rely on different assumptions to impose structure on the unknown exposure-response surface: the ACE- and DRF-DLNM reduce the complexity related to time lags, and the additive and single-index structures reduce the complexity across multiple exposures. 
These model assumptions yield different inferences, and incorrect assumptions lead to biased estimates and invalid inference \citep{pan2025estimating}.
In our analysis of all-cause mortality in Ontario, the ACE-DLNM structures yield null overall effects while the additive DRF-DLNM produces a positive overall effect; 
the single-index ACE-DLNM yields null effects of PM$_{2.5}$ while the two additive DLNMs show negative effects of PM$_{2.5}$ (see Figure \ref{fig:app_rr}). 
Despite common practice of assuming a single model \textit{a priori} and basing conclusions on that choice, there is often little prior knowledge favouring one of several subtly different DLNM parametrizations. 
This motivates a twofold strategy. First, we need methods for identifying the best fitting model. 
Second, as different models may be misspecified but capture different aspects of the underlying exposure-response surface, 
we aim to combine inferences across models to borrow strength and obtain results that are robust to misspecifications.

To investigate daily air pollution and mortality counts in Ontario, between 2001 and 2015, we develop a unified framework for multiple exposure DLNMs for assessing non-linear and time-lagged associations between multiple exposures and a health outcome. 
This framework integrates model specification, estimation, selection, and stacking, requiring four novel contributions.  
First, we propose the single-index ACE-DLNM and single-index DRF-DLNM for multiple pollutants, applicable to general outcome types including the mortality counts in the motivating application. 
Second, we develop an estimation and inference approach based on penalized splines for the two proposed single-index DLNMs as well as the additive ACE- and DRF-DLNM. This approach scales to large datasets like our motivating data collected over 15 years across 20 census divisions in Ontario, totaling 106,346 observations. 
Third, we derive a new AIC to select among these four DLNMs fitted within the proposed penalized regression framework, where existing AICs are not directly applicable. Fourth, we extend the model stacking approach \citep{yao2018using} to combine these DLNMs for inference on a common estimand of interest such as relative mortality reduction in our analysis, and develop an implementation based on a one-step Newton update in \citet{wood2024neighbourhood}, which allows the stacking approach to scale to our large dataset. 
In our analysis, the single-index DLNMs and additive DLNMs produce different estimates, and stacking combines them, yielding results distinct from either model.

\section{Data, Notation, and Estimand}
\subsection{Ontario Air Pollution and Mortality Data}
\label{s:data}

In this study, we quantify the associations between a mixture of PM$_{2.5}$, O$_3$ and NO$_2$ and all-cause and cause-specific mortality counts in Ontario, Canada.

The data cover January 1, 2001, to December 31, 2015 and are aggregated by geographic and administrative units called Census Divisions (CD; \citealp{statcan_CD}). 
We obtained data for 20 CDs with available health and air pollution data, covering about 84\% of the population in Ontario. 
Among these, 15 CDs have available data beginning on January 1, 2001 while the remaining 5 CDs start slightly later (see supplementary Table S1). 
From each CD’s starting date to December 31, 2015, there are no missing daily observations. In total, 106,346 observations are modeled in our analysis. 

The mortality counts were collected from the Canadian Vital Statistics--Death database \citep{statcan2023} and aggregated by CD and date. 
We model three mortality counts: all-cause non-accidental (A00-R99), hereafter termed all-cause; circulatory (I00-I99); and respiratory (J00-J99), classified according to the International Classification of Diseases code \citep{WHO_ICD10}. 
The National Air Pollution Surveillance program \citep{NAPS} provided hourly air pollutant concentrations, including fine particulate matter less than 2.5 micrometres in size (PM$_{2.5}$), ozone (O$_3$) and nitrogen dioxide (NO$_2$). 
Daily O$_3$ concentrations were defined as the daily maximum of 8-hour rolling averages, and daily PM$_{2.5}$ and NO$_2$ concentrations were defined as the daily averages, following \citet{shin2022sex} and the Canadian Ambient Air Quality Standards \citep{CAAQS}. 
We also collected the daily mean temperature for each CD \citep{ECCC}. 

A data summary is provided in supplementary Table S1. 
Mortality rates are low for each outcome; for example, the daily average respiratory mortality rate is 0.1 per 100,000 people in Toronto, the most populous CD in Ontario. Supplementary Figures S1--S3 indicate seasonality and correlation in concentrations of the three pollutants. The low mortality rates and correlated pollutants complicate inference, motivating us to propose approaches that achieve statistical efficiency and computational scalability.

\subsection{Notation}
Let $Y_{it}$ be the response for group $i \in \left\{1, \ldots,N\right\}$ observed at time $t \in \mathcal{T}_i =  \left\{1, \ldots, n_i\right\}$, and let $n =\sum_{i=1}^N n_i$ be the total number of observations. 
We consider $M$ exposures $\boldsymbol{x}_{it} = (x_{it1}, \ldots, x_{itM})^\top$. 
We assume an underlying exposure process $\boldsymbol{X}_i = (X_{i1}, \ldots, X_{iM})^\top$, where each $X_{im}: \mathbb{R} \rightarrow \mathbb{R}$, and $x_{itm}$ is the realization of $X_{im}$ at time $t$. Let $\boldsymbol{z}_{it} \in \mathbb{R}^p$ be a vector of covariates. 
In the motivating analysis, index $i$ refers to a CD ($N = 20$), $Y_{it}$ is the mortality count for CD $i$ on day $t$, $\boldsymbol{x}_{it}$ is the daily concentrations of PM$_{2.5}$, O$_3$ and NO$_2$ ($M = 3$), and $\boldsymbol{z}_{it}$ contains calendar time, seasonality, day of week and temperature.

We denote the association between response at time $t$ and exposures by $s(\boldsymbol{X}_i, t)$. 
This relationship is complex due to the three challenges: (i) multiple exposures, where correlated pollutants (PM$_{2.5}$, O$_3$ and NO$_2$) may be jointly associated with mortality; (ii) time lags, such that mortality at time $t$ is influenced not only by contemporaneous exposure but also prior exposures (e.g., up to two weeks), a structure referred to as a distributed lag; and (iii) non-linearity, where the effects of these pollutants vary across concentration levels. 
Imposing assumptions yields several different parameterizations, which we introduce in Section \ref{s:models}.

\subsection{Estimand of Interest}
Different model assumptions parametrize the relationship $s(\boldsymbol{X}_i, t)$ in distinct ways with different interpretations, as we discuss in Section \ref{s:models}. We consider a common estimand, referred to as the relative mortality reduction: 
\begin{equation}
\Delta(\alpha) = 1-\frac{\sum_{i=1}^N \sum_{t \in \mathcal{T}_i}\hat{\mu}_{it}\left\{(1-\alpha) \boldsymbol{x}_i, \boldsymbol{z}_{it}\right\}}{\sum_{i=1}^N \sum_{t \in \mathcal{T}_i} \hat{\mu}_{it}\left\{\boldsymbol{x}_i, \boldsymbol{z}_{it}\right\}},
\label{eq:est}
\end{equation}
where $\hat{\mu}_{it}\left\{(1-\alpha) \boldsymbol{x}_i, \boldsymbol{z}_{it}\right\}$ is the estimated expected death count for CD $i$ at time $t$ under exposure levels $(1-\alpha) \boldsymbol{x}_i$ and covariates $\boldsymbol{z}_{it}$. 
Here $\boldsymbol{x}_i$ %
is the observed levels of PM$_{2.5}$, O$_3$, and NO$_2$ for CD $i$ across the study period, and $(1-\alpha)\boldsymbol{x}_i$ corresponds to the levels obtained by reducing the observed levels by a proportion $\alpha$. 
In this analysis, we set $\alpha = 0.1$, representing a 10\% reduction. The estimand $\Delta(10\%)$ measures the overall effect and is interpreted as the relative reduction in the expected number of deaths under a 10\% reduction in all exposure levels relative to the observed levels. This estimand has a natural policy interpretation: it quantifies the changes in mortality associated with a policy that reduces 10\% of air pollutant concentrations. 
We define analogous estimands to quantify individual associations by reducing one exposure by 10\% while holding the others at their observed levels. 
We do not use rate ratios based on two fixed exposure levels, such as the 75\% and 25\% percentiles of observed concentrations. 
Temporal trends in exposure levels make it difficult to choose two levels that consistently represent high and low concentrations over time, and the results are sensitive to the choice (see supplementary Appendix H.4).

\section{Multiple Exposure DLNMs}
\label{s:models}

We assume that $Y_{it} | \boldsymbol{X}_{i}, \boldsymbol{z}_{it}$ follows a distribution with finite variance and a density that is 
twice continuously differentiable with respect to the mean parameter $\mu_{it} \in \mathcal{S} \subseteq \mathbb{R}$ 
and additional unknown parameters $\boldsymbol{\theta}$. This class includes distributions in the exponential family. 
In our analysis of mortality counts, we use a negative binomial distribution with mean $\mu_{it} \in \mathcal{S} = (0,\infty)$ and unknown dispersion parameter $\theta > 0$. 

We consider the following general model structure: 
\begin{equation}
    g(\mu_{it}) = s(\boldsymbol{X}_i, t) + \sum_{j=1}^p h_j (z_{itj}),
    \label{eq:general}
\end{equation}
where $g: \mathcal{S} \rightarrow \mathbb{R}$ is a smooth, monotone link function, and we use log-link for the negative binomial distribution $g(\mu_{it}) = \log (\mu_{it})$. 
The $p$ unknown smooth functions $\boldsymbol{h} = (h_1, \ldots, h_p)$ can represent non-linear smooth functions, linear components and random effects \citep{wood2016smoothing}. 
The association between outcome at time $t$ and exposures is captured by the unknown surface $s(\boldsymbol{X}_i, t)$. 
Because of multiple exposures, time lags, and non-linearity mentioned above, the form of $s(\boldsymbol{X}_i, t)$ is complex. Imposing further assumptions leads to several different DLNM parameterizations.

\subsection{Existing Models: Additive DLNMs}
\label{ss:existingDLNM}
The ACE-DLNM and DRF-DLNM make different assumptions about lagged effects of a time-varying exposure, and both have been extended to incorporate additive effects of multiple exposures \citep{pan2025estimating,gasparrini2017penalized}.

The additive ACE-DLNM \citep{pan2025estimating} specifies 
\begin{equation}
    s(\boldsymbol{X}_i, t) = \sum_{m=1}^{M} f_m \left\{\int_{0}^{L_m} w_m(l) X_{im}(t-l) dl\right\}, 
    \label{eq:additive-ace-dlnm}
\end{equation}
where $\int_{0}^{L_m} w_m(l) X_{im}(t-l) dl$ is the adaptive cumulative exposure (ACE) at time $t$, with $w_m(\cdot)$ denoting
an unknown smooth lag weight function, and $L_m$ is the (user-defined) maximum lag. 
In our analysis, we set $L_m = 15$, and the integral includes the same-day exposure and exposures 1 to 14 days prior. 
The association between ACE and response is captured by the unknown smooth function $f_m(\cdot)$, for $m = 1, \ldots, M$. The ACE-DLNM replaces the underlying exposure process $X_{im}$ with an interpolating spline fit to the observed exposure level $x_{itm}, t \in  \mathcal{T}_m$, which enables exact evaluation of the integral. 

The additive DRF-DLNM \citep{gasparrini2017penalized} specifies 
\begin{equation}
    s(\boldsymbol{X}_i, t) = \sum_{m=1}^M \int_{0}^{L_m} \psi_m\left\{X_{im}(t-l), l\right\} dl. 
    \label{eq:additive-drf-dlnm}
\end{equation}
where $\psi_m(\cdot, \cdot)$ is a bivariate smooth function capturing the exposure-response association that varies by lag. 
The integral aggregates the lag-specific associations and is approximated by a discrete summation.

Both additive models require fitting a separate surface for each exposure, which is challenging when the effect sizes are small and multiple exposures are highly correlated.

\subsection{Proposed Models: Single-Index DLNMs}
\label{ss:proposedDLNM}
We propose two single-index DLNMs, extending both the ACE-DLNM and DRF-DLNM, respectively.
The models reduce the dimensionality of multiple exposures by introducing a linear multi-exposure index, and capture the joint effects through a single function. 

We first define the single index for CD $i$ and day $t$ as $E_i(t; \boldsymbol{\alpha}) = \sum_{m = 1}^M \alpha_m X_{im}(t)$, 
where $\boldsymbol{\alpha} = (\alpha_1, \ldots, \alpha_M)^\top$ are the index weights quantifying the relative contribution of each individual exposure to the overall effect. 
The relationship between the outcome and the single index can be modeled using the ACE-DLNM or DRF-DLNM reparameterization. 

Our proposed single-index ACE-DLNM is 
\begin{equation}
  s(\boldsymbol{X}_i, t) = f \left\{\int_{0}^{L} w(l) E_{i}(t-l;\boldsymbol{\alpha}) dl\right\}. 
  \label{eq:si-ace-dlnm}
\end{equation}
We denote $E^L_{i}(t;\boldsymbol{\alpha}, w) = \int_{0}^{L} w(l) E_{i}(t-l;\boldsymbol{\alpha}) dl$ as the cumulative index, aggregating the single index over lags from 0 to the maximum lag $L$. We set $L = 15$ in the analysis to allow the effect of the mixture to last for up to two weeks. The unknown smooth function $w(\cdot)$ captures the relative importance of index lags, and the unknown function $f(\cdot)$ models the association between the cumulative index and the response. The underlying exposure processes are replaced by interpolating splines for exact evaluation of the integral as in additive ACE-DLNMs.

Our proposed single-index DRF-DLNM is specified as
\begin{equation}
  s(\boldsymbol{X}_i, t) = \int_{0}^{L} \psi\left\{E_{i}(t-l; \boldsymbol{\alpha}), l\right\} dl,
  \label{eq:si-drf-dlnm}
\end{equation}
where the bivariate smooth function $\psi(\cdot, \cdot)$ models the lag-varying association between the single index and response. 
The model approximates the integral by discrete summation, as in the existing additive DRF-DLNMs. 
The bivariate function can be interpreted through slices along either dimension. 
For a fixed lag $l$, the slice $\psi(\cdot, l)$ captures the association between outcome and single index at lag $l$. For a fixed index level $E$, the slice $\psi(E, \cdot)$ captures the lag structure at the specific index level. 

Both single-index DLNMs yield mixture-level interpretations through the single index, but the interpretation of $\psi$ in the single-index DRF-DLNM differs from that of $f$ and $w$ in the single-index ACE-DLNM: $\psi(\cdot, l)$ characterizes the association with the index at specific lags while $f(\cdot)$ characterizes the association with the cumulative index; $\psi(E, \cdot)$ measures lag contributions to the overall effect while $w(\cdot)$ measures them to the cumulative index. 
On the other hand, the additive DLNMs yield exposure-specific interpretations. 
These differences in interpretations motivate us to consider common estimands such as the relative mortality reduction (Equation \ref{eq:est}). In our analysis, however, the multiple models yield different estimates of the relative mortality reduction, raising the question of how to draw inference in practice. We will address this question in Section \ref{s:selection-averaging}.

Introducing index weights reduces the number of non-linear functions that need to be estimated, but the nested structure of the index weights inside the unknown association functions presents challenges for estimation. 
In Section \ref{s:est}, we describe a unified estimation framework for any of the models described in this section.

\subsection{Identifiability}

Parameter identifiability is a general concern in index models. A common strategy is to constrain the magnitude and sign of index weights: $\boldsymbol{\alpha}^\top \boldsymbol{\alpha} = 1$ and $\alpha_1 > 0$ (e.g., \citealp{yu2002penalized}). %
However, the latter arbitrarily forces the effect of the first exposure to be non-null ($\alpha_1 > 0$). 
This might not hold in practice; in our analysis we find null effects of the first exposure PM$_{2.5}$ on respiratory and all-cause mortality under both single-index DLNMs. 
We instead propose the constraint: $\boldsymbol{1}_M^\top \boldsymbol{\alpha} > 0$ where $\boldsymbol{1}_M$ is the unit vector of length $M$. 
This constrains the sign of the overall index while allowing any individual weight to be zero, and can be applied broadly to any index model.

To facilitate an efficient fitting, we propose a reparameterization of the constrained index weights to an unconstrained parameter space. 
Consider a more general form of the constraints: $\boldsymbol{\alpha}^\top \boldsymbol{\alpha} = 1$ and $\boldsymbol{c}^\top \boldsymbol{\alpha}> 0$, where $\boldsymbol{c}$ is a non-zero vector of length $M$. This encompasses both our preferred constraint $\mathbf{1}_M^\top \boldsymbol{\alpha} > 0$ ($\boldsymbol{c} = \mathbf{1}_M$) and the common $\alpha_1 > 0$ ($\boldsymbol{c} = \mathbf{e}_1= [1,0,\ldots,0]^{\top}$).
We satisfy these by mapping $\boldsymbol{\alpha}$ from an unconstrained parameter $\boldsymbol{\alpha}^* \in \mathbb{R}^{M-1}$ as follows: 
\begin{equation}
    \boldsymbol{\alpha} = \frac{\mathbf{B}_{\boldsymbol{\alpha}} \left[1, \boldsymbol{\alpha}^{*\top}\right]^\top}{\left(\left[1, \boldsymbol{\alpha}^{*\top}\right] \mathbf{B}_{\boldsymbol{\alpha}}^\top \mathbf{B}_{\boldsymbol{\alpha}} \left[1, \boldsymbol{\alpha}^{*\top}\right]^\top\right)^{1/2}},
    \label{eq:rapa}
\end{equation}
where $\mathbf{B}_{\boldsymbol{\alpha}} = \left[\boldsymbol{c}, \mathbf{Q}^+_{\boldsymbol{\alpha}}\right]$. 
The matrix $\mathbf{Q}^+_{\boldsymbol{\alpha}}$ is obtained by removing the first column of $\mathbf{Q}_{\boldsymbol{\alpha}}$ from the QR decomposition: $\mathbf{Q}_{\boldsymbol{\alpha}} \mathbf{R}_{\boldsymbol{\alpha}} = \boldsymbol{c}$. The proof is in supplementary Appendix B. 
This generalizes the reparameterization of \citet{yu2002penalized} which is obtained as a special case when $\boldsymbol{c} = \mathbf{e}_1$; see details in supplementary Appendix B.

We adopt standard identifiability constraints for all other parameters. The lag weight function $w$ satisfies $\int_{0}^{L} w(l) > 0$ and $\int_{0}^{L} w(l)^2 = 1$ \citep{pan2025estimating}, allowing for mortality displacement \citep{zanobetti2000generalized} and facilitating computation via reparameterization. 
The functions $f$, $\psi$ and $\boldsymbol{h}$ are subject to linear constraints used in generalized additive models \citep{wood2017generalized}.

\section{Estimation and Inference}
\label{s:est}

This section develops a unified estimation and inference framework for the proposed single-index ACE- and DRF-DLNMs; the framework also accommodates the additive ACE- and DRF-DLNMs. 
Within this framework, all four multiple exposure DLNMs in Section \ref{s:models} are estimated based on a penalized log-likelihood with a common general form and a shared optimization approach. 
This framework provides valid uncertainty quantification and scales to large datasets such as our analysis with 106,346 observations, unlike the ad-hoc manner in \citet{wood2017generalized} (Chapter 7.4.1) that conducts inference conditional on fixed index weights estimated via derivative-free methods.

We represent unknown smooth functions by cubic B-spline basis expansions: $f(x) = \sum_{q=1}^{d^f} b^f_{q}(x)\beta^f_{q}$, $w(x) = \sum_{q=1}^{d^w} b^w_{q}(x)\beta^w_{q}$, $\psi(x,l) = \sum_{s = 1}^{d^{\psi_x}} \sum_{r = 1}^{d^{\psi_l}} b^{\psi_x}_s(x) b^{\psi_l}_r(l) \beta^\psi_{sr}$, and $h_j(x) = \sum_{q = 1}^{d^{h_j}} b_{jq}^{h}(x) \beta^{h}_{jq}$. 
Here $b_q^f$, $b_q^w$, $b^{\psi_x}_s$, $b^{\psi_l}_r$ and $b_{jq}^{h}$ are known cubic B-spline functions.
We denote the unknown spline coefficients as $\boldsymbol{\beta}^f = (\beta^f_q)_{q=1,\ldots,d^f}$, $\boldsymbol{\beta}^w = (\beta^w_q)_{q=1,\ldots,d^w}$, 
$\boldsymbol{\beta}^\psi = (\beta^\psi_{sr})_{s=1,\ldots,d^{\psi_x}, r = 1,\ldots,d^{\psi_l}}$, 
and $\boldsymbol{\beta}^{\boldsymbol{h}} = (\boldsymbol{\beta}_j^{h})_{j=1,\ldots,p}$ where $\boldsymbol{\beta}_j^{h} = (\beta^{h}_{j q})_{q = 1,\ldots,d^{h_j}}$. Quadratic smoothness penalties are imposed on the spline coefficients. 
For a univariate smooth function, the penalty takes the form $\mathcal{P}(\boldsymbol{\beta}; \lambda) = (1/2) \lambda \boldsymbol{\beta}^\top \mathbf{S} \boldsymbol{\beta}$, %
where $\lambda$ is an unknown smoothing parameter to be estimated and $\mathbf{S}$ is a known penalty matrix \citep{wood2017p}. 
The penalty for the bivariate smooth function $\psi$ is $\mathcal{P}(\boldsymbol{\beta}^{\psi}; \boldsymbol{\lambda}^{\psi}) = (1/2) \boldsymbol{\beta}^{\psi \top} (\lambda^{\psi_x} \mathbf{S}^{\psi_x} \otimes \mathbf{I}_{d^{\psi_x}} + \lambda^{\psi_l} \mathbf{I}_{d^{\psi_l}} \otimes \mathbf{S}^{\psi_l} ) \boldsymbol{\beta}^{\psi}$ where $\mathbf{I}_d$ is the identity matrix with dimension $d$ and $\otimes$ denotes the Kronecker product. %

The basis functions of $f$ and $\psi$ are defined through their knot sequences, 
which depend on the ranges of 
$E^{L}_i(t; \boldsymbol{\alpha}, w)=\int_{0}^{L} w(l) E_i(t-l; \boldsymbol{\alpha}) dl$ and $E_i(t; \boldsymbol{\alpha}) = \boldsymbol{\alpha}^\top \boldsymbol{X}_i(t)$, and rely on the unknown $\boldsymbol{\alpha}$ and $w$. 
Without well-defined ranges, the splines cannot be constructed, yielding challenges in inference. 
To address this, we pre-specify their ranges that do not depend on any unknown quantities, using the Cauchy-Schwarz inequality. 
Specifically, the range of $E_i^{L}(t; \boldsymbol{\alpha}, w)$ over all $t$ and $i$ is $\left[-\bar{E}^{L}, \bar{E}^{L}\right]$ where $\bar{E}^{L} = \max_{i,t} \sqrt{\sum_{m=1}^M \int_0^L X_{im}(t-l)^2 d l}$, and the range of $E_i(t; \boldsymbol{\alpha})$ over all $t$ and $i$ is $\left[-\bar{E}, \bar{E}\right]$ where $\bar{E} = \max_{i,t} \sqrt{\sum_{m = 1}^M X_{im}(t)^2}$. Details are provided in supplementary Appendix C.

To construct a unified estimation framework, we partition the parameters in both the single-index ACE-DLNM and the single-index DRF-DLNM into three groups: (i) inner parameters $\boldsymbol{\phi}$: the parameters defining $E^{L}_i(t; \boldsymbol{\alpha}, w)$ or $E_i(t; \boldsymbol{\alpha})$ inside the association functions $f$ or $\psi$; (ii) outer parameters $\boldsymbol{\gamma}$: the spline coefficients for the association function $f$ or $\psi$ and the covariate functions $\boldsymbol{h}$; and (iii) tuning parameters including $\boldsymbol{\lambda}$ and $\boldsymbol{\theta}$. 
The smooth penalty is denoted as as a general form $\mathcal{P}(\boldsymbol{\phi}, \boldsymbol{\gamma}; \boldsymbol{\lambda})$. 
Specifically, for the single-index ACE-DLNM, we have $\boldsymbol{\phi} = (\boldsymbol{\alpha}^{*\top}, \boldsymbol{\beta}^{w*\top})^\top$ where $\boldsymbol{\beta}^{w*}$ is the unconstrained parameter
mapped to $\boldsymbol{\beta}^{w}$ to ensure the identifiability constraints on $w$ \citep{pan2025estimating}, 
$\boldsymbol{\gamma} = (\boldsymbol{\beta}^{f\top}, \boldsymbol{\beta}^{\boldsymbol{h\top}})^\top$, $\boldsymbol{\lambda} = (\lambda^w, \lambda^f, \boldsymbol{\lambda}^{h\top})^\top$ where $\boldsymbol{\lambda}^{h} = (\lambda^{h}_1, \ldots, \lambda^{h}_p)^\top$, and $\mathcal{P}(\boldsymbol{\phi}, \boldsymbol{\gamma}; \boldsymbol{\lambda}) = \mathcal{P}(\boldsymbol{\beta}^{w*}; \lambda^w) + \mathcal{P}(\boldsymbol{\beta}^{f}; \lambda^f) + \mathcal{P}(\boldsymbol{\beta}^{\boldsymbol{h}}; \boldsymbol{\lambda}^{h})$.
For the single-index DRF-DLNM, we have $\boldsymbol{\phi} = \boldsymbol{\alpha}^{*}$, $\boldsymbol{\gamma} = (\boldsymbol{\beta}^{\psi\top}, \boldsymbol{\beta}^{\boldsymbol{h}\top})^\top$, $\boldsymbol{\lambda} = (\lambda^{\psi_x}, \lambda^{\psi_l}, \boldsymbol{\lambda}^{h\top})^\top$ and $\mathcal{P}(\boldsymbol{\phi}, \boldsymbol{\gamma}; \boldsymbol{\lambda}) = \mathcal{P}(\boldsymbol{\beta}^{\psi}; \lambda^{\psi_x}, \lambda^{\psi_l}) + \mathcal{P}(\boldsymbol{\beta}^{\boldsymbol{h}}; \boldsymbol{\lambda}^{h})$. 
This framework incorporates the additive ACE- and DRF-DLNMs by replacing $w$, $f$ and $\phi$ with exposure-specific functions $w_m$, $f_m$ and $\psi_m$. %

The penalized log-likelihood can be written in a general form:
$
\mathcal{L}(\boldsymbol{\phi}, \boldsymbol{\gamma}; \boldsymbol{\lambda}, \boldsymbol{\theta}) = l(\boldsymbol{\phi}, \boldsymbol{\gamma}; \boldsymbol{\theta}) - \mathcal{P}(\boldsymbol{\phi}, \boldsymbol{\gamma}; \boldsymbol{\lambda}),
$
where the log-likelihood $l(\boldsymbol{\phi}, \boldsymbol{\gamma}; \boldsymbol{\theta}) = \sum_{i=1}^N \sum_{t \in \mathcal{T}_i} \log p(Y_{it} | \boldsymbol{x}_{it}, \boldsymbol{z}_{it}; \boldsymbol{\phi}, \boldsymbol{\gamma}, \boldsymbol{\theta})$.

We aim to maximize the penalized log-likelihood to obtain the estimators. The optimization problem is complicated by the lag structure, multiple smoothing parameters and unknown index weights.  
\citet{pan2025estimating} developed an efficient algorithm to address the lag structure and smoothing parameter estimation for single-exposure ACE-DLNMs. We extend this to the general form of $\mathcal{L}(\boldsymbol{\phi}, \boldsymbol{\gamma}; \boldsymbol{\lambda}, \boldsymbol{\theta})$, in which the inner parameters $\boldsymbol{\phi}$ incorporate both index weights parameters and spline coefficients for smooth functions, and the penalty terms can involve multiple smoothing parameters (e.g. $\mathcal{P}(\boldsymbol{\gamma}^{\psi}; \lambda^{\psi_x}, \lambda^{\psi_l})$ in the single-index DRF-DLNM). 
We estimate $\boldsymbol{\phi}$ and $\boldsymbol{\gamma}$ by maximizing the profile likelihood, and $\boldsymbol{\lambda}$ and $\boldsymbol{\theta}$ by maximizing the marginal likelihood. See supplementary Appendix D for details. 
The estimates are denoted by $\widehat{\boldsymbol{\phi}}$, $\widehat{\boldsymbol{\gamma}}$, $\widehat{\boldsymbol{\lambda}}$ and $\widehat{\boldsymbol{\theta}}$.

Because the index weights $\boldsymbol{\alpha}$ and the spline coefficients for $w$ are subject to identifiability constraints and are mapped from the inner parameters $\boldsymbol{\phi}$ through a non-linear reparameterization, we propose a sampling-based approach to compute confidence intervals. 
We first draw $R$ samples of the unconstrained $\boldsymbol{\phi}$ and $\boldsymbol{\gamma}$ from $\mathcal{N}\left((\widehat{\boldsymbol{\phi}}^\top, \widehat{\boldsymbol{\gamma}}^\top)^\top, \boldsymbol{\mathcal{H}}^{-1}\right)$, 
where $\boldsymbol{\mathcal{H}}$ is the negative Hessian of $\mathcal{L}$ with respect to $(\boldsymbol{\phi}, \boldsymbol{\gamma})$ evaluated at the estimates. 
For each sample, we map the unconstrained inner parameters to the constrained parameters. 
Confidence intervals for function values and index weights are then constructed using the sample quantiles of their estimates across the $R$ samples. 
In this procedure, the parameters are jointly sampled, with the Hessian taken with respect to all inner and outer parameters, unlike \citet{wood2017generalized} (Chapter 7.4.1), where inference treats index weights as fixed.

\section{Model Selection and Stacking}
\label{s:selection-averaging}

Due to the complexity of $s(\boldsymbol{X}_i, t)$ involving multiple exposures, time lags and non-linearity, 
explicitly specifying and fitting the true $s(\boldsymbol{X}_i, t)$ is infeasible. 
The models described in Section \ref{s:models}, including the proposed single-index ACE- and DRF-DLNMs and the additive ACE- and DRF-DLNMs, collected in the model set $\mathcal{M} = \left\{M_1, \ldots, M_4\right\}$, rely on restrictive simplifications of the true model. The true model is unlikely to be in $\mathcal{M}$; the setting is $\mathcal{M}$-open \citep{bernardo1994bayesian}.

Each candidate model yields inferences about model-specific quantities with distinct interpretations, and results in different estimates for the common estimand (see Section \ref{s:application}). 
This raises a challenge: how should we draw inferences when candidate models disagree? 
One strategy is to select a single model and draw inferences from it. In practice, however, there may be no prior knowledge on which simplification is appropriate. To address this, Section \ref{ss:AIC} proposes an Akaike information criterion (AIC) for these misspecified models. 
Beyond selecting a single model, Section \ref{ss:MA} extends a model stacking approach in the $\mathcal{M}$-open case to combine inferences on a common estimand across DLNMs, borrowing strength while accounting for model uncertainty, and proposes a scalable implementation.

\subsection{AIC}
\label{ss:AIC}

Existing AIC cannot be directly applied to the $\mathcal{M}$-open DLNMs considered here. 
\citet{wood2016smoothing} proposed an AIC for general smooth models. 
We extend this to account for both the smoothness penalty and model misspecification arising from restrictive simplifications of the true model. 
The proposed AIC is: $\mathrm{AIC}_{\text{cond}} = -2 l(\widehat{\boldsymbol{u}}) + \mathrm{tr}\left[(\widehat{\boldsymbol{\mathcal{I}}} + \mathbf{S^{\boldsymbol{\lambda}}})^{-1} \widehat{\mathbf{K}}\right]$, where $\boldsymbol{u}=(\boldsymbol{\phi}^\top, \boldsymbol{\gamma}^\top, \boldsymbol{\theta}^\top)^\top$ denotes the parameters that determine the log-likelihood, $\widehat{\boldsymbol{u}}(\boldsymbol{\lambda})$ denotes the estimator of $\boldsymbol{u}$ for a given $\boldsymbol{\lambda}$, and $\widehat{\boldsymbol{u}} = \widehat{\boldsymbol{u}}(\widehat{\boldsymbol{\lambda}})$. 
The matrix $\widehat{\boldsymbol{\mathcal{I}}}$ is the Hessian of the negative log-likelihood with respect to $\boldsymbol{u}$ evaluated at $\widehat{\boldsymbol{u}}$, and $\widehat{\mathbf{K}} = \sum_{i=1}^N \sum_{t \in \mathcal{T}_i} \left[\partial \log p(Y_{it} | \boldsymbol{x}_{it}, \boldsymbol{z}_{it};\widehat{\boldsymbol{u}})/\partial\boldsymbol{u} \right] \left[\partial \log p(Y_{it} | \boldsymbol{x}_{it}, \boldsymbol{z}_{it};\widehat{\boldsymbol{u}})/\partial\boldsymbol{u} \right]^\top$. The derivation is provided in See supplementary Appendix F.

We refer to this proposed criterion as conditional AIC because it fixes $\boldsymbol{\lambda}$ at $\widehat{\boldsymbol{\lambda}}$ without accounting for uncertainty in estimating $\boldsymbol{\lambda}$. 
An adjusted AIC that incorporates this uncertainty is proposed in supplementary Appendix F.3. 
It requires intractable third-order derivatives and is more computationally expensive than the conditional AIC; we develop an efficient computational approach. 
In the simulation (Section \ref{s:simC}), the conditional and adjusted AIC yield broadly similar results, in line with findings in \citet{wood2016smoothing}. We therefore use the conditional AIC in the data analysis.

\subsection{Model Stacking}
\label{ss:MA}
For $\mathcal{M}$-open cases, \citet{yao2018using} proposed a Bayesian stacking approach to combine Bayesian predictive distributions. 
Combining penalized regression models for general estimands has not been investigated in the Bayesian stacking literature. 
In this section, we extend the stacking approach for penalized regression models including DLNMs, and develop a scalable implementation using a one-step Newton update \citep{wood2024neighbourhood}.

Consider an estimand of interest $\Delta$, such as the relative mortality reduction, 
which can be estimated from any candidate model in $\mathcal{M} = \left\{M_1, \ldots, M_K\right\}$. 
We focus on the case where all candidate models are penalized regression models, such as the single-index and additive DLNMs fitted within our proposed framework. 
We adopt a pseudo-Bayesian perspective, which is widely used in penalized regression models \citep{wood2016smoothing}. 
The model-weighted posterior distribution of $\Delta$ is
$p(\Delta | \mathcal{D}) = \sum_{k=1}^K \xi_k p(\Delta | \mathcal{D}, M_k)$, 
where $p(\Delta | \mathcal{D}, M_k)$ is the (pseudo-)posterior distribution of $\Delta$ under model $M_k$ based on the observed dataset $\mathcal{D}$, and $\boldsymbol{\xi} = (\xi_1, \ldots, \xi_K)^\top$ is the model-averaging weights, satisfying $\sum_{k=1}^K \xi_k = 1$ and $\xi_k \geq 0$. 
Classic Bayesian model averaging chooses $\boldsymbol{\xi}$ based on the posterior model probabilities but is not suitable for $\mathcal{M}$-open cases. 
We instead use a Bayesian stacking \citep{yao2018using} designed for $\mathcal{M}$-open cases, and determine $\boldsymbol{\xi}$ by solving the following cross-validation-based optimization problem:
$$
\max_{\boldsymbol{\xi}} \sum_{s=1}^{n} \log \sum_{k=1}^K \xi_k p(Y_s | \mathcal{D}^{(-s)}, M_k). 
$$
Here we denote $\mathcal{D}^{(-s)}$ as the dataset omitting the $s$-th observation and $Y_s$ is the corresponding response. 
Let $\boldsymbol{u}_k$ be the parameters that determine the log-likelihood function for model $M_k$. We evaluate
$p(Y_s | \mathcal{D}^{(-s)}, M_k) = \int p\left(Y_s | \boldsymbol{u}_k, M_k\right) p(\boldsymbol{u}_k | \mathcal{D}^{(-s)}, M_k) d \boldsymbol{u}_k$, 
where $p\left(Y_s | \boldsymbol{u}_k, M_k\right)$ is the likelihood of $Y_s$ given parameters $\boldsymbol{u}_k$ under $M_k$, and $p\left(\boldsymbol{u}_k | \mathcal{D}^{(-s)}, M_k\right)$ is the pseudo-posterior distribution of $\boldsymbol{u}_k$ based on data $\mathcal{D}^{(-s)}$ under $M_k$. 
To simplify notation, we omit the subscript $k$, but note that the parameter $\boldsymbol{u}$ and the corresponding gradient and Hessian differ across models. 
As is conventional in penalized regression models \citep{wood2016smoothing}, we fix $\boldsymbol{\lambda}$ at their estimate and approximate the pseudo-posterior distribution of $\boldsymbol{u}$ by 
$$
\boldsymbol{u}\ |\ \mathcal{D}^{(-s)} \sim N\left(\widehat{\boldsymbol{u}}^{(-s)}, \left(\boldsymbol{\mathcal{H}}^{(-s)}\right)^{-1}\right),
$$
where $\widehat{\boldsymbol{u}}^{(-s)}$ is the estimate of $\boldsymbol{u}$ based on the data $\mathcal{D}^{(-s)}$
and $\boldsymbol{\mathcal{H}}^{(-s)}$ is the Hessian of the negative penalized log-likelihood for data $\mathcal{D}^{(-s)}$ evaluated at $\widehat{\boldsymbol{u}}^{(-s)}$.

Obtaining the estimates $\widehat{\boldsymbol{u}}^{(-s)}$ for $s = 1,\ldots,n$ requires fitting each model $n$ times. 
This is computationally infeasible for the complex DLNMs in the data analysis in Section~\ref{s:application} where $n = 106,346$. 
To overcome the computational burden, we propose a scalable implementation that incorporates the one-step Newton update of \citet{wood2024neighbourhood}.
Instead of refitting $n$ times, we fit each model a single time using the full data $\mathcal{D}$ to obtain the estimate $\widehat{\boldsymbol{u}}$, and approximate $\widehat{\boldsymbol{u}}^{(-s)}$ via the one-step Newton update:
$$
\widehat{\boldsymbol{u}}^{(-s)} \approx \widehat{\boldsymbol{u}} - \left(\boldsymbol{\mathcal{H}}(\widehat{\boldsymbol{u}}) - \widehat{\boldsymbol{\mathcal{I}}}^{(s)}(\widehat{\boldsymbol{u}})\right)^{-1} \boldsymbol{g}^{(s)}(\widehat{\boldsymbol{u}}),
$$
where $\boldsymbol{\mathcal{H}}(\widehat{\boldsymbol{u}})$ is the Hessian of the negative penalized log-likelihood using $\mathcal{D}$ evaluated at $\widehat{\boldsymbol{u}}$ and does not vary across updates, 
and $\widehat{\boldsymbol{\mathcal{I}}}^{(s)}(\widehat{\boldsymbol{u}})$ and $\boldsymbol{g}^{(s)}(\widehat{\boldsymbol{u}})$ are the Hessian and gradient, respectively, of the negative log-likelihood for the $s$-th observation evaluated at $\widehat{\boldsymbol{u}}$.
With this approximation, we only need to update the estimate $n$ times, each involving just a single observation. 
The $n$ updates are separate %
and can be parallelized to further improve the computational efficiency. 
We implement the method in {\tt Cpp} with {\tt OpenMP} \citep{openmp} to enable parallel computing.

\section{Simulation} 
\label{s:simC}

We conduct three simulation studies. 
The first two assess the performance of the developed estimation and inference framework for fitting the proposed single-index ACE-DLNM and single-index DRF-DLNM, respectively. We observe low bias and near-nominal coverage for the model-specific quantities; see supplementary Appendices G.1 and G.2. 
This section presents the simulation that evaluates the performance of model selection and stacking.

We consider a data generating mechanism that encompasses both single-index DLNMs as special cases but cannot be fitted using existing approaches. 
The exposure data are daily standardized PM$_{2.5}$, O$_3$ and NO$_2$ concentrations in Waterloo, Ontario, starting from January 1, 2001, with a sample size of $n$=1,000. 
The count outcome $Y_t$ is generated from a single-index ACE-DLNM with a negative binomial distribution with mean such that $\log(\mu_t) = s(\boldsymbol{X}, t) + h(t)$ where $h(t) = 0.5 ~\mathrm{sin}(t/150)$ represents a seasonal trend, and a dispersion parameter $\theta = 8$. 
We set index weights $\boldsymbol{\alpha} = (1, 0.5, 0.3)^\top/c $ such that $c$ ensures $\boldsymbol{\alpha}^\top \boldsymbol{\alpha} = 1$ and a non-linear lag weight $w(\cdot)$ %
(see Type (i) in supplementary Figure S4). The true exposure-response surface is defined as
$$
s(\boldsymbol{X}, t) = f_{\text{ACE}} \left\{\int_{0}^{15} w(l) f_{\text{DRF}}\left(\sum_{m=1}^3 \alpha_m X_{m}(t-l)\right)d l\right\}. 
$$
The single-index ACE-DLNM models a non-linear $f_{\text{ACE}}$ but constrains $f_{\text{DRF}}$ to be linear; whereas the single-index DRF-DLNM allows a non-linear $f_{\text{DRF}}$ but constrains $f_{\text{ACE}}$ to be linear. 
Each model is correct when its corresponding constrained function is truly linear, and the setting is $\mathcal{M}$-open when both functions are non-linear. 
We specify each function ($f_{\text{ACE}}$ and $f_{\text{DRF}}$) as a weighted average of the form $(1-\omega) z\beta + \omega \eta(z)$, where $\eta$ is a non-linear function constructed using a sine function and $\omega \in \{0, 0.25, 0.5,0.75, 1\}$, yielding a total of 25 combinations. 
The weight $\omega$ controls the degree of nonlinearity: $\omega = 0$ yields a linear function and $\omega = 1$ yields a highly non-linear function (see supplementary Figure S6).

We estimate $\Delta(10\%)$ in Equation \ref{eq:est} via: 
(i) the single-index ACE-DLNM;
(ii) the single-index DRF-DLNM;
(iii) the model with lowest conditional AIC; 
(iv) the model with lowest adjusted AIC (supplementary Appendix F.3); and
(v) model stacking.

\begin{figure}[htbp]
  \centering
  \begin{subfigure}{\textwidth}
    \centering
    \includegraphics[width=0.8\textwidth]{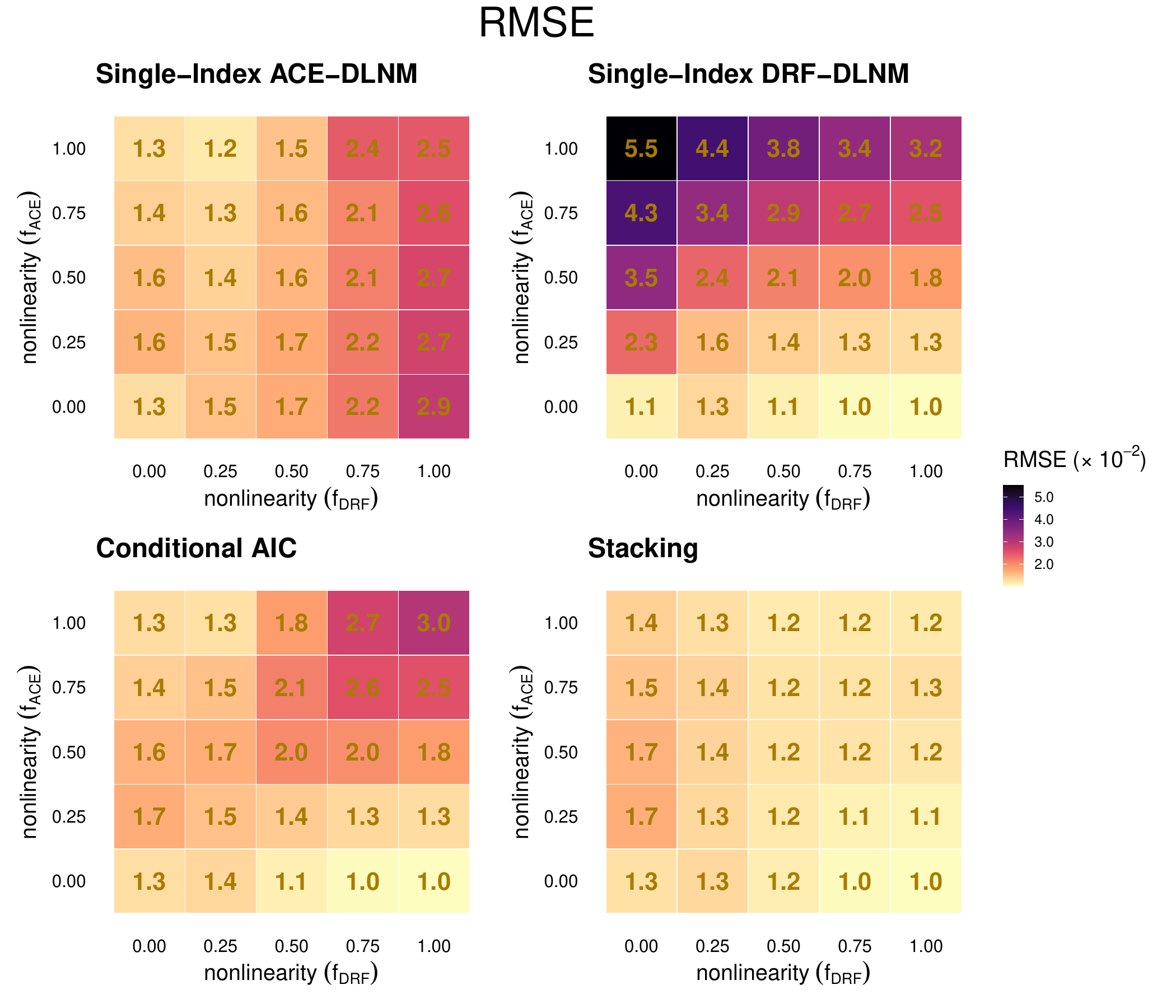}
  \end{subfigure}

  \begin{subfigure}{\textwidth}
    \centering
    \includegraphics[width=0.8\textwidth]{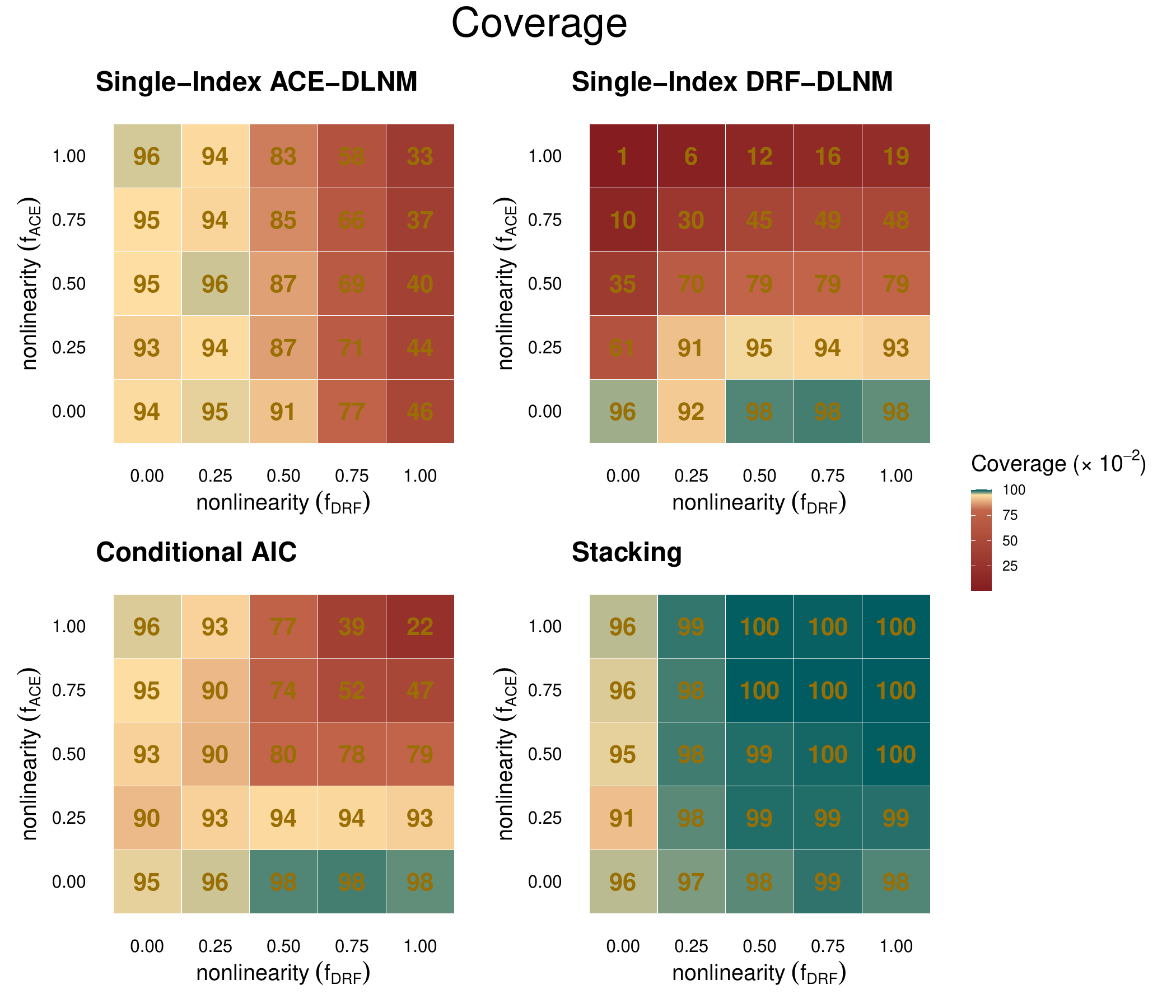}
  \end{subfigure}
  \caption{Results of Simulation C. The RMSE and coverage of the quantity of interest $\Delta(10\%)$ are reported. The nonlinearity of $f_{\text{ACE}}$ and $f_{\text{DRF}}$ are varied in the rows and columns.}
  \label{fig:simC-heatmap}
\end{figure}

Figure \ref{fig:simC-heatmap} shows heatmaps of RMSE and coverage across 1,000 replicates, with the degree of nonlinearity of $f_{\text{ACE}}$ and $f_{\text{DRF}}$ varying along the rows and columns. %
The heatmaps for bias are in supplementary Figure S9. Results for adjusted AIC are similar to those of conditional AIC and are shown in supplementary Figures S7--S9.
When one function is linear (x- or y-axis at 0), the corresponding model performs well with smaller RMSEs and coverages closer to 95\%; increasing the nonlinearity of $f_{\text{DRF}}$ ($f_{\text{ACE}}$) leads to worse performance of the single-index ACE-DLNM (DRF-DLNM). 
The model with lowest AIC behaves similarly to the best model in each setting, supporting its use in model selection. 
When both models are misspecified (top-right corner), however, selecting a single model yields poor performance. 

\begin{figure}[!htb]
  \centering
  \begin{subfigure}{\textwidth}
    \centering
    \includegraphics[width=\textwidth]{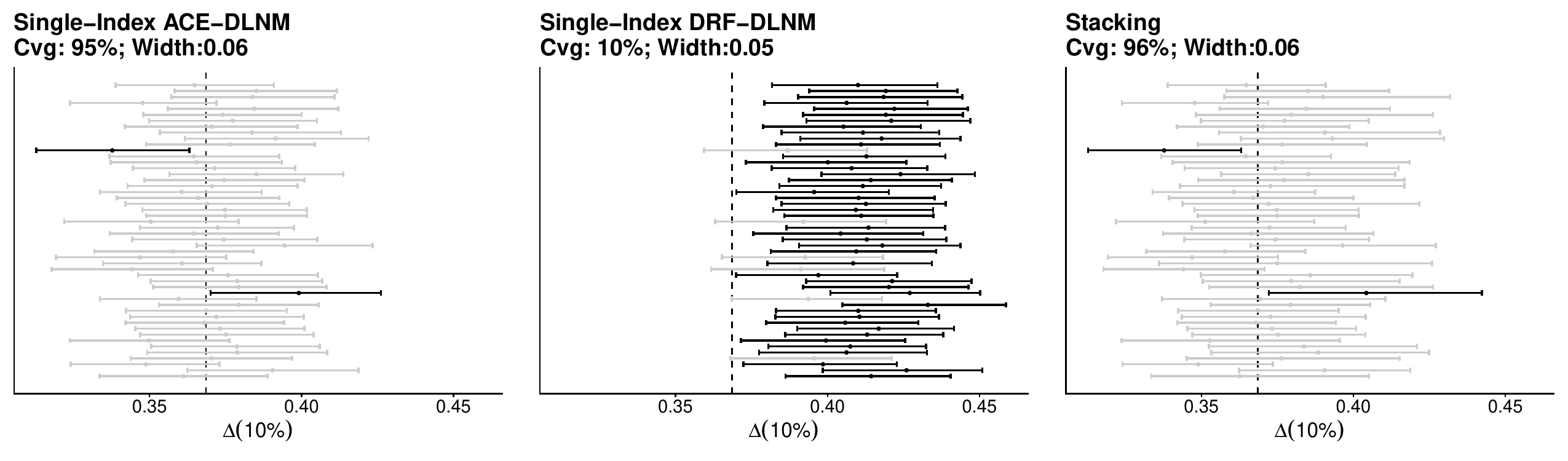}
    \caption{The nonlinearity of $f_{\text{ACE}}$ is 0.75 and the nonlinearity of $f_{\text{DRF}}$ is 0.}
    \label{fig:simC-interval-a}
  \end{subfigure}
  \begin{subfigure}{\textwidth}
    \centering
    \includegraphics[width=\textwidth]{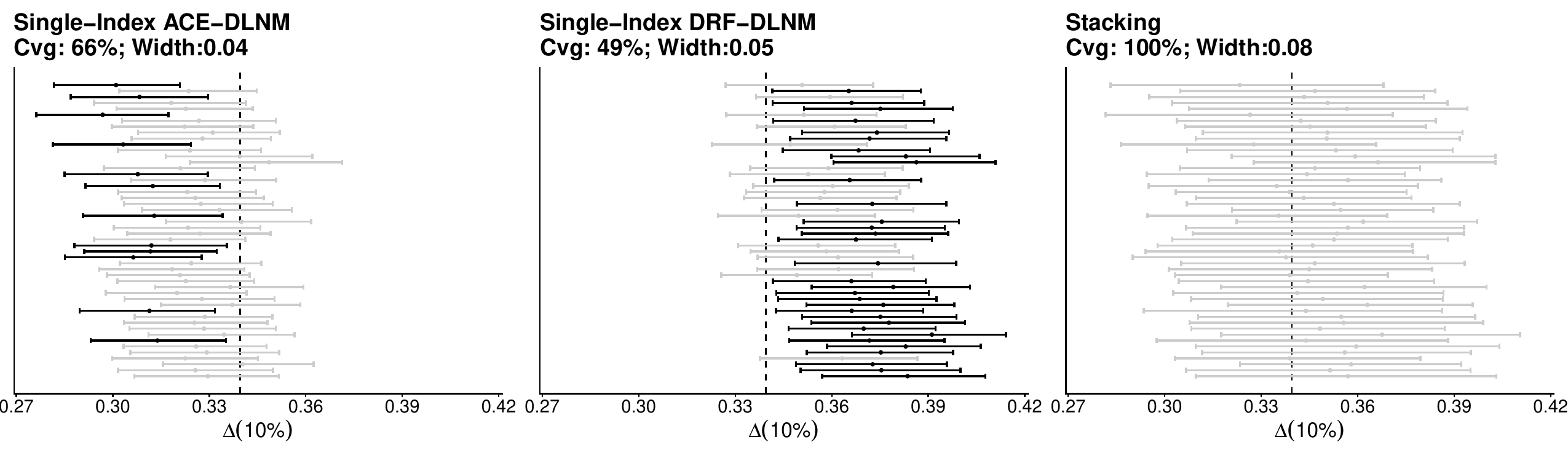}
    \caption{The nonlinearity of $f_{\text{ACE}}$ is 0.75 and the nonlinearity of $f_{\text{DRF}}$ is 0.75.}
    \label{fig:simC-interval-b}
  \end{subfigure}
  \caption{Confidence intervals from the first 50 replicates. In panel (a) the single-index DRF-DLNM is misspecified and the single-index ACE-DLNM is correct, and in panel (b) both are misspecified. The intervals that do not cover the true value (vertical dashed line) are highlighted. The captions report the coverage and the average width of the confidence intervals across the 1,000 replicates.}
  \label{fig:simC-interval}
\end{figure}

The stacking approach leads to similar results to the model with lowest AIC when one of the candidate models is correct (x- or y-axis at 0). When both models are misspecified, though, the stacking generally outperforms the model with lowest AIC and candidate models in terms of RMSE and bias. In these settings, stacking does result in over-coverage, however. 
To further investigate this, Figure \ref{fig:simC-interval} shows individual confidence intervals from the first 50 simulation results in two representative settings: when one model is correct [panel (a)], and when neither is correct [panel (b)]. 
Intervals that do not cover the true value (vertical dashed line) are highlighted. 
Stacking results in similar intervals to those of the correct model in panel (a). %
When both models are misspecified (panel [b]), the candidate models and the model with lowest AIC perform poorly with coverage of 66\%, 49\% and 52\%, respectively. 
Stacking achieves 100\% coverage, with confidence interval widths roughly double those of the candidate models, yet these intervals remain useful rather than uninformatively wide. 
In such misspecified cases, a conservative interval estimate is more desirable than a too optimistic one in practice. 
Overall, stacking tends to be close to the correctly specified model when one exists, and yields less biased point estimates and conservative confidence intervals when both models are misspecified. 

We observe opposite bias patterns between the two models: the single-index ACE-DLNM tends to under-estimate while the single-index DRF-DLNM tends to over-estimate (see, e.g., Figure \ref{fig:simC-interval-b}). 
We provide an additional simulation where both models tend to under-estimate in supplementary Appendix F.4. 
In this scenario, model selection and stacking yield estimates close to the best-fitting model.

\section{Associations between Air Pollution and Mortality in Ontario, Canada}
\label{s:application}
Using the proposed unified framework for multiple exposure DLNMs, we analyze the associations between air pollution and mortality in 20 CDs in Ontario from 2001 to 2015, considering three air pollutants (PM$_{2.5}$, O$_3$ and NO$_2$) and three outcomes (all-cause, circulatory and respiratory mortality counts) described in Section \ref{s:data}. 
For each of the outcomes, we assume a negative binomial distribution and fit the following model:
$$
\begin{aligned}
    \log(\mu_{it}) = s(\boldsymbol{X}_i, t)  + h_1(t) + h_2(\text{Month}_t)
    + h_3(\text{Temp}_{it}) + \sum_{p=1}^6 \beta^{\text{DOW}}_p\text{DOW}_{pt} + \delta_i + \log (\text{population}_{it}),
\end{aligned}
$$
where the exposure-response surface $s(\boldsymbol{X}_i, t)$ is represented using one of four DLNMs fitted with the proposed estimation framework: single-index ACE-DLNM, single-index DRF-DLNM, additive ACE-DLNM, and additive DRF-DLNM.
In these models, we use a two-week maximum lag. Long-term trend and seasonality are captured by the cubic B-spline $h_1(t)$ for time $t$ and the cyclic cubic B-spline $h_2(\text{Month}_t)$ for month, respectively. 
The temperature association is modeled by the cubic B-spline $h_3(\text{Temp}_{it})$. The $\log (\text{population}_{it})$ acts as an offset. We include the indicator variables for the day of week (DOW). 
A CD-specific random intercept $\delta_i$ is included to allow heterogeneity in baseline mortality rates. 
We quantify associations without making causal claims and acknowledge potential confounding. 
Supplementary Appendices H.1--H.2 present the estimated parameters and functions specific to each model. %
Randomized quantile residual diagnostics \citep{dunn1996randomized} suggest adequate model fit for each candidate model (supplementary Appendix H.3).

We quantify the association between air pollution and mortality using the relative mortality reduction (Equation \ref{eq:est}), which is the relative reduction in the expected number of deaths under a 10\% reduction in all exposure levels relative to the observed levels. Supplementary Appendix H.7 presents a sensitivity analysis in which reduced exposure levels falling below the CD-specific annual minimum are replaced by that minimum; the results are identical in our analysis. 

\begin{figure}[htbp]
    \centering
    \includegraphics[width=\linewidth]{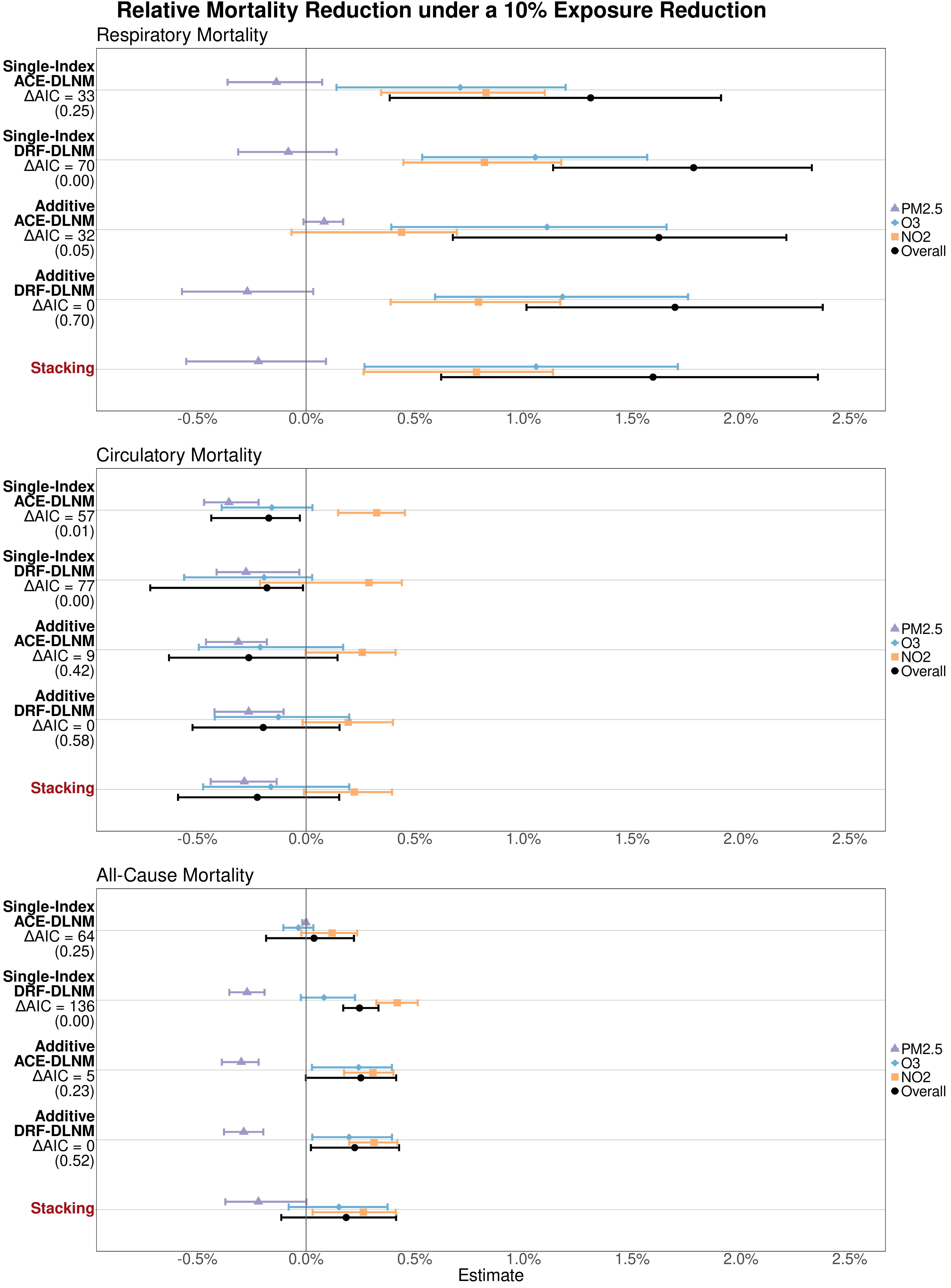}
    \caption{Relative mortality reduction under a 10\% exposure reduction for three types of mortality. Point estimates and 95\% confidence intervals are shown. The overall effect corresponds to a 10\% reduction in all pollutants, and the individual effects correspond to a 10\% reduction in one pollutant holding the others at observed levels. $\Delta$AIC is the difference in AIC relative to the best-fitting model, and the stacking weights are shown in brackets.}
    \label{fig:app_rr}
\end{figure}

We estimate the relative mortality reduction based on the four DLNMs. We select the best-fitting model using the conditional AIC, and apply stacking to combine the four DLNMs. 
Figure \ref{fig:app_rr} reports estimated relative mortality reduction for respiratory mortality, circulatory mortality, and all-cause mortality, based on the four DLNMs and the stacking. Point estimates and 95\% confidence intervals are displayed for the overall effects as well as individual pollutants. 
We report $\Delta$AIC as the difference of AIC relative to the best-fitting model and the stacking weights. %

AIC prefers the additive DRF-DLNM for the three outcomes. The stacking approach generally assigns a large weight to the AIC-selected model, but still weights other candidate models. The weighting pattern varies by outcome. 
For respiratory mortality, the four models yield broadly comparable results, and stacking assigns weights of 0.70, 0.25 and 0.05 to the additive DRF-DLNM, the single-index ACE-DLNM and the additive ACE-DLNM, respectively.
For circulatory mortality, the two additive DLNMs achieve nearly equal AICs, with the additive DRF-DLNM slightly lower. 
While AIC selects the model alone, the stacking, in contrast, balances both additive DLNMs with weights of 0.58 and 0.42. 
For all-cause mortality, these DLNMs exhibit larger differences. The two ACE-DLNMs find null overall effects, while the two DRF-DLNMs find significant positive overall effects; 
meanwhile, the single-index ACE-DLNM yields null PM$_{2.5}$ and NO$_2$ effects, while the other three DLNMs yield negative PM$_{2.5}$ effects and positive NO$_2$ effects. 
Stacking combines the single-index ACE-DLNM (0.25), additive ACE-DLNM (0.23) and additive DRF-DLNM (0.52) and yields null overall and PM$_{2.5}$ effects but a positive NO$_2$ effect. 
In the remainder of this section, we focus on stacking estimates.

The overall effect estimates indicate that simultaneously reducing all three air pollutants by 10\% is associated with a decrease in respiratory mortality by 1.60\% (95\% CI: [0.62\%, 2.35\%]), and the estimated reductions in respiratory mortality counts are 1287 (95\% CI: [498, 1896]).
The reductions in circulatory and all-cause mortality are not significant. 
Although respiratory mortality is relatively low compared to other causes, for example, in Toronto, the daily average count is 3.7 for respiratory versus 13.1 for circulatory, it exhibits a stronger estimated association with air pollution.

For the individual effects of reducing a single pollutant by 10\% while keeping the others at their observed levels, we find heterogeneous associations across PM$_{2.5}$, O$_3$, and NO$_2$. 
PM$_{2.5}$ exhibits surprising negative associations with circulatory mortality, which may be related to unmeasured confounding and warrants further investigation. 
The estimated effects of PM$_{2.5}$ on respiratory and all-cause mortality are near null. For all-cause mortality, we observe different patterns among the models. The additive models, which fit separate curves for each pollutant, tend to yield negative associations for PM$_{2.5}$, distinct from other pollutants. In contrast, the single-index ACE-DLNM shrinks the index weights of PM$_{2.5}$ toward zero, leading to null effects and contributing to the stacking estimate indicating no association. 
For O$_3$, the estimated associations are positive for all-cause and respiratory mortality and null for circulatory mortality.
NO$_2$ is estimated to have significant positive associations with respiratory and all-cause mortality, aligning with the estimated index weights from the single-index models, which assign the largest index weights to NO$_2$. 
This suggests that NO$_2$ may be the primary contributor to the adverse effects of air pollution on mortality.

\begin{figure}[htbp]
    \centering
    \includegraphics[width=\linewidth]{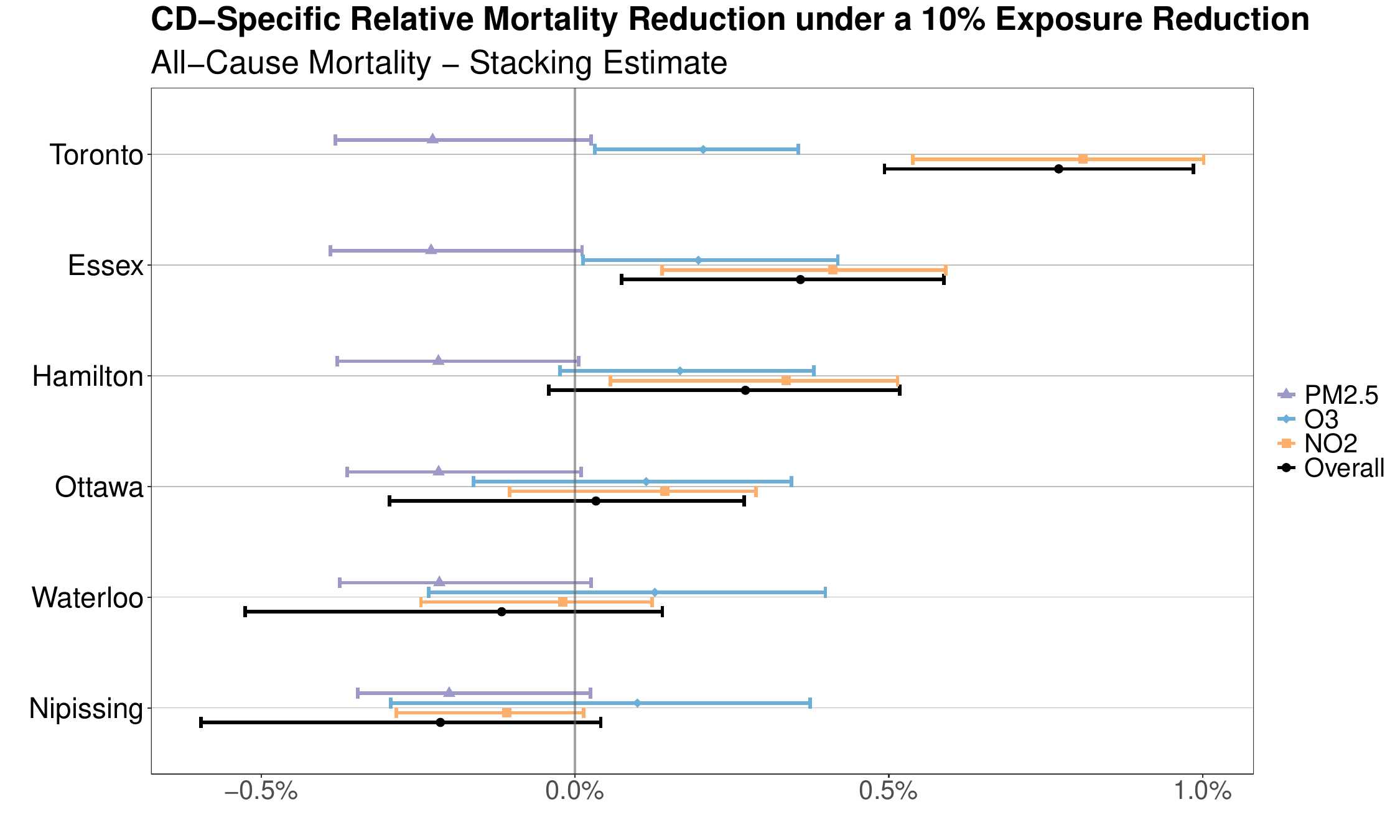}
    \caption{CD-specific relative all-cause mortality reduction under a 10\% exposure reduction in five representative CDs. Point estimates and 95\% confidence intervals from model stacking are shown. The overall effect corresponds to a 10\% reduction in all pollutants, and the individual effects correspond to a 10\% reduction in one pollutant holding the others at observed levels. }
    \label{fig:app_rr_CD}
\end{figure}

We report the CD-specific relative mortality reduction, obtained by restricting summation in Equation \ref{eq:est} to a given CD. 
This quantity measures the relative reduction under a 10\% reduction in exposure levels for that specific CD. 
Figure \ref{fig:app_rr_CD} shows the results for all-cause mortality for six representative CDs, and the full results are provided in supplementary Appendix H.5. 
Because of the non-linear exposure-response relationships, we observe heterogeneity in the overall and NO$_2$ associations on all-cause mortality across CDs with different levels of air pollution. 
Most CDs exhibit null effects, consistent with Ontario-level estimates; Ottawa, Waterloo, and Nipissing (the least populous CD in this analysis) are presented as examples. 
In contrast, only Toronto and Essex display significant positive overall associations, with NO$_2$ contributing most; significant positive NO$_2$ associations are observed only in Toronto, Essex and Hamilton.
A similar pattern is observed for circulatory mortality: Toronto, Essex and Hamilton are the only three CDs with significant positive NO$_2$ associations, although the overall associations are null. 
In our analysis, these three CDs have the highest NO$_2$ concentrations, which are primarily from transportation, industry, and power generation. Toronto is the largest city in Canada, Essex is adjacent to Detroit, and Hamilton is a major industrial center in Ontario.

We examine temporal patterns in the associations between air pollution and mortality using year-specific relative mortality reduction, based on the four DLNMs fitted to all CDs during the entire study period. 
The year-specific estimates are obtained by restricting the summation in Equation \ref{eq:est} to each year, measuring the relative reduction under a 10\% reduction in exposure levels for the study population in that year.
The results for all-cause mortality are shown in Figure \ref{fig:app_rr_year}, and those for circulatory and respiratory mortality are provided in supplementary Appendix H.6. 
For all-cause mortality, we observe that the overall mixture effect and the individual effect of NO$_2$ decrease over time: both peak in 2001 and decline to insignificance after 2005 and 2007, respectively. O$_3$ effects are significant between 2001 and 2003 and slightly decline to null after 2005, and PM$_{2.5}$ effects remain relatively stable. 
Circulatory mortality and respiratory mortality exhibit similar decreasing trends in overall and NO$_2$ effects and generally stable patterns for PM$_{2.5}$ and O$_3$. 
These patterns can be explained by the non-linear exposure-response relationships and the observed decline in NO$_2$ concentration. A 10\% reduction applied to the higher NO$_2$ levels in 2001 (median: 14.7 parts per billion, ppb, 25\% percentile: 9.48 ppb and 75\% percentile: 21.3 ppb) yields an estimated 0.90\% reduction (95\% CI: [0.68\%, 1.09\%]) in all-cause mortality, while the 10\% reduction applied to lower NO$_2$ levels in 2015 (median: 6.21 ppb, 25\% percentile: 3.83 ppb and 75\% percentile: 10.3 ppb) yields an estimated -0.01\% reduction (95\% CI: [-0.23\%, 0.14\%]).

\begin{figure}[htbp]
    \centering
    \includegraphics[width=1\linewidth]{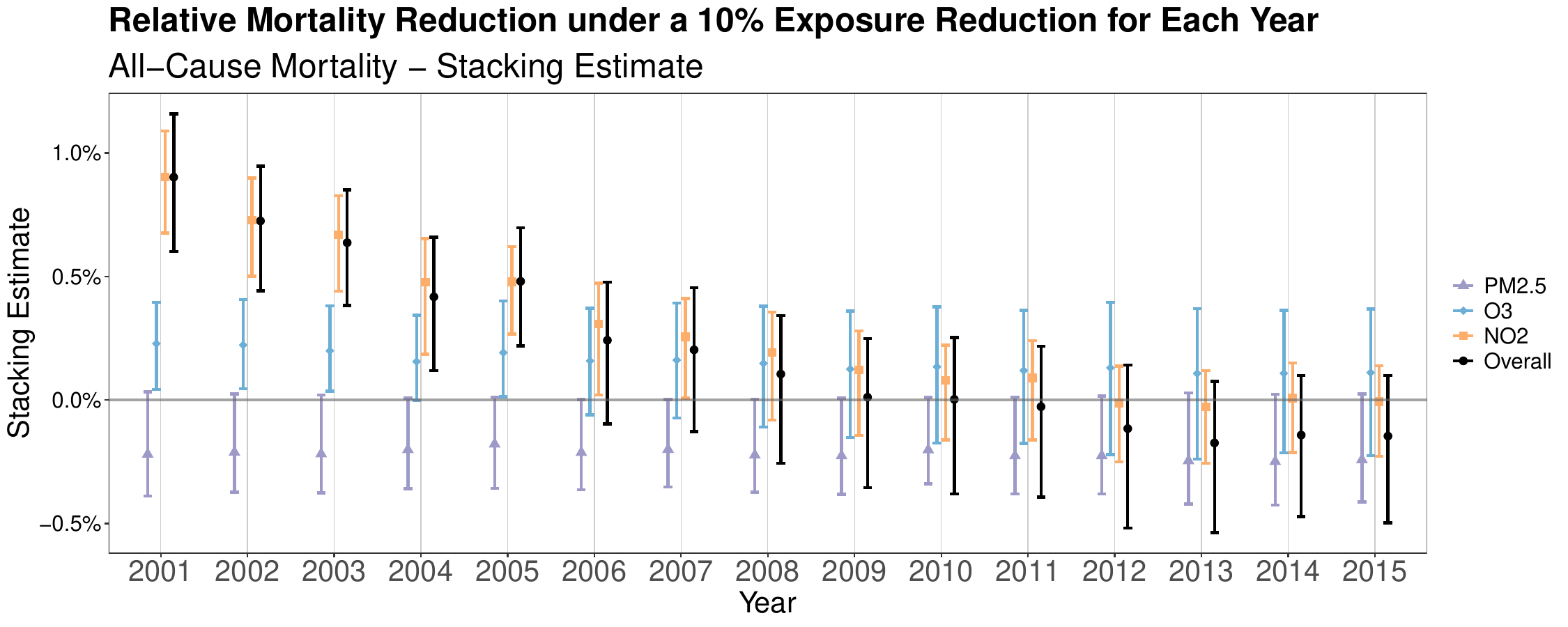}
    \caption{Year-specific relative all-cause mortality reduction under a 10\% exposure reduction. Point estimates and 95\% confidence intervals from model stacking are shown. The overall effect corresponds to a 10\% reduction in all pollutants, and the individual effects correspond to a 10\% reduction in one pollutant holding the others at observed levels. }
    \label{fig:app_rr_year}
\end{figure}

\section{Discussion}
In this paper we analyze the association between a mixture of multiple air pollutants and mortality in Ontario, Canada, covering 20 CDs from 2001 to 2015, with 106,346 observations in total. 
To simultaneously address time-lagged associations, non-linear associations, and multiple pollutants, we propose a unified framework for multiple exposure DLNMs that integrates model specification, estimation, selection and stacking. 
We highlight the practical use of model selection and stacking, which removes the need to assume a model \textit{a priori} as is common in the literature despite no guarantee of correct specification. 
Beyond the Canadian context, this framework is broadly applicable to studies involving multiple exposures measured repeatedly over time. 

Many existing studies focus primarily on PM$_{2.5}$, including Canadian \citep{shin2024PM2.5} and 
global studies \citep{liu2019ambient} as noted by \citet{huang2023multi}. 
By including O$_3$ and NO$_2$ and accounting for non-linear and lagged effects, we find that the associations between PM$_{2.5}$ and mortality tend to be null, consistent with the findings of \citet{huang2023multi} in four Canadian cities. PM$_{2.5}$ concentrations in Canada are among the lowest worldwide \citep{health_canada_2021_air_pollution_report}, which may contribute to these null associations. 
We find that NO$_2$ is significantly associated with all three types of mortality, suggesting it may be a main risk factor, and O$_3$ has a significant association with respiratory mortality. This finding aligns with a study in 47 Canadian cities by \citet{huang2023multi}.
Beyond these pollutant-specific associations, we provide estimates for overall associations of multiple air pollutants. 
Our results suggest that respiratory mortality shows the strongest association with air pollution. This is consistent with studies of particulate matter reporting stronger associations with respiratory mortality than with other outcomes \citep{liu2019ambient, chen2012association}. 
Our year-specific analyses suggest that the effects of air pollution on mortality decrease over time. This pattern is consistent with the findings of \citet{pan2026temporal} and the Air Health Trends Indicator (AHTI, \citealp{AHTI}), both of which are based on single-pollutant models, while we provide evidence considering multiple air pollutants simultaneously. 
Our analysis assumes a common exposure-response relationship across years, and allowing this relationship to vary over time, as in \citet{pan2026temporal} and the AHTI \citep{AHTI}, is an avenue for future research.

We acknowledge some limitations of this study. 
First, we use pollutant concentrations measured at the CD level and may not accurately represent individual exposures, leading to exposure measurement error. 
Second, our population-level estimates are subject to ecological bias \citep{greenland_ecological_1989} and may not reflect individual-level associations. 
Third, we observe a negative association between PM$_{2.5}$ and circulatory mortality. This may be due to potential confounding, such as socioeconomic status, as we quantify associations without making causal claims. 
Lastly, our analysis assumes no remaining spatial correlation among CDs; although no lack of model fit is observed, future work will explore potential spatial dependencies.

\section*{Supplementary Materials}
The R package is available at \url{https://github.com/tianyi-pan/mDLNM}. Code to reproduce the results in this paper is available at \url{https://github.com/tianyi-pan/mDLNM-paper-code}.

\bibliography{refs}

\clearpage
\appendix
\renewcommand{\appendixname}{}

\def\spacingset#1{\renewcommand{\baselinestretch}%
{#1}\small\normalsize} \spacingset{1}

\renewcommand{\thesection}{\Alph{section}}
\renewcommand{\thefigure}{S\arabic{figure}}
\renewcommand{\thetable}{S\arabic{table}}

\renewcommand{\theequation}{S.\arabic{equation}}

\part{Supplementary Materials}

\renewcommand{\ptctitle}{}
\parttoc

\section{Data}

\begin{table}[H]
\centering
\caption{Summary statistics for population (in thousands), daily mortality (all-cause, circulatory, respiratory), air pollution (PM$_{2.5}$, O$_3$, NO$_2$), temperature, and start dates of available data for 20 census divisions from 2001 to 2015. Mortality data for individuals aged under 1 year are not available, and the corresponding population is excluded. Rates are daily counts per 100,000 people. Population values are annual averages, and all other values are daily averages.}
\makebox[\textwidth][c]{%
\resizebox{1.15\textwidth}{!}{
\begin{tabular}{@{}lrrrrrrrrrrrr@{}}
\toprule
   &            & \multicolumn{2}{c}{All-cause} & \multicolumn{2}{c}{Circulatory} & \multicolumn{2}{c}{Respiratory} & \multicolumn{3}{c}{Pollutants}   &             &            \\ 
\cmidrule(lr){3-4} \cmidrule(lr){5-6} \cmidrule(lr){7-8} \cmidrule(lr){9-11}
Census division & \makecell[t]{Population\\($\times 10^3$)} & No. & Rate & No. & Rate & No. & Rate &
\makecell[t]{PM$_{2.5}$ \\ ($\mu g\,m^{-3}$)} &
\makecell[t]{O$_3$ \\ (ppb)} &
\makecell[t]{NO$_2$ \\ (ppb)} &
\makecell[t]{Temp. \\ ($^\circ$C)} &
Data start \\ 
\midrule
Toronto        & 2633 & 42.8 & 1.6 & 13.1 & 0.5 & 3.7 & 0.1 & 8.0 & 33.2 & 18.6 & 9.2 & Jan 01, 2001 \\
Peel           & 1235 & 12.4 & 1.0 & 3.7 & 0.3 & 1.0 & 0.1 & 7.6 & 35.5 & 13.5 & 8.7 & Jan 01, 2001 \\
York           & 966  & 10.3 & 1.1 & 3.1 & 0.3 & 0.9 & 0.1 & 6.8 & 38.3 & 8.6 & 7.9 & Jan 01, 2001 \\
Ottawa         & 866  & 13.6 & 1.6 & 4.3 & 0.5 & 1.2 & 0.1 & 6.4 & 31.8 & 11.2 & 6.9 & Jan 01, 2001 \\
Durham         & 592  & 8.6 & 1.5 & 2.6 & 0.4 & 0.7 & 0.1 & 7.1 & 34.9 & 9.9 & 7.6 & Jan 01, 2001 \\
Hamilton       & 524  & 10.9 & 2.1 & 3.4 & 0.7 & 1.0 & 0.2 & 8.8 & 35.4 & 14.3 & 8.4 & Jan 01, 2001 \\
Waterloo       & 499  & 7.9 & 1.6 & 2.6 & 0.5 & 0.6 & 0.1 & 7.7 & 37.2 & 9.7 & 7.5 & Jan 01, 2001 \\
Halton         & 473  & 6.9 & 1.5 & 2.0 & 0.4 & 0.6 & 0.1 & 7.8 & 35.8 & 12.9 & 9.0 & Jan 01, 2001 \\
Middlesex      & 440  & 8.4 & 1.9 & 2.6 & 0.6 & 0.8 & 0.2 & 8.2 & 36.6 & 10.8 & 8.4 & Jan 01, 2001 \\
Simcoe         & 439  & 8.6 & 2.0 & 2.7 & 0.6 & 0.8 & 0.2 & 6.8 & 36.5 & 10.8 & 7.1 & Dec 01, 2001 \\
Niagara        & 437  & 10.6 & 2.4 & 3.7 & 0.9 & 0.9 & 0.2 & 7.8 & 36.5 & 11.0 & 9.4 & Jan 01, 2001 \\
Essex          & 398  & 7.7 & 1.9 & 2.8 & 0.7 & 0.6 & 0.2 & 8.9 & 36.5 & 15.1 & 10.0 & Jan 01, 2001 \\
Peterborough   & 136  & 3.5 & 2.5 & 1.1 & 0.8 & 0.3 & 0.3 & 6.4 & 37.3 & 6.4 & 7.0 & Feb 28, 2001 \\
Brant          & 137  & 3.1 & 2.2 & 0.9 & 0.7 & 0.3 & 0.2 & 7.6 & 39.0 & 6.9 & 8.4 & Jan 02, 2004 \\
Thunder Bay    & 151  & 3.5 & 2.3 & 1.2 & 0.8 & 0.3 & 0.2 & 5.1 & 32.7 & 9.0 & 2.8 & Jun 22, 2001 \\
Greater Sudbury& 163  & 3.7 & 2.3 & 1.2 & 0.7 & 0.3 & 0.2 & 4.5 & 34.8 & 7.1 & 4.7 & Jun 28, 2004 \\
Lambton        & 130  & 3.2 & 2.4 & 1.2 & 0.9 & 0.2 & 0.2 & 10.6 & 37.1 & 11.0 & 8.9 & Jan 01, 2001 \\
Algoma         & 119  & 3.2 & 2.7 & 1.0 & 0.9 & 0.3 & 0.2 & 5.9 & 36.3 & 6.4 & 4.0 & Jan 01, 2001 \\
Haldimand-Norfolk & 110 & 2.6 & 2.3 & 1.0 & 0.9 & 0.2 & 0.2 & 7.5 & 41.4 & 5.4 & 8.7 & Jan 01, 2001 \\
Nipissing      & 86  & 2.0 & 2.3 & 0.7 & 0.8 & 0.2 & 0.2 & 5.1 & 35.7 & 7.5 & 4.5 & Jan 01, 2001 \\
\bottomrule
\end{tabular}
}
}
\end{table}

\begin{figure}[H]
    \centering
    \includegraphics[width=\linewidth]{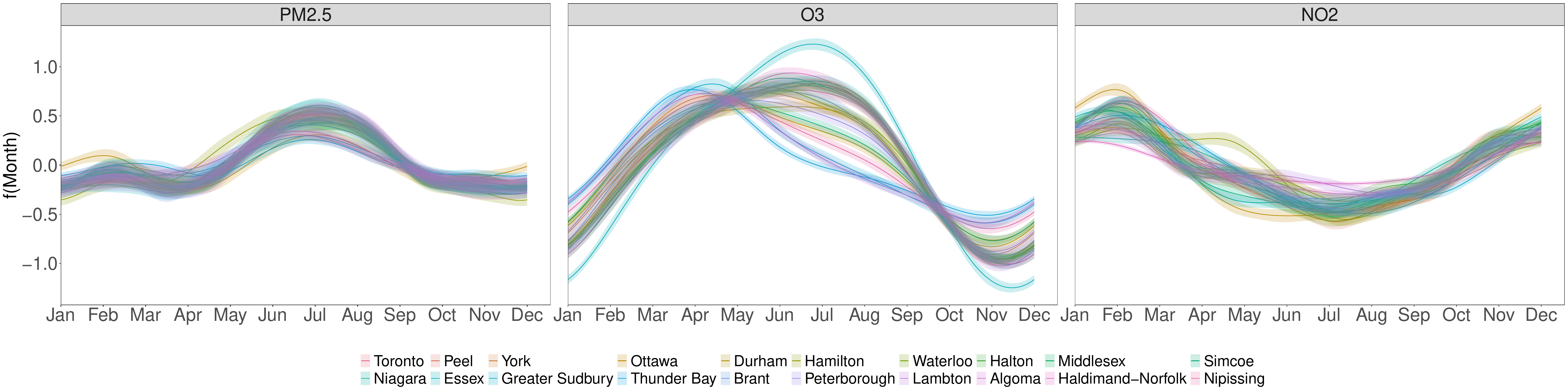}
    \caption{Seasonalities in PM$_{2.5}$, O$_3$ and NO$_2$ concentrations across 20 CDs from 2001 to 2015.}
    \label{fig:fmonth}
\end{figure}

\begin{figure}[H]
    \centering
    \includegraphics[width=\linewidth]{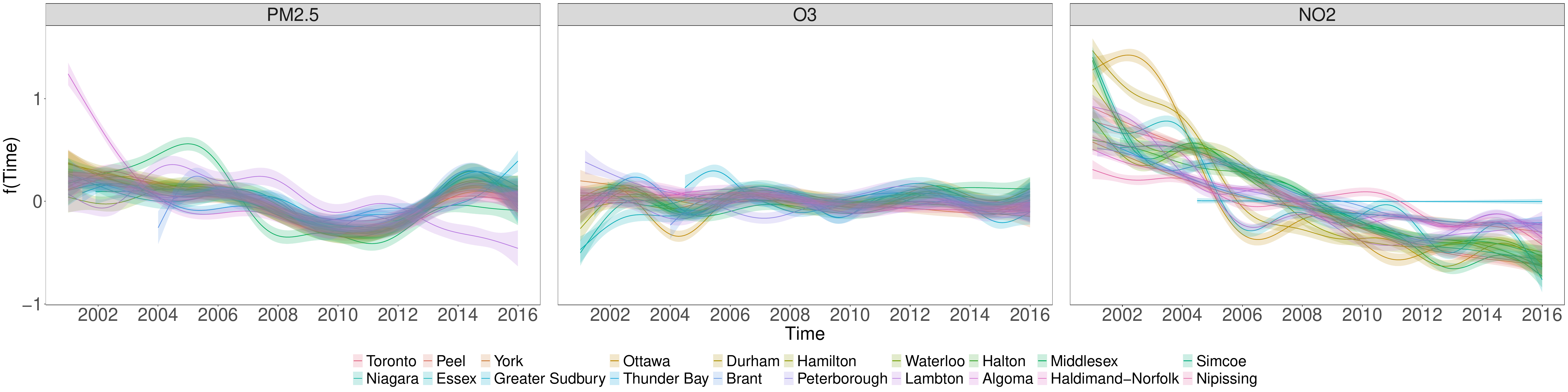}
    \caption{Long-term trends in PM$_{2.5}$, O$_3$ and NO$_2$ concentrations across 20 CDs from 2001 to 2015.}
    \label{fig:fday}
\end{figure}

\begin{figure}[H]
    \centering
    \includegraphics[width=\linewidth]{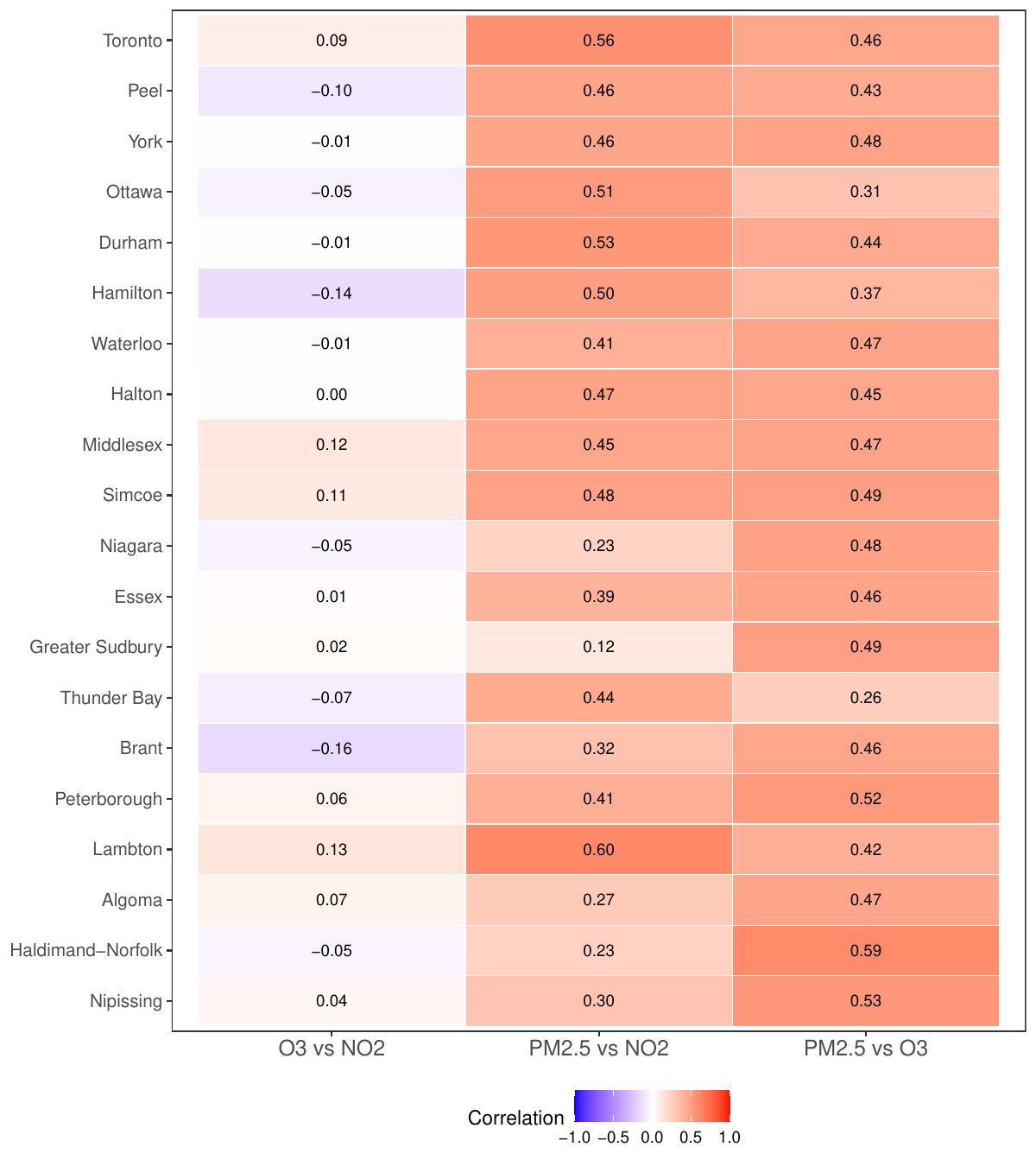}
    \caption{Pairwise correlation between PM$_{2.5}$, O$_3$, and NO$_2$ by CD. Correlations are calculated using residuals from the generalized additive models after removing long-term trends and seasonality. Colors indicate the sign and magnitude of correlations, with values shown in the figure.}
    \label{fig:corr}
\end{figure}

We fit generalized additive models to regress a standardized pollutant concentrations on a cyclic smooth function of month to capture the seasonality and a smooth function of time to account for long-term trends, separately for each CD and air pollutant. The estimated curves are shown in Figures \ref{fig:fmonth} and \ref{fig:fday}. 
We calculate the pairwise correlation between PM$_{2.5}$, O$_3$, and NO$_2$ using the model residuals after removing temporal trends and seasonality, and the results are shown in Figure \ref{fig:corr}. 

\section{Identifiability Reparameterization}
We introduce a general constraint $\boldsymbol{\alpha}^\top \boldsymbol{\alpha} = 1$ and $\boldsymbol{c}^\top \boldsymbol{\alpha}> 0$ in Section 3.4. This appendix shows that the general constraint is satisfied by mapping $\boldsymbol{\alpha}$ from an unconstrained parameter $\boldsymbol{\alpha}^* \in \mathbb{R}^{M-1}$ as follows: 
\begin{equation}
\boldsymbol{\alpha} = \frac{\mathbf{B}_{\boldsymbol{\alpha}} \left[1, \boldsymbol{\alpha}^{*\top}\right]^\top}{\left(\left[1, \boldsymbol{\alpha}^{*\top}\right] \mathbf{B}_{\boldsymbol{\alpha}}^\top \mathbf{B}_{\boldsymbol{\alpha}} \left[1, \boldsymbol{\alpha}^{*\top}\right]^\top\right)^{1/2}}, 
\label{eq:repa}
\end{equation}
where $\mathbf{B}_{\boldsymbol{\alpha}} = \left[\boldsymbol{c}, \mathbf{Q}^+_{\boldsymbol{\alpha}}\right]$ with $\mathbf{e}_1 = \left[1,0,\cdots,0\right]^\top$. 
The matrix $\mathbf{Q}^+_{\boldsymbol{\alpha}}$ is obtained by removing the first column of $\mathbf{Q}_{\boldsymbol{\alpha}}$ from the QR decomposition: $\mathbf{Q}_{\boldsymbol{\alpha}} \mathbf{R}_{\boldsymbol{\alpha}} = \boldsymbol{c}$.

Because the vector $\boldsymbol{c}$ is non-zero, 
the QR decomposition yields $\mathbf{R}_{\boldsymbol{\alpha}} = [r, 0,\cdots,0]^\top$ with length $M$ where $r \neq 0$, and an orthogonal matrix $\mathbf{Q}_{\boldsymbol{\alpha}} = [\mathbf{q}_{\boldsymbol{\alpha}}^1, \mathbf{Q}_{\boldsymbol{\alpha}}^+]$. 
Therefore, we can rewrite $\mathbf{B}_{\boldsymbol{\alpha}} = [r \mathbf{q}_{\boldsymbol{\alpha}}^1, \mathbf{Q}_{\boldsymbol{\alpha}}^+]$. It follows that $\mathbf{B}_{\boldsymbol{\alpha}}$ has full column rank, and therefore 
$\left[1, \boldsymbol{\alpha}^{*\top}\right]^\top \mathbf{B}_{\boldsymbol{\alpha}}^\top \mathbf{B}_{\boldsymbol{\alpha}} \left[1, \boldsymbol{\alpha}^{*\top}\right] > 0$. 

For the orthogonal matrix $\mathbf{Q}^+_{\boldsymbol{\alpha}}$, we have 
$$\boldsymbol{c}^\top \mathbf{Q}^+_{\boldsymbol{\alpha}} =  
[r, 0,\cdots,0] 
\begin{bmatrix}
\mathbf{q}_{\boldsymbol{\alpha}}^{1\top}\\
\mathbf{Q}_{\boldsymbol{\alpha}}^{+\top}
\end{bmatrix} 
\mathbf{Q}_{\boldsymbol{\alpha}}^+
= r \mathbf{q}_{\boldsymbol{\alpha}}^{1\top} \mathbf{Q}_{\boldsymbol{\alpha}}^+ = \boldsymbol{0}. 
$$
Consequently, $\boldsymbol{c}^\top \mathbf{B}_{\boldsymbol{\alpha}} = [\boldsymbol{c}^\top \boldsymbol{c}, \boldsymbol{c}^\top \mathbf{Q}^+_{\boldsymbol{\alpha}}] = [\boldsymbol{c}^\top \boldsymbol{c}, \boldsymbol{0}]$ where $\boldsymbol{c}^\top \boldsymbol{c} > 0$. 
Finally, we have 
$$
\boldsymbol{c}^\top\boldsymbol{\alpha} = 
\frac{\boldsymbol{c}^\top\mathbf{B}_{\boldsymbol{\alpha}} \left[1, \boldsymbol{\alpha}^{*\top}\right]^\top}{\left(\left[1, \boldsymbol{\alpha}^{*\top}\right] \mathbf{B}_{\boldsymbol{\alpha}}^\top \mathbf{B}_{\boldsymbol{\alpha}} \left[1, \boldsymbol{\alpha}^{*\top}\right]^\top\right)^{1/2}} = 
\frac{\boldsymbol{c}^\top \boldsymbol{c}}{\left(\left[1, \boldsymbol{\alpha}^{*\top}\right]\mathbf{B}_{\boldsymbol{\alpha}}^\top \mathbf{B}_{\boldsymbol{\alpha}} \left[1, \boldsymbol{\alpha}^{*\top}\right]^\top\right)^{1/2}} > 0. 
$$
The constraint $\boldsymbol{\alpha}^\top \boldsymbol{\alpha} = 1$ is satisfied by construction in Equation~\ref{eq:repa}, which completes the proof.

Next, we argue that the reparameterization used in literature, e.g., \citet{yu2002penalized, yu2017penalised}, is a special case of our proposed approach when $\boldsymbol{c} = \mathbf{e}_1$. 
Under $\boldsymbol{c} = \mathbf{e}_1$, the QR decomposition can be simplified to $\mathbf{e}_1 = \mathbf{Q}_{\boldsymbol{\alpha}} \mathbf{R}_{\boldsymbol{\alpha}}$ where $\mathbf{Q}_{\boldsymbol{\alpha}} = \mathbf{I}_M$ and $\mathbf{R}_{\boldsymbol{\alpha}} = \mathbf{e}_1$. 
It follows that $\mathbf{B}_{\boldsymbol{\alpha}} = [r \mathbf{q}_{\boldsymbol{\alpha}}^1, \mathbf{Q}_{\boldsymbol{\alpha}}^+] = \mathbf{I}_M$. Plugging this into Equation \ref{eq:repa}, we have $\boldsymbol{\alpha} = \left[1, \boldsymbol{\alpha}^{*\top}\right]^\top / \left(\left[1, \boldsymbol{\alpha}^{*\top}\right]^\top \left[1, \boldsymbol{\alpha}^{*\top}\right]\right)^{1/2}$ which is the reparameterization used in literature. 

\section{Ranges of Index and Cumulative Index}

This Appendix Section provides details of the ranges of index $E_i(t; \boldsymbol{\alpha})$ and cumulative index $E_i^{L}(t; \boldsymbol{\alpha}, w)$. 
By the Cauchy-Schwarz inequality, for a given time $t$ and exposure $\boldsymbol{X}_i$ for CD $i$, we have
\begin{equation}
E_i(t; \boldsymbol{\alpha})^2 = [\boldsymbol{\alpha}^\top \boldsymbol{X}_i(t)]^2 \le \left(\sum_{m = 1}^M \alpha_m^2 \right)\left(\sum_{m = 1}^M X_{im}(t)^2\right). 
\label{eq:CSindex}
\end{equation}
We impose the identifiability constraint $\boldsymbol{\alpha}^\top \boldsymbol{\alpha}$ described in Section 3.4. It follows that the range of index $E_i(t; \boldsymbol{\alpha})$ is 
$$
- \sqrt{\sum_{m = 1}^M X_{im}(t)^2} \le E_i(t; \boldsymbol{\alpha}) \le \sqrt{\sum_{m = 1}^M X_{im}(t)^2}. 
$$
For the cumulative index $E_i^{L}(t; \boldsymbol{\alpha}, w)$, the Cauchy-Schwarz inequality yields
$$
E_i^{L}(t; \boldsymbol{\alpha}, w)^2 =  \left(\int_{0}^{L} w(l) E_i(t-l; \boldsymbol{\alpha}) dl\right)^2 \le \int_{0}^{L} w(l)^2 dl  \int_{0}^{L} E_i(t-l; \boldsymbol{\alpha})^2 dl. 
$$
Since $\int_{0}^{L} w(l)^2 dl = 1$ described in Section 3.4, and $E_i(t-l; \boldsymbol{\alpha})^2 \le \sum_{m = 1}^M X_{im}(t-l)^2$ from Equation \ref{eq:CSindex} using the Cauchy-Schwarz inequality, we obtain 
$$
E_i^{L}(t; \boldsymbol{\alpha}, w)^2 \le \int_{0}^{L} \sum_{m = 1}^M X_{im}(t-l)^2 dl.
$$
Therefore, for a given time $t$ and exposure $\boldsymbol{X}_i$ for CD $i$, we have 
$$
-\sqrt{\sum_{m=1}^M \int_0^L X_{im}(t-l)^2 d l} \le E_i^{L}(t; \boldsymbol{\alpha}, w) \le \sqrt{\sum_{m=1}^M \int_0^L X_{im}(t-l)^2 d l}. 
$$
\citet{pan2025estimating} used a similar strategy for the range of $\int_{0}^{L} w(l) X(t-l) dl$. Here we extend this by replacing $X(t-l)$ with $E_i(t-l; \boldsymbol{\alpha})$, which requires an additional application of the Cauchy-Schwarz inequality to account for the unknown parameter $\boldsymbol{\alpha}$.

It follows that the range of $E_i^{L}(t; \boldsymbol{\alpha}, w)$ over all $t$ and $i$ can be specified as $\left[-\bar{E}^{L}, \bar{E}^{L}\right]$ where $\bar{E}^{L} = \max_{i,t} \sqrt{\sum_{m=1}^M \int_0^L X_{im}(t-l)^2 d l}$, 
and the range of $E_i(t; \boldsymbol{\alpha})$ over all $t$ and $i$ is $\left[-\bar{E}, \bar{E}\right]$ where $\bar{E} = \max_{i,t} \sqrt{\sum_{m = 1}^M X_{im}(t)^2}$.

\section{Optimization}
We aim to maximize the penalized log-likelihood to obtain the estimators. 
The optimization problem is complicated by the lag structure, multiple smoothing parameters and unknown index weights.  
\citet{pan2025estimating} developed an algorithm to address the lag structure and smoothing parameter estimation for single-exposure ACE-DLNMs. 
However, in both single-index ACE- and DRF-DLNMs, the unknown index weights induce nested model structures, which require more complex optimization procedures. 

We extend the approach in \citet{pan2025estimating} for a general form of $\mathcal{L}(\boldsymbol{\phi}, \boldsymbol{\gamma}; \boldsymbol{\lambda}, \boldsymbol{\theta})$, in which the inner parameters $\boldsymbol{\phi}$ incorporate both index weights parameters and spline coefficients for smooth functions, and the penalty terms can involve multiple smoothing parameters (e.g. $\mathcal{P}(\boldsymbol{\gamma}^{\psi}; \lambda^{\psi_x}, \lambda^{\psi_l})$ in the single-index DRF-DLNM). We describe the approach below.

Given $\boldsymbol{\lambda}$ and $\boldsymbol{\theta}$, the inner and outer parameters are estimated by maximizing profile log-likelihood: 
$\boldsymbol{Q}(\boldsymbol{\phi}; \boldsymbol{\lambda}, \boldsymbol{\theta}) = \mathcal{L}(\boldsymbol{\phi}, \widehat{\boldsymbol{\gamma}}(\boldsymbol{\phi}; \boldsymbol{\lambda}, \boldsymbol{\theta}); \boldsymbol{\lambda}, \boldsymbol{\theta})$, 
where $\widehat{\boldsymbol{\gamma}}(\boldsymbol{\phi}; \boldsymbol{\lambda}, \boldsymbol{\theta}) = \arg\max_{\boldsymbol{\gamma}} \mathcal{L}(\boldsymbol{\phi}, \boldsymbol{\gamma}; \boldsymbol{\lambda}, \boldsymbol{\theta})$. We denote the estimator as $\widehat{\boldsymbol{\phi}}(\boldsymbol{\lambda}, \boldsymbol{\theta}) = \arg\max_{\boldsymbol{\phi}} \boldsymbol{Q}(\boldsymbol{\phi}; \boldsymbol{\lambda}, \boldsymbol{\theta})$ and $\widehat{\boldsymbol{\gamma}}(\boldsymbol{\lambda}, \boldsymbol{\theta}) = \widehat{\boldsymbol{\gamma}}(\widehat{\boldsymbol{\phi}}(\boldsymbol{\lambda}, \boldsymbol{\theta}); \boldsymbol{\lambda}, \boldsymbol{\theta})$. We find the analytical gradients and Hessian of $\mathcal{L}(\boldsymbol{\phi}, \boldsymbol{\gamma}; \boldsymbol{\lambda}, \boldsymbol{\theta})$ and $\boldsymbol{Q}(\boldsymbol{\phi}; \boldsymbol{\lambda}, \boldsymbol{\theta})$ using implicit differentiation techniques; see supplementary Sections \ref{ss:deL} and \ref{ss:deQ} for details. 
These analytical derivatives facilitate efficient optimization using the Q-Newton method \citep{truong2023fast} with a step-halving strategy \citep{pan2025estimating}.

We estimate $\boldsymbol{\lambda}$ and $\boldsymbol{\theta}$ by maximizing the Laplace approximate marginal likelihood (LAML; \citealp{wood2016smoothing}). The log-LAML is 
$$
\mathcal{L}^*_{\text{LA}}(\boldsymbol{\lambda}, \boldsymbol{\theta}) =
\mathcal{L}(\widehat{\boldsymbol{\phi}}\left(\boldsymbol{\lambda}, \boldsymbol{\theta}\right), \widehat{\boldsymbol{\gamma}}\left(\boldsymbol{\lambda}, \boldsymbol{\theta}\right); \boldsymbol{\lambda}, \boldsymbol{\theta})
- \frac{1}{2} \log \left\{\mathrm{det} \boldsymbol{\mathcal{H}}\left(\widehat{\boldsymbol{\phi}}\left(\boldsymbol{\lambda}, \boldsymbol{\theta}\right), \widehat{\boldsymbol{\gamma}}\left(\boldsymbol{\lambda}, \boldsymbol{\theta}\right); \boldsymbol{\lambda}, \boldsymbol{\theta}\right)\right\} + 
\frac{1}{2}\log|\mathbf{S}^{\boldsymbol{\lambda}}|_{+} + \mathrm{C},
$$
where $\boldsymbol{\mathcal{H}}\left(\widehat{\boldsymbol{\phi}}\left(\boldsymbol{\lambda}, \boldsymbol{\theta}\right), \widehat{\boldsymbol{\gamma}}\left(\boldsymbol{\lambda}, \boldsymbol{\theta}\right); \boldsymbol{\lambda}, \boldsymbol{\theta}\right)$ is the negative Hessian of $\mathcal{L}$ with respect to $(\boldsymbol{\phi}, \boldsymbol{\gamma})$ evaluated at $\left(\widehat{\boldsymbol{\phi}}\left(\boldsymbol{\lambda}, \boldsymbol{\theta}\right), \widehat{\boldsymbol{\gamma}}\left(\boldsymbol{\lambda}, \boldsymbol{\theta}\right)\right)$, and constant $\mathrm{C} = M_p\log(2\pi)/2$ where $M_p$ is the number of zero eigenvalues in $\mathbf{S}^{\boldsymbol{\lambda}}$. The $\log|\mathbf{S}^{\boldsymbol{\lambda}}|_{+}$ is the product of the positive eigenvalues of $\mathbf{S}^{\boldsymbol{\lambda}}$, which is the block diagonal matrix with each block being the penalty matrix multiplied by the corresponding smoothing parameters. 
We maximize $\mathcal{L}^*_{\text{LA}}(\boldsymbol{\lambda}, \boldsymbol{\theta})$ using a BFGS algorithm, a Newton-type method that requires the gradient. 
We derive the analytical gradients of $\mathcal{L}(\widehat{\boldsymbol{\phi}}\left(\boldsymbol{\lambda}, \boldsymbol{\theta}\right), \widehat{\boldsymbol{\gamma}}\left(\boldsymbol{\lambda}, \boldsymbol{\theta}\right); \boldsymbol{\lambda}, \boldsymbol{\theta})$ and $\log|\mathbf{S}^{\boldsymbol{\lambda}}|_{+}$, where multiple smoothing-parameter blocks in $\mathbf{S}^{\boldsymbol{\lambda}}$ require additional steps beyond \citet{pan2025estimating}. 
Deriving the analytical gradient of $\log \left\{\mathrm{det} \boldsymbol{\mathcal{H}}\left(\widehat{\boldsymbol{\phi}}\left(\boldsymbol{\lambda}, \boldsymbol{\theta}\right), \widehat{\boldsymbol{\gamma}}\left(\boldsymbol{\lambda}, \boldsymbol{\theta}\right); \boldsymbol{\lambda}, \boldsymbol{\theta}\right)\right\}$ is challenging for this complicated general model form, and hence we compute it via automatic differentiation (AD) %
using {\tt CppAD} \citep{CppAD} in {\tt C++} and implicit differentiation. See supplementary Section \ref{ss:deLAML} for details. 
Unlike the forward-mode AD used for single-exposure ACE-DLNM \citet{pan2025estimating}, we employ reverse-mode AD
to handle the large number of parameters in our models. 
Finally, we obtain $\widehat{\boldsymbol{\phi}} = \widehat{\boldsymbol{\phi}}(\widehat{\boldsymbol{\lambda}}, \widehat{\boldsymbol{\theta}})$ and $\widehat{\boldsymbol{\gamma}} = \widehat{\boldsymbol{\gamma}}(\widehat{\boldsymbol{\lambda}}, \widehat{\boldsymbol{\theta}})$.

\section{Derivative Computation}
\label{s:de}
\subsection{\texorpdfstring{Derivative of $\mathcal{L}
(\boldsymbol{\phi}, \boldsymbol{\gamma}; \boldsymbol{\lambda}, \boldsymbol{\theta})$}{Derivative of L(phi, gamma; lambda, theta)}} \label{ss:deL}
We consider the model with only exposure-response surface $s(\boldsymbol{X}_i, t)$. Including $h_j$ for covariates is straightforward as in the regular generalized additive models. 
Using the general form with inner, outer, and tuning parameters, we write the model as 
$$
g(\mu_{it}) = \sum_{j = 1}^{d^{\boldsymbol{\gamma}}} \boldsymbol{\gamma}_j B_j\left(\sum_{m=1}^M \alpha_m D_m(t) \right),
$$
where $B_j$ is constructed from the basis function for the basis functions of $f$ and $\psi$ for ACE-DLNM and DRF-DLNM structures; specifically: $B_j(x) = b^w_{j}(x)$ for ACE-DLNM structure, and $B_j(\boldsymbol{x}) = \sum_l b^{\psi_x}_s(x_l) b^{\psi_l}_r(l)$. For the term inside $B_j$, we have $D_{m}(t) = \sum_{q=1}^{d^w}\beta^w\int_{0}^L  b^w_{q}(l) X_m(t-l)dl$ for ACE-DLNM, and $D_{m}(t)= [X_m(t), \cdots, X_m(t-L+1)]^\top$ for DRF-DLNM. 
The derivatives of $\mathcal{L}(\boldsymbol{\phi}, \boldsymbol{\gamma}; \boldsymbol{\lambda}, \boldsymbol{\theta})$ require the derivatives of the B-spline which is calculated using the recursive formula \citep{wood2017generalized}, as well as $d \boldsymbol{\alpha}/d\boldsymbol{\alpha}^*$. We define $\tilde{\boldsymbol{\alpha}}$ as $[1, \boldsymbol{\alpha}^{*\top}]^\top$. According to Equation 6 in Section 3, we have 
$$
\frac{d \boldsymbol{\alpha}^{w}}{d \tilde{\boldsymbol{\alpha}}} = \mathbf{B}_{\boldsymbol{\alpha}} \left[\left(\tilde{\boldsymbol{\alpha}}^\top \mathbf{B}_{\boldsymbol{\alpha}}^\top \mathbf{B}_{\boldsymbol{\alpha}} \tilde{\boldsymbol{\alpha}}\right)^{-\frac{1}{2}} \mathbf{I} - \tilde{\boldsymbol{\alpha}} \tilde{\boldsymbol{\alpha}}^\top \mathbf{B}_{\boldsymbol{\alpha}}^\top \mathbf{B}_{\boldsymbol{\alpha}}\left(\tilde{\boldsymbol{\alpha}}^\top \mathbf{B}_{\boldsymbol{\alpha}}^\top \mathbf{B}_{\boldsymbol{\alpha}} \tilde{\boldsymbol{\alpha}}\right)^{-\frac{3}{2}}\right]. 
$$
Then, $d \boldsymbol{\alpha}/d\boldsymbol{\alpha}^*$ is obtained by removing the first column of $d \boldsymbol{\alpha}^{w}/d \tilde{\boldsymbol{\alpha}}$.

\subsection{\texorpdfstring{Derivative of $\boldsymbol{Q}(\boldsymbol{\phi}; \boldsymbol{\lambda}, \boldsymbol{\theta})$}{Derivative of Q(phi; lambda, theta)}}
\label{ss:deQ}
Recall that $\boldsymbol{Q}(\boldsymbol{\phi}; \boldsymbol{\lambda}, \boldsymbol{\theta}) = \mathcal{L}(\boldsymbol{\phi}, \widehat{\boldsymbol{\gamma}}(\boldsymbol{\phi}; \boldsymbol{\lambda}, \boldsymbol{\theta}); \boldsymbol{\lambda}, \boldsymbol{\theta})$. 
The gradient is given by, 
$$
\begin{aligned}
    \frac{\partial \mathbf{Q}(\boldsymbol{\phi}; \boldsymbol{\lambda}, \boldsymbol{\theta})}{\partial \boldsymbol{\phi}} =& \frac{\partial \mathcal{L}\left(\boldsymbol{\phi},\boldsymbol{\gamma};  \boldsymbol{\lambda}, \boldsymbol{\theta}\right)}{\partial \boldsymbol{\phi}} \Bigr|_{\boldsymbol{\gamma} = \widehat{\boldsymbol{\gamma}}(\boldsymbol{\phi}; \boldsymbol{\lambda}, \boldsymbol{\theta})} +\frac{\partial \widehat{\boldsymbol{\gamma}}(\boldsymbol{\phi}; \boldsymbol{\lambda}, \boldsymbol{\theta})}{\partial \boldsymbol{\phi}} \frac{\partial \mathcal{L}\left(\boldsymbol{\phi},\widehat{\boldsymbol{\gamma}}(\boldsymbol{\phi}; \boldsymbol{\lambda}, \boldsymbol{\theta}); \boldsymbol{\lambda}, \boldsymbol{\theta}\right)}{\partial \boldsymbol{\gamma}},
\end{aligned}
$$
where $\partial \mathcal{L}\left(\boldsymbol{\phi},\widehat{\boldsymbol{\gamma}}(\boldsymbol{\phi}; \boldsymbol{\lambda}, \boldsymbol{\theta}); \boldsymbol{\lambda}, \boldsymbol{\theta}\right)/\partial \boldsymbol{\gamma} = 0$ since $\widehat{\boldsymbol{\gamma}}(\boldsymbol{\phi}; \boldsymbol{\lambda}, \boldsymbol{\theta})$ is obtained by maximizing $\mathcal{L}$. 

The Hessian is given by, 
$$
\begin{aligned}
    \frac{\partial^2 \mathbf{Q}(\boldsymbol{\phi}; \boldsymbol{\lambda}, \boldsymbol{\theta})}{\partial \boldsymbol{\phi} \partial \boldsymbol{\phi}^\top} =& 
    \frac{\partial^2 \mathcal{L}\left(\boldsymbol{\phi},\boldsymbol{\gamma};  \boldsymbol{\lambda}, \boldsymbol{\theta}\right)}{\partial \boldsymbol{\phi}\partial \boldsymbol{\phi}^\top} \Bigr|_{\boldsymbol{\gamma} = \widehat{\boldsymbol{\gamma}}(\boldsymbol{\phi}; \boldsymbol{\lambda}, \boldsymbol{\theta})}
    +\left(\frac{\partial \widehat{\boldsymbol{\gamma}}(\boldsymbol{\phi}; \boldsymbol{\lambda}, \boldsymbol{\theta})}{\partial \boldsymbol{\phi}}\right)^\top \frac{\partial^2 \mathcal{L}\left(\boldsymbol{\phi},\boldsymbol{\gamma};  \boldsymbol{\lambda}, \boldsymbol{\theta}\right)}{\partial \boldsymbol{\gamma} \partial \boldsymbol{\phi}^\top} \Bigr|_{\boldsymbol{\gamma} = \widehat{\boldsymbol{\gamma}}(\boldsymbol{\phi}; \boldsymbol{\lambda}, \boldsymbol{\theta})}, 
\end{aligned}
$$
where $\partial \widehat{\boldsymbol{\gamma}}(\boldsymbol{\phi}; \boldsymbol{\lambda}, \boldsymbol{\theta})/\partial \boldsymbol{\phi}$ is calculated using implicit differentiation. Specifically, it is solved by 
$$
\begin{aligned}
    \frac{\partial^2 \mathcal{L}\left(\boldsymbol{\phi},\widehat{\boldsymbol{\gamma}}(\boldsymbol{\phi}; \boldsymbol{\lambda}, \boldsymbol{\theta}); \boldsymbol{\lambda}, \boldsymbol{\theta}\right)}{\partial \boldsymbol{\gamma} \partial \boldsymbol{\gamma}^\top} \frac{\partial \widehat{\boldsymbol{\gamma}}(\boldsymbol{\phi}; \boldsymbol{\lambda}, \boldsymbol{\theta})}{\partial \boldsymbol{\phi}} = -\frac{\partial^2 \mathcal{L}\left(\boldsymbol{\phi},\boldsymbol{\gamma};  \boldsymbol{\lambda}, \boldsymbol{\theta}\right)}{\partial \boldsymbol{\gamma} \partial \boldsymbol{\phi}^\top} \Bigr|_{\boldsymbol{\gamma} = \widehat{\boldsymbol{\gamma}}(\boldsymbol{\phi}; \boldsymbol{\lambda}, \boldsymbol{\theta})}, 
\end{aligned}
$$
which is from differentiating $\partial \mathcal{L}\left(\boldsymbol{\phi},\widehat{\boldsymbol{\gamma}}(\boldsymbol{\phi}; \boldsymbol{\lambda}, \boldsymbol{\theta}); \boldsymbol{\lambda}, \boldsymbol{\theta}\right)/\partial \boldsymbol{\gamma} = 0$ with respect to $\boldsymbol{\phi}$.

\subsection{\texorpdfstring{Derivative of $\mathcal{L}_{\mathrm{LA}}^*(\boldsymbol{\lambda}, \boldsymbol{\theta})$}{Derivative of L-LA star(lambda, theta)}}
\label{ss:deLAML}
The log-LAML described in Section 4.2 is
$$
\begin{aligned}
\mathcal{L}^*_{\text{LA}}(\boldsymbol{\lambda}, \boldsymbol{\theta}) =&
\mathcal{L}(\widehat{\boldsymbol{\phi}}\left(\boldsymbol{\lambda}, \boldsymbol{\theta}\right), \widehat{\boldsymbol{\gamma}}\left(\boldsymbol{\lambda}, \boldsymbol{\theta}\right); \boldsymbol{\lambda}, \boldsymbol{\theta}) \\
&- \frac{1}{2} \log \left\{\mathrm{det} \boldsymbol{\mathcal{H}}\left(\widehat{\boldsymbol{\phi}}\left(\boldsymbol{\lambda}, \boldsymbol{\theta}\right), \widehat{\boldsymbol{\gamma}}\left(\boldsymbol{\lambda}, \boldsymbol{\theta}\right); \boldsymbol{\lambda}, \boldsymbol{\theta}\right)\right\} + 
\frac{1}{2}\log|\mathbf{S}^{\boldsymbol{\lambda}}|_{+} + \frac{M_p}{2} \log(2\pi),
\end{aligned}
$$

The first-order derivative of $\mathcal{L}(\widehat{\boldsymbol{\phi}}\left(\boldsymbol{\lambda}, \boldsymbol{\theta}\right), \widehat{\boldsymbol{\gamma}}\left(\boldsymbol{\lambda}, \boldsymbol{\theta}\right); \boldsymbol{\lambda}, \boldsymbol{\theta})$ is calculated using implicit differentiation which requires the derivatives of $\mathcal{L}(\boldsymbol{\phi}, \boldsymbol{\gamma}; \boldsymbol{\lambda}, \boldsymbol{\theta})$ in Supplementary Materials Section \ref{ss:deL}, as well as $\partial \widehat{\boldsymbol{\phi}}\left(\boldsymbol{\lambda}, \boldsymbol{\theta}\right) / \partial [\log\boldsymbol{\lambda}, \log\boldsymbol{\theta}]$ and $\partial \widehat{\boldsymbol{\gamma}}\left(\boldsymbol{\lambda}, \boldsymbol{\theta}\right) / \partial [\log\boldsymbol{\lambda}, \log\boldsymbol{\theta}]$. 

We differentiate 
$\partial \mathcal{L}(\widehat{\boldsymbol{\phi}}\left(\boldsymbol{\lambda}, \boldsymbol{\theta}\right), \widehat{\boldsymbol{\gamma}}\left(\boldsymbol{\lambda}, \boldsymbol{\theta}\right); \boldsymbol{\lambda}, \boldsymbol{\theta}) / \partial [\boldsymbol{\phi}, \boldsymbol{\gamma}] = 0$
with respect to $[\log\boldsymbol{\lambda}, \log\boldsymbol{\theta}]$ to obtain $\partial \widehat{\boldsymbol{\phi}}\left(\boldsymbol{\lambda}, \boldsymbol{\theta}\right) / \partial [\log\boldsymbol{\lambda}, \log\boldsymbol{\theta}]$ and $\partial \widehat{\boldsymbol{\gamma}}\left(\boldsymbol{\lambda}, \boldsymbol{\theta}\right) / \partial [\log\boldsymbol{\lambda}, \log\boldsymbol{\theta}]$; see details in Web Appendix F.3 in \citet{pan2025estimating}.

The multiple smoothing-parameter blocks in $\mathbf{S}^{\boldsymbol{\lambda}}$ complicate the computation. Consider the matrix $\mathbf{S}^{\boldsymbol{\lambda}} = \lambda^{\psi_x} \mathbf{S}^{\psi_x} \otimes \mathbf{I}_{d^{\psi_x}} + \lambda^{\psi_l} \mathbf{I}_{d^{\psi_l}} \otimes \mathbf{S}^{\psi_l}$ in the single-index DRF-DLNM. We reparameterize the spline coefficients using the eigen-decomposition, yielding full rank penalty matrices $\mathbf{S}^{\psi_x}$ and $\mathbf{S}^{\psi_l}$ to improve numerical stability, following \citet{wood2017generalized}. 
Then we have 
$$
\begin{aligned}
\frac{\partial \mathbf{S}^{\boldsymbol{\lambda}}}{\partial \log \lambda^{\psi_x}} &= \lambda^{\psi_x} \mathrm{tr}\left\{(\mathbf{S}^{\boldsymbol{\lambda}})^{-1}(\mathbf{S}^{\psi_x} \otimes \mathbf{I}_{d^{\psi_x}})\right\},\\ 
\frac{\partial \mathbf{S}^{\boldsymbol{\lambda}}}{\partial \log \lambda^{\psi_l}} &= \lambda^{\psi_l} \mathrm{tr}\left\{(\mathbf{S}^{\boldsymbol{\lambda}})^{-1} (\mathbf{I}_{d^{\psi_l}} \otimes \mathbf{S}^{\psi_l})\right\}. 
\end{aligned}
$$

Deriving the analytical gradient of $\log \left\{\mathrm{det} \boldsymbol{\mathcal{H}}\left(\widehat{\boldsymbol{\phi}}\left(\boldsymbol{\lambda}, \boldsymbol{\theta}\right), \widehat{\boldsymbol{\gamma}}\left(\boldsymbol{\lambda}, \boldsymbol{\theta}\right); \boldsymbol{\lambda}, \boldsymbol{\theta}\right)\right\}$ is challenging for this complicated general model form. 
We therefore obtain the log-determinant via an LU decomposition,
computed as the sum of the logarithms of the diagonal elements.
We adopt reverse-mode automatic differentiation, implemented via {\tt CppAD} \citep{CppAD} in {\tt C++}, to calculate the derivative of $\log \left\{\mathrm{det} \boldsymbol{\mathcal{H}}\left(\boldsymbol{\phi}, \boldsymbol{\gamma}; \boldsymbol{\lambda}, \boldsymbol{\theta}\right)\right\}$ with respect to $[\boldsymbol{\phi}, \boldsymbol{\gamma}, \boldsymbol{\lambda}, \boldsymbol{\theta}]$ and then the gradient is obtained using implicit differentiation.

\section{AIC}
\subsection{Derivation}
Let the true model be $\varphi^*$. 
We fit a candidate model $\varphi(y; \boldsymbol{u})$, and denote $\widehat{\boldsymbol{u}}$ as the value of $\boldsymbol{u}$ that maximizes the penalized log-likelihood $\mathcal{L}(\boldsymbol{u}; \boldsymbol{\lambda}) = l(\boldsymbol{u}) - 1/2 \boldsymbol{u}^\top \mathbf{S}^{\boldsymbol{\lambda}} \boldsymbol{u}$ where $l(\boldsymbol{u}) = \sum_{i=1}^n \varphi(y_i; \boldsymbol{u})$. 

Denote $D(\varphi_{\boldsymbol{u}}, \varphi^*)$ as the Kullback-Leibler (KL) divergence, 
$$
D(\varphi_{\boldsymbol{u}}, \varphi^*) = \int \log \left\{\frac{\varphi^*(y)}{\varphi(y; \boldsymbol{u})}\right\} \varphi^*(y) dy. 
$$
Let $\boldsymbol{u}^*$ be a value of $\boldsymbol{u}$ that minimizes $D(\varphi_{\boldsymbol{u}}, \varphi^*)$, i.e. $\boldsymbol{u}^* = \arg \min_{\boldsymbol{u}} D(\varphi_{\boldsymbol{u}}, \varphi^*)$. 
The KL divergence comparing at $\boldsymbol{u} = \widehat{\boldsymbol{u}}$ is given by
$$
D(\varphi_{\widehat{\boldsymbol{u}}}, \varphi^*) = \int \log \left\{\frac{\varphi^*(y)}{\varphi(y; \widehat{\boldsymbol{u}})} \right\} \varphi^*(y) dy \ge D(\varphi_{\boldsymbol{u}^*}, \varphi^*). 
$$
We use $\mathrm{E}^*$ to represent the expectation with respect to the true model $\varphi^*$. Taking the expectation with respect to $\widehat{\boldsymbol{u}}$, we have 
\begin{equation}
\begin{aligned}
\mathrm{E}^* \left\{D(\varphi_{\widehat{\boldsymbol{u}}}, \varphi^*) \right\} & = \mathrm{E}^* \left[\int \log \left\{\frac{\varphi^*(y)}{\varphi(y; \widehat{\boldsymbol{u}})} \right\} \varphi^*(y) dy\right] \\
& = \int \log \left\{\varphi^*(y) \right\} \varphi^*(y) dy -  \mathrm{E}^*\left\{ \int \log \left\{\varphi(y; \widehat{\boldsymbol{u}}) \right\} \varphi^*(y) dy\right\}. 
\end{aligned}
\label{eq:AIC1}
\end{equation}
Taking the Taylor expansion of $\log \varphi(y; \widehat{\boldsymbol{u}})$ around $\boldsymbol{u}^*$, we have 
$$
\begin{aligned}
\log \varphi(y; \widehat{\boldsymbol{u}}) \approx \log \varphi(y; \boldsymbol{u}^*) + (\widehat{\boldsymbol{u}} - \boldsymbol{u}^*)^\top \frac{\partial \log \varphi(y; \boldsymbol{u}^*)}{\partial \boldsymbol{u}} +  \frac{1}{2}(\widehat{\boldsymbol{u}} - \boldsymbol{u}^*)^\top \frac{\partial^2 \log \varphi(y; \boldsymbol{u}^*)}{\partial \boldsymbol{u} \partial \boldsymbol{u}^\top} (\widehat{\boldsymbol{u}} - \boldsymbol{u}^*)
\end{aligned}
$$
Since $\boldsymbol{u}^*$ minimizes $D(\varphi_{\boldsymbol{u}}, \varphi^*)$, we have $\int \left\{\partial \log \varphi(y; \boldsymbol{u}^*)/\partial \boldsymbol{u}\right\} \varphi^*(y) dy = 0$. Therefore, 
\begin{equation}
\begin{aligned}
    \mathrm{E}^*\left\{ \int \log \left\{\varphi(y; \widehat{\boldsymbol{u}}) \right\} \varphi^*(y) dy\right\} & \approx \int \log \left\{\varphi(y; \boldsymbol{u}^*)\right\}  \varphi^*(y) dy - \frac{1}{2n}\mathrm{E}^* \left\{(\widehat{\boldsymbol{u}} - \boldsymbol{u}^*)^\top \boldsymbol{\mathcal{I}}^* (\widehat{\boldsymbol{u}} - \boldsymbol{u}^*)\right\},
\end{aligned}
\label{eq:AIC2}
\end{equation}
where 
$$
\boldsymbol{\mathcal{I}}^* = -n \int \frac{\partial^2 \log \varphi(y; \boldsymbol{u}^*)}{\partial \boldsymbol{u} \partial \boldsymbol{u}^\top} \varphi^*(y) dy
$$
Plugging Equation \ref{eq:AIC2} into Equation \ref{eq:AIC1} leads to
\begin{equation}
\begin{aligned}
    \mathrm{E}^* \left\{D(\varphi_{\widehat{\boldsymbol{u}}}, \varphi^*) \right\} &\approx D(\varphi_{\boldsymbol{u}^*}, \varphi^*) + \frac{1}{2n}\mathrm{E}^* \left\{(\widehat{\boldsymbol{u}} - \boldsymbol{u}^*)^\top \boldsymbol{\mathcal{I}}^* (\widehat{\boldsymbol{u}} - \boldsymbol{u}^*)\right\}%
\end{aligned}
\label{eq:ED}
\end{equation}

We aim to choose the candidate model $\varphi$ minimizing $\mathrm{E}^* \left\{D(\varphi_{\widehat{\boldsymbol{u}}}, \varphi^*) \right\}$, an unknown quantity to be estimated. Considering the Taylor expansion of $l(\boldsymbol{u}^*)$, we have
$$
l(\boldsymbol{u}^*) \approx l(\widehat{\boldsymbol{u}}) + (\boldsymbol{u}^* - \widehat{\boldsymbol{u}})^\top \frac{\partial l(\widehat{\boldsymbol{u}})}{\partial \boldsymbol{u}} + \frac{1}{2} (\boldsymbol{u}^* - \widehat{\boldsymbol{u}})^\top \frac{\partial^2 l(\widehat{\boldsymbol{u}})}{\partial \boldsymbol{u} \partial \boldsymbol{u}^\top}(\boldsymbol{u}^* - \widehat{\boldsymbol{u}})
$$

As $\widehat{\boldsymbol{u}}$ maximizes the penalized log-likelihood, 
$$
0=\frac{\partial \mathcal{L}(\widehat{\boldsymbol{u}})}{\partial \boldsymbol{u}} = \frac{\partial l(\widehat{\boldsymbol{u}})}{\partial \boldsymbol{u}} - \mathbf{S}^{\boldsymbol{\lambda}} \widehat{\boldsymbol{u}}.
$$
Hence,
$$
l(\boldsymbol{u}^*) \approx l(\widehat{\boldsymbol{u}}) + (\boldsymbol{u}^* - \widehat{\boldsymbol{u}})^\top \mathbf{S}^{\boldsymbol{\lambda}} \widehat{\boldsymbol{u}} - \frac{1}{2} (\boldsymbol{u}^* - \widehat{\boldsymbol{u}})^\top \widehat{\boldsymbol{\mathcal{I}}} (\boldsymbol{u}^* - \widehat{\boldsymbol{u}})
$$
where $\widehat{\boldsymbol{\mathcal{I}}} = -\frac{\partial^2 l(\widehat{\boldsymbol{u}})}{\partial \boldsymbol{u} \partial \boldsymbol{u}^\top}$ is the Hessian of the negative log-likelihood evaluated at $\widehat{\boldsymbol{u}}$. It follows that 
$$
\begin{aligned}
\mathrm{E}^* \left\{-l(\widehat{\boldsymbol{u}})\right\} \approx & - \mathrm{E}^* \left\{l(\boldsymbol{u}^*)\right\} + \mathrm{E}^* \left\{(\boldsymbol{u}^* - \widehat{\boldsymbol{u}})^\top \mathbf{S}^{\boldsymbol{\lambda}} \widehat{\boldsymbol{u}}\right\} -
\frac{1}{2} \mathrm{E}^* \left\{(\boldsymbol{u}^* - \widehat{\boldsymbol{u}})^\top \widehat{\boldsymbol{\mathcal{I}}} (\boldsymbol{u}^* - \widehat{\boldsymbol{u}})\right\} \\
= & n D(\varphi_{\boldsymbol{u}^*}, \varphi^*) - n \int \log \left\{\varphi^*(y)\right\} \varphi^*(y) dy \\ &+ \mathrm{E}^* \left\{(\boldsymbol{u}^* - \widehat{\boldsymbol{u}})^\top \mathbf{S}^{\boldsymbol{\lambda}} \widehat{\boldsymbol{u}}\right\} - \frac{1}{2} \mathrm{E}^* \left\{(\boldsymbol{u}^* - \widehat{\boldsymbol{u}})^\top \widehat{\boldsymbol{\mathcal{I}}} (\boldsymbol{u}^* - \widehat{\boldsymbol{u}})\right\},
\end{aligned}
$$
which leads to 
$$
\begin{aligned}
n D(\varphi_{\boldsymbol{u}^*}, \varphi^*) \approx & \mathrm{E}^* \left\{-l(\widehat{\boldsymbol{u}})\right\} +  n \int \log \left\{\varphi^*(y)\right\} \varphi^*(y) dy  \\ 
& - \mathrm{E}^* \left\{(\boldsymbol{u}^* - \widehat{\boldsymbol{u}})^\top \mathbf{S}^{\boldsymbol{\lambda}} \widehat{\boldsymbol{u}}\right\} + \frac{1}{2} \mathrm{E}^* \left\{(\boldsymbol{u}^* - \widehat{\boldsymbol{u}})^\top \widehat{\boldsymbol{\mathcal{I}}} (\boldsymbol{u}^* - \widehat{\boldsymbol{u}})\right\}. 
\end{aligned}
$$
Plugging this into Equation \ref{eq:ED} yields
$$
\begin{aligned}
n \mathrm{E}^* \left\{D(\varphi_{\widehat{\boldsymbol{u}}}, \varphi^*) \right\} \approx& \mathrm{E}^* \left\{-l(\widehat{\boldsymbol{u}})\right\} + n \int \log \left\{\varphi^*(y)\right\} \varphi^*(y) dy  - 
\mathrm{E}^* \left\{(\boldsymbol{u}^* - \widehat{\boldsymbol{u}})^\top \mathbf{S}^{\boldsymbol{\lambda}} \widehat{\boldsymbol{u}}\right\} \\ 
&+ \frac{1}{2} \mathrm{E}^* \left\{(\boldsymbol{u}^* - \widehat{\boldsymbol{u}})^\top \widehat{\boldsymbol{\mathcal{I}}} (\boldsymbol{u}^* - \widehat{\boldsymbol{u}})\right\}  + \frac{1}{2}\mathrm{E}^* \left\{(\widehat{\boldsymbol{u}} - \boldsymbol{u}^*)^\top \boldsymbol{\mathcal{I}}^* (\widehat{\boldsymbol{u}} - \boldsymbol{u}^*)\right\}. 
\end{aligned}
$$
Following \citet{wood2016smoothing} and \citet{davison2003statistical}, the last two terms can be estimated by 
$$
\mathrm{tr}\left[\mathrm{E}^* \left\{(\widehat{\boldsymbol{u}} - \boldsymbol{u}^*) (\widehat{\boldsymbol{u}} - \boldsymbol{u}^*)^\top\right\}\widehat{\boldsymbol{\mathcal{I}}}\right]. 
$$
Ignoring the constant term $n \int \log \left\{\varphi^*(y)\right\} \varphi^*(y) dy$, we obtain the following AIC
$$
\mathrm{AIC} = -2 l(\widehat{\boldsymbol{u}}) + 2 \mathrm{tr}\left[\mathrm{E}^* \left\{\widehat{\boldsymbol{u}}(\widehat{\boldsymbol{u}} - \boldsymbol{u}^*)^\top \right\} \mathbf{S}^{\boldsymbol{\lambda}} \right]+ 2 \mathrm{tr}\left[\mathrm{E}^* \left\{(\widehat{\boldsymbol{u}} - \boldsymbol{u}^*)^\top (\widehat{\boldsymbol{u}} - \boldsymbol{u}^*)\right\}\widehat{\boldsymbol{\mathcal{I}}}\right]. 
$$
The second term extends \citet{wood2016smoothing} to adjust for the non-zero gradient of the log-likelihood evaluated at $\widehat{\boldsymbol{u}}$ induced by the smoothness penalties, i.e. $\partial l(\widehat{\boldsymbol{u}})/\partial \boldsymbol{u} = \mathbf{S}^{\boldsymbol{\lambda}} \widehat{\boldsymbol{u}}$. 
Note that
$$
\begin{aligned}
    \mathrm{E}^* \left\{\widehat{\boldsymbol{u}}(\widehat{\boldsymbol{u}} - \boldsymbol{u}^*)^\top \right\} = \mathrm{E}^* \left\{(\widehat{\boldsymbol{u}} - \boldsymbol{u}^*)(\widehat{\boldsymbol{u}} - \boldsymbol{u}^*)^\top \right\} + \boldsymbol{u}^*(\mathrm{E}^* (\widehat{\boldsymbol{u}}) - \boldsymbol{u}^*)^\top, 
\end{aligned}
$$
where $\mathrm{E}^* (\widehat{\boldsymbol{u}}) \approx \left(\widehat{\boldsymbol{\mathcal{I}}} + \mathbf{S}^{\boldsymbol{\lambda}}\right)^{-1} \widehat{\boldsymbol{\mathcal{I}}} \boldsymbol{u}^*$ and following Lemma 1 of \citet{wood2016smoothing} we can approximate $\mathrm{tr}\left[\mathrm{E}^* \left\{\widehat{\boldsymbol{u}}(\widehat{\boldsymbol{u}} - \boldsymbol{u}^*)^\top \right\} \mathbf{S}^{\boldsymbol{\lambda}} \right]$ by
$$
\begin{aligned}
    \mathrm{tr}\left[\mathrm{E}^* \left\{\widehat{\boldsymbol{u}}(\widehat{\boldsymbol{u}} - \boldsymbol{u}^*)^\top \right\} \mathbf{S}^{\boldsymbol{\lambda}} \right] 
    \approx \mathrm{tr}\left[\mathrm{E}^* \left\{(\widehat{\boldsymbol{u}} - \boldsymbol{u}^*) (\widehat{\boldsymbol{u}} - \boldsymbol{u}^*)^\top\right\}\mathbf{S}^{\boldsymbol{\lambda}}\right] + 
    \mathrm{tr}\left[\widehat{\boldsymbol{\mathcal{I}}} \left(\widehat{\boldsymbol{\mathcal{I}}} + \mathbf{S}^{\boldsymbol{\lambda}}\right)^{-1} - \mathrm{I}\right]. 
\end{aligned}
$$
Therefore, the AIC score is given by 
\begin{equation}
\begin{aligned}
\mathrm{AIC} =& -2 l(\widehat{\boldsymbol{u}}) + 2\mathrm{tr}\left[\mathrm{E}^* \left\{(\widehat{\boldsymbol{u}} - \boldsymbol{u}^*)^\top (\widehat{\boldsymbol{u}} - \boldsymbol{u}^*)\right\}\mathbf{S}^{\boldsymbol{\lambda}}\right] \\ 
    &+ 2\mathrm{tr}\left[\widehat{\boldsymbol{\mathcal{I}}} \left(\widehat{\boldsymbol{\mathcal{I}}} + \mathbf{S}^{\boldsymbol{\lambda}}\right)^{-1} - \mathrm{I}\right]+ 2\mathrm{tr}\left[\mathrm{E}^* \left\{(\widehat{\boldsymbol{u}} - \boldsymbol{u}^*)^\top (\widehat{\boldsymbol{u}} - \boldsymbol{u}^*)\right\}\widehat{\boldsymbol{\mathcal{I}}}\right]. 
\end{aligned}
\label{eq:AIC}
\end{equation}
This AIC can be viewed as the negative log-likelihood penalized by three trace terms. The first term corresponds to the effective number of parameters and is the only adjustment term in the AIC of \citet{wood2016smoothing}. 
The last two terms extend \citet{wood2016smoothing} to penalize for the non-zero gradient of the log-likelihood evaluated at $\widehat{\boldsymbol{u}}$ induced by smoothness penalties. 

The unknown term $\mathrm{E}^* \left\{(\widehat{\boldsymbol{u}} - \boldsymbol{u}^*)^\top (\widehat{\boldsymbol{u}} - \boldsymbol{u}^*)\right\}$ can be estimated in different ways, leading to different AICs. 
The conditional AIC proposed in Section 5.1 and Section \ref{ss:condAIC} estimates the quantity by fixing the smoothing parameters at their point estimates. To account for the smoothing parameter uncertainty, in Section \ref{ss:adjustAIC}, we propose an adjusted AIC, as well as an efficient computational approach to address the computational challenge arising from the intractable third-order derivatives.

\subsection{Conditional AIC}
\label{ss:condAIC}

The unknown term $\mathrm{E}^* \left\{(\widehat{\boldsymbol{u}} - \boldsymbol{u}^*) (\widehat{\boldsymbol{u}} - \boldsymbol{u}^*)^\top\right\}$ can be decomposed as
$(\mathrm{E}^*(\widehat{\boldsymbol{u}}) - \boldsymbol{u}^*) (\mathrm{E}^*(\widehat{\boldsymbol{u}}) - \boldsymbol{u}^*)^\top   +  \mathrm{E}^* \left\{(\widehat{\boldsymbol{u}} - \mathrm{E}^*(\widehat{\boldsymbol{u}})) (\widehat{\boldsymbol{u}} - \mathrm{E}^*(\widehat{\boldsymbol{u}}))^\top\right\} $. 
We estimate first component by
$(\widehat{\boldsymbol{\mathcal{I}}} + \mathbf{S^{\boldsymbol{\lambda}}})^{-1} - (\widehat{\boldsymbol{\mathcal{I}}} + \mathbf{S^{\boldsymbol{\lambda}}})^{-1} \widehat{\boldsymbol{\mathcal{I}}} (\widehat{\boldsymbol{\mathcal{I}}} + \mathbf{S^{\boldsymbol{\lambda}}})^{-1}$ (Lemma 1; \citealp{wood2016smoothing}). 
Unlike \citet{wood2016smoothing}, we account for model misspecification \citep{davison2003statistical} by estimating the second term $\mathrm{E}^* \left\{(\widehat{\boldsymbol{u}} - \mathrm{E}^*(\widehat{\boldsymbol{u}})) (\widehat{\boldsymbol{u}} - \mathrm{E}^*(\widehat{\boldsymbol{u}}))^\top\right\}$ using $(\widehat{\boldsymbol{\mathcal{I}}} + \mathbf{S^{\boldsymbol{\lambda}}})^{-1} \widehat{\mathbf{K}} (\widehat{\boldsymbol{\mathcal{I}}} + \mathbf{S^{\boldsymbol{\lambda}}})^{-1}$,  where
$$
\widehat{\mathbf{K}} = \sum_{i=1}^N \sum_{t \in \mathcal{T}_i} \left[\partial \log p(Y_{it} | \boldsymbol{x}_{it}, \boldsymbol{z}_{it};\widehat{\boldsymbol{u}})/\partial\boldsymbol{u} \right] \left[\partial \log p(Y_{it} | \boldsymbol{x}_{it}, \boldsymbol{z}_{it};\widehat{\boldsymbol{u}})/\partial\boldsymbol{u} \right]^\top.
$$
Plugging this estimator into Equation \ref{eq:AIC} yields the proposed AIC in a simplified form,
\begin{equation}
\mathrm{AIC}_{\text{cond}} = -2 l(\widehat{\boldsymbol{u}}) + \mathrm{tr}\left[(\widehat{\boldsymbol{\mathcal{I}}} + \mathbf{S^{\boldsymbol{\lambda}}})^{-1} \widehat{\mathbf{K}}\right].
\label{eq:AICcond}
\end{equation}

\subsection{Adjusted AIC}
\label{ss:adjustAIC}

The conditional AIC does not account for the uncertainty in estimating the smoothing parameters. 
We take the first-order Taylor expansion of $\boldsymbol{u}$ around $\log \widehat{\boldsymbol{\lambda}}$ following \citet{wood2016smoothing}. Under this approximation, we estimate $\mathrm{E}^* \left\{(\widehat{\boldsymbol{u}} - \mathrm{E}^*\widehat{\boldsymbol{u}}) (\widehat{\boldsymbol{u}} - \mathrm{E}^*\widehat{\boldsymbol{u}})^\top\right\}$ as
$$
(\widehat{\boldsymbol{\mathcal{I}}} + \mathbf{S}^{\boldsymbol{\lambda}})^{-1} \widehat{\mathbf{K}} (\widehat{\boldsymbol{\mathcal{I}}} + \mathbf{S}^{\boldsymbol{\lambda}})^{-1} + \mathbf{V}' + \mathbf{V}'',
$$ 
where $\mathbf{V}' = \mathbf{J} \mathbf{V}_{\log \boldsymbol{\boldsymbol{u}}} \mathbf{J}^\top$, $\mathbf{J} = d \boldsymbol{\widehat{\boldsymbol{u}}} / d \log \boldsymbol{\lambda}$, and $\mathbf{V}_{\log \boldsymbol{\lambda}}$ is the covariance matrix of $\log \widehat{\boldsymbol{\lambda}}$ estimated by the inverse Hessian of the negative LAML; and $\mathbf{V}_{j m}^{\prime \prime}=\sum_i \sum_l \sum_k \frac{\partial \mathbf{R}_{i j}}{\partial \log \boldsymbol{\lambda}_k} V_{\log \boldsymbol{\lambda}, k l} \frac{\partial \mathbf{R}_{i m}}{\partial \log \boldsymbol{\lambda}_l}$ where 
$\mathbf{R}^\top \mathbf{R} = (\widehat{\boldsymbol{\mathcal{I}}} + \mathbf{S^{\boldsymbol{\lambda}}})^{-1} \widehat{\mathbf{K}} (\widehat{\boldsymbol{\mathcal{I}}} + \mathbf{S^{\boldsymbol{\lambda}}})^{-1}$. 
The first correction term $\mathbf{V}'$ requires the Jacobian matrix $\mathbf{J}$, which can be computed using the implicit differentiation technique described in Appendix F.3 of \citet{pan2025estimating}. 
The second correction term involving the derivative of $\mathbf{R}$ is computationally intensive. 
In single-index DLNMs, the complex model structure makes this derivative intractable, and we compute it by automatic differentiation using {\tt CppAD} library \citep{CppAD}. 
To avoid the computationally expensive differentiation for the decomposition of the entire sandwich form, we decompose the positive semi-definite matrix $\widehat{\mathbf{K}}$ first: 
$\widehat{\mathbf{K}} = \mathbf{P}^\top \mathbf{U}^\top \mathbf{D} \mathbf{U} \mathbf{P}$ where $\mathbf{P}$ is a permutation matrix, $\mathbf{U}$ is an upper triangular matrix, and $\mathbf{D}$ is a diagonal matrix with non-negative entries. Then we formulate $\mathbf{R}$ as $\mathbf{R} = \mathbf{D_+}^{1/2} \mathbf{U} \mathbf{P} (\widehat{\boldsymbol{\mathcal{I}}} + \mathbf{S^{\boldsymbol{\lambda}}})^{-1}$,
where $\mathbf{D_+}^{1/2}$ is a diagonal matrix with the square root of the positive entries in $\mathbf{D}$. After the decomposition, we obtain the derivative of $\mathbf{R}$ by differentiating the factor $\mathbf{D_+}^{1/2} \mathbf{U} \mathbf{P}$ and the Hessian $\widehat{\boldsymbol{\mathcal{I}}}$, which are less expensive than differentiating the decomposition of the entire sandwich form.

\clearpage
\section{Simulations}

\subsection{Simulation A: Single-Index ACE-DLNM}
\label{ss:simA}
\begin{figure}[htbp]
  \centering
  \begin{subfigure}[b]{0.45\linewidth}
    \includegraphics[width=\linewidth]{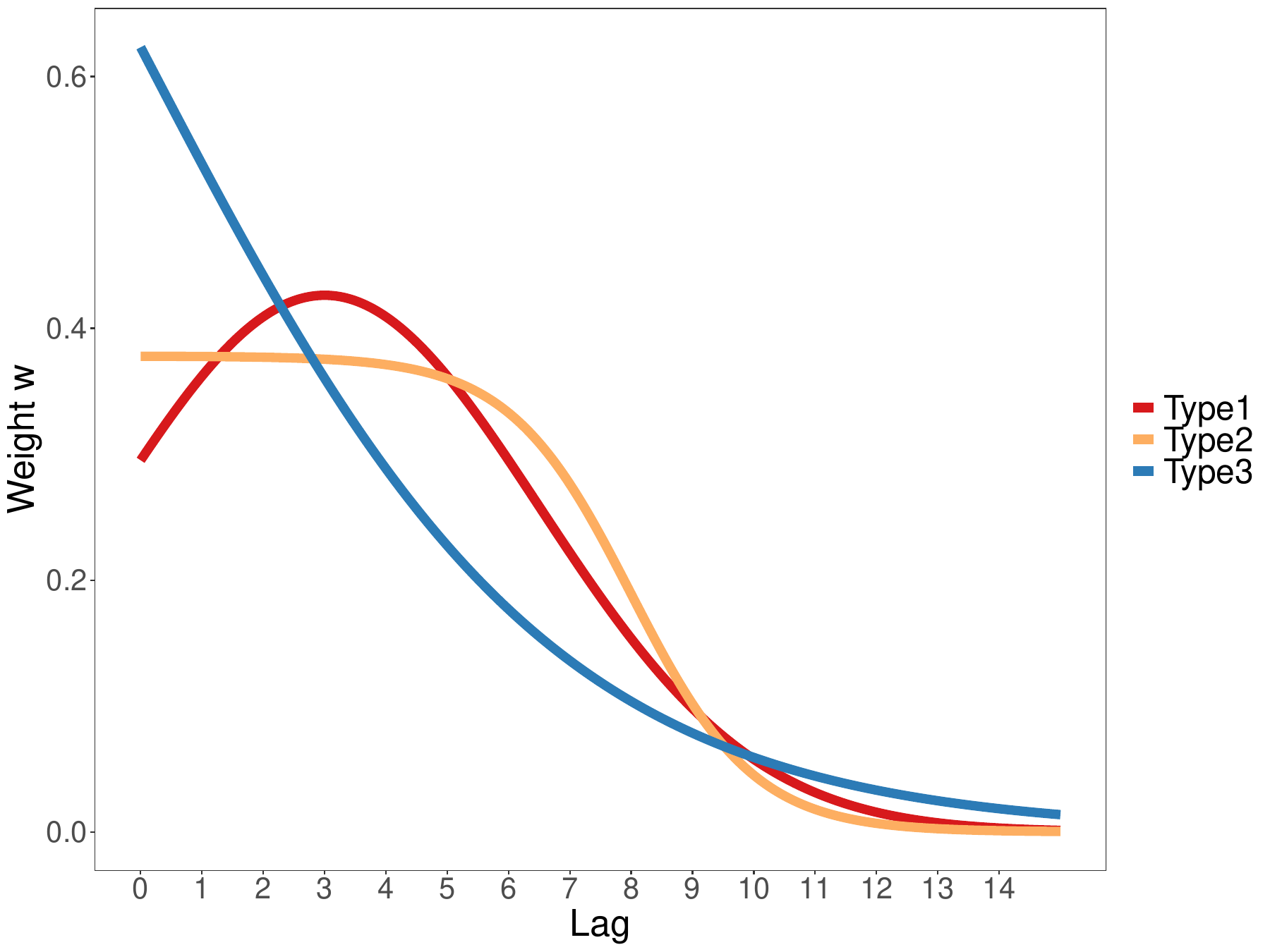}
    \caption{True $w$}
  \end{subfigure}
  \hspace{0.05\linewidth}
  \begin{subfigure}[b]{0.45\linewidth}
    \includegraphics[width=\linewidth]{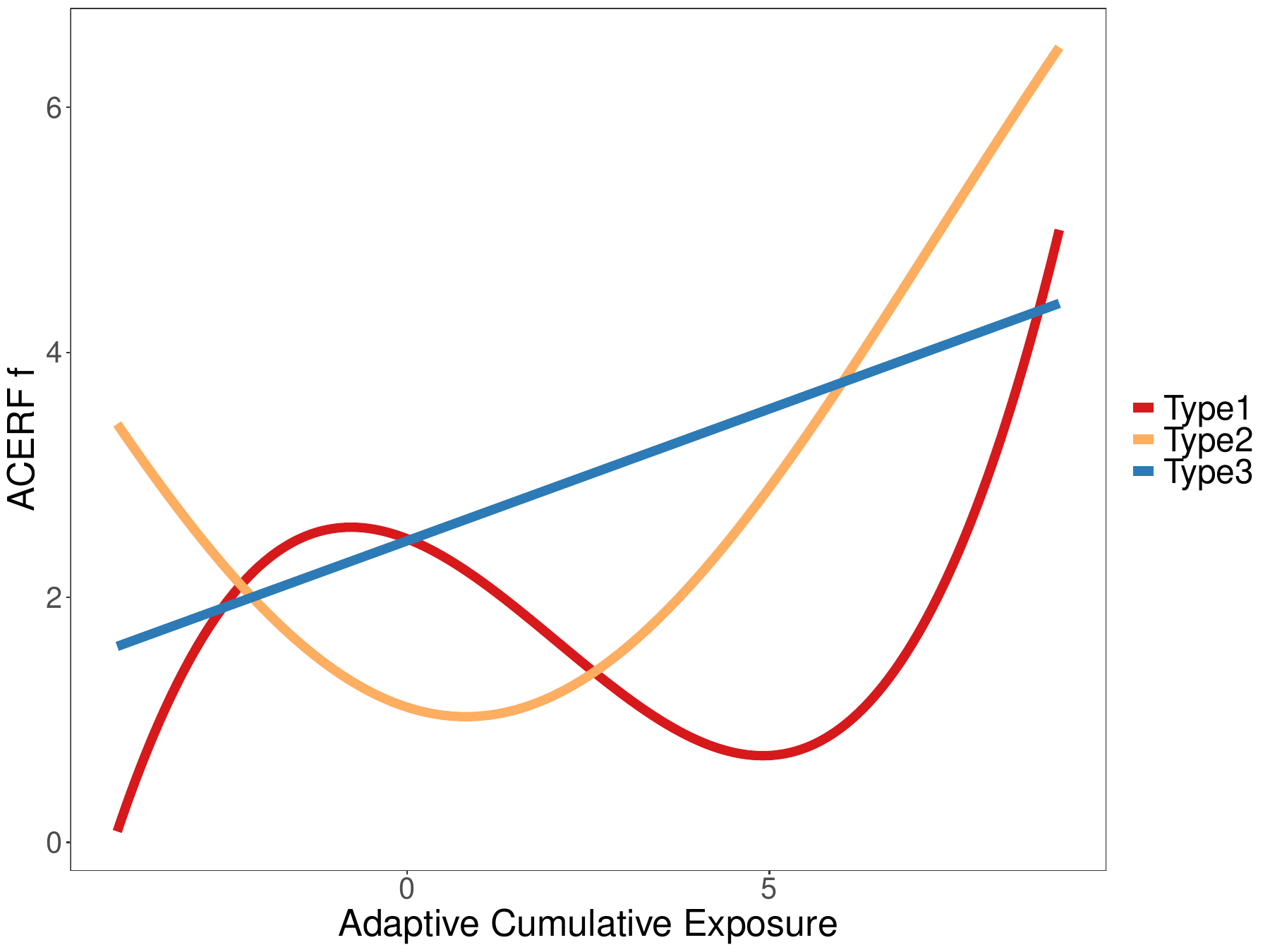}
    \caption{True $f$}
  \end{subfigure}
  \caption{The true weight functions $w$ and association functions $f$ used in Simulation A.}
  \label{fig:simA-wf}
\end{figure}

This simulation evaluates the estimation framework for the single-index ACE-DLNM, focusing on estimation of weight function $w$, the association function $f$, and the index weights $\boldsymbol{\alpha}$.
The exposure data are daily standardized PM$_{2.5}$, O$_3$ and NO$_2$ concentrations in Waterloo, Ontario, starting from January 1, 2001, with a sample size of $n$=1,000 or 2,000. 
The count outcome $Y_t$ is generated from a single-index ACE-DLNM with a negative binomial distribution with mean $\log(\mu_t) = f \left\{\int_{0}^{15} w(l) E(t-l;\boldsymbol{\alpha}) dl\right\} + h(t)$ and a dispersion parameter $\theta = 8$. 
The true index weights are set as $\boldsymbol{\alpha} = (1, 0.5, 0.3)^\top/c $ such that $c$ ensures $\boldsymbol{\alpha}^\top \boldsymbol{\alpha} = 1$. 
We consider three versions of $w$ and $f$ shown in Figure \ref{fig:simA-wf}. We included the seasonal trend $h(t) = 0.5 ~\mathrm{sin}(t/150)$. 

Table \ref{tab:simA-f} reports the root mean square error (RMSE), 95\% confidence interval coverage (Cvg) and average confidence interval widths (Width) for $w$ and $f$ evaluated on grids of evenly spaced values over 5,000 replicates.
In all scenarios, RMSEs and average widths decrease as the sample size increases, indicating improved estimation accuracy. 
The empirical coverages are close to the nominal level of 95\%. 
The results for index weights $\boldsymbol{\alpha}$ are presented in Table \ref{tab:simA-alpha}: the bias decreases with increasing sample size, and the empirical coverage of the confidence intervals is close to the nominal level.

\begin{table}[htbp]
    \centering
    \caption{Results of Simulation A. The RMSE, 95\% confidence intervals coverage (Cvg) and average widths (Width) for $w$ and $f$ with the sample size $n = 1,000$ and $2,000$ are reported.}
    \label{tab:simA-f}
    \begin{tabular}{@{}cccccccccc@{}}
\toprule
     & & & \multicolumn{3}{c}{$w$}                 &  & \multicolumn{3}{c}{$f$}                 \\ \cmidrule(lr){4-6} \cmidrule(l){8-10} 
Type of $w$ & Type of $f$ & $n$ & RMSE        & Cvg        & Width        &  & RMSE        & Cvg        & Width        \\ \midrule
Type i & Type i  & 1000  & 0.030       & 0.928      & 0.085      &  & 0.179       & 0.929      & 0.573      \\
            &          & 2000  & 0.013       & 0.969      & 0.057      &  & 0.100       & 0.937      & 0.322      \\
            & Type ii & 1000  & 0.018       & 0.951      & 0.061      &  & 0.119       & 0.943      & 0.477      \\
            &               & 2000  & 0.015       & 0.962      & 0.048      &  & 0.081       & 0.949      & 0.275      \\
            & Type iii & 1000  & 0.024       & 0.927      & 0.090      &  & 0.032       & 0.928      & 0.119      \\
            &               & 2000  & 0.017       & 0.963      & 0.073      &  & 0.021       & 0.935      & 0.083      \\[0.5em]
Type ii & Type i  & 1000  & 0.017       & 0.957      & 0.071      &  & 0.110       & 0.948      & 0.393      \\
            &               & 2000  & 0.014       & 0.964      & 0.056      &  & 0.090       & 0.939      & 0.299      \\
            & Type ii & 1000  & 0.023       & 0.938      & 0.059      &  & 0.126       & 0.935      & 0.384      \\
            &               & 2000  & 0.012       & 0.968      & 0.047      &  & 0.061       & 0.952      & 0.215      \\
            & Type iii & 1000  & 0.023       & 0.929      & 0.088      &  & 0.030       & 0.949      & 0.114      \\
            &               & 2000  & 0.017       & 0.955      & 0.071      &  & 0.021       & 0.944      & 0.080      \\[0.5em]
Type iii  & Type i & 1000  & 0.017       & 0.963      & 0.059      &  & 0.131       & 0.941      & 0.453      \\
            &               & 2000  & 0.009       & 0.977      & 0.044      &  & 0.093       & 0.933      & 0.304      \\
            & Type ii & 1000  & 0.011       & 0.971      & 0.049      &  & 0.091       & 0.948      & 0.437      \\
            &               & 2000  & 0.009       & 0.974      & 0.039      &  & 0.061       & 0.948      & 0.228      \\
            & Type iii & 1000  & 0.018       & 0.965      & 0.077      &  & 0.029       & 0.947      & 0.118      \\
            &               & 2000  & 0.013       & 0.975      & 0.059      &  & 0.021       & 0.937      & 0.084      \\ \bottomrule
\end{tabular}
\end{table}

\begin{table}[htbp]
\centering
\caption{Results of Simulation A. Three weight functions and three association functions are considered. 
The bias and 95\% confidence interval coverage (Cvg) for index weights $\boldsymbol{\alpha}$ with sample sizes $n = 1,000$ and $2,000$ are reported.}
\label{tab:simA-alpha}
\begin{tabular}{@{}cccccccccccc@{}}
\toprule
          &            &      & \multicolumn{2}{c}{PM$_{2.5}$ (0.86)} &  & \multicolumn{2}{c}{O$_3$ (0.43)} &  & \multicolumn{2}{c}{NO$_2$ (0.26)} \\ 
\cmidrule(lr){4-5} \cmidrule(lr){7-8} \cmidrule(l){10-11} 
$w$         & $f$          & $n$    & Bias    & Cvg   &  & Bias   & Cvg   &  & Bias    & Cvg   \\ 
\midrule
Type (i)  & Type (i)   & 1000 & 0.045 & 0.882 &  & 0.021 & 0.899 &  & 0.011 & 0.892 \\
          &            & 2000 & 0.009 & 0.928 &  & 0.003 & 0.930 &  & 0.001 & 0.948 \\
          & Type (ii)  & 1000 & 0.015 & 0.947 &  & 0.006 & 0.932 &  & 0.000 & 0.937 \\
          &            & 2000 & 0.010 & 0.950 &  & 0.008 & 0.931 &  & 0.000 & 0.953 \\
          & Type (iii) & 1000 & 0.002 & 0.959 &  & 0.002 & 0.956 &  & 0.014 & 0.960 \\
          &            & 2000 & 0.002 & 0.961 &  & 0.000 & 0.965 &  & 0.001 & 0.970 \\
          \\
Type (ii) & Type (i)   & 1000 & 0.009 & 0.952 &  & 0.005 & 0.954 &  & 0.005 & 0.953 \\
          &            & 2000 & 0.007 & 0.945 &  & 0.004 & 0.943 &  & 0.000 & 0.960 \\
          & Type (ii)  & 1000 & 0.006 & 0.943 &  & 0.005 & 0.933 &  & 0.011 & 0.926 \\
          &            & 2000 & 0.002 & 0.944 &  & 0.003 & 0.944 &  & 0.001 & 0.945 \\
          & Type (iii) & 1000 & 0.002 & 0.967 &  & 0.002 & 0.959 &  & 0.012 & 0.974 \\
          &            & 2000 & 0.004 & 0.961 &  & 0.002 & 0.961 &  & 0.001 & 0.970 \\
          \\
Type (iii)& Type (i)   & 1000 & 0.018 & 0.939 &  & 0.012 & 0.943 &  & 0.007 & 0.953 \\
          &            & 2000 & 0.003 & 0.952 &  & 0.002 & 0.958 &  & 0.002 & 0.961 \\
          & Type (ii)  & 1000 & 0.009 & 0.951 &  & 0.002 & 0.950 &  & 0.005 & 0.947 \\
          &            & 2000 & 0.002 & 0.949 &  & 0.000 & 0.947 &  & 0.000 & 0.954 \\
          & Type (iii) & 1000 & 0.004 & 0.966 &  & 0.000 & 0.961 &  & 0.010 & 0.973 \\
          &            & 2000 & 0.005 & 0.958 &  & 0.003 & 0.962 &  & 0.001 & 0.961 \\
\bottomrule
\end{tabular}
\end{table}

\clearpage
\subsection{Simulation B: Single-Index DRF-DLNM}
\label{ss:simB}

\begin{figure}[htbp]
    \centering
    \begin{subfigure}{0.32\textwidth}
        \includegraphics[width=\linewidth]{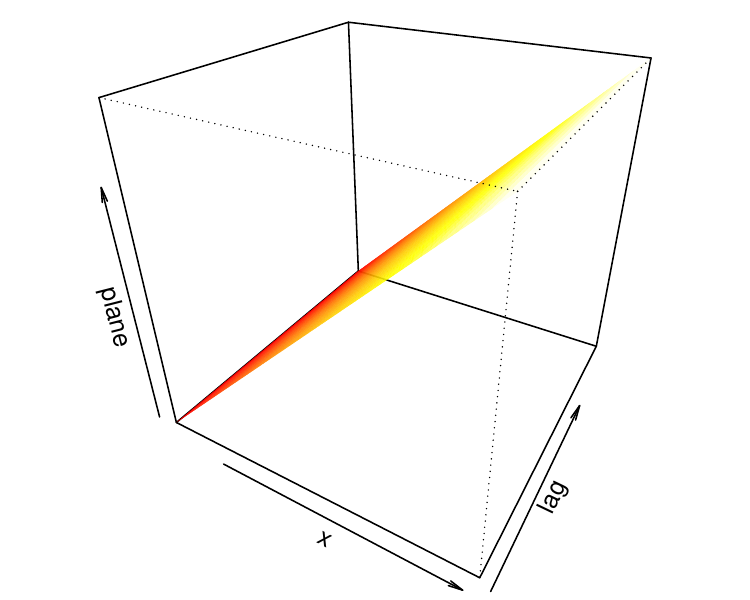}
        \caption{Scenario 1 ($\psi$ plane)}
    \end{subfigure}
    \begin{subfigure}{0.32\textwidth}
        \includegraphics[width=\linewidth]{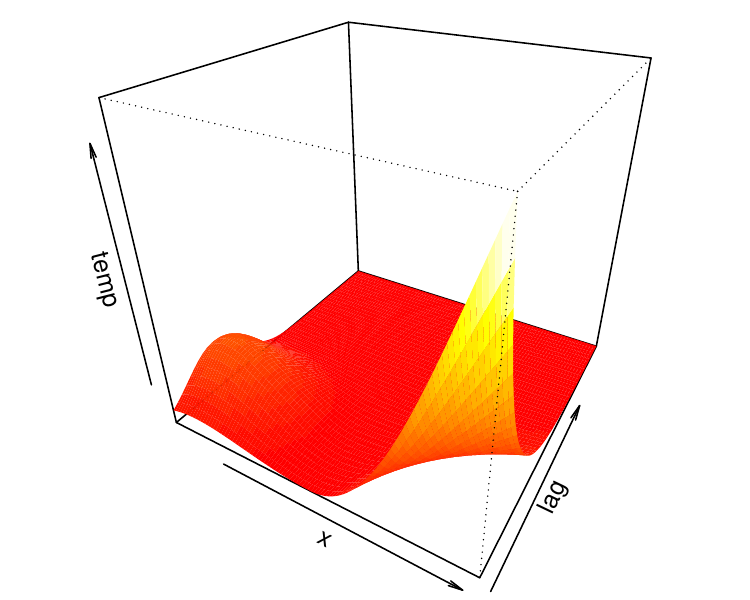}
        \caption{Scenario 2 ($\psi$ temperature)}
    \end{subfigure}
    \begin{subfigure}{0.32\textwidth}
        \includegraphics[width=\linewidth]{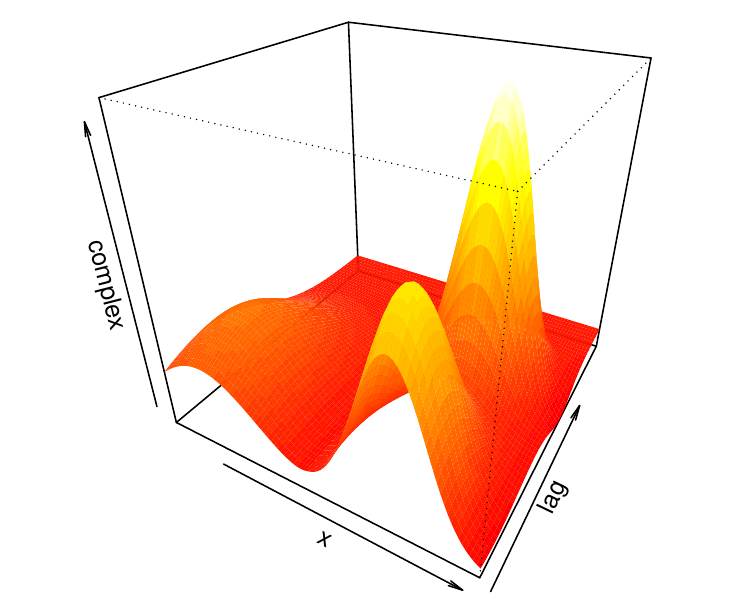}
        \caption{Scenario 3 ($\psi$ complex)}
    \end{subfigure}

    \caption{The true functions $\psi$ used in Simulation B.}
    \label{fig:simB-psi}
\end{figure}

In this simulation, we evaluate the performance of the proposed single-index DRF-DLNM under different functions $\psi$. 
This simulation uses the same exposure data as in Simulation A, and 
generates $Y_t$ from a single-index DRF-DLNM: $\log(\mu_t) = \int_{0}^{15} \psi\left\{E(t-l; \boldsymbol{\alpha}), l\right\} dl + h(t)$, where $\boldsymbol{\alpha}$, $h(t)$ and $\theta$ are the same as in Simulation A. We consider three $\psi$ functions, with shapes following \citet{gasparrini2017penalized}: plane $\psi$ (Scenario (i)), temperature $\psi$ (Scenario (ii)) and complex $\psi$ (Scenario (iii)); visualized in Figure \ref{fig:simB-psi}.

Table \ref{tab:simB} reports the root mean square error (RMSE), 95\% confidence intervals coverage (Cvg) and average confidence interval width (Width) for the $\psi$ functions evaluated on evenly spaced grids, as well as the bias and coverage for the index weights $\boldsymbol{\alpha}$, across 5,000 replicates. %
For the surface $\psi$, we observe decreases in RMSE and average widths of the confidence intervals as the sample size increases; and empirical coverages are close to the nominal level, with slightly lower coverage in some scenarios, a pattern also observed for the single-exposure DRF-DLNM \citep{gasparrini2017penalized}. 
For the index weights, the biases are small and the empirical coverages are close to 95\%. 
Overall, these results indicate that the proposed framework provides accurate estimation and reliable uncertainty quantification.

\begin{table}[htbp]
  \caption{Results of Simulation B: performance of the single-index DRF-DLNM. The RMSE, 95\% confidence intervals coverage (Cvg) and average widths (Width) for $\psi$, and bias and coverage for index weights $\boldsymbol{\alpha}$, with sample size $n = 1,000$ and $2,000$, are reported.}
  \label{tab:simB}
\centering
\resizebox{\textwidth}{!}{
\begin{tabular}{@{}ccccccccccccccc@{}}
\toprule
&      & \multicolumn{3}{c}{$\psi$} 
& & \multicolumn{2}{c}{PM$_{2.5}$ (0.86)} 
& & \multicolumn{2}{c}{O$_3$ (0.43)} 
& & \multicolumn{2}{c}{NO$_2$ (0.26)} \\ 
\cmidrule(lr){3-5}
\cmidrule(lr){7-8}
\cmidrule(lr){10-11}
\cmidrule(l){13-14}
 
& $n$
& RMSE   & Cvg   & Width 
&  & Bias    & Cvg   
&  & Bias   & Cvg   
&  & Bias    & Cvg   \\ 
\midrule
Scenario (i) 
& 1000 & 0.010 & 0.958 & (0.036)
& & 0.005 & 0.960
& & 0.002 & 0.958
& & 0.004 & 0.972 \\
& 2000 & 0.008 & 0.962 & (0.025)
& & 0.004 & 0.970
& & 0.002 & 0.972
& & 0.002 & 0.981 \\
\\
Scenario (ii)
& 1000 & 0.057 & 0.973 & (0.203)
& & 0.000 & 0.962
& & 0.005 & 0.972
& & 0.001 & 0.955 \\
& 2000 & 0.050 & 0.968 & (0.168)
& & 0.003 & 0.954
& & 0.001 & 0.954
& & 0.001 & 0.951 \\
\\
Scenario (ii)
& 1000 & 0.094 & 0.923 & (0.276)
& & 0.002 & 0.960
& & 0.003 & 0.971
& & 0.018 & 0.895 \\
& 2000 & 0.064 & 0.932 & (0.195)
& & 0.003 & 0.944
& & 0.004 & 0.942
& & 0.002 & 0.938 \\
\bottomrule
\end{tabular}
}
\end{table}

\subsection{Simulation C: Model Selection and Stacking}

\begin{figure}[H]
  \centering
  \begin{subfigure}[b]{0.49\linewidth}
    \includegraphics[width=\linewidth]{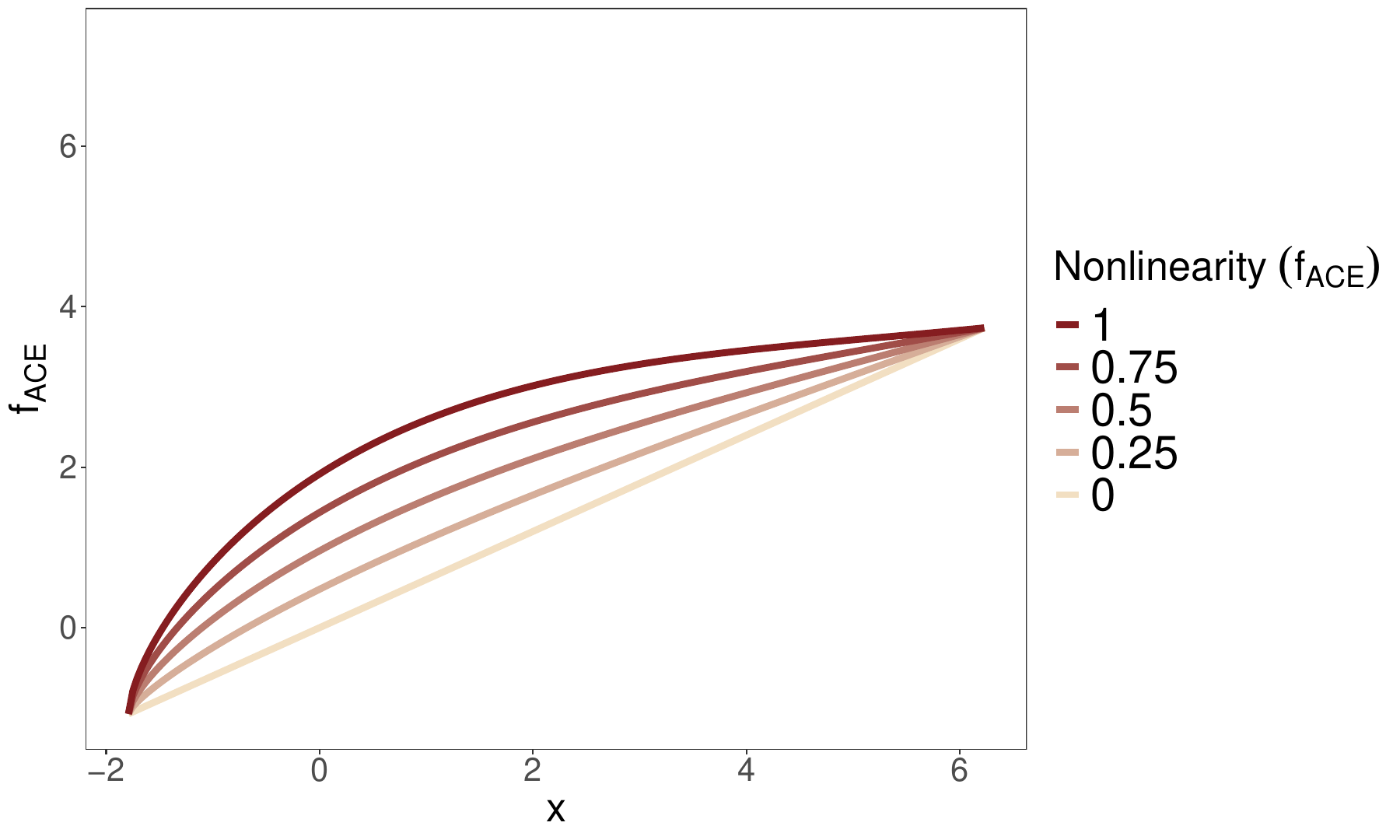}
    \caption{$f_{\text{ACE}}$}
  \end{subfigure}
  \begin{subfigure}[b]{0.49\linewidth}
    \includegraphics[width=\linewidth]{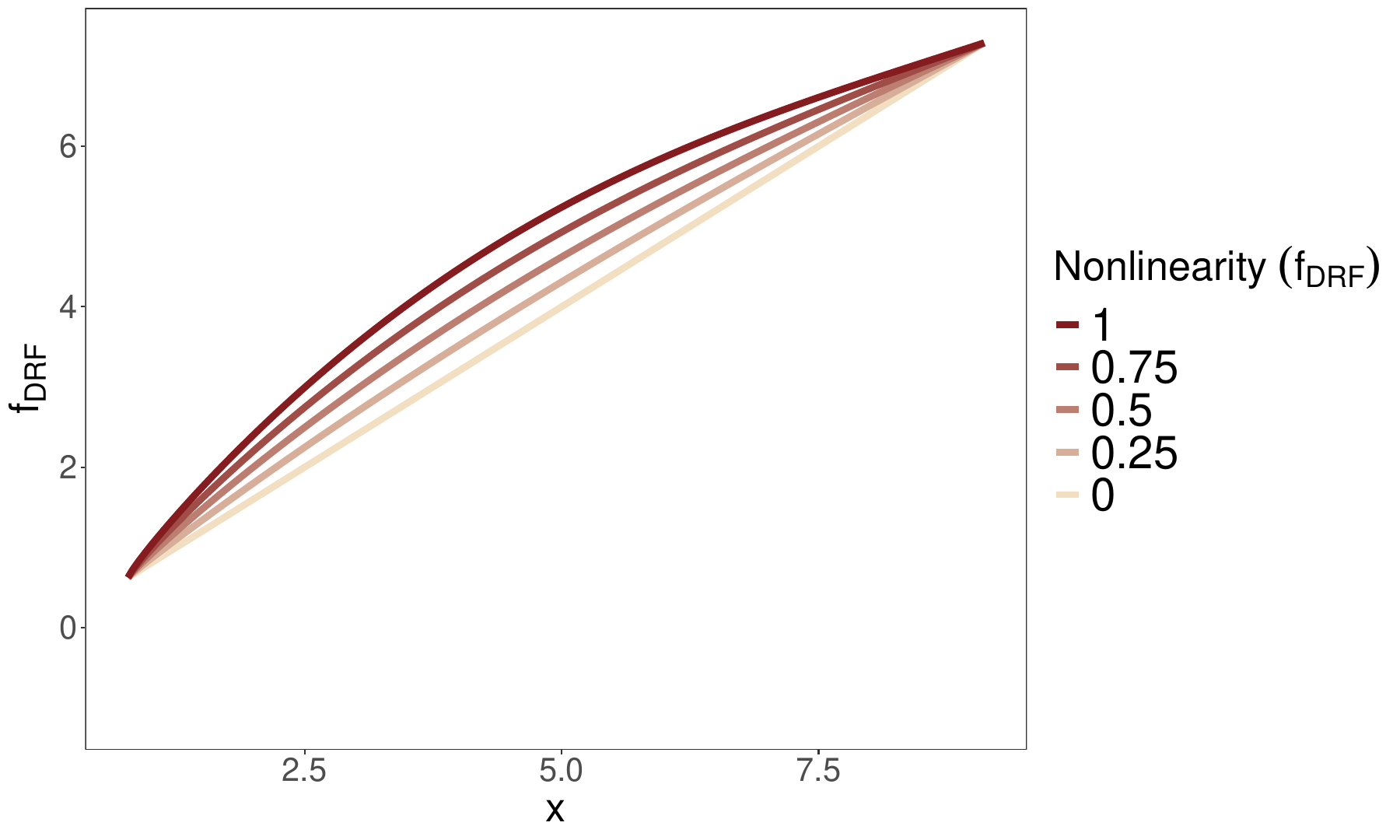}
    \caption{$f_{\text{DRF}}$}
  \end{subfigure}
  \caption{Shapes of $f_{\text{ACE}}$ and $f_{\text{DRF}}$ in Simulation C.}
\end{figure}

\begin{figure}[H]
    \centering
    \includegraphics[width=\linewidth]{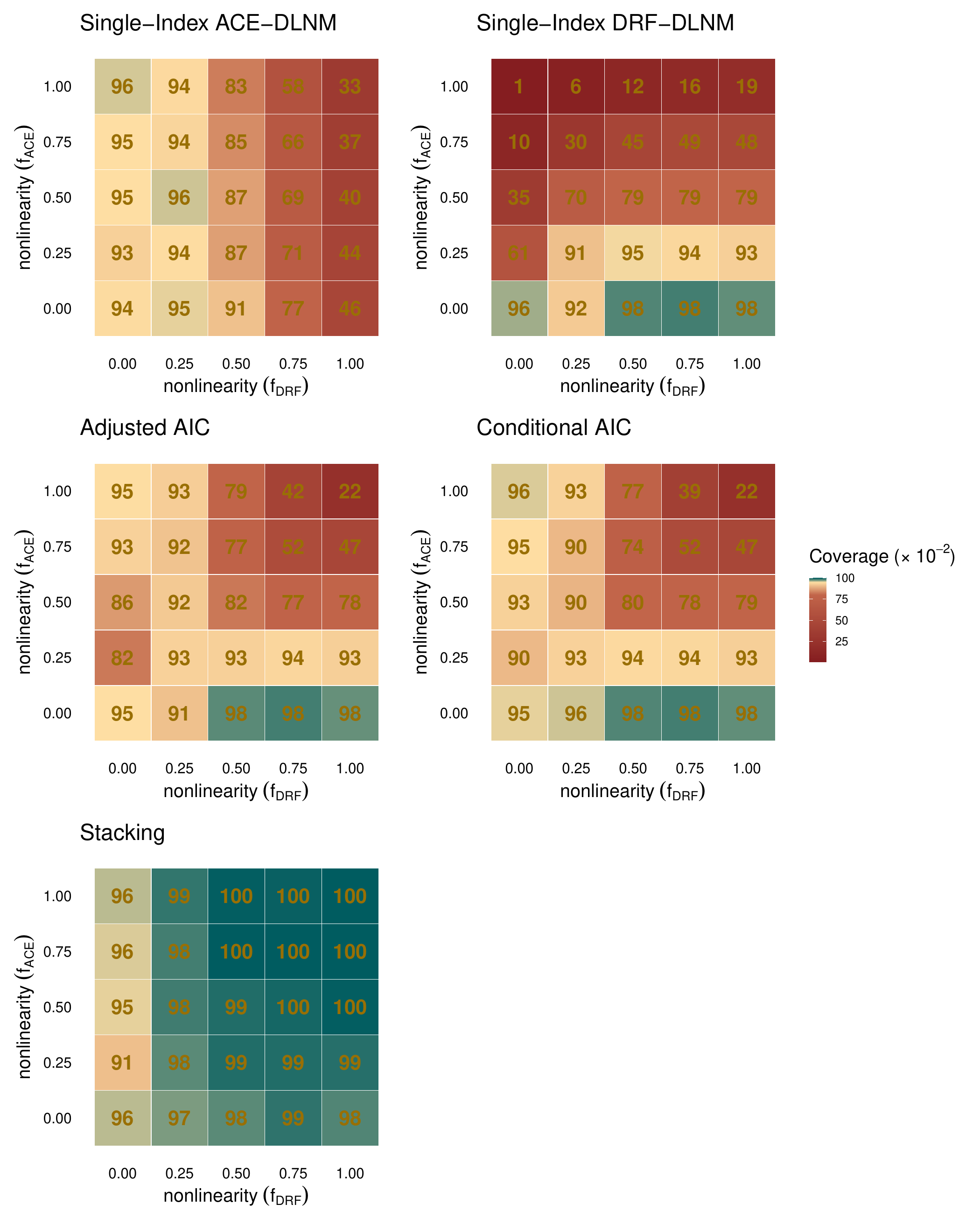}
    \caption{Results of Simulation C: performance of the model selection and stacking approaches. The Coverage of the quantity of interest $\Delta(10\%)$ are reported for different approaches. The nonlinearity of $f_{\text{ACE}}$ and $f_{\text{DRF}}$ are varied in the rows and columns.}
\end{figure}

\begin{figure}[H]
    \centering
    \includegraphics[width=\linewidth]{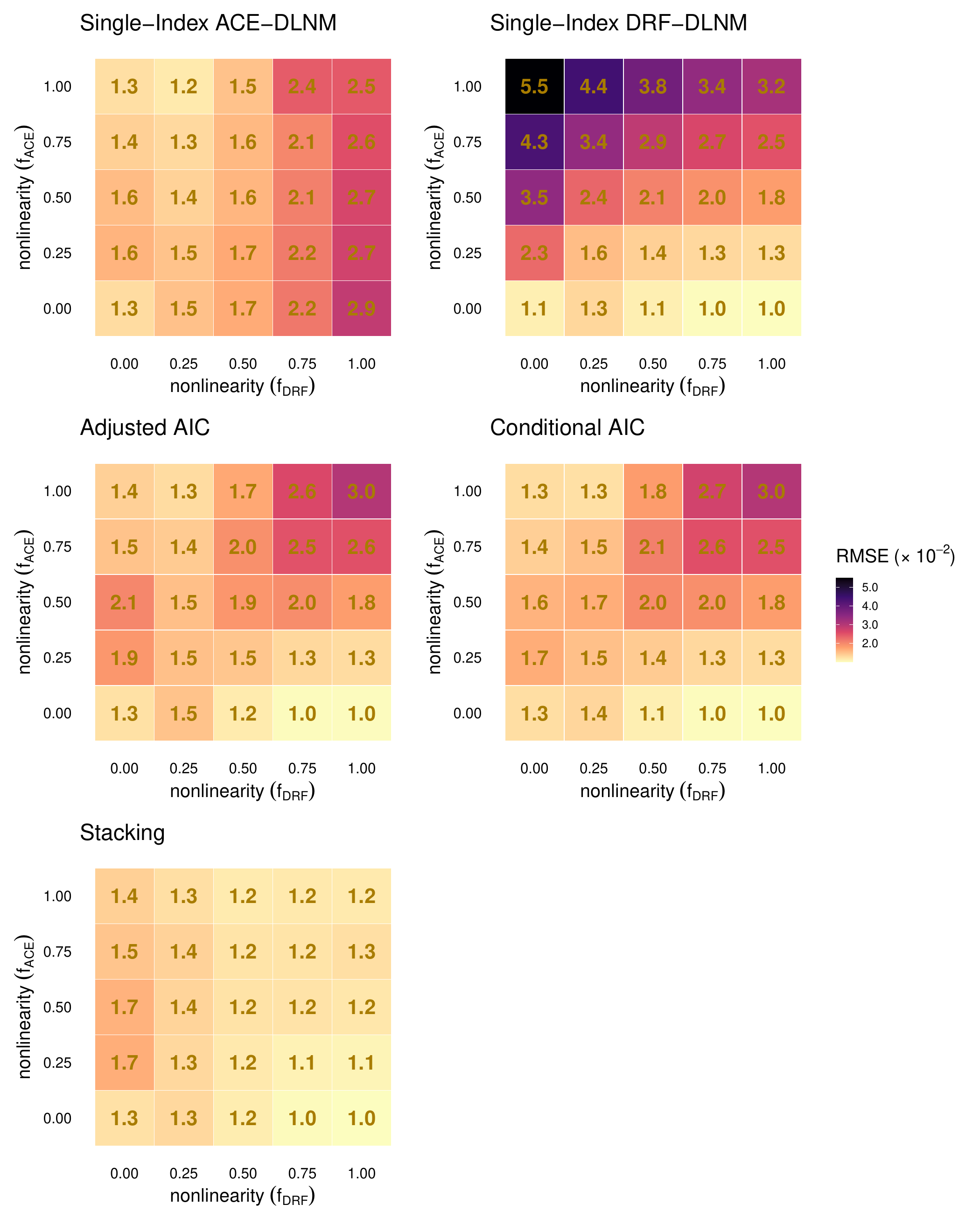}
    \caption{Results of Simulation C: performance of the model selection and stacking approaches. The RMSE of the quantity of interest $\Delta(10\%)$ are reported for different approaches. The nonlinearity of $f_{\text{ACE}}$ and $f_{\text{DRF}}$ are varied in the rows and columns.}
\end{figure}

\begin{figure}[H]
    \centering
    \includegraphics[width=\linewidth]{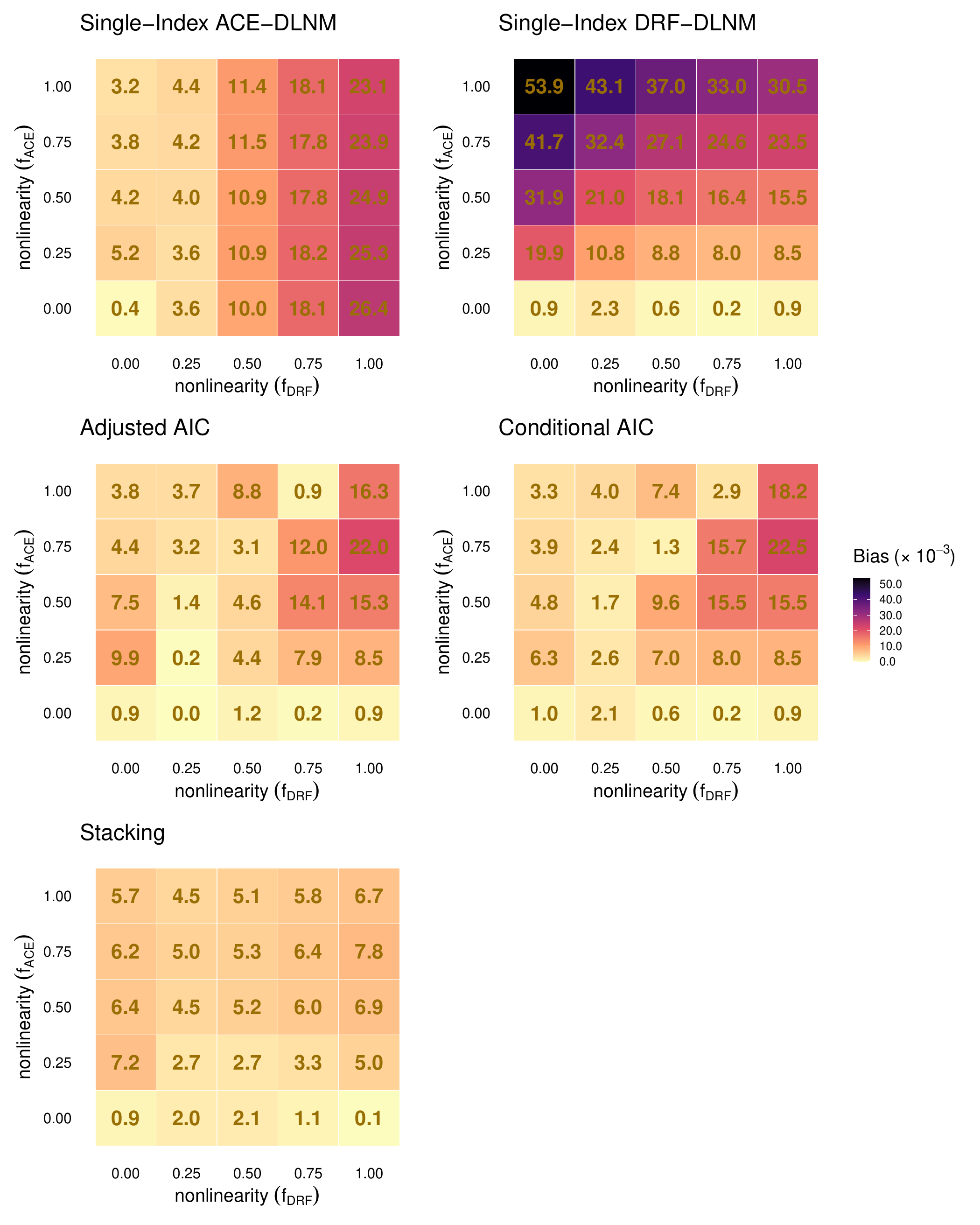}
    \caption{Results of Simulation C: performance of the model selection and stacking approaches. The Bias of the quantity of interest $\Delta(10\%)$ are reported for different approaches. The nonlinearity of $f_{\text{ACE}}$ and $f_{\text{DRF}}$ are varied in the rows and columns.}
\end{figure}

\subsection{Supplementary Simulation}
\begin{figure}[H]
  \centering
  \begin{subfigure}[b]{0.45\linewidth}
    \includegraphics[width=\linewidth]{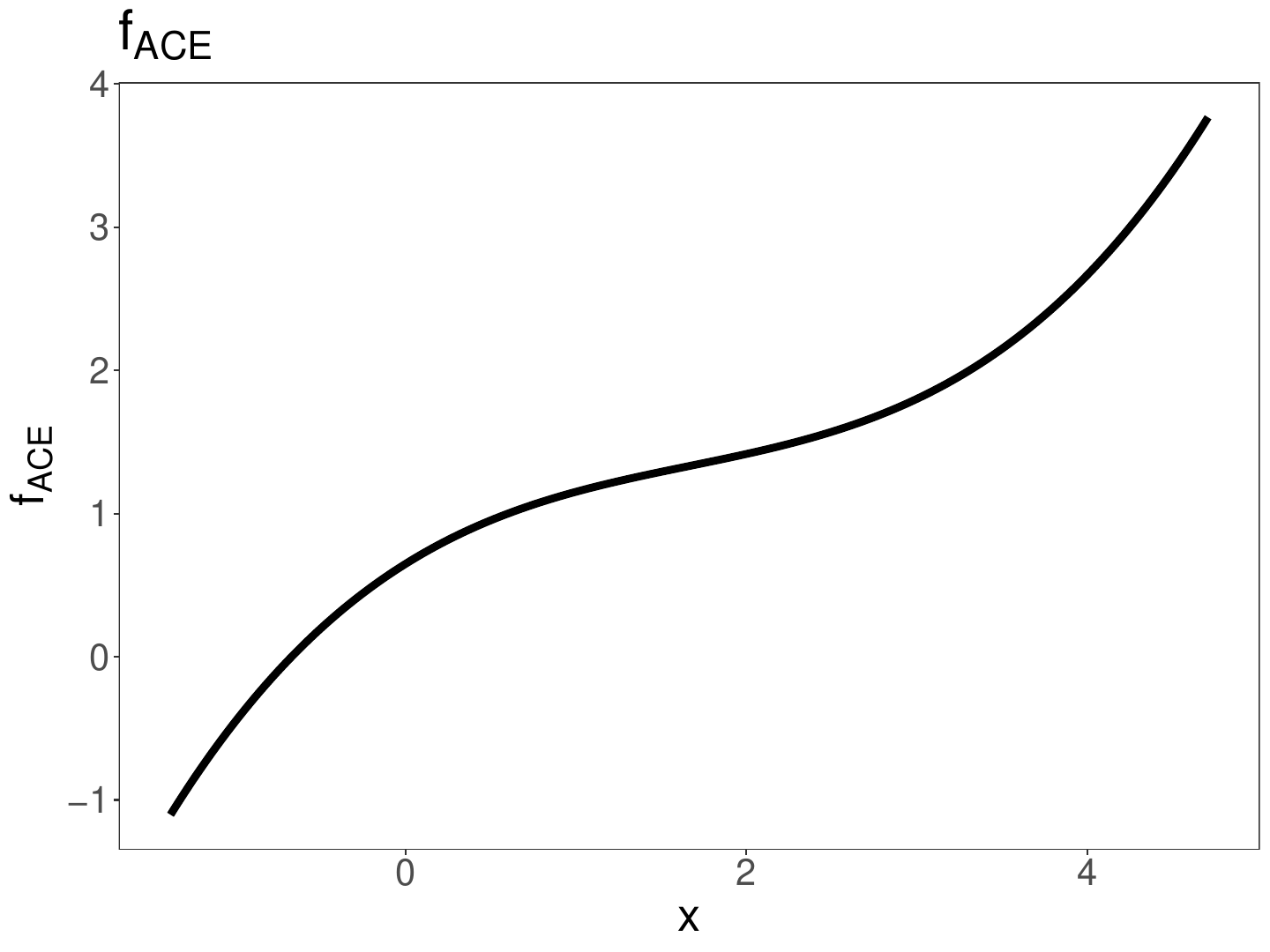}
    \caption{$f_{\text{ACE}}$}
  \end{subfigure}
  \hspace{0.05\linewidth}
  \begin{subfigure}[b]{0.45\linewidth}
    \includegraphics[width=\linewidth]{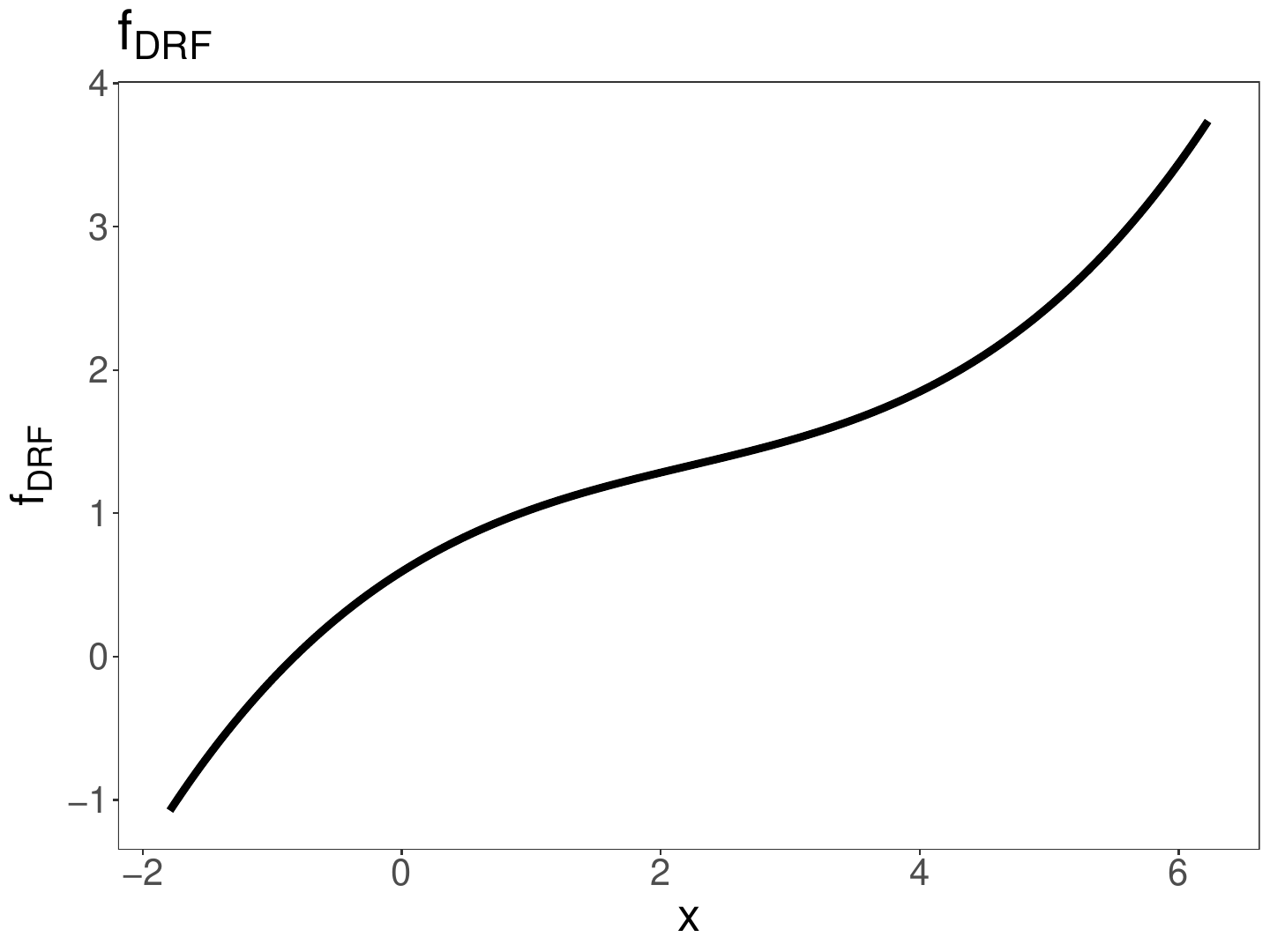}
    \caption{$f_{\text{DRF}}$}
  \end{subfigure}
  \caption{Shapes of $f_{\text{ACE}}$ and $f_{\text{DRF}}$ in Supplementary Simulation. }
  \label{fig:supp-stacking-f}
\end{figure}

We conduct a simulation study to investigate the performance of AIC and stacking when the single-index ACE-DLNM and single-index DRF-DLNM are biased in the same direction. We generate data from the same model as Simulation C in Section 6.3, but specify $f_{\text{ACE}}$ and $f_{\text{DRF}}$ as cubic functions shown in Figure \ref{fig:supp-stacking-f}. 

\begin{figure}[H]
    \centering
    \includegraphics[width=\linewidth]{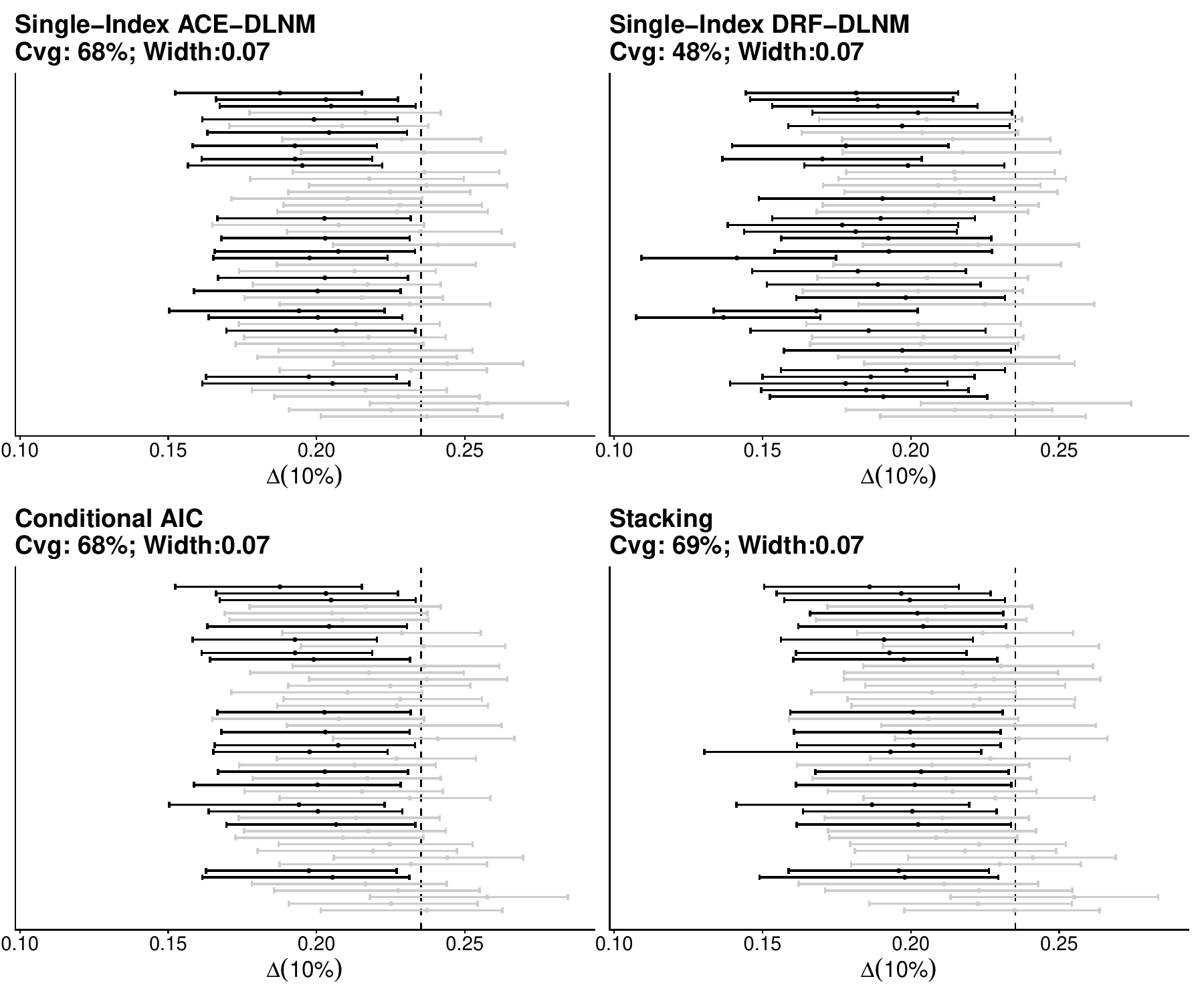}
    \caption{The first 50 simulation results in each setting. The true values are shown as vertical dashed lines, and intervals that do not cover the true value are highlighted. The figure caption reports the empirical coverage and the average width of the confidence intervals across the 1,000 simulation replicates.}
    \label{fig:supp-stacking-cvg}
\end{figure}

We estimate the quantity of interest as in Simulation C, using the single-index ACE-DLNM, the single-index DRF-DLNM and model selection and stacking. The single-index ACE- and DRF-DLNM lead to bias of 0.02 and 0.04. The AIC and stacking perform similarly as the best fitting model with bias of 0.02 and 0.02. 
Figure \ref{fig:supp-stacking-cvg} shows the confidence intervals from the first 50 simulation results. 
The true values are shown as vertical dashed lines, and intervals that do not cover the true value are highlighted. We also indicate the empirical coverage and the average width of the confidence intervals across the 1,000 replicates in the figure caption. We observe that both model selection and stacking yield estimates close to the best fitting candidate model.

\section{Application Results}

\subsection{Model Estimates}
We present the estimates of index weights and functions for exposure-response surfaces for the four models. 
For single-index ACE-DLNM, the association $f$ centred at the estimated cumulative index corresponding to all exposures are at their 25\% quantiles. 
For additive ACE-DLNM, each association $f_m$ centred at the estimated adaptive cumulative exposures corresponding to the exposure at its 25\% quantile. 
For single-index DRF-DLNM, we report the point estimates and confidence intervals for the bivariate surface $\psi$ and selected slices. Specifically, we show the univariate curves $\psi(\text{Index}=Q^E_2,\text{Lag})$ and $\psi(\text{Index}=Q^E_3,\text{Lag})$ representing the relative importance of lags under some fixed index, where the index are fixed at $Q^E_2$ and $Q^E_3$ that denote the estimated index values when exposures are at their 50\% and 75\% quantiles, respectively; and we show the curves $\psi(\text{Index},\text{Lag}=0)$, $\psi(\text{Index},\text{Lag}=3)$, and $\psi(\text{Index},\text{Lag}=15)$ representing the association between response and the index at the specific lag. For additive DRF-DLNM, we present the surfaces and slices in the same manner for each exposure separately.

The index weights measure the relative contributions of individual pollutants to the overall effects. 
The two single-index DLNMs yield broadly similar estimates for respiratory mortality: NO$_2$ has the largest index weight, followed by O$_3$, while PM$_{2.5}$ has non-significant weights. 
For all-cause mortality, a similar pattern is observed for the single-index ACE-DLNM, while the single-index DRF-DLNM yields a negative PM$_{2.5}$ weight and a non-significant O$_3$ weight. The AIC prefers the single-index ACE-DLNM.  
The results suggest that NO$_2$ contributes most to the joint effects of air pollution on mortality in Ontario, and O$_3$ also has a relatively large contribution, particularly to respiratory mortality, whereas PM$_{2.5}$ tends to have null effects. 
The lag weights from the single-index ACE-DLNM suggest that the importance of lagged exposures is generally constant over the two-week lag period for respiratory and all-cause mortality. The patterns of lag importance from single-index DRF-DLNM can be illustrated from the slices of $\psi$ at fixed index: we show lag patterns at estimated index when exposures at their 25\% and 75\% quantiles, and observe slightly decreasing trends for respiratory mortality. 
The results highlight the need to consider lagged effects when assessing associations between air pollution and mortality. 
These models exhibit non-linearity in association functions, demonstrating the importance of allowing non-linear associations in this analysis; however, these functions are based on distinct reparameterizations with different interpretations. 
The association function $f$ in single-index ACE-DLNM can be interpreted as the log rate ratio of the cumulative index relative to the specific cumulative index when all exposures are at their 25\% quantiles. In the additive ACE-DLNM, each association function $f_m$ represents the log rate ratio for the adaptive cumulative exposure relative to the value when that exposure is set at its 25\% quantile. 
The sliced $\psi$ can be interpreted as the log rate ratio of the index at a specific lag relative to the specific index when exposures are at their 25\% quantiles. The additive DRF-DLNM has analogous interpretations for each individual exposure rather than for the index.

\subsubsection{Respiratory Mortality}

\begin{figure}[H]

    \centering
    \includegraphics[width=\linewidth]{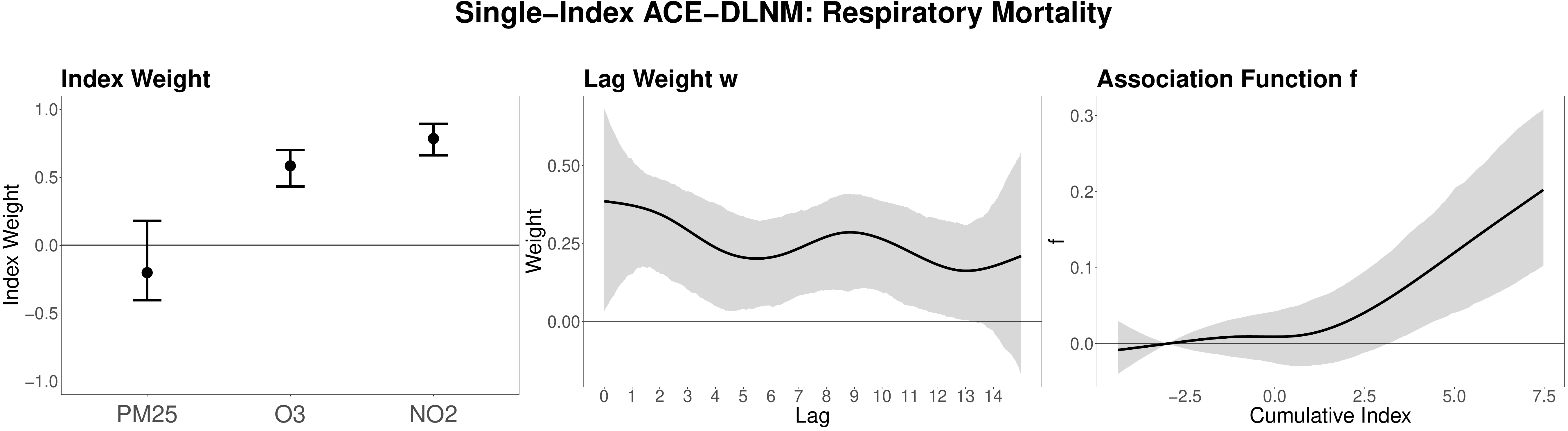}    
    \caption{Estimated index weights and functions from the single-index ACE-DLNM for respiratory mortality. }
\end{figure}

\begin{figure}[H]
    \centering
    \includegraphics[width=\linewidth]{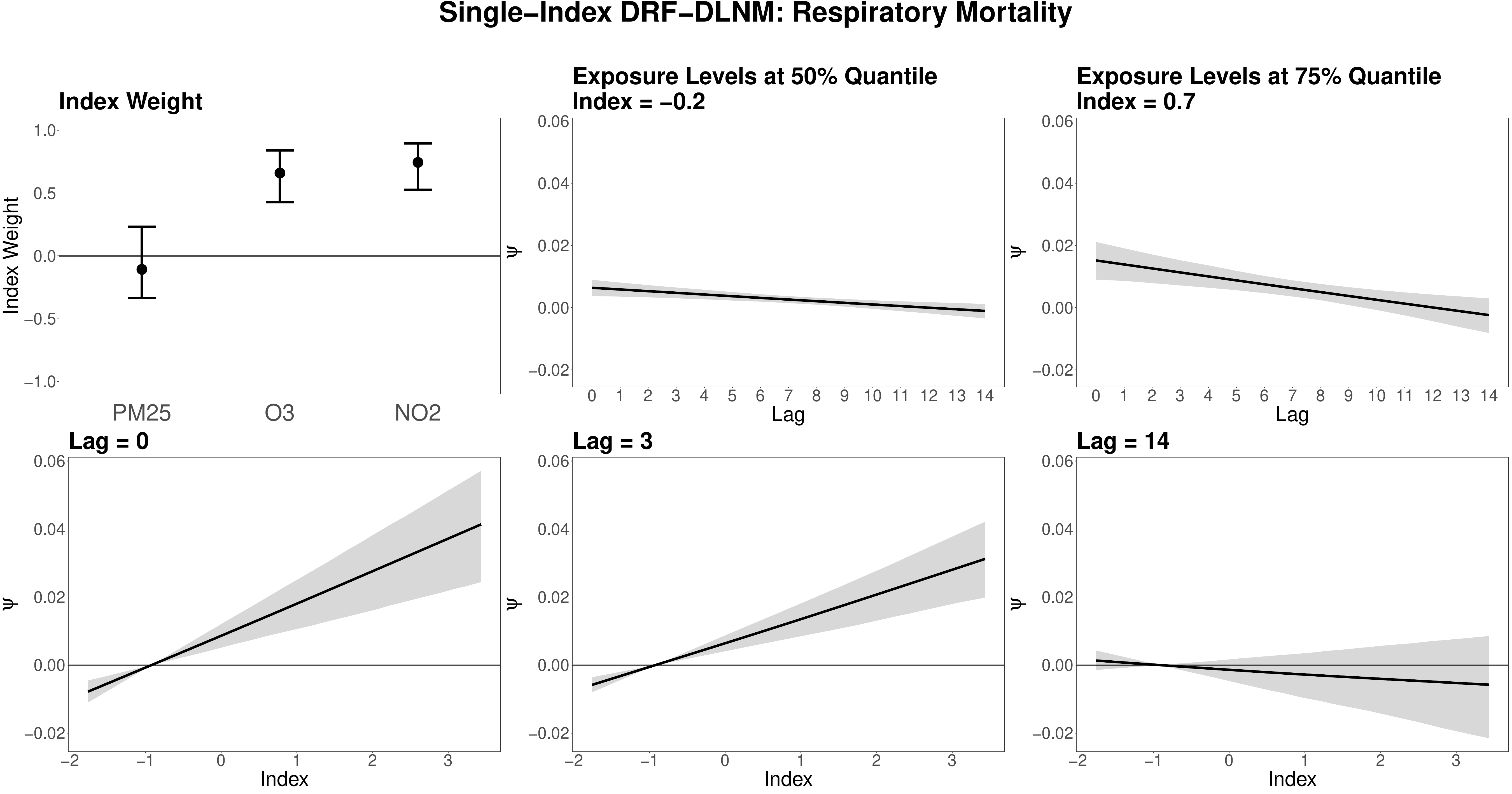}
    \caption{Estimated index weights and functions from the single-index DRF-DLNM for respiratory mortality. The function $\psi$ is shown in selected slices of the bivariate surface $\psi$, by fixing the index at estimated values under the exposures are at their 25\% and 75\% quantiles, and fixing the lag at 0, 3, and 14. }
\end{figure}

\begin{figure}[H]
  \centering
  \makebox[\textwidth][c]{
  \begin{subfigure}[t]{0.33\textwidth}
    \centering
    \includegraphics[width=\linewidth]{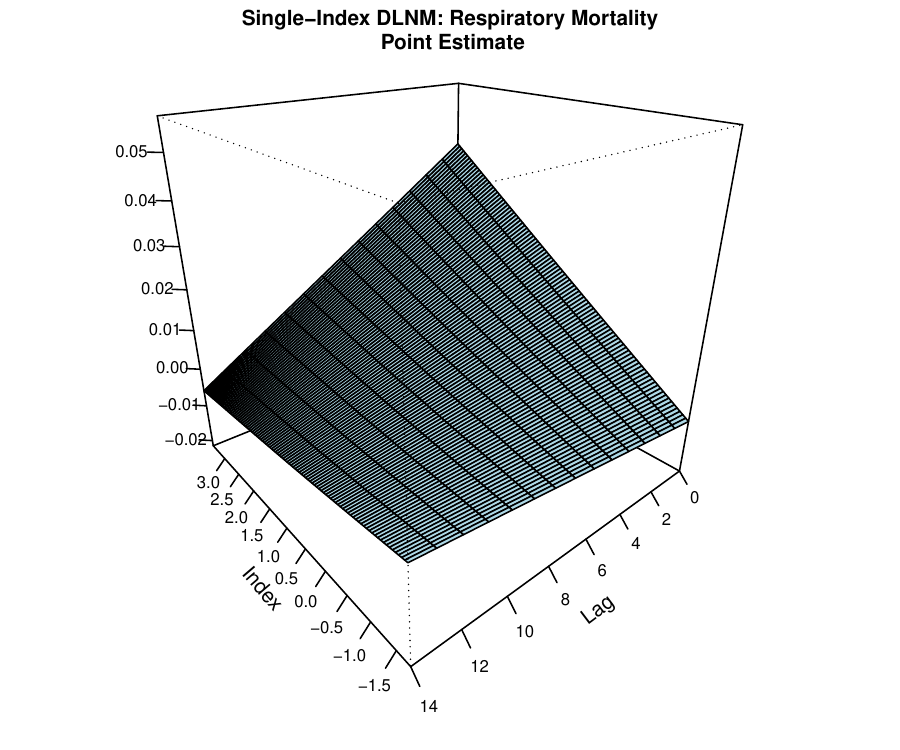}
  \end{subfigure}
  }
  \vspace{0.5em}
  
  \begin{subfigure}[t]{0.33\textwidth}
    \centering
    \includegraphics[width=\linewidth]{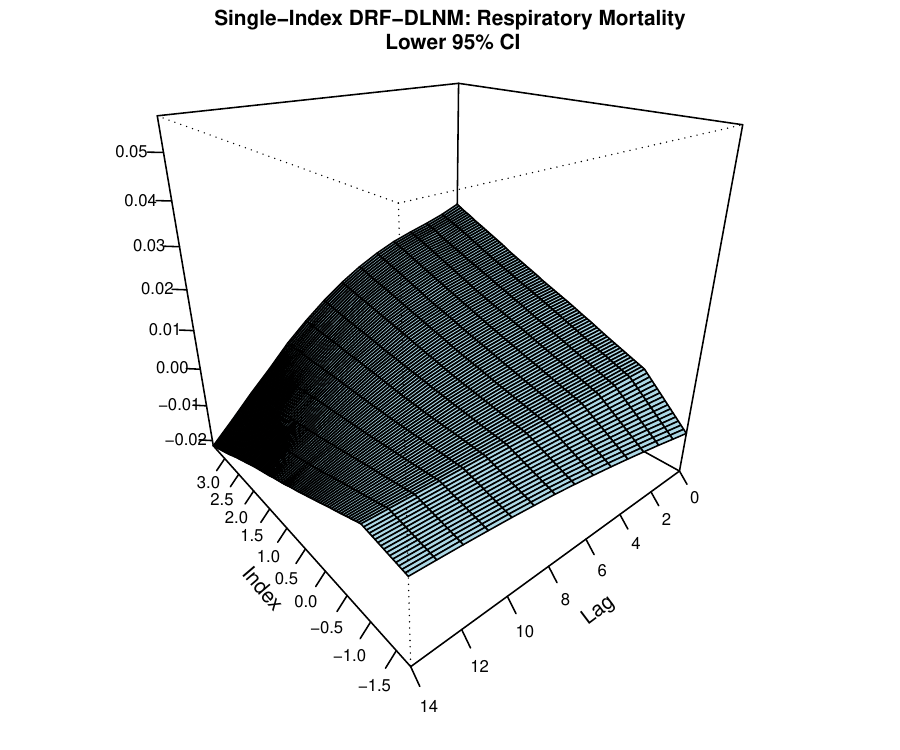}
  \end{subfigure}
  \begin{subfigure}[t]{0.33\textwidth}
    \centering
    \includegraphics[width=\linewidth]{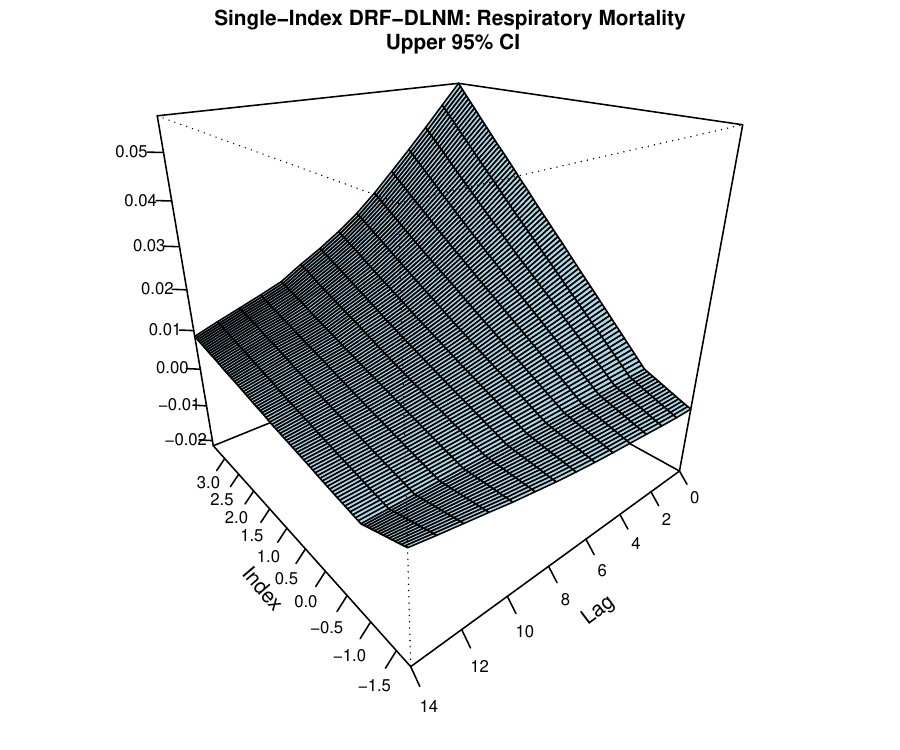}
  \end{subfigure}
  \caption{Estimated $\psi$ from the single-index DRF-DLNM for respiratory mortality.}
\end{figure}

\begin{figure}[H]
    \centering
    \includegraphics[width=\linewidth]{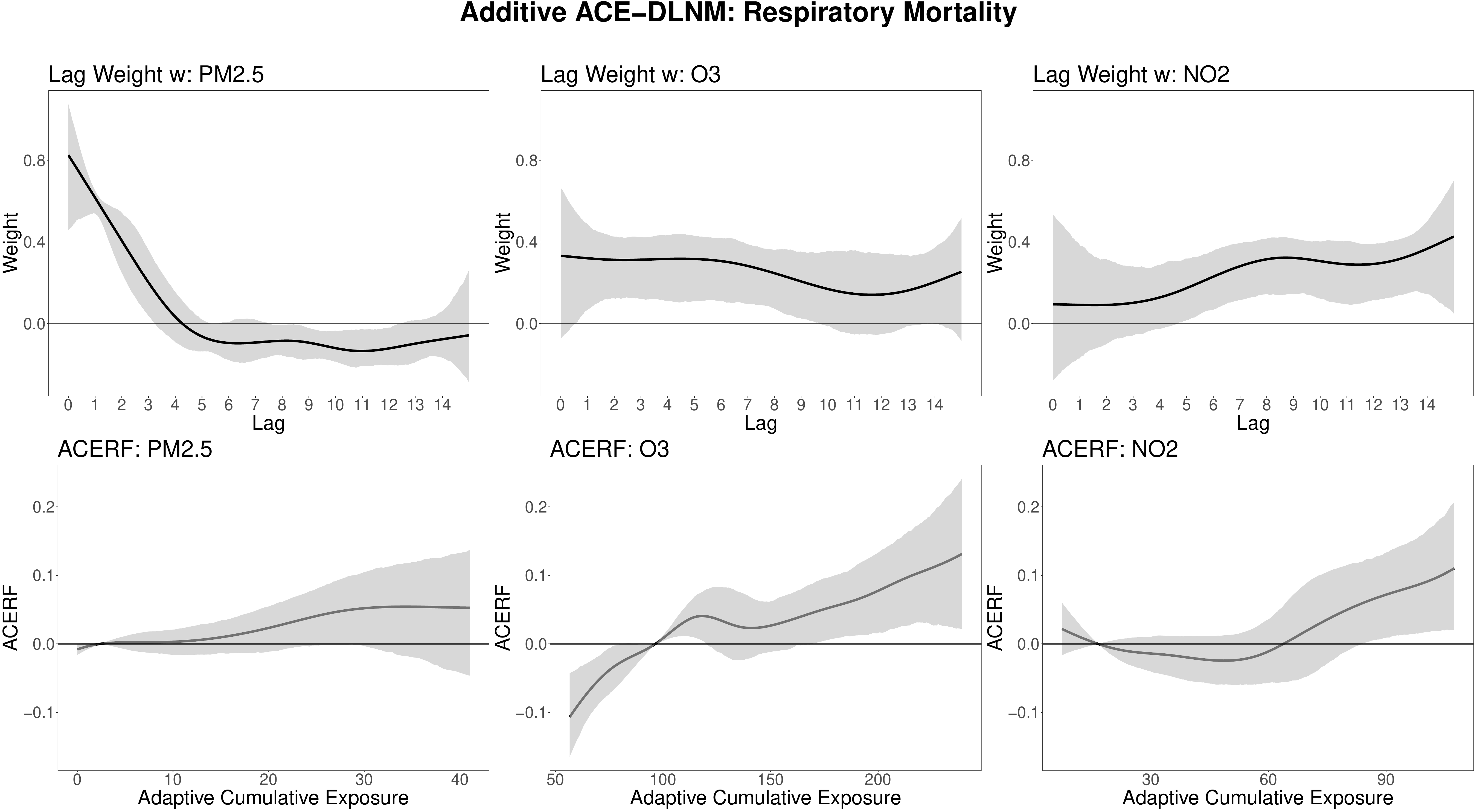}
    \caption{Estimated index weights and functions from the additive ACE-DLNM for respiratory mortality.}
\end{figure}

\begin{figure}[H]
    \centering
    \includegraphics[width=\linewidth]{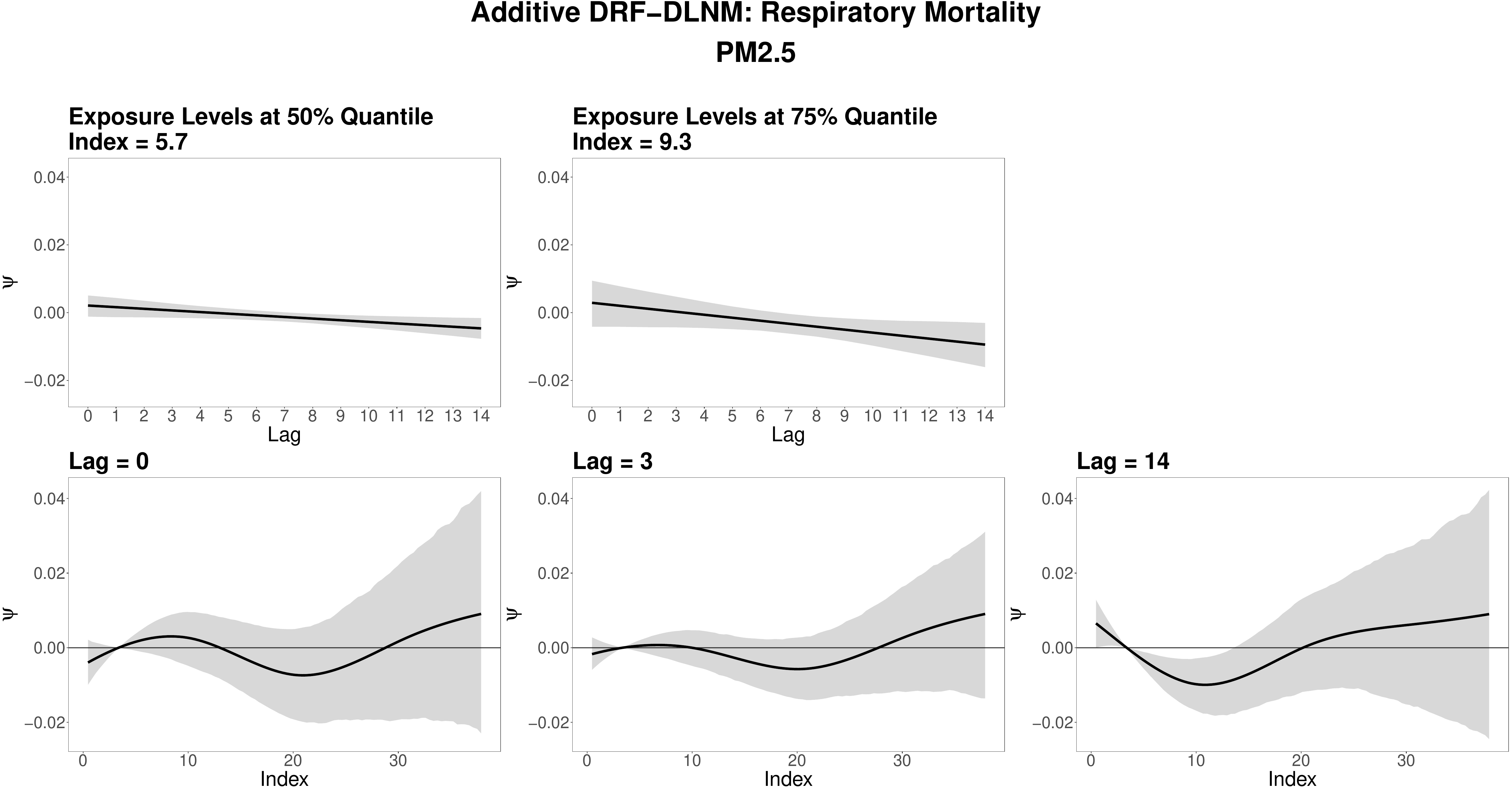}
    \caption{Estimated index weights and functions from the additive DRF-DLNM for PM$_{2.5}$ and respiratory mortality. The function $\psi$ is shown in selected slices of the bivariate surface $\psi$, by fixing the index at estimated values under the exposures are at their 25\% and 75\% quantiles, and fixing the lag at 0, 3, and 14. }
\end{figure}

\begin{figure}[H]
  \centering
  \makebox[\textwidth][c]{
  \begin{subfigure}[t]{0.33\textwidth}
    \centering
    \includegraphics[width=\linewidth]{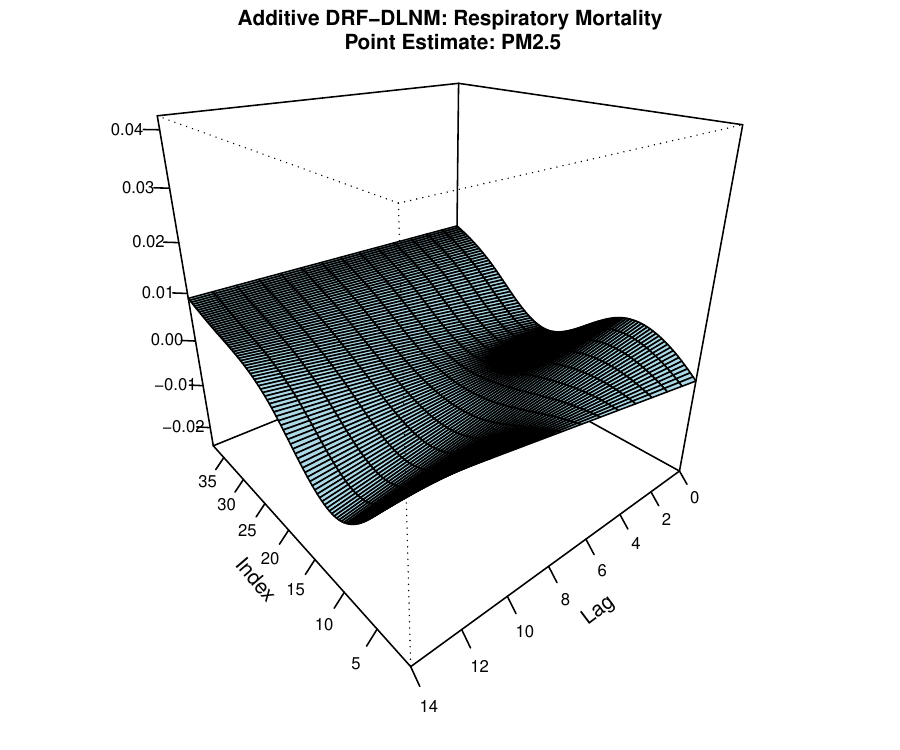}
  \end{subfigure}
  }
  \vspace{0.5em}
  
  \begin{subfigure}[t]{0.33\textwidth}
    \centering
    \includegraphics[width=\linewidth]{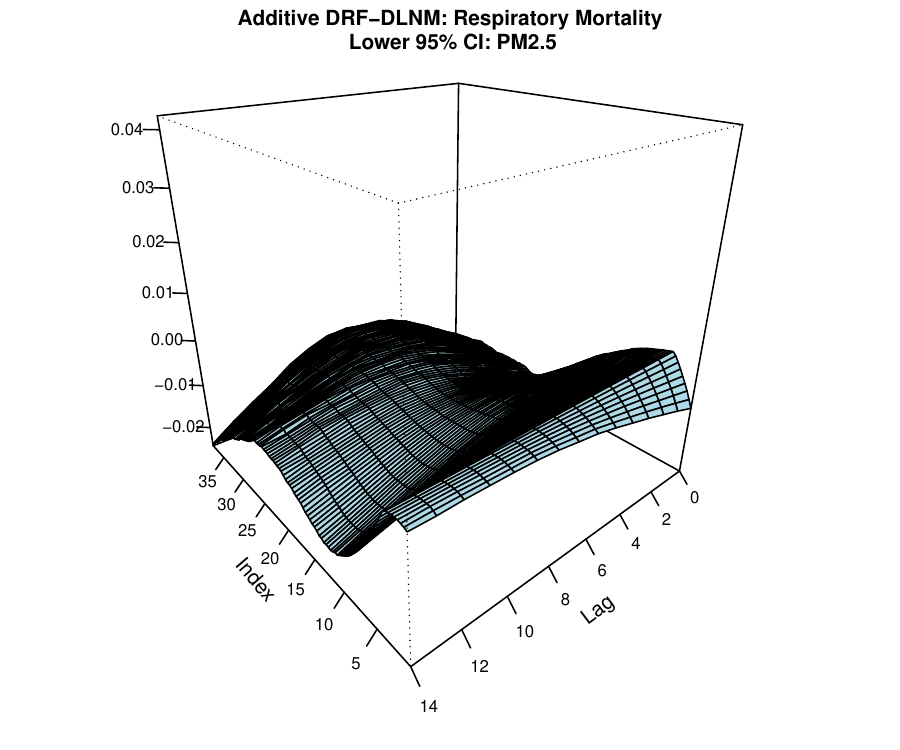}
  \end{subfigure}
  \begin{subfigure}[t]{0.33\textwidth}
    \centering
    \includegraphics[width=\linewidth]{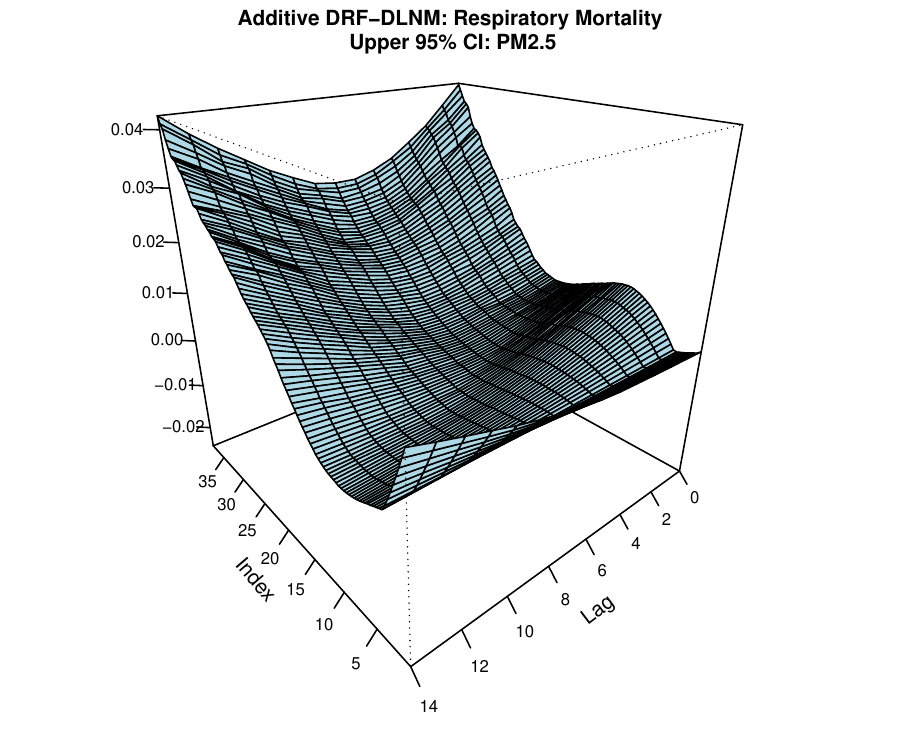}
  \end{subfigure}
  \caption{Estimated $\psi$ from the additive DRF-DLNM for PM$_{2.5}$ and respiratory mortality.}
\end{figure}

\begin{figure}[H]
    \centering
    \includegraphics[width=\linewidth]{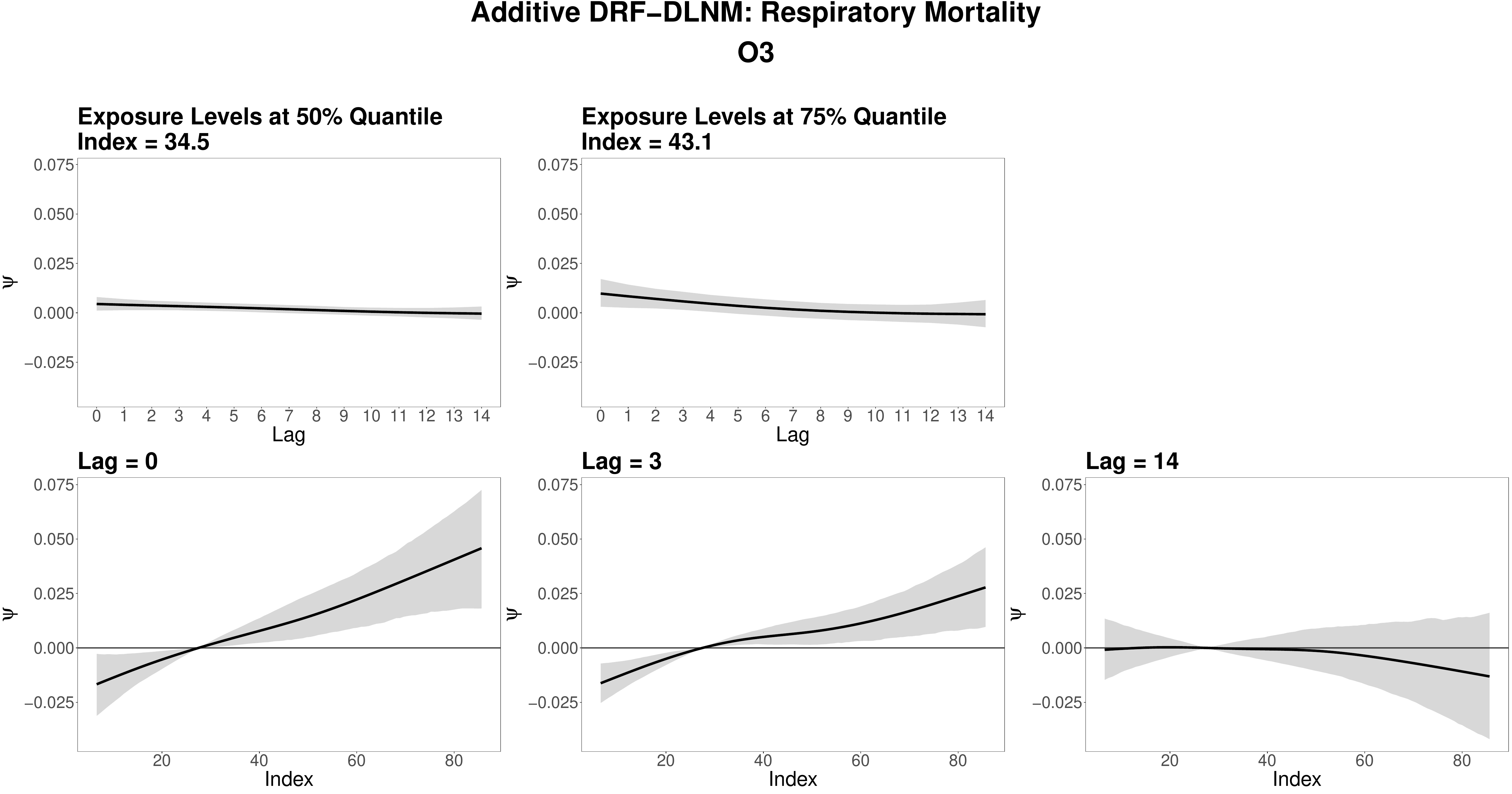}
    \caption{Estimated index weights and functions from the additive DRF-DLNM for O$_3$ and respiratory mortality. The function $\psi$ is shown in selected slices of the bivariate surface $\psi$, by fixing the index at estimated values under the exposures are at their 25\% and 75\% quantiles, and fixing the lag at 0, 3, and 14. }
\end{figure}

\begin{figure}[H]
  \centering
  \makebox[\textwidth][c]{
  \begin{subfigure}[t]{0.33\textwidth}
    \centering
    \includegraphics[width=\linewidth]{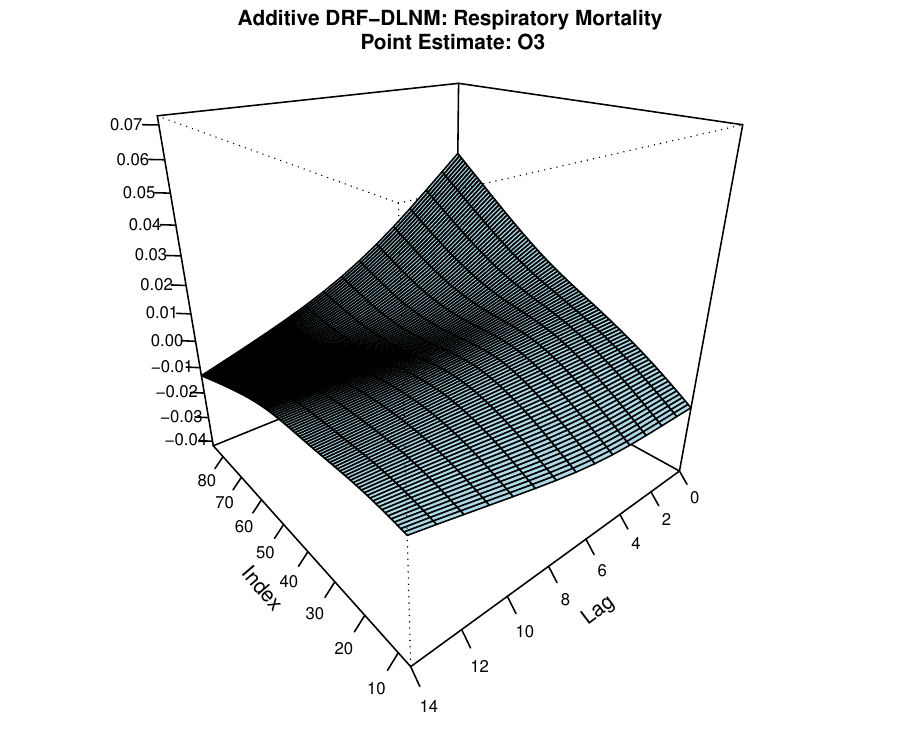}
  \end{subfigure}
  }
  \vspace{0.5em}
  
  \begin{subfigure}[t]{0.33\textwidth}
    \centering
    \includegraphics[width=\linewidth]{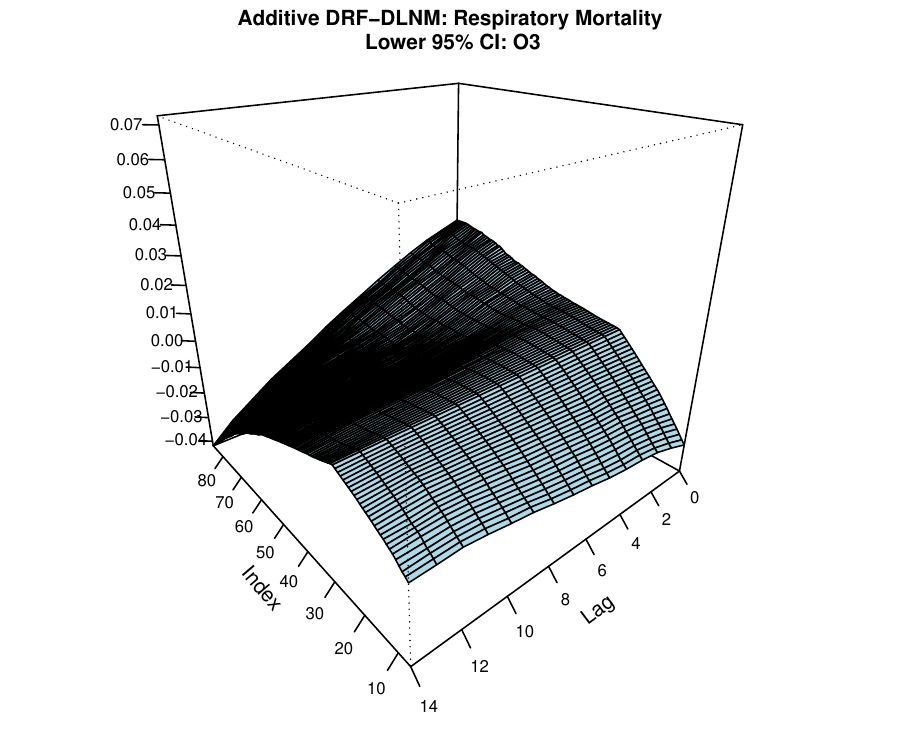}
  \end{subfigure}
  \begin{subfigure}[t]{0.33\textwidth}
    \centering
    \includegraphics[width=\linewidth]{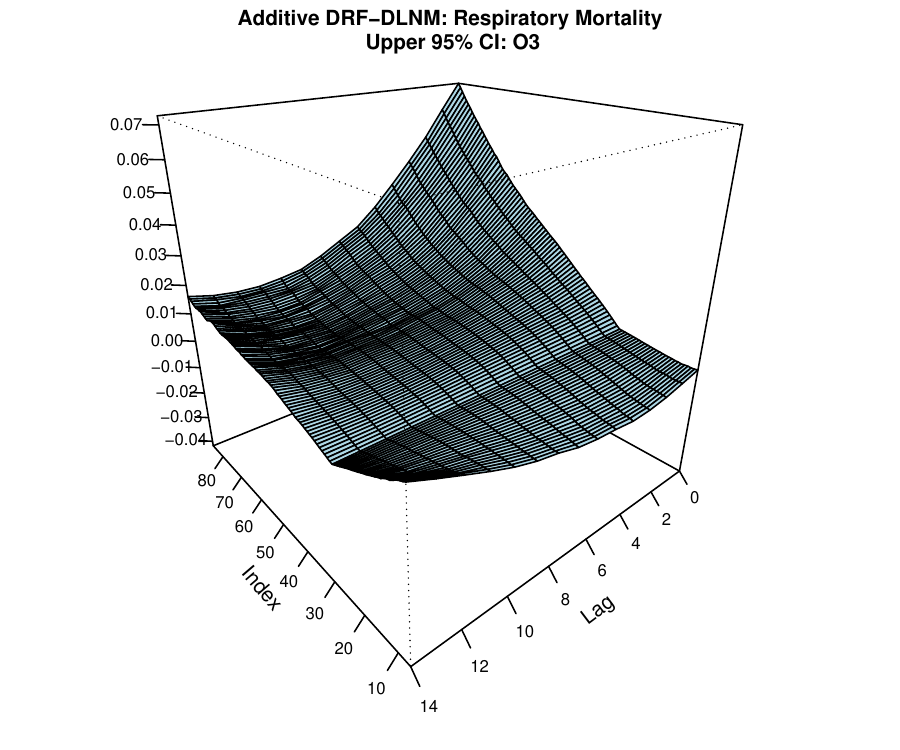}
  \end{subfigure}
  \caption{Estimated $\psi$ from the additive DRF-DLNM for O$_3$ and respiratory mortality.}
\end{figure}

\begin{figure}[H]
    \centering
    \includegraphics[width=\linewidth]{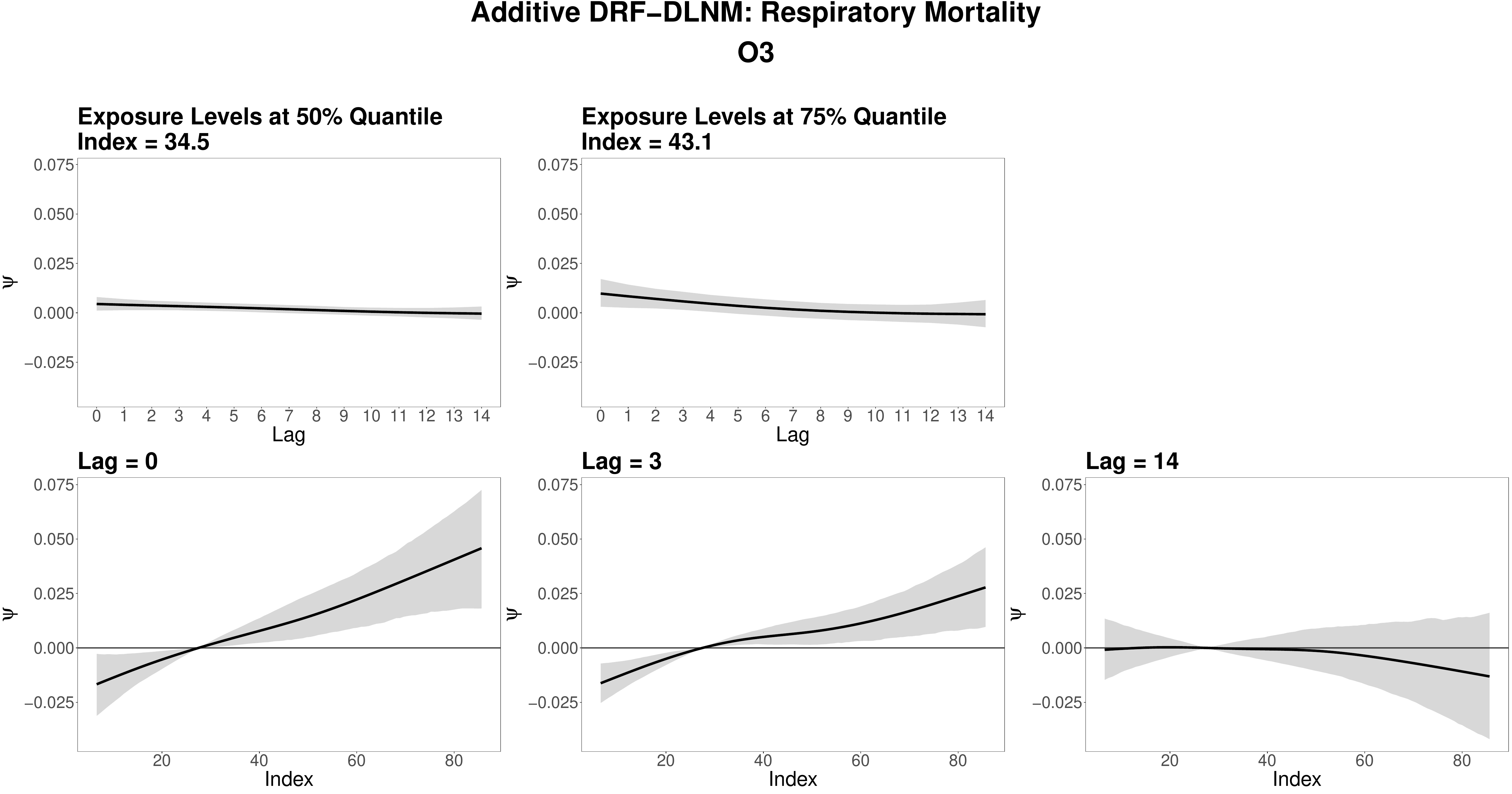}
    \caption{Estimated index weights and functions from the additive DRF-DLNM for O$_3$ and respiratory mortality. The function $\psi$ is shown in selected slices of the bivariate surface $\psi$, by fixing the index at estimated values under the exposures are at their 25\% and 75\% quantiles, and fixing the lag at 0, 3, and 14. }
\end{figure}

\begin{figure}[H]
  \centering
  \makebox[\textwidth][c]{
  \begin{subfigure}[t]{0.33\textwidth}
    \centering
    \includegraphics[width=\linewidth]{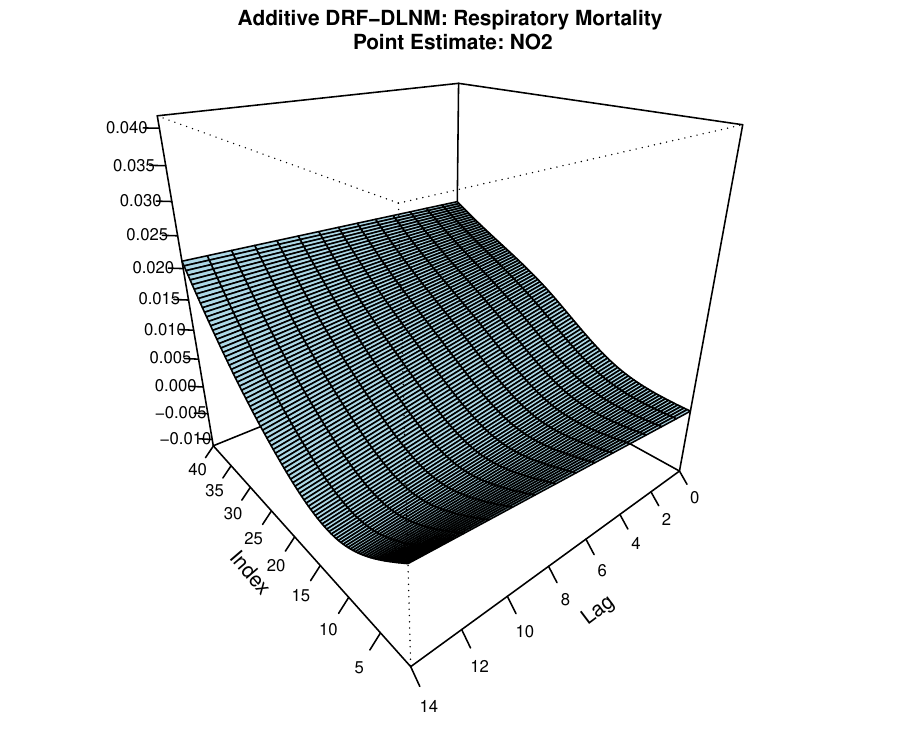}
  \end{subfigure}
  }
  \vspace{0.5em}
  
  \begin{subfigure}[t]{0.33\textwidth}
    \centering
    \includegraphics[width=\linewidth]{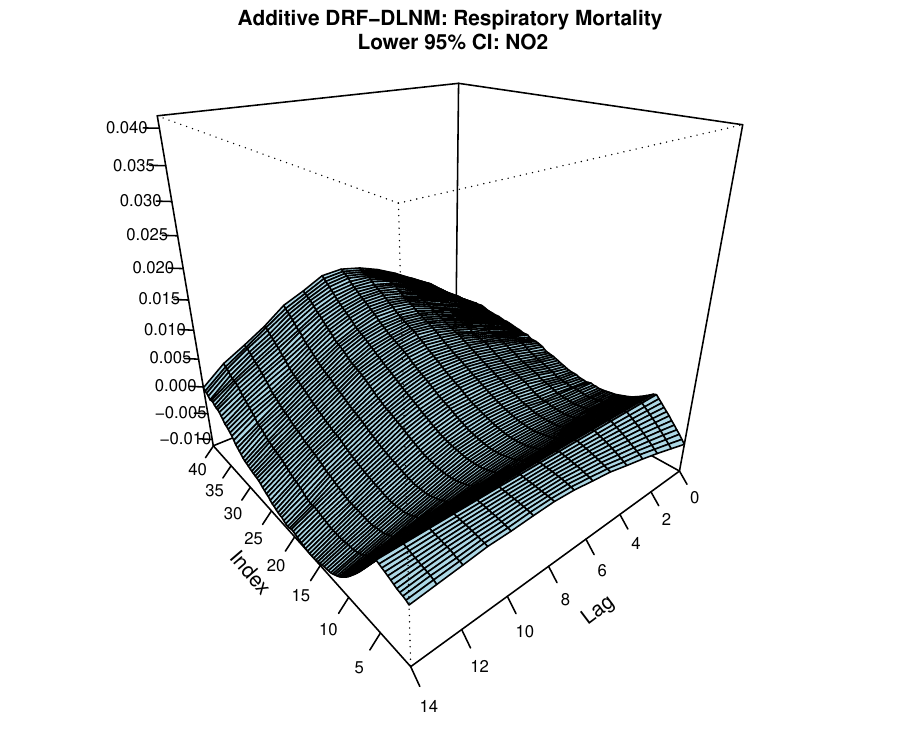}
  \end{subfigure}
  \begin{subfigure}[t]{0.33\textwidth}
    \centering
    \includegraphics[width=\linewidth]{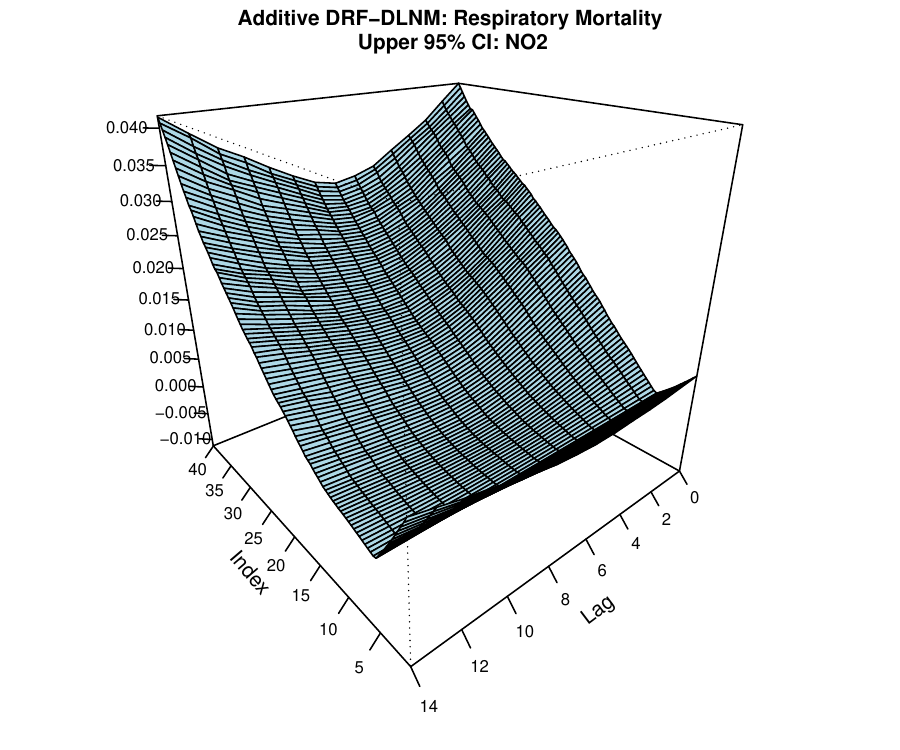}
  \end{subfigure}
  \caption{Estimated $\psi$ from the additive DRF-DLNM for NO$_2$ and respiratory mortality.}
\end{figure}

\subsubsection{Circulatory Mortality}

\begin{figure}[H]

    \centering
    \includegraphics[width=\linewidth]{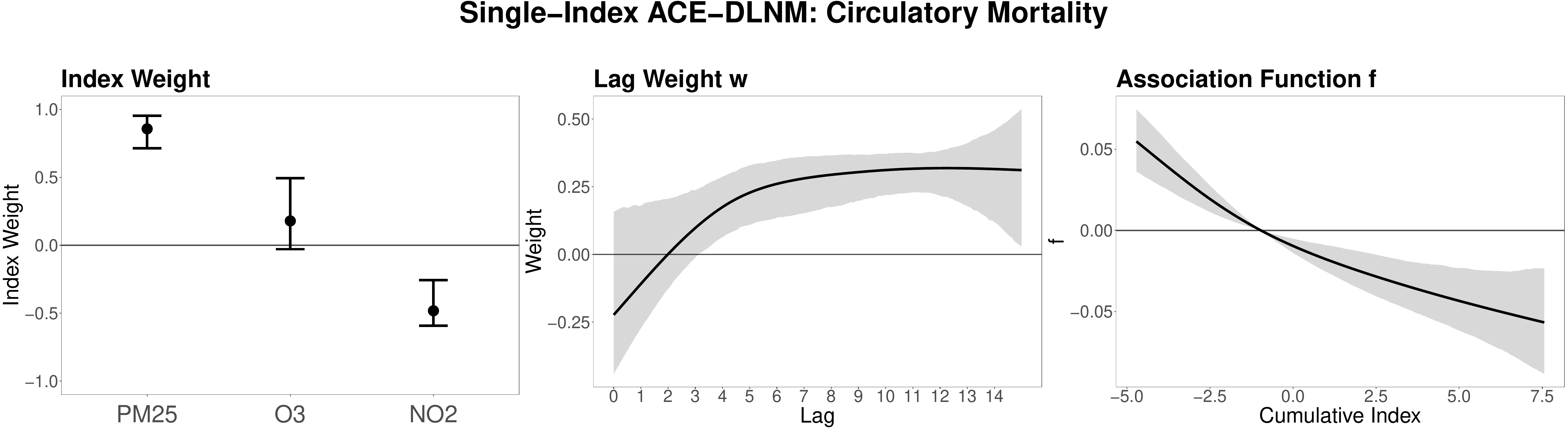}    
    \caption{Estimated index weights and functions from the single-index ACE-DLNM for circulatory mortality. }
\end{figure}

\begin{figure}[H]
    \centering
    \includegraphics[width=\linewidth]{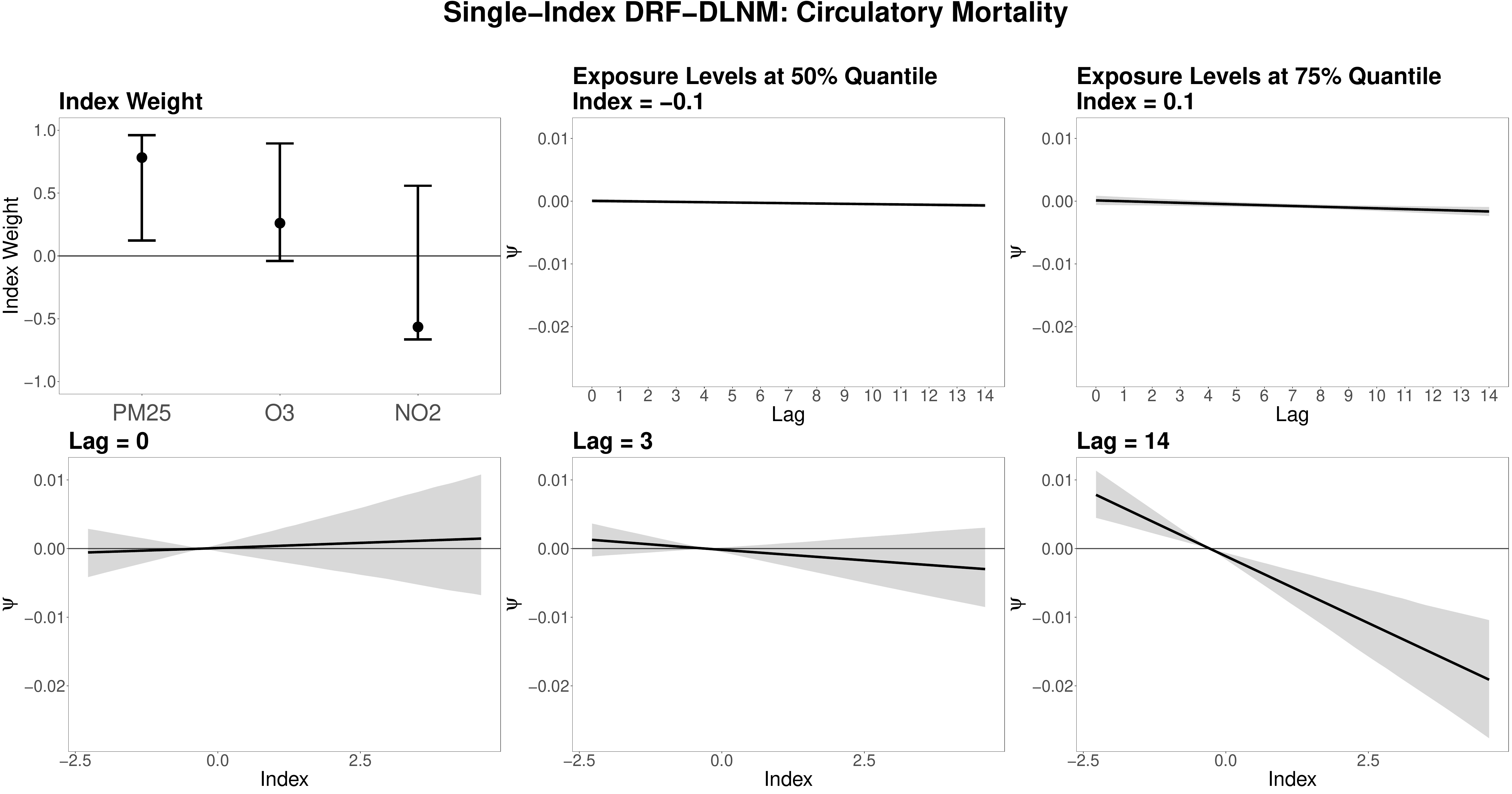}
    \caption{Estimated index weights and functions from the single-index DRF-DLNM for circulatory mortality. The function $\psi$ is shown in selected slices of the bivariate surface $\psi$, by fixing the index at estimated values under the exposures are at their 25\% and 75\% quantiles, and fixing the lag at 0, 3, and 14. }
\end{figure}

\begin{figure}[H]
  \centering
  \makebox[\textwidth][c]{
  \begin{subfigure}[t]{0.33\textwidth}
    \centering
    \includegraphics[width=\linewidth]{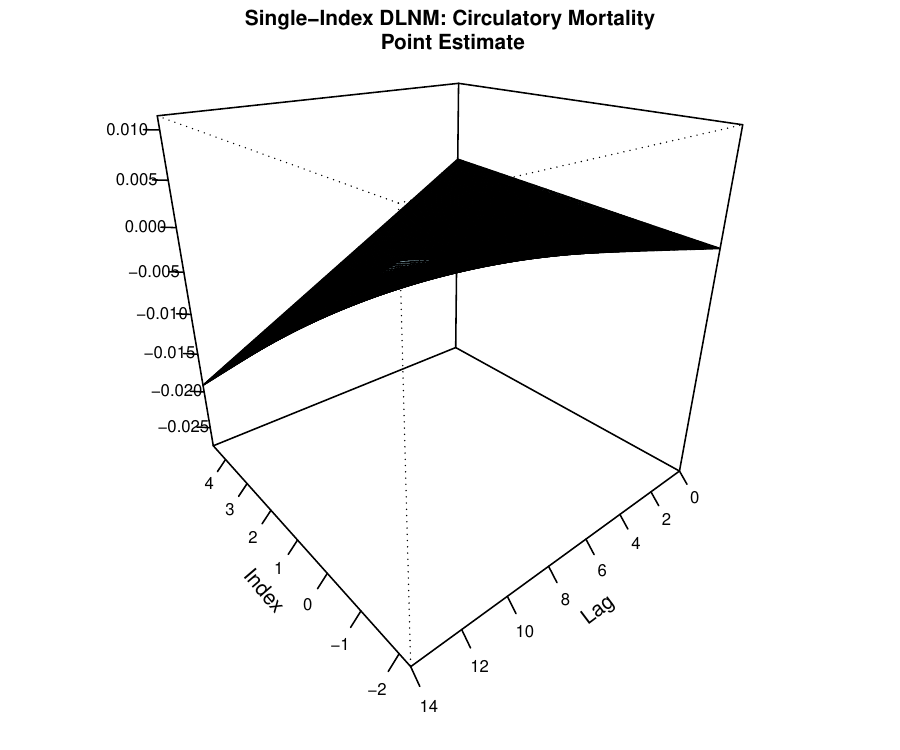}
  \end{subfigure}
  }
  \vspace{0.5em}
  
  \begin{subfigure}[t]{0.33\textwidth}
    \centering
    \includegraphics[width=\linewidth]{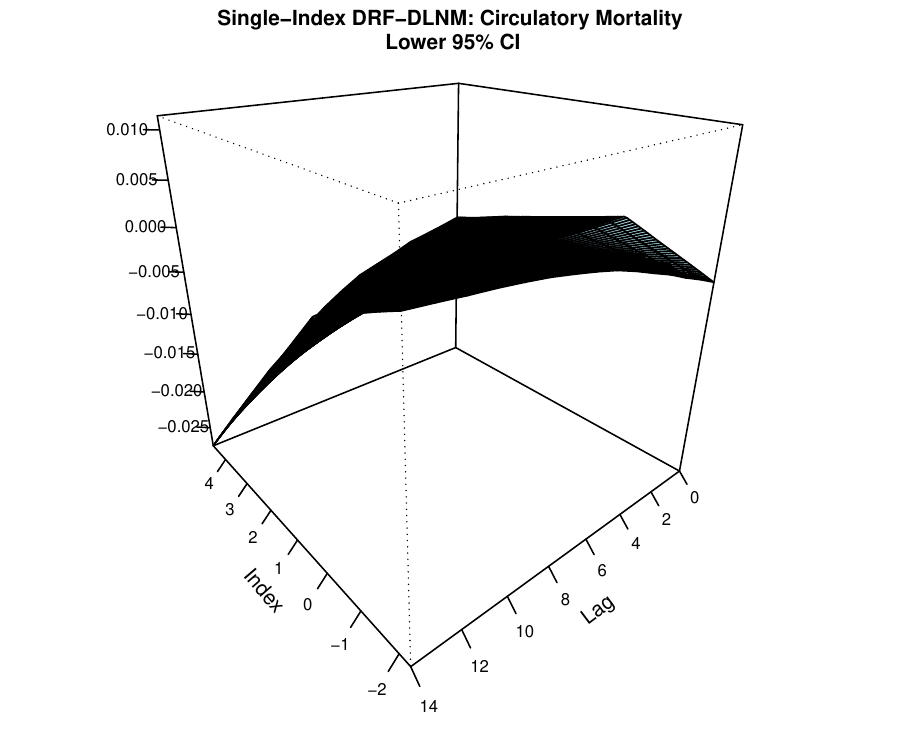}
  \end{subfigure}
  \begin{subfigure}[t]{0.33\textwidth}
    \centering
    \includegraphics[width=\linewidth]{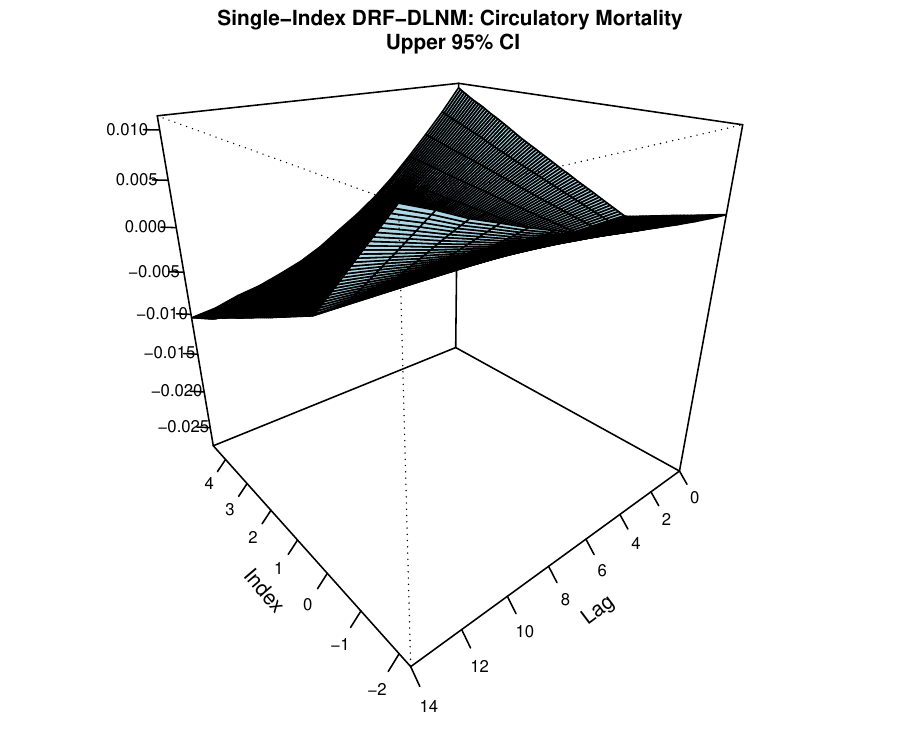}
  \end{subfigure}
  \caption{Estimated $\psi$ from the single-index DRF-DLNM for circulatory mortality.}
\end{figure}

\begin{figure}[H]
    \centering
    \includegraphics[width=\linewidth]{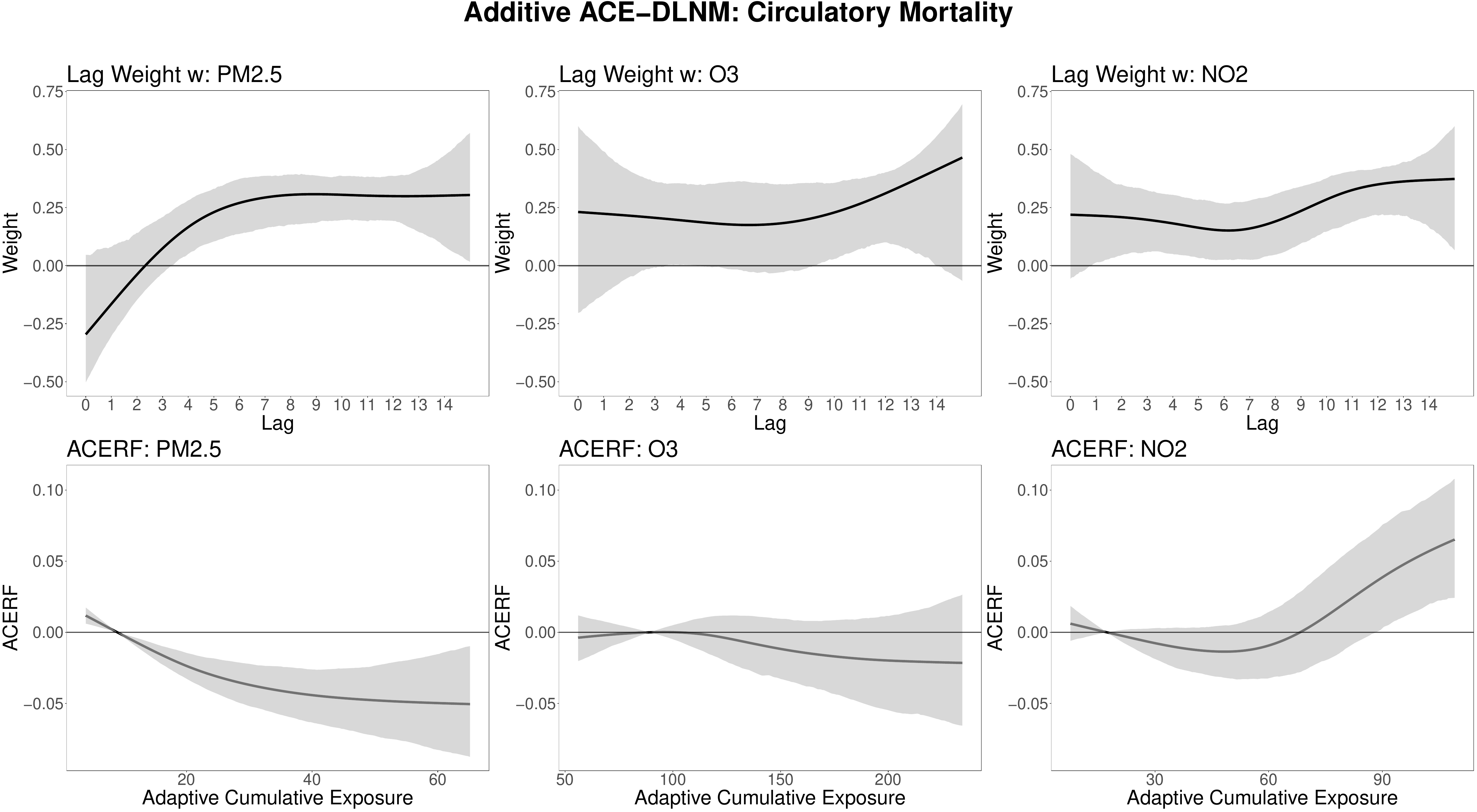}
    \caption{Estimated index weights and functions from the additive ACE-DLNM for circulatory mortality.}
\end{figure}

\begin{figure}[H]
    \centering
    \includegraphics[width=\linewidth]{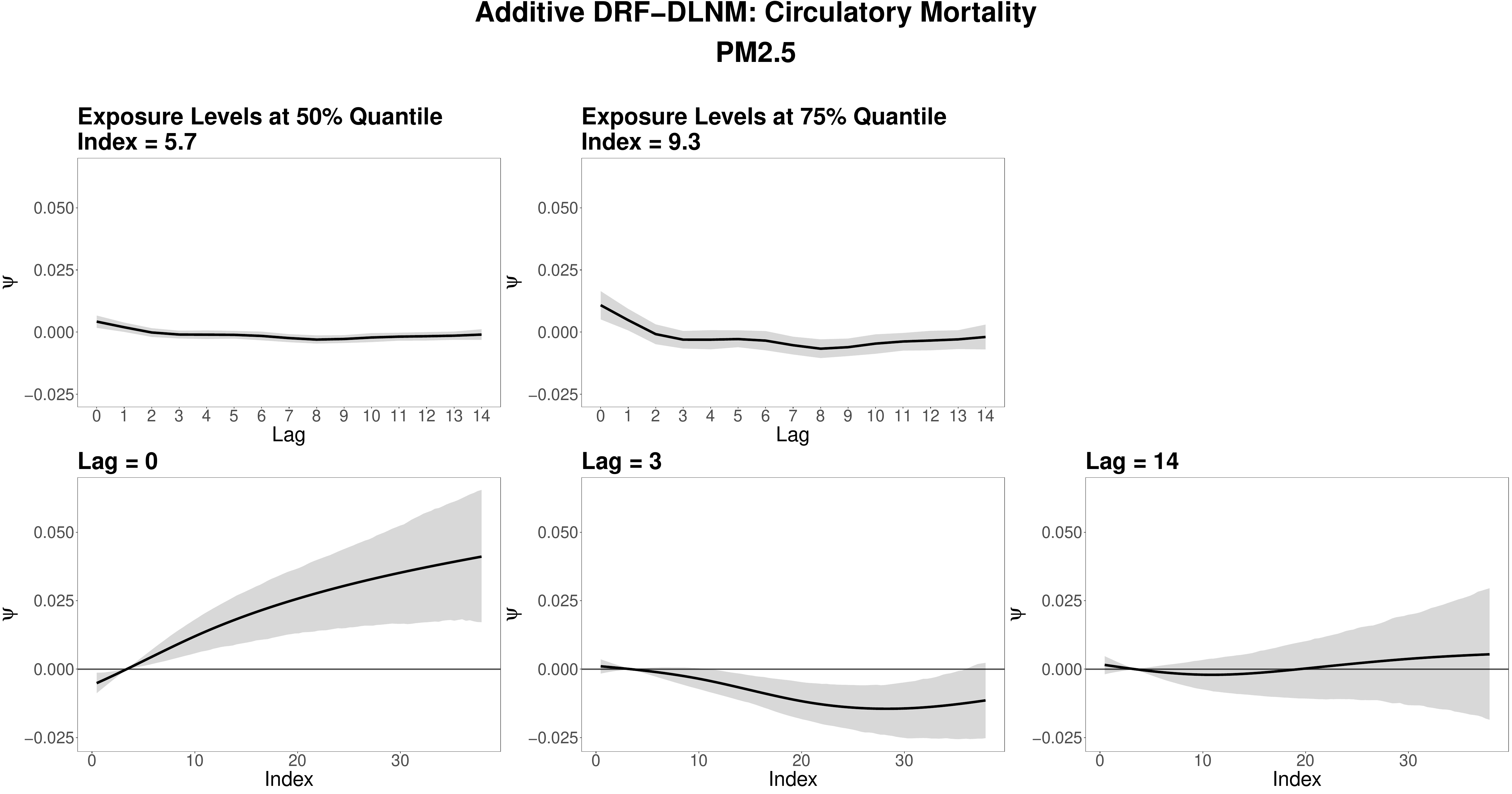}
    \caption{Estimated index weights and functions from the additive DRF-DLNM for PM$_{2.5}$ and circulatory mortality. The function $\psi$ is shown in selected slices of the bivariate surface $\psi$, by fixing the index at estimated values under the exposures are at their 25\% and 75\% quantiles, and fixing the lag at 0, 3, and 14. }
\end{figure}

\begin{figure}[H]
  \centering
  \makebox[\textwidth][c]{
  \begin{subfigure}[t]{0.33\textwidth}
    \centering
    \includegraphics[width=\linewidth]{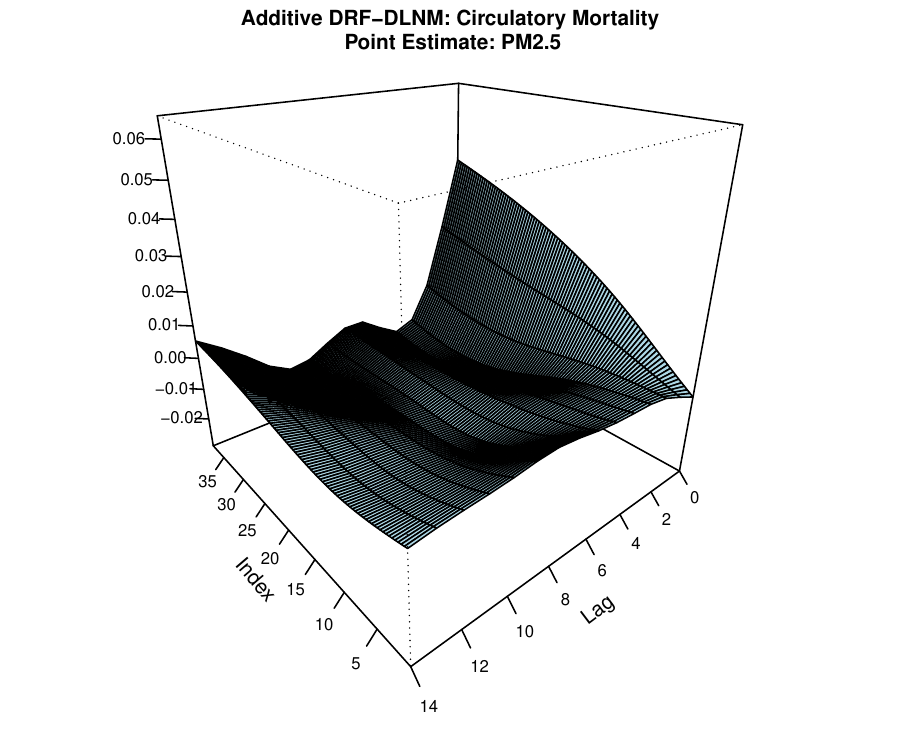}
  \end{subfigure}
  }
  \vspace{0.5em}
  
  \begin{subfigure}[t]{0.33\textwidth}
    \centering
    \includegraphics[width=\linewidth]{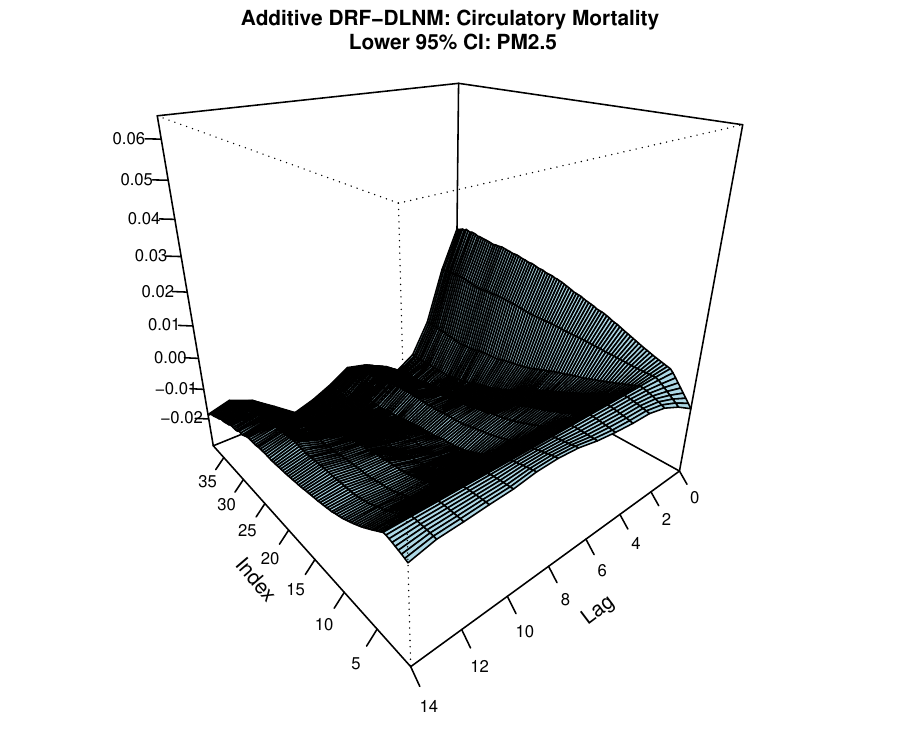}
  \end{subfigure}
  \begin{subfigure}[t]{0.33\textwidth}
    \centering
    \includegraphics[width=\linewidth]{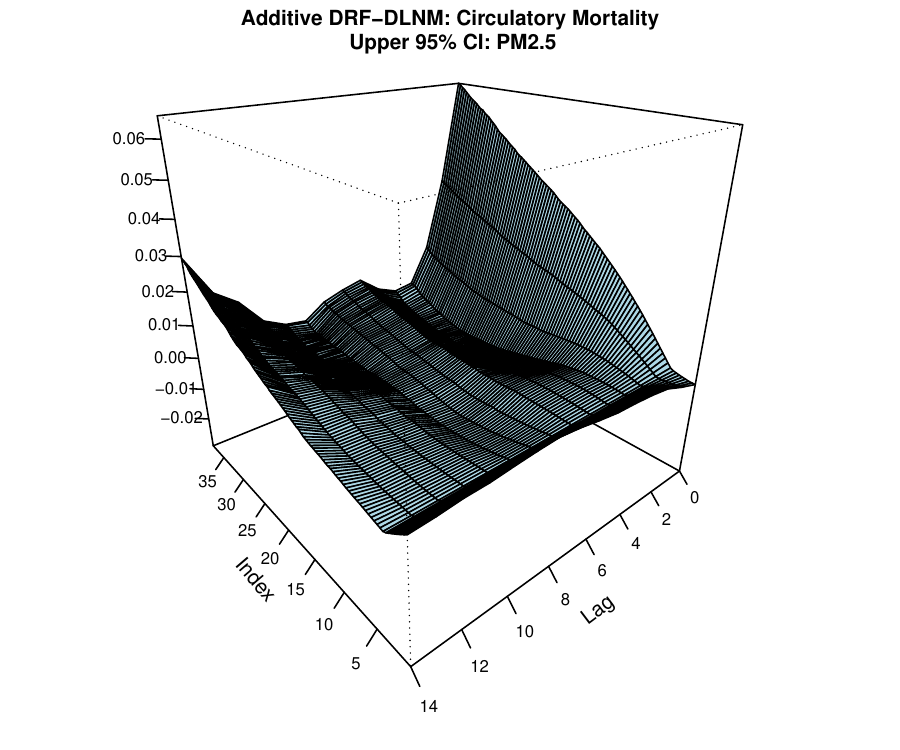}
  \end{subfigure}
  \caption{Estimated $\psi$ from the additive DRF-DLNM for PM$_{2.5}$ and circulatory mortality.}
\end{figure}

\begin{figure}[H]
    \centering
    \includegraphics[width=\linewidth]{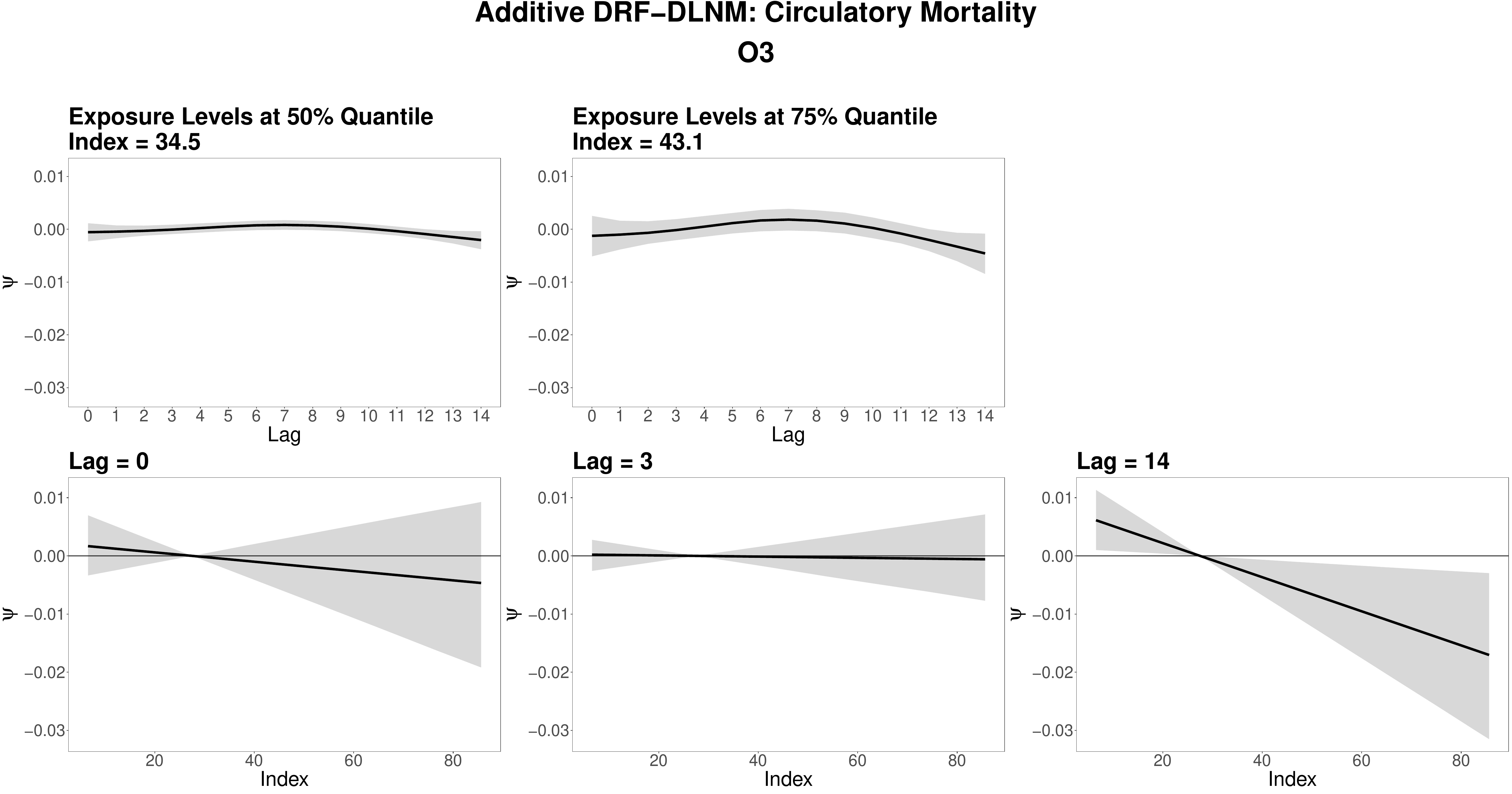}
    \caption{Estimated index weights and functions from the additive DRF-DLNM for O$_3$ and circulatory mortality. The function $\psi$ is shown in selected slices of the bivariate surface $\psi$, by fixing the index at estimated values under the exposures are at their 25\% and 75\% quantiles, and fixing the lag at 0, 3, and 14. }
\end{figure}

\begin{figure}[H]
  \centering
  \makebox[\textwidth][c]{
  \begin{subfigure}[t]{0.33\textwidth}
    \centering
    \includegraphics[width=\linewidth]{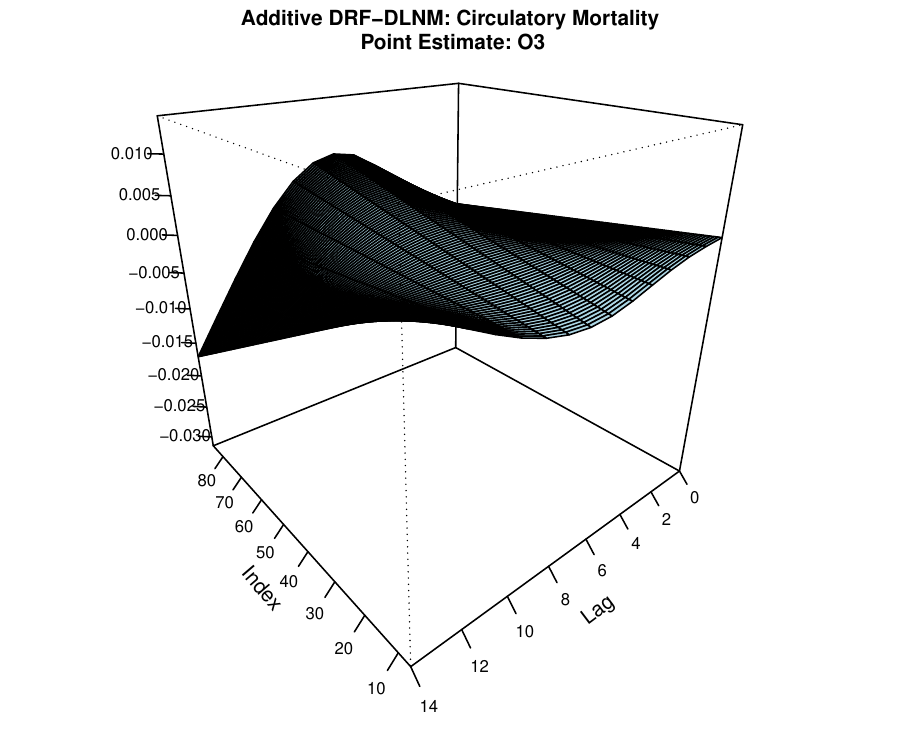}
  \end{subfigure}
  }
  \vspace{0.5em}
  
  \begin{subfigure}[t]{0.33\textwidth}
    \centering
    \includegraphics[width=\linewidth]{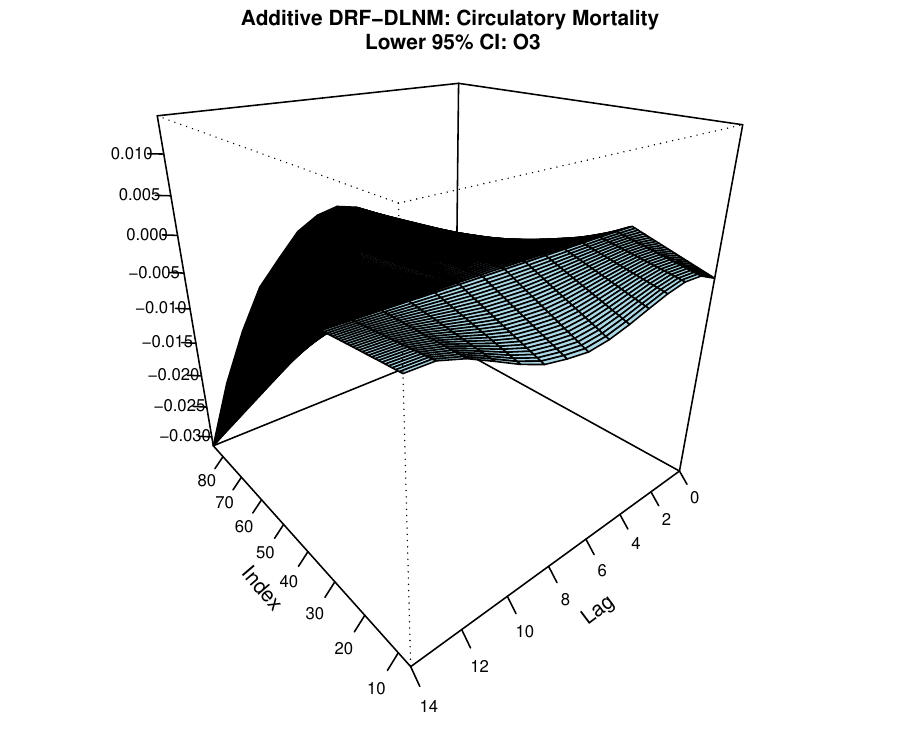}
  \end{subfigure}
  \begin{subfigure}[t]{0.33\textwidth}
    \centering
    \includegraphics[width=\linewidth]{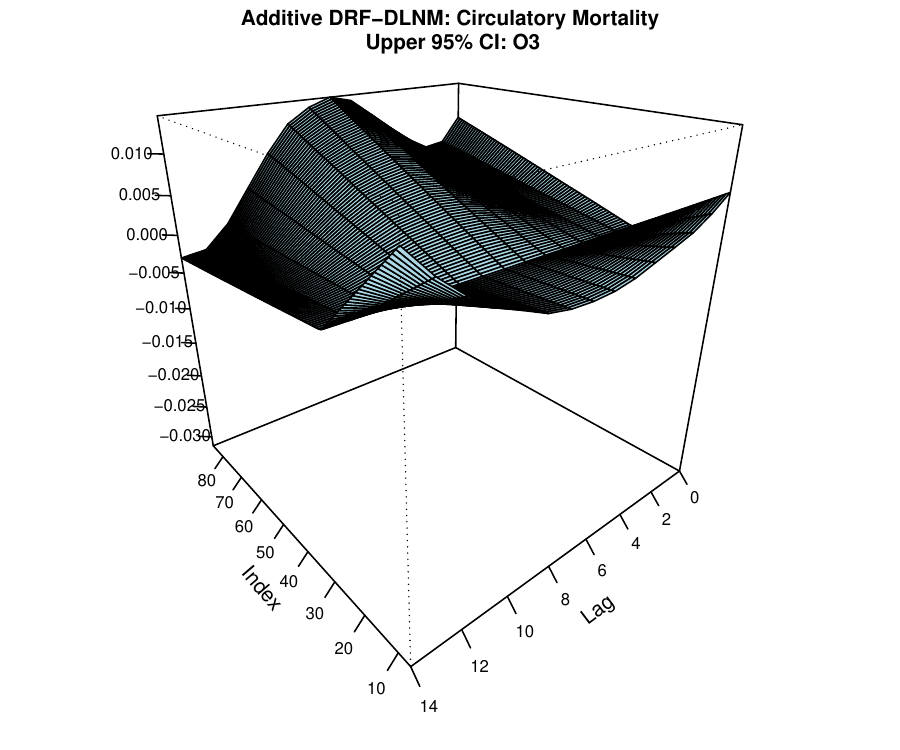}
  \end{subfigure}
  \caption{Estimated $\psi$ from the additive DRF-DLNM for O$_3$ and circulatory mortality.}
\end{figure}

\begin{figure}[H]
    \centering
    \includegraphics[width=\linewidth]{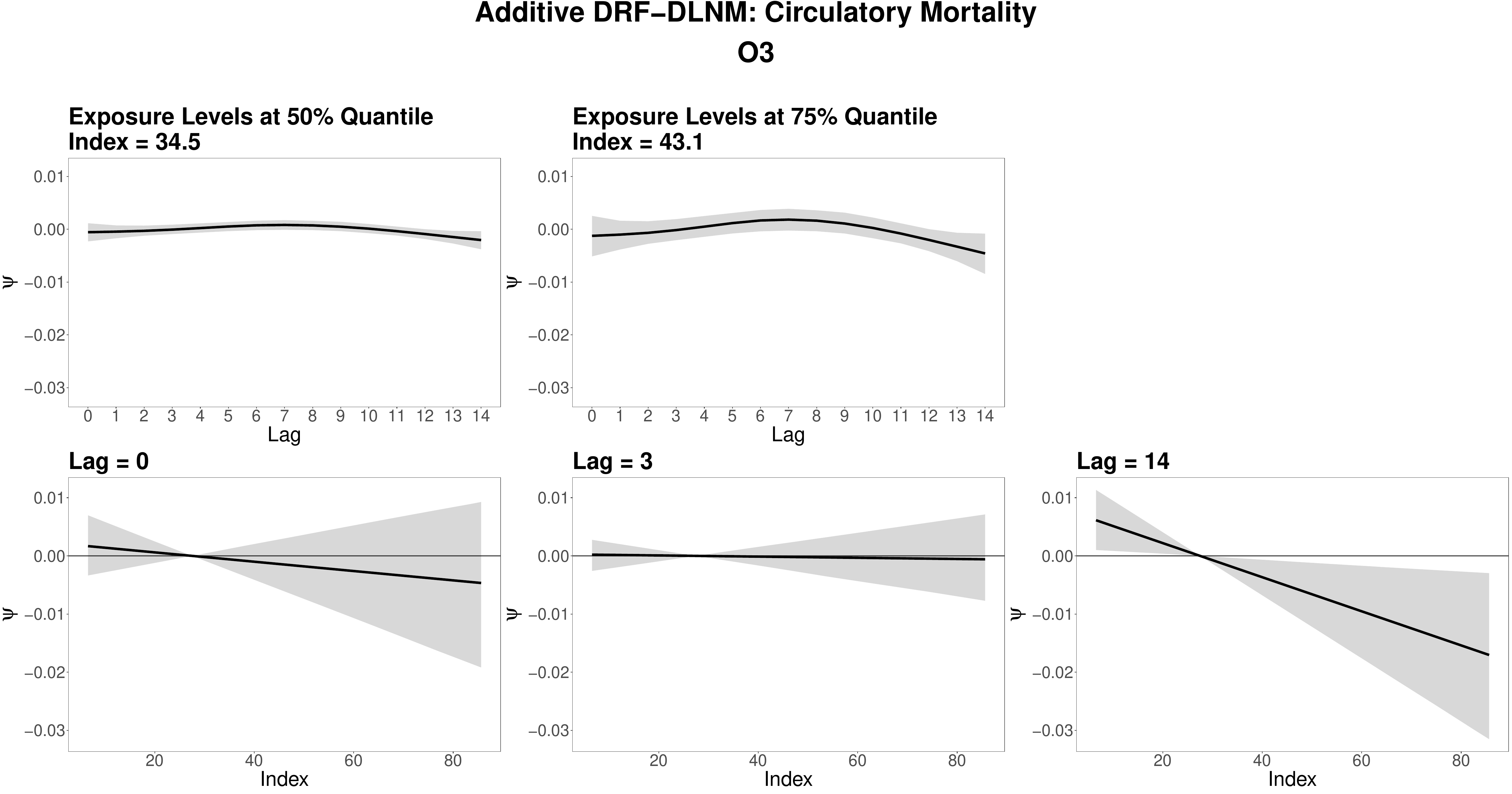}
    \caption{Estimated index weights and functions from the additive DRF-DLNM for O$_3$ and circulatory mortality. The function $\psi$ is shown in selected slices of the bivariate surface $\psi$, by fixing the index at estimated values under the exposures are at their 25\% and 75\% quantiles, and fixing the lag at 0, 3, and 14. }
\end{figure}

\begin{figure}[H]
  \centering
  \makebox[\textwidth][c]{
  \begin{subfigure}[t]{0.33\textwidth}
    \centering
    \includegraphics[width=\linewidth]{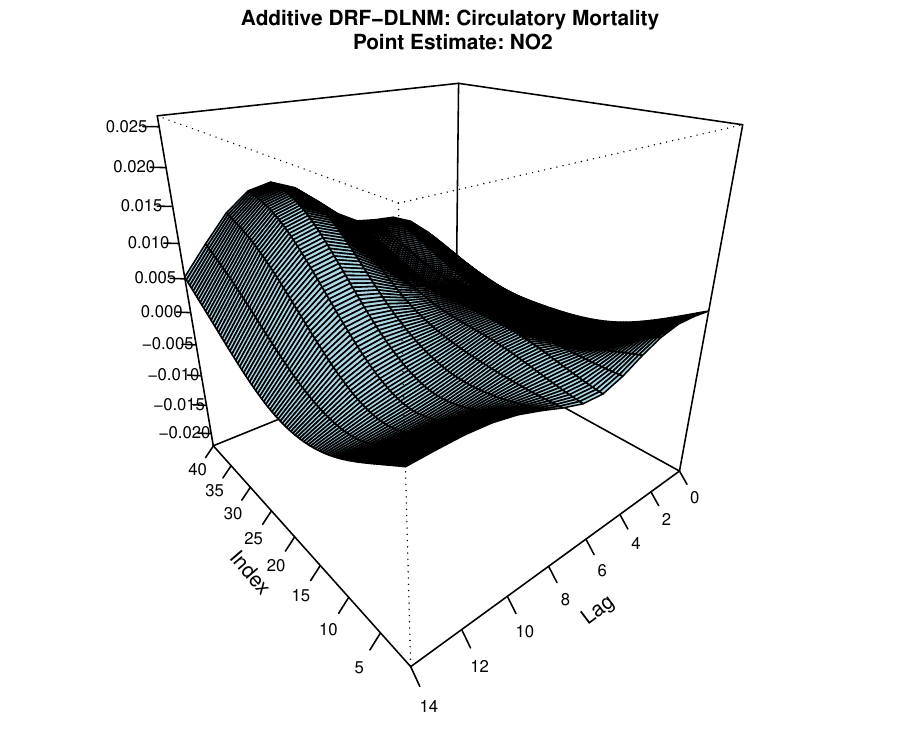}
  \end{subfigure}
  }
  \vspace{0.5em}
  
  \begin{subfigure}[t]{0.33\textwidth}
    \centering
    \includegraphics[width=\linewidth]{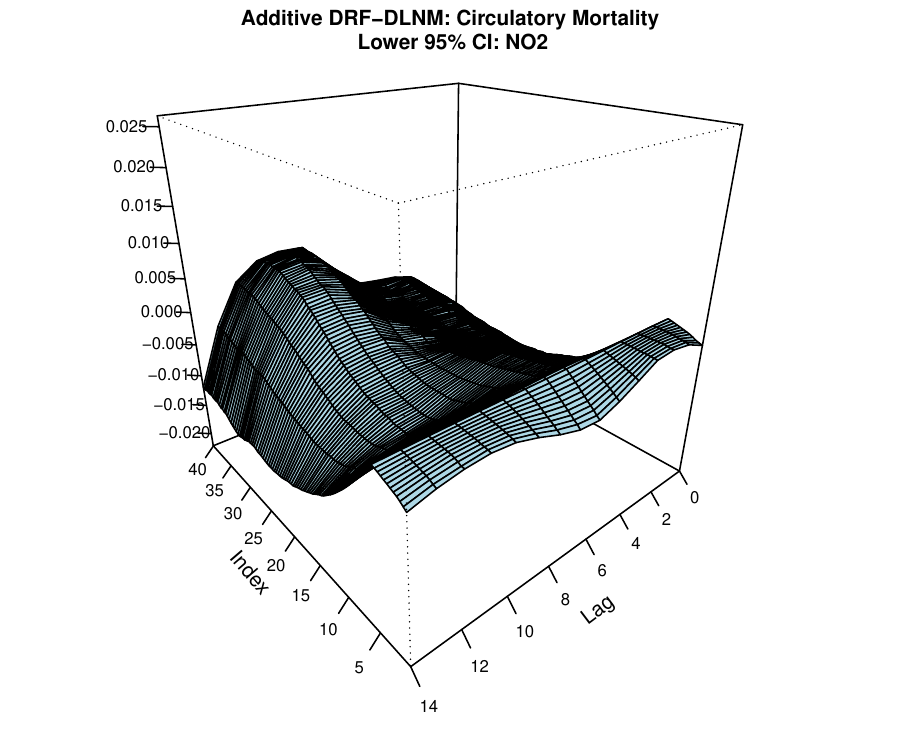}
  \end{subfigure}
  \begin{subfigure}[t]{0.33\textwidth}
    \centering
    \includegraphics[width=\linewidth]{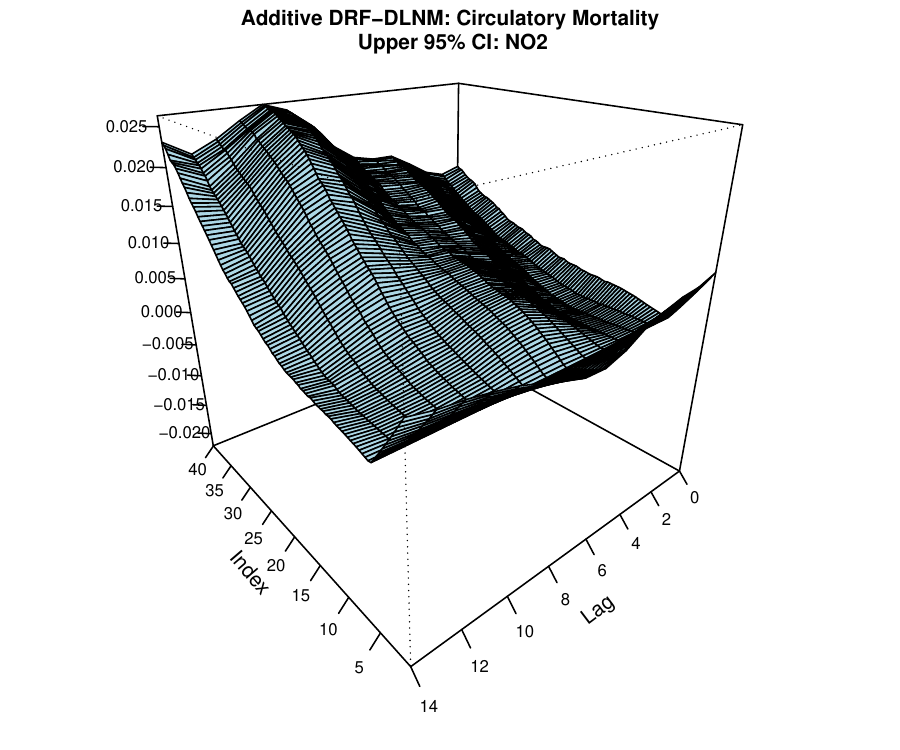}
  \end{subfigure}
  \caption{Estimated $\psi$ from the additive DRF-DLNM for NO$_2$ and circulatory mortality.}
\end{figure}

\subsubsection{All-Cause Mortality}

\begin{figure}[H]

    \centering
    \includegraphics[width=\linewidth]{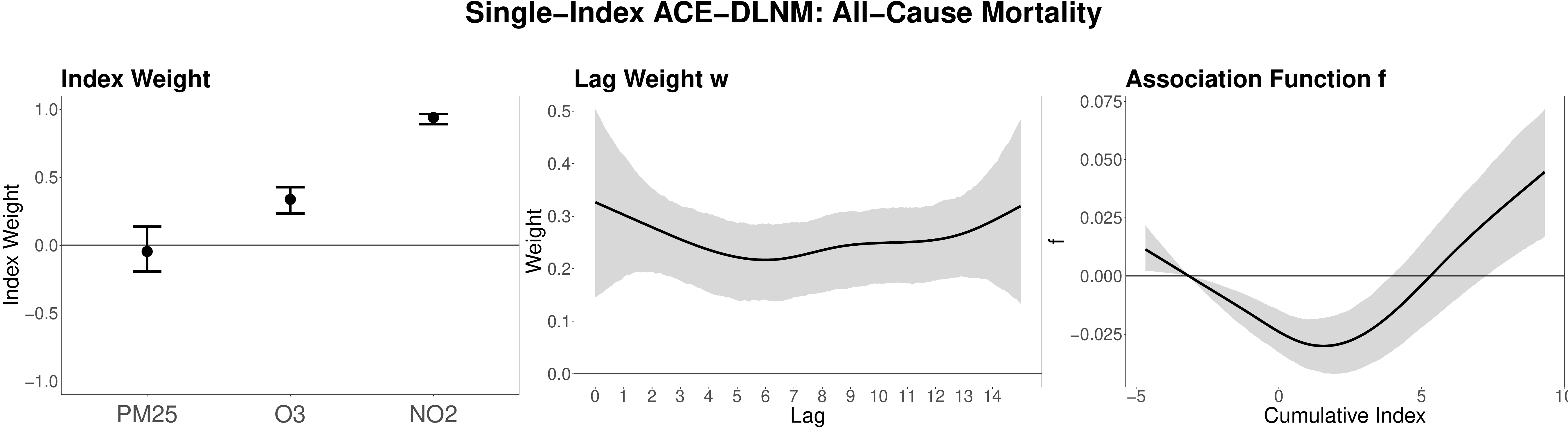}    
    \caption{Estimated index weights and functions from the single-index ACE-DLNM for all-cause mortality. }
\end{figure}

\begin{figure}[H]
    \centering
    \includegraphics[width=\linewidth]{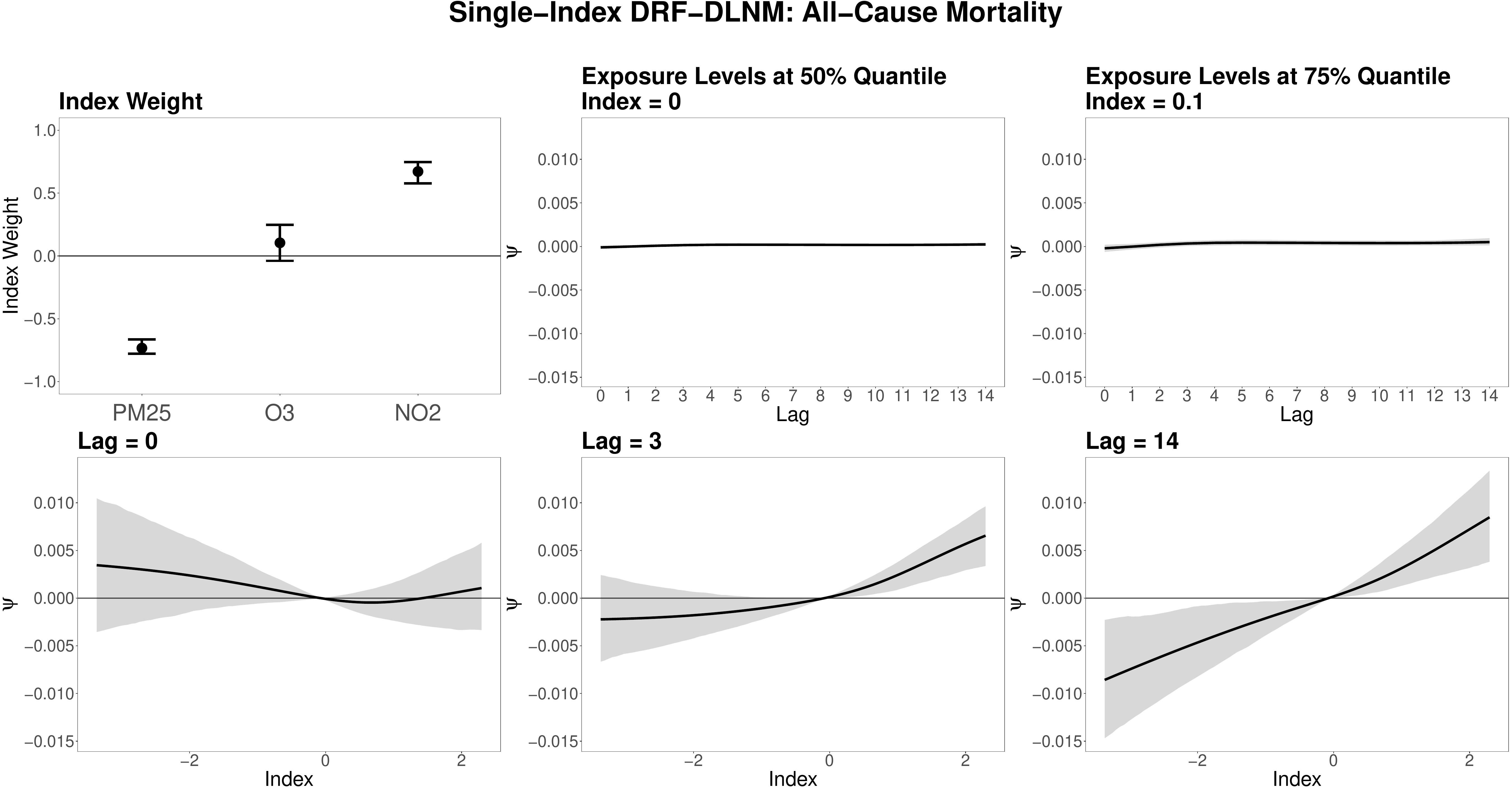}
    \caption{Estimated index weights and functions from the single-index DRF-DLNM for all-cause mortality. The function $\psi$ is shown in selected slices of the bivariate surface $\psi$, by fixing the index at estimated values under the exposures are at their 25\% and 75\% quantiles, and fixing the lag at 0, 3, and 14. }
\end{figure}

\begin{figure}[H]
  \centering
  \makebox[\textwidth][c]{
  \begin{subfigure}[t]{0.33\textwidth}
    \centering
    \includegraphics[width=\linewidth]{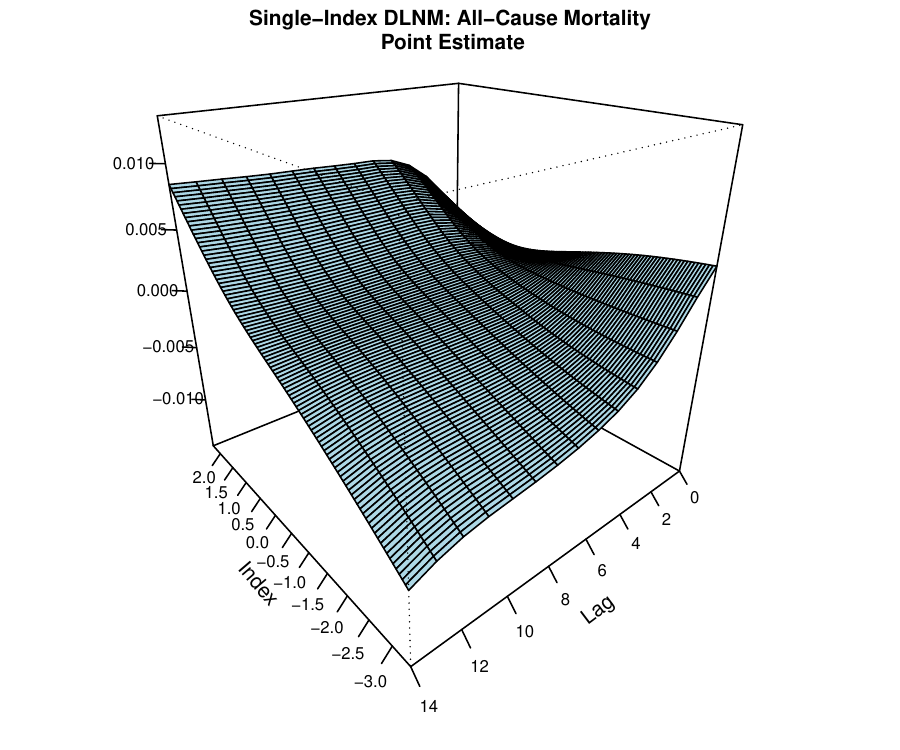}
  \end{subfigure}
  }
  \vspace{0.5em}
  
  \begin{subfigure}[t]{0.33\textwidth}
    \centering
    \includegraphics[width=\linewidth]{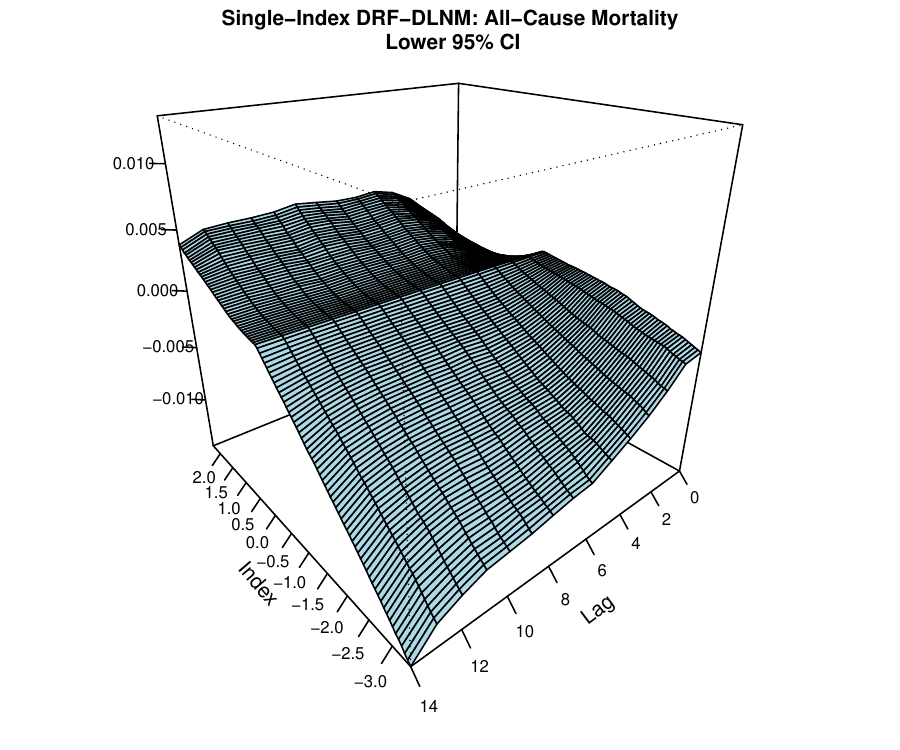}
  \end{subfigure}
  \begin{subfigure}[t]{0.33\textwidth}
    \centering
    \includegraphics[width=\linewidth]{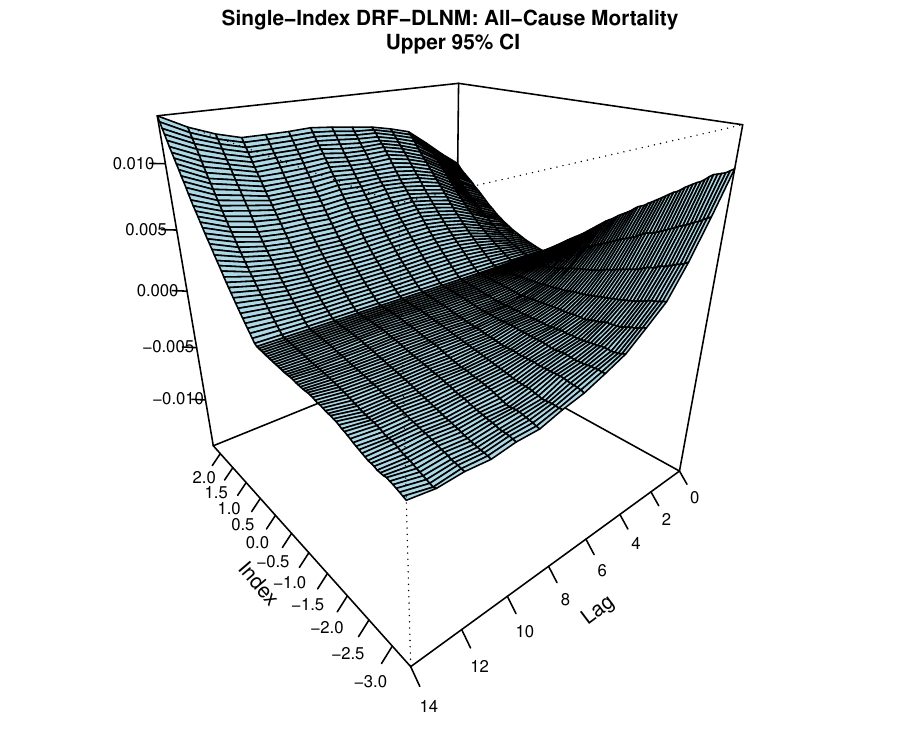}
  \end{subfigure}
  \caption{Estimated $\psi$ from the single-index DRF-DLNM for all-cause mortality.}
\end{figure}

\begin{figure}[H]
    \centering
    \includegraphics[width=\linewidth]{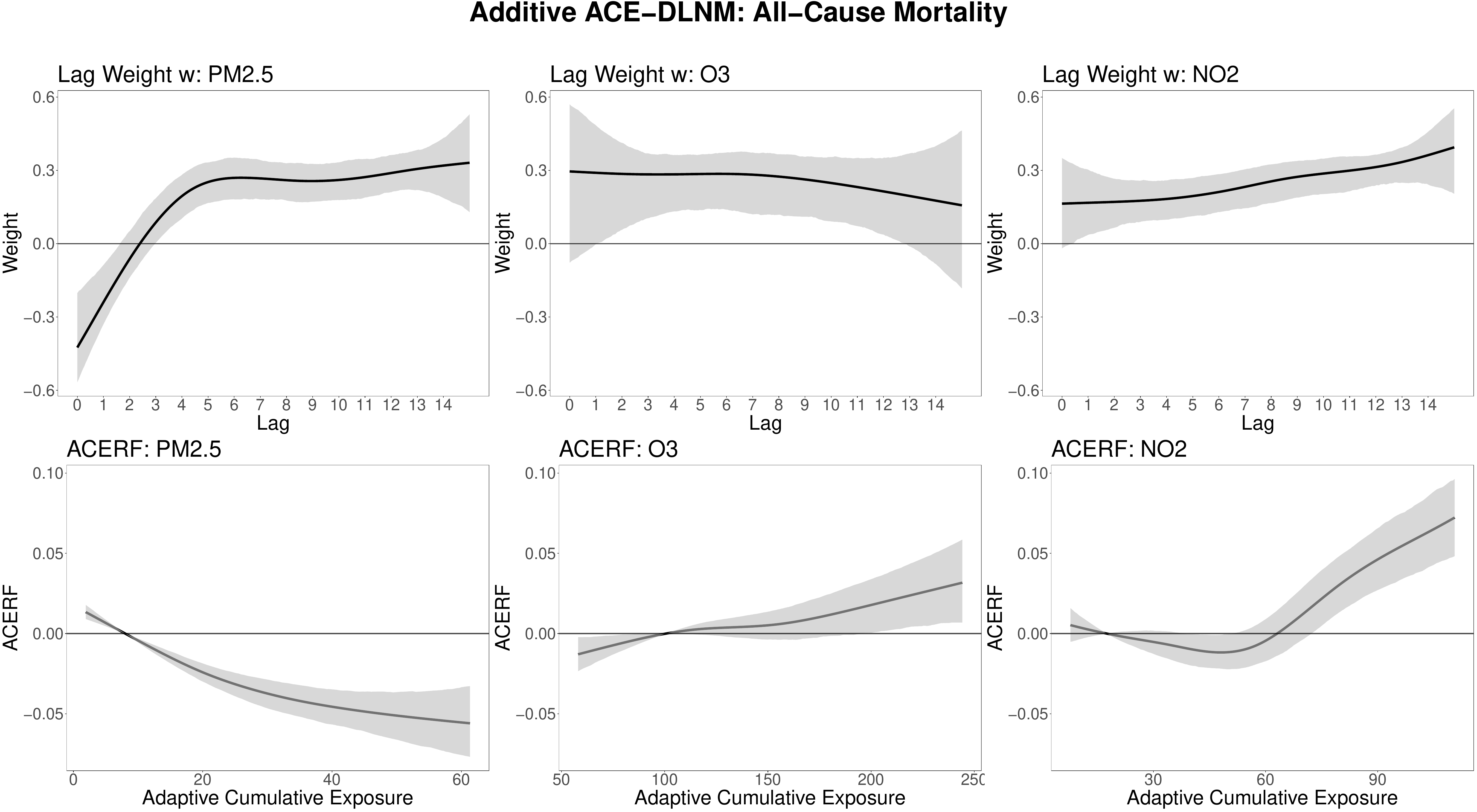}
    \caption{Estimated index weights and functions from the additive ACE-DLNM for all-cause mortality.}
\end{figure}

\begin{figure}[H]
    \centering
    \includegraphics[width=\linewidth]{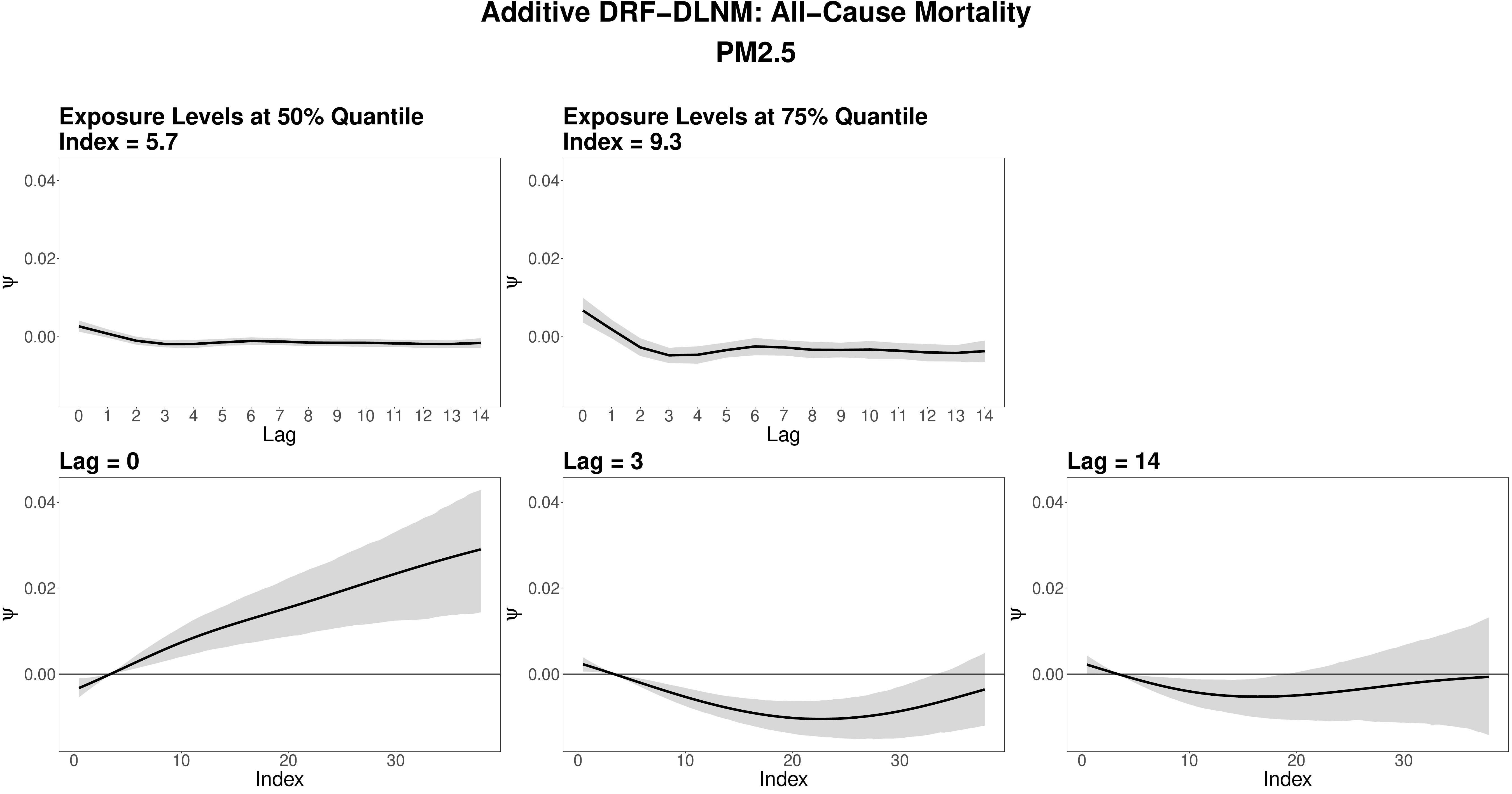}
    \caption{Estimated index weights and functions from the additive DRF-DLNM for PM$_{2.5}$ and all-cause mortality. The function $\psi$ is shown in selected slices of the bivariate surface $\psi$, by fixing the index at estimated values under the exposures are at their 25\% and 75\% quantiles, and fixing the lag at 0, 3, and 14. }
\end{figure}

\begin{figure}[H]
  \centering
  \makebox[\textwidth][c]{
  \begin{subfigure}[t]{0.33\textwidth}
    \centering
    \includegraphics[width=\linewidth]{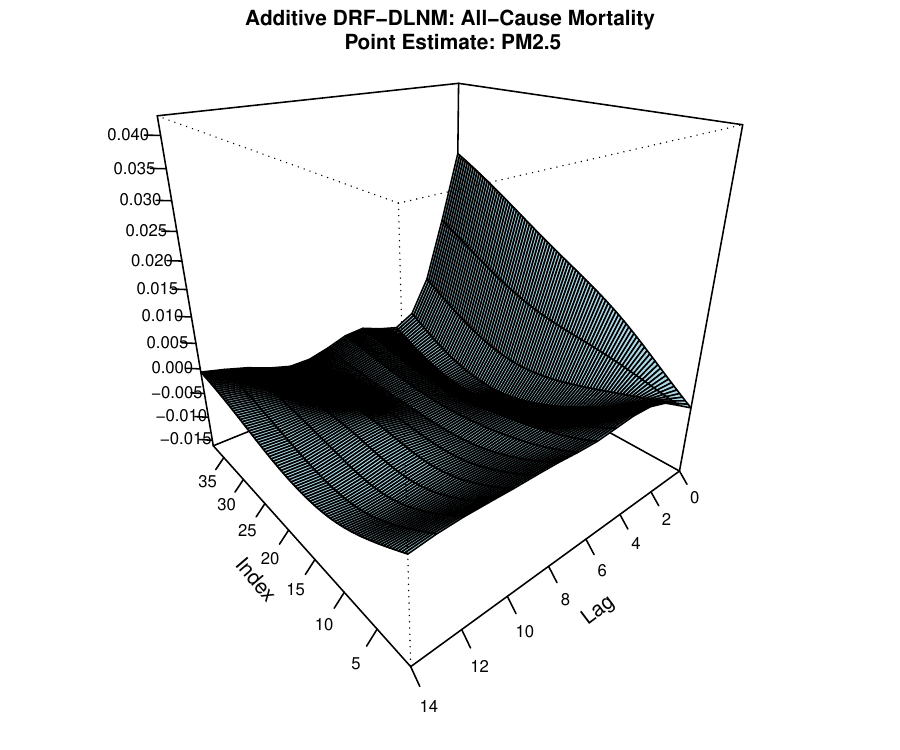}
  \end{subfigure}
  }
  \vspace{0.5em}
  
  \begin{subfigure}[t]{0.33\textwidth}
    \centering
    \includegraphics[width=\linewidth]{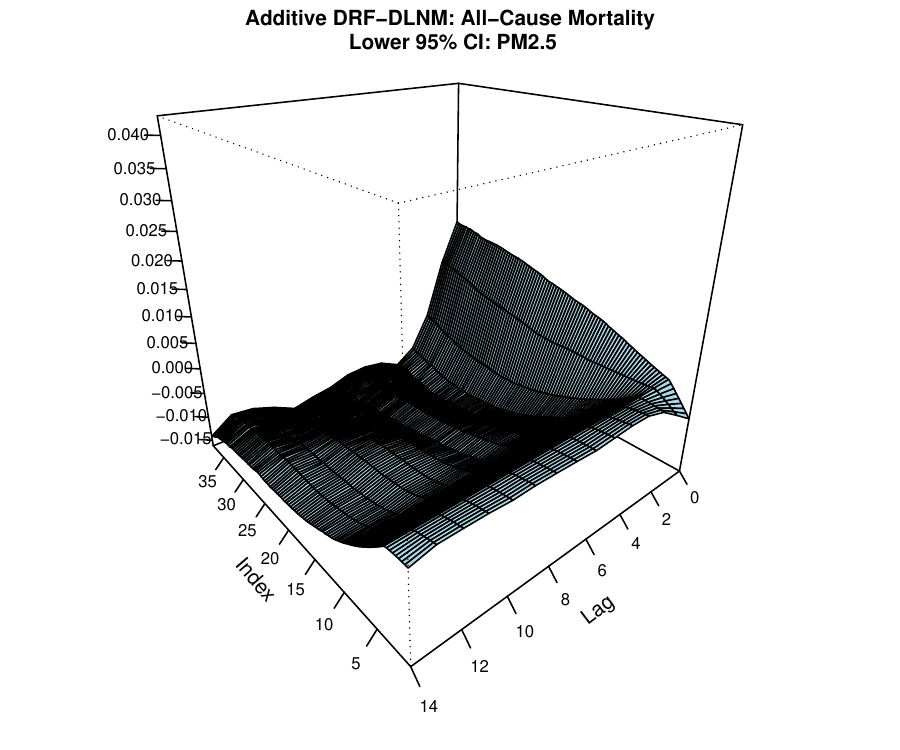}
  \end{subfigure}
  \begin{subfigure}[t]{0.33\textwidth}
    \centering
    \includegraphics[width=\linewidth]{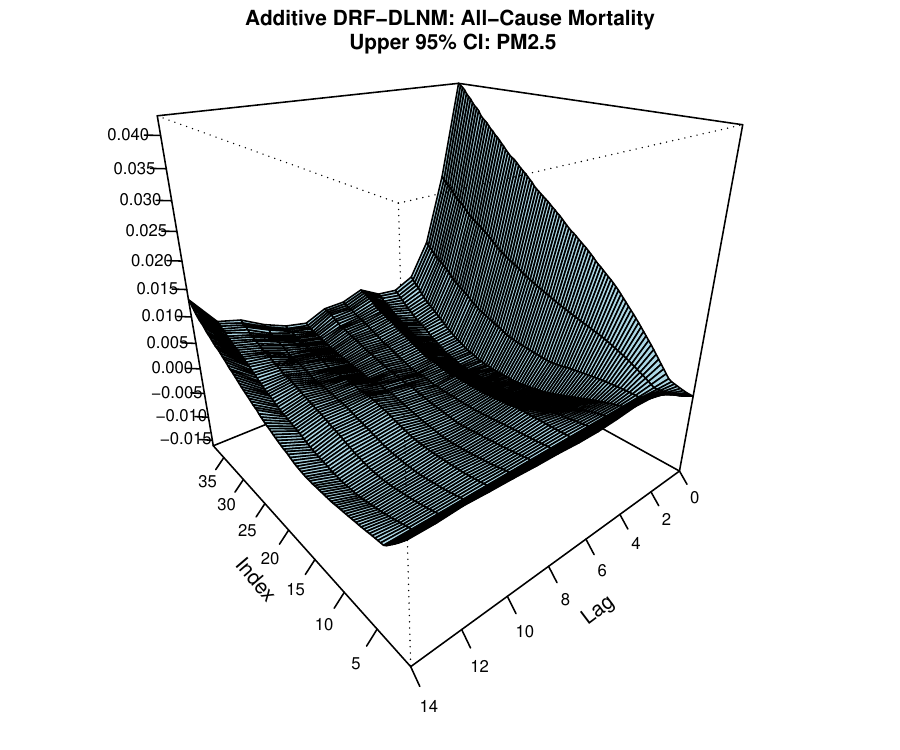}
  \end{subfigure}
  \caption{Estimated $\psi$ from the additive DRF-DLNM for PM$_{2.5}$ and all-cause mortality.}
\end{figure}

\begin{figure}[H]
    \centering
    \includegraphics[width=\linewidth]{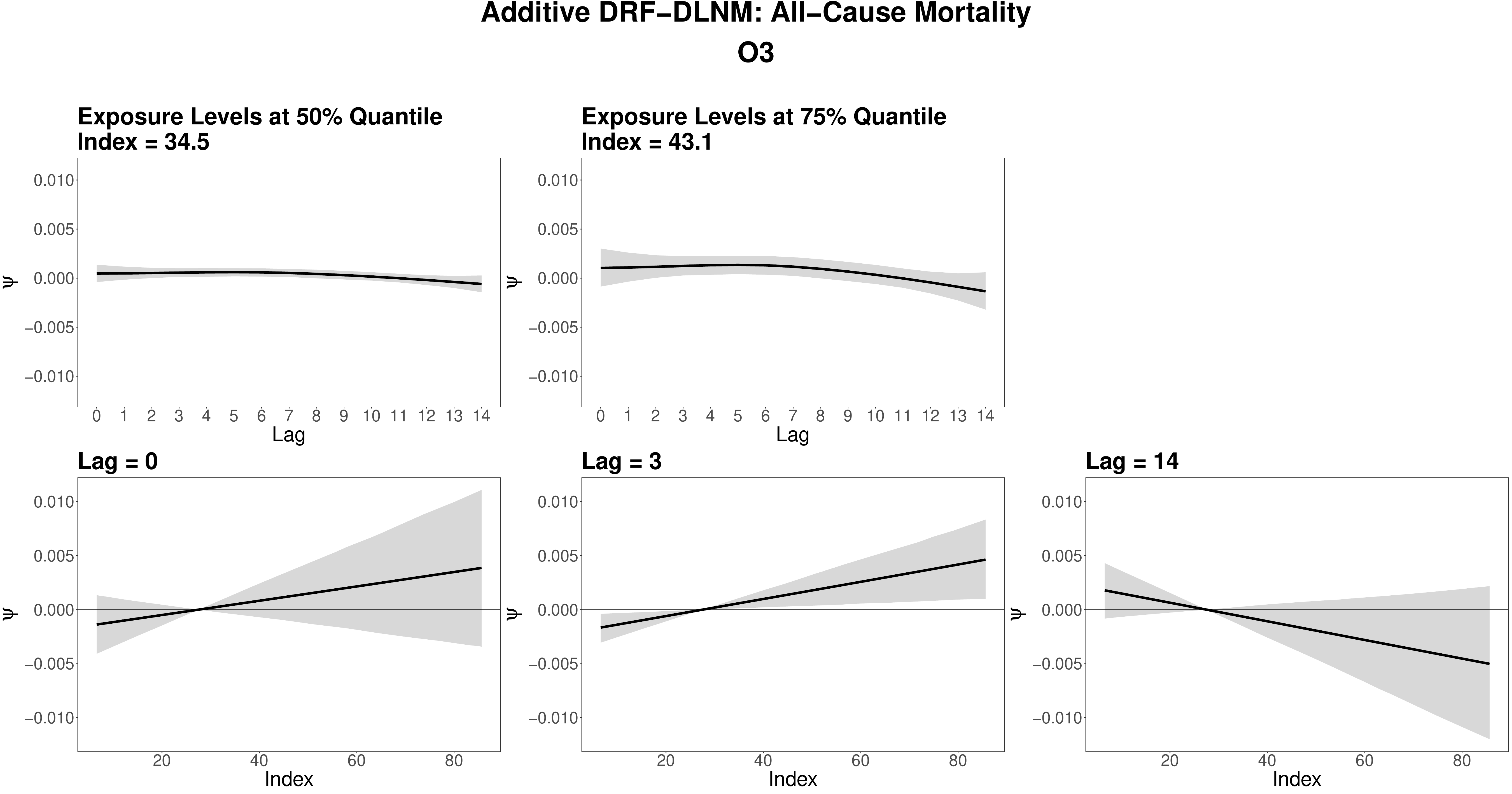}
    \caption{Estimated index weights and functions from the additive DRF-DLNM for O$_3$ and all-cause mortality. The function $\psi$ is shown in selected slices of the bivariate surface $\psi$, by fixing the index at estimated values under the exposures are at their 25\% and 75\% quantiles, and fixing the lag at 0, 3, and 14. }
\end{figure}

\begin{figure}[H]
  \centering
  \makebox[\textwidth][c]{
  \begin{subfigure}[t]{0.33\textwidth}
    \centering
    \includegraphics[width=\linewidth]{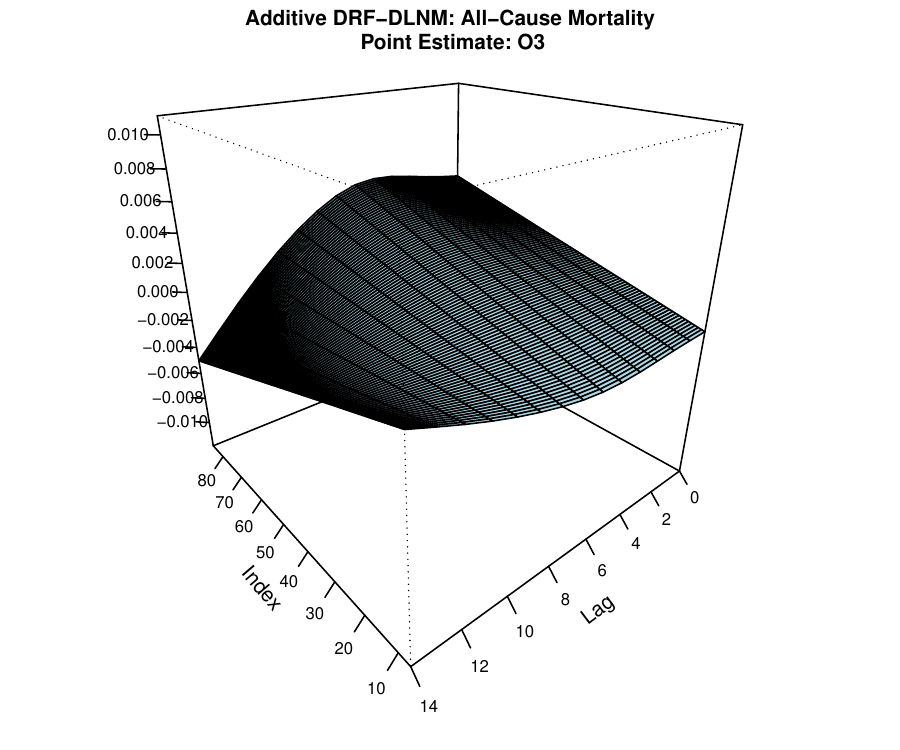}
  \end{subfigure}
  }
  \vspace{0.5em}
  
  \begin{subfigure}[t]{0.33\textwidth}
    \centering
    \includegraphics[width=\linewidth]{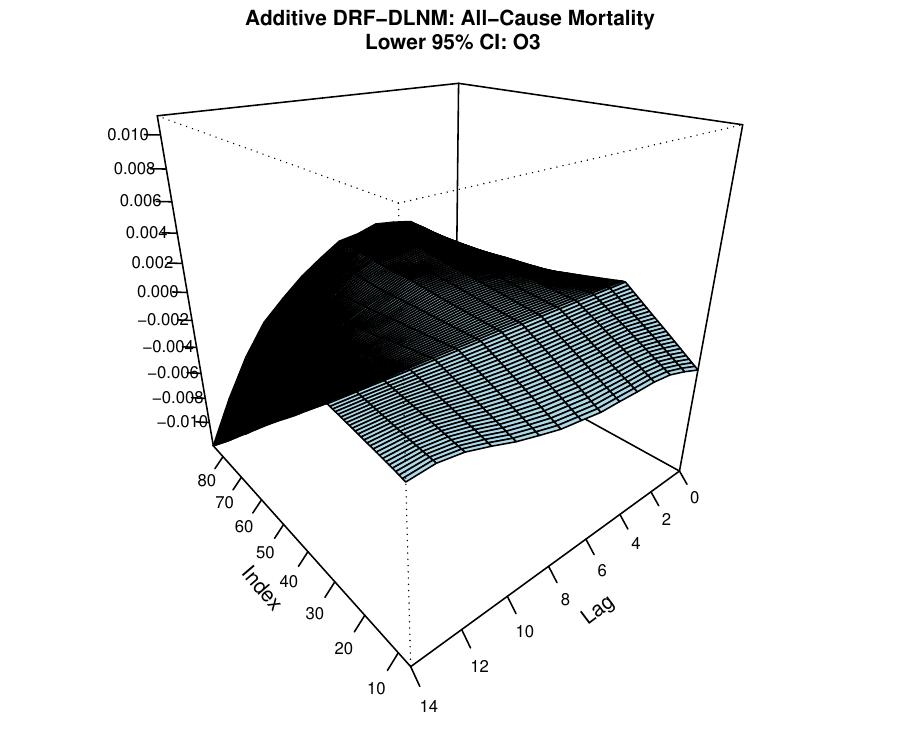}
  \end{subfigure}
  \begin{subfigure}[t]{0.33\textwidth}
    \centering
    \includegraphics[width=\linewidth]{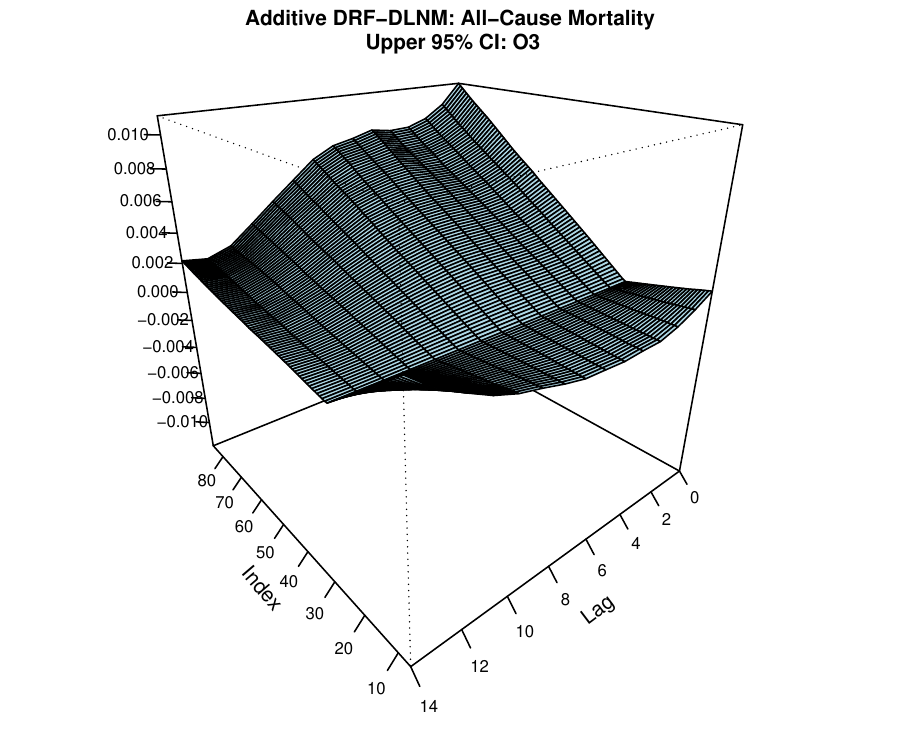}
  \end{subfigure}
  \caption{Estimated $\psi$ from the additive DRF-DLNM for O$_3$ and all-cause mortality.}
\end{figure}

\begin{figure}[H]
    \centering
    \includegraphics[width=\linewidth]{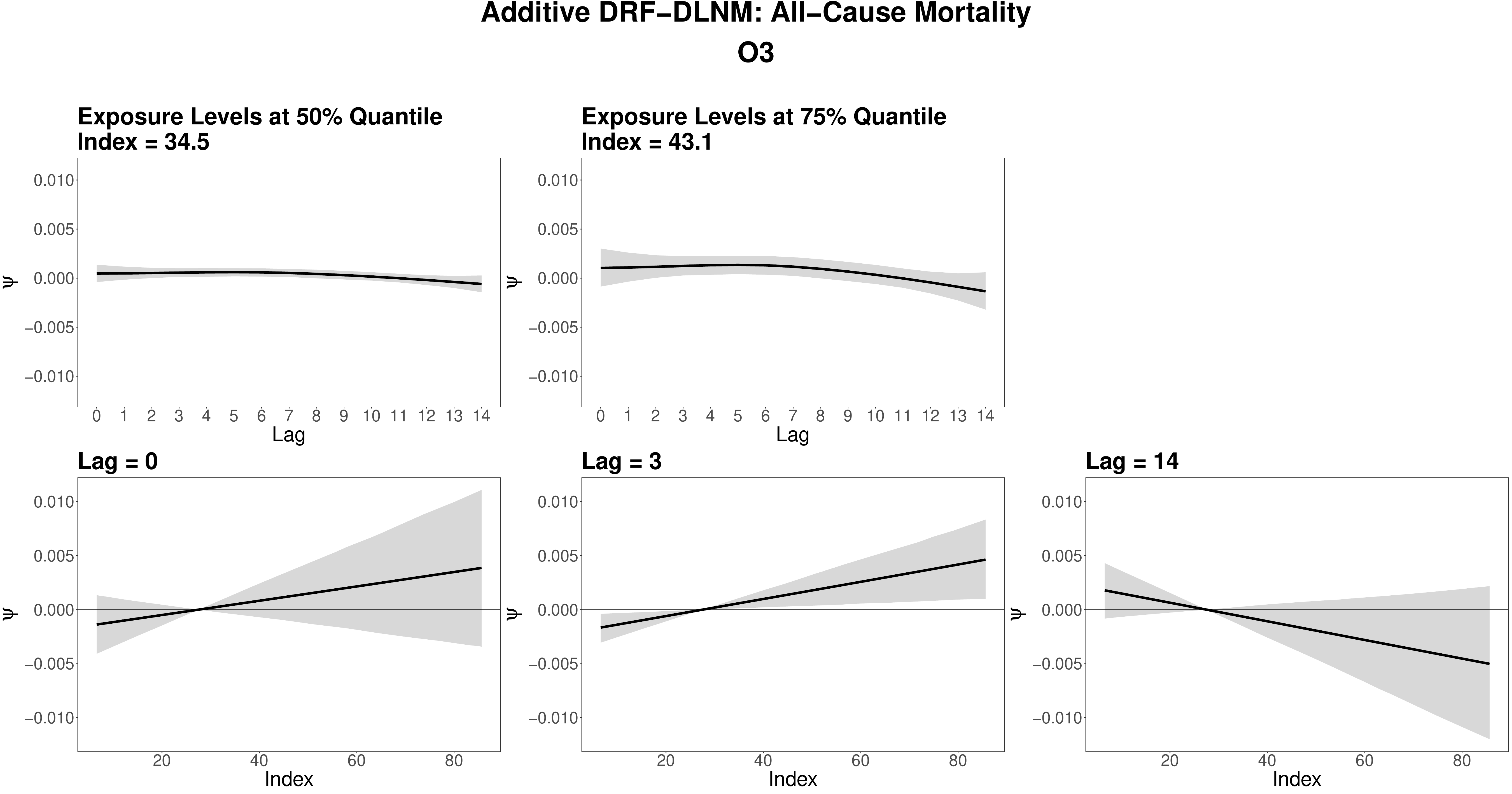}
    \caption{Estimated index weights and functions from the additive DRF-DLNM for O$_3$ and all-cause mortality. The function $\psi$ is shown in selected slices of the bivariate surface $\psi$, by fixing the index at estimated values under the exposures are at their 25\% and 75\% quantiles, and fixing the lag at 0, 3, and 14. }
\end{figure}

\begin{figure}[H]
  \centering
  \makebox[\textwidth][c]{
  \begin{subfigure}[t]{0.33\textwidth}
    \centering
    \includegraphics[width=\linewidth]{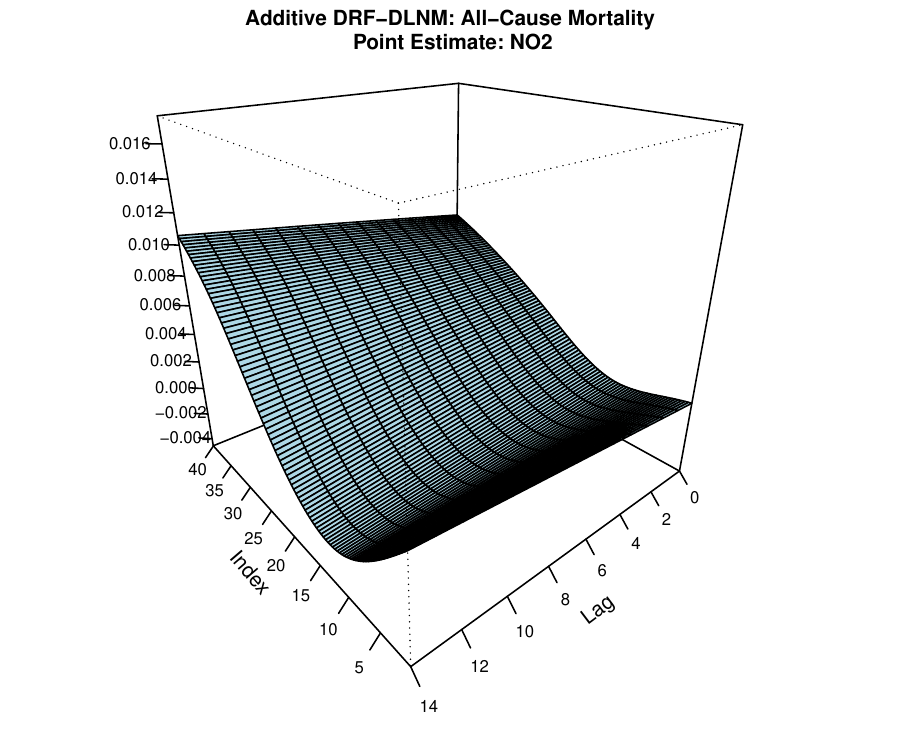}
  \end{subfigure}
  }
  \vspace{0.5em}
  
  \begin{subfigure}[t]{0.33\textwidth}
    \centering
    \includegraphics[width=\linewidth]{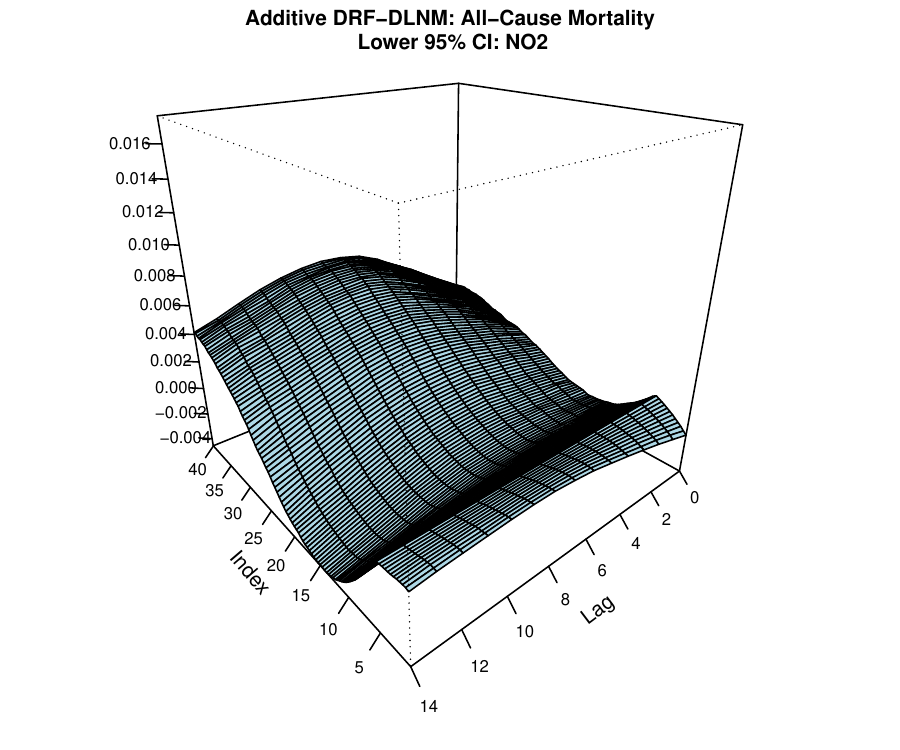}
  \end{subfigure}
  \begin{subfigure}[t]{0.33\textwidth}
    \centering
    \includegraphics[width=\linewidth]{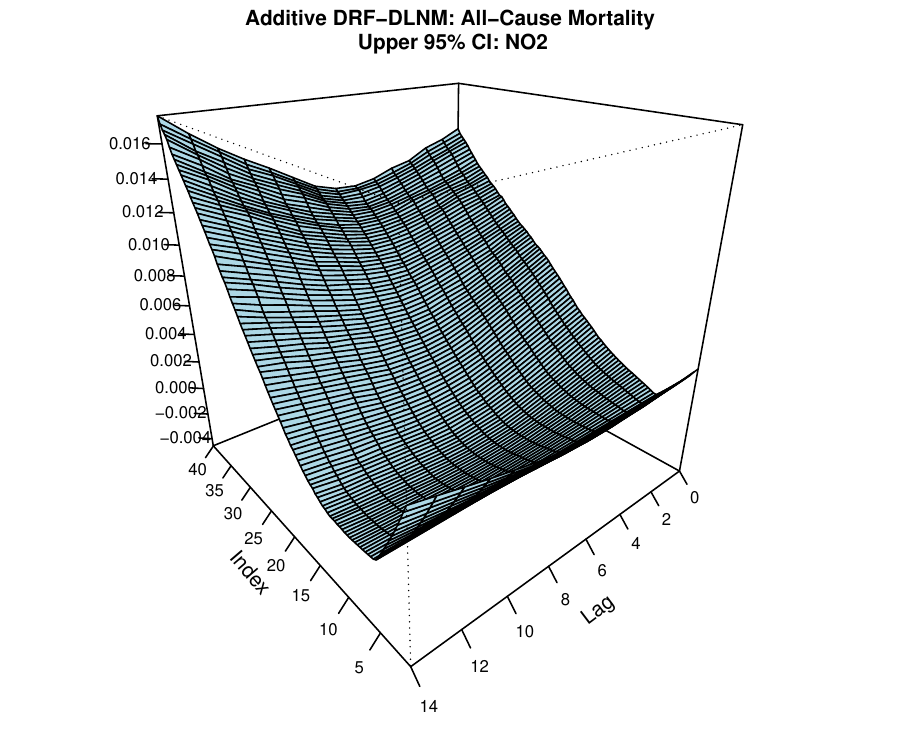}
  \end{subfigure}
  \caption{Estimated $\psi$ from the additive DRF-DLNM for NO$_2$ and all-cause mortality.}
\end{figure}

\subsection{Additional Results}

\subsubsection{Respiratory Mortality}
\begin{figure}[H]
  \centering
  \begin{subfigure}[t]{0.32\textwidth}
    \includegraphics[width=\linewidth]{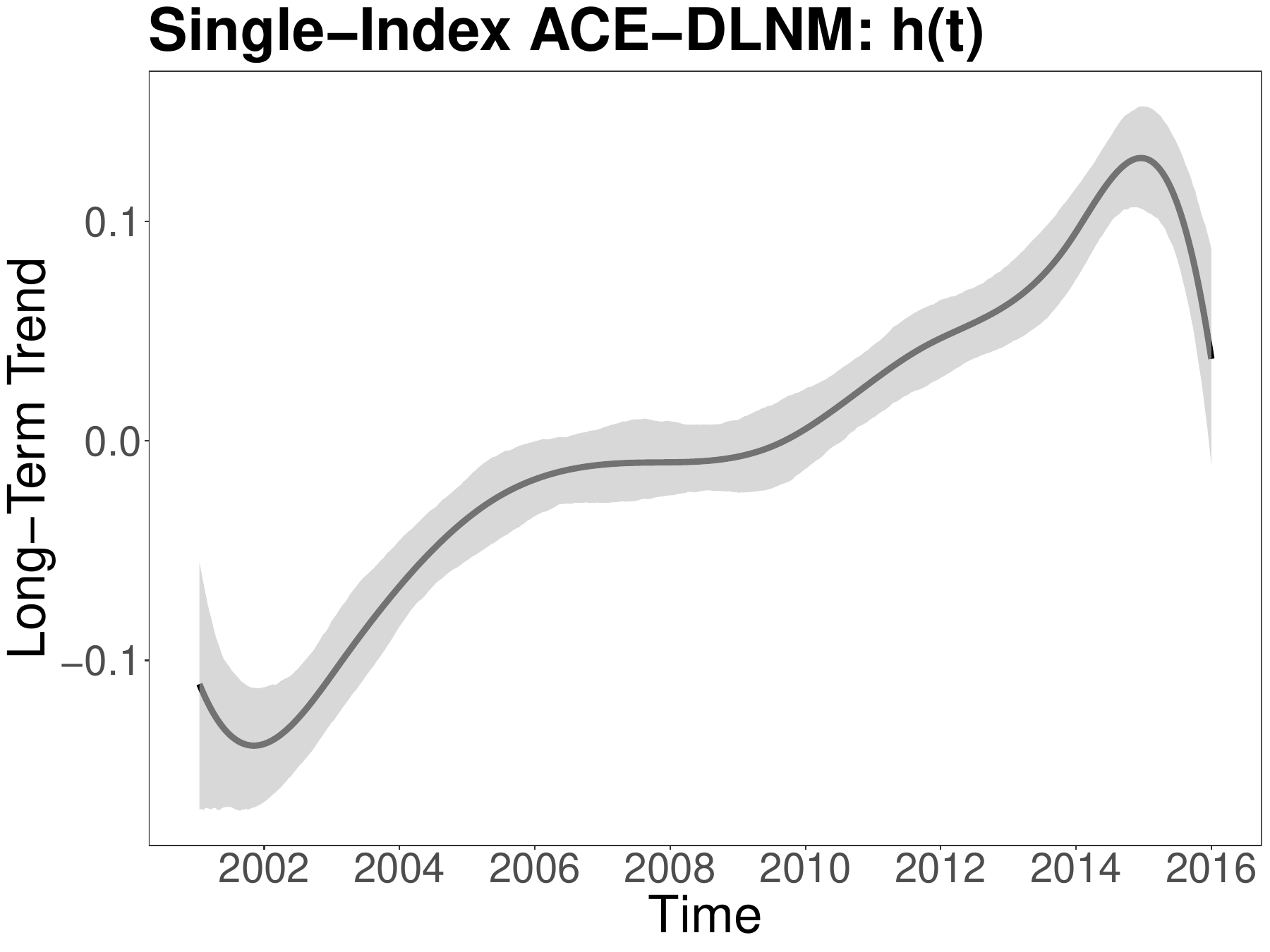}
  \end{subfigure}
  \begin{subfigure}[t]{0.32\textwidth}
    \includegraphics[width=\linewidth]{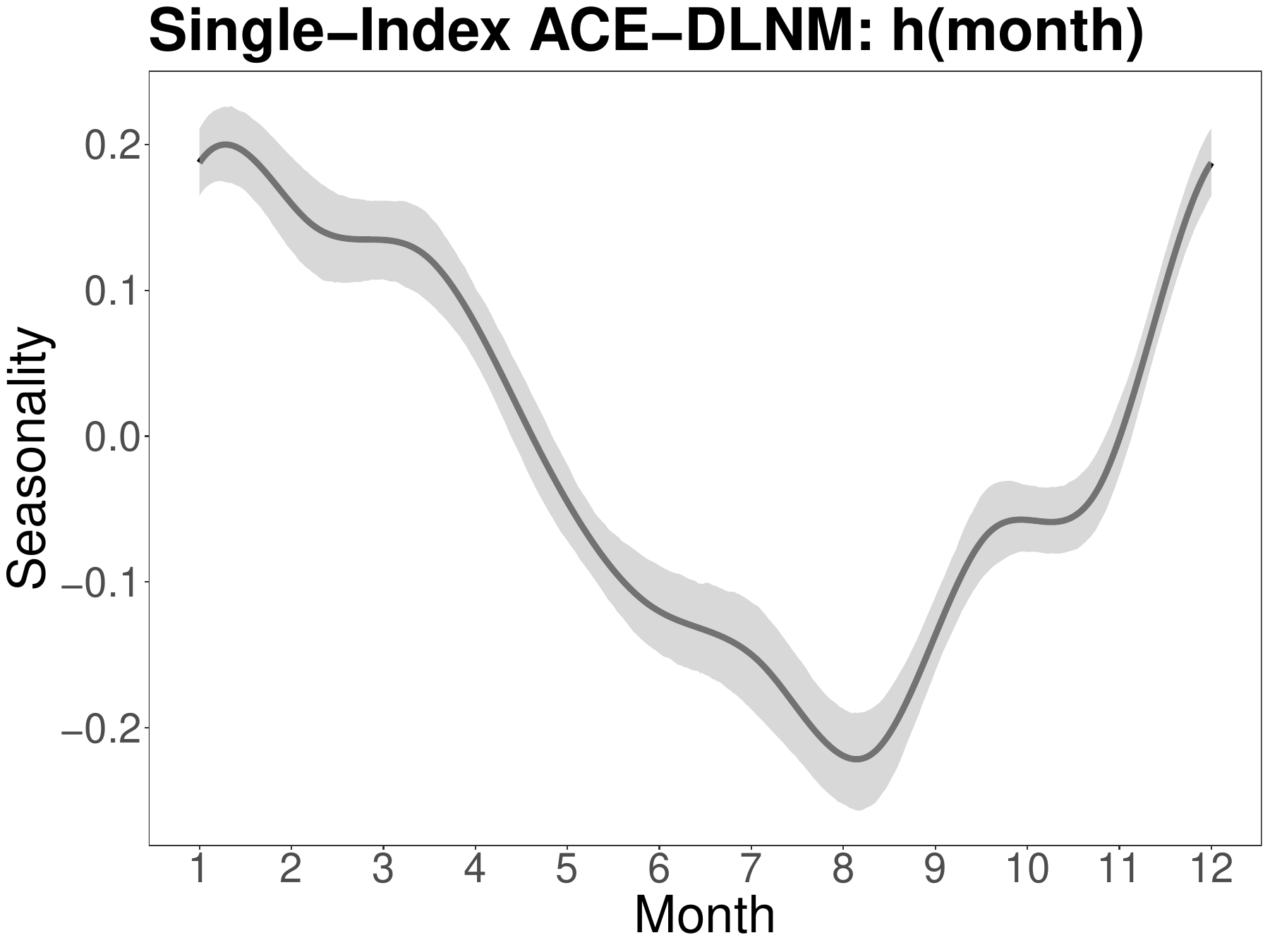}
  \end{subfigure}
  \begin{subfigure}[t]{0.32\textwidth}
    \includegraphics[width=\linewidth]{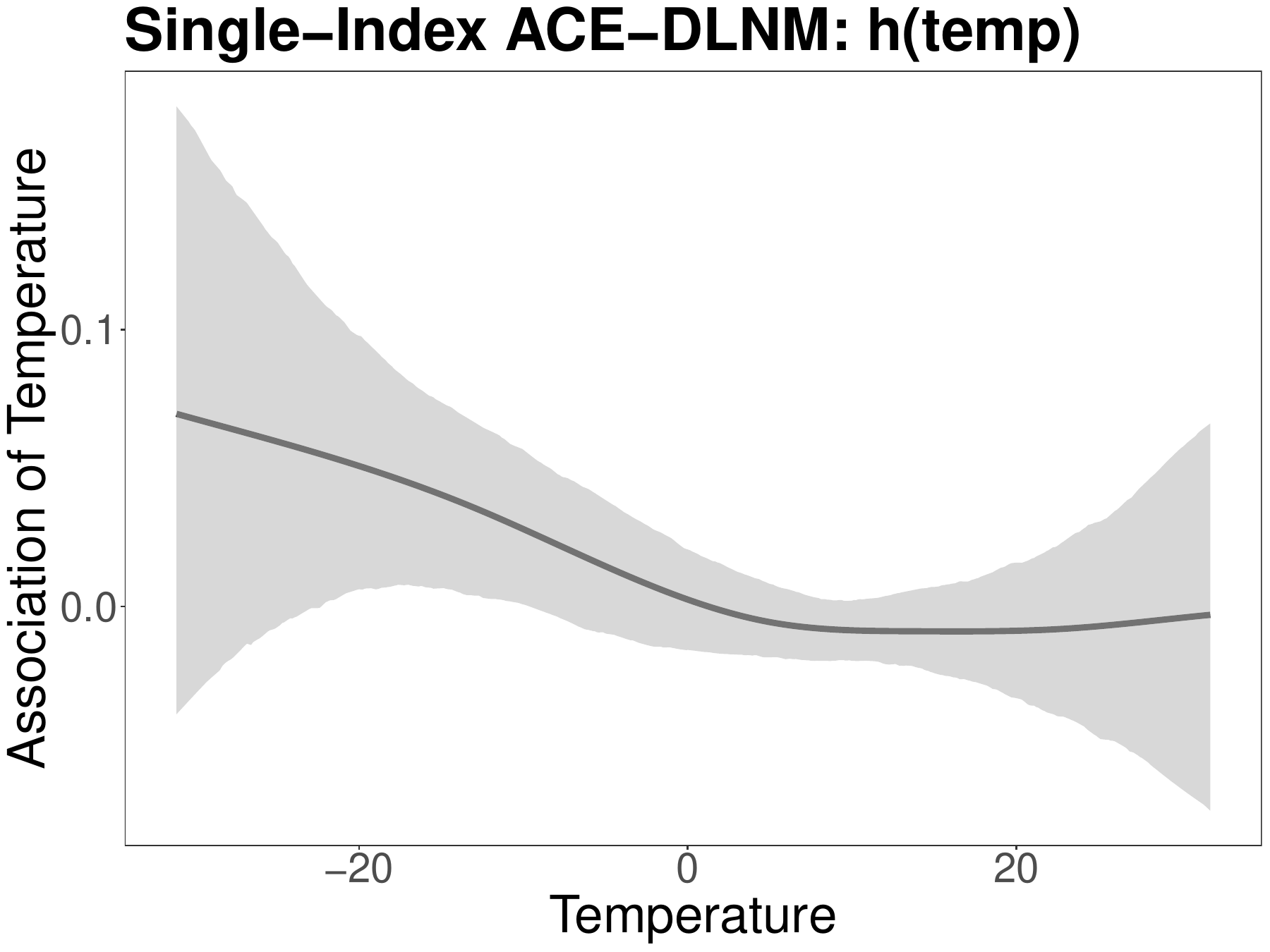}
  \end{subfigure}

  \medskip

  \begin{subfigure}[t]{0.32\textwidth}
    \includegraphics[width=\linewidth]{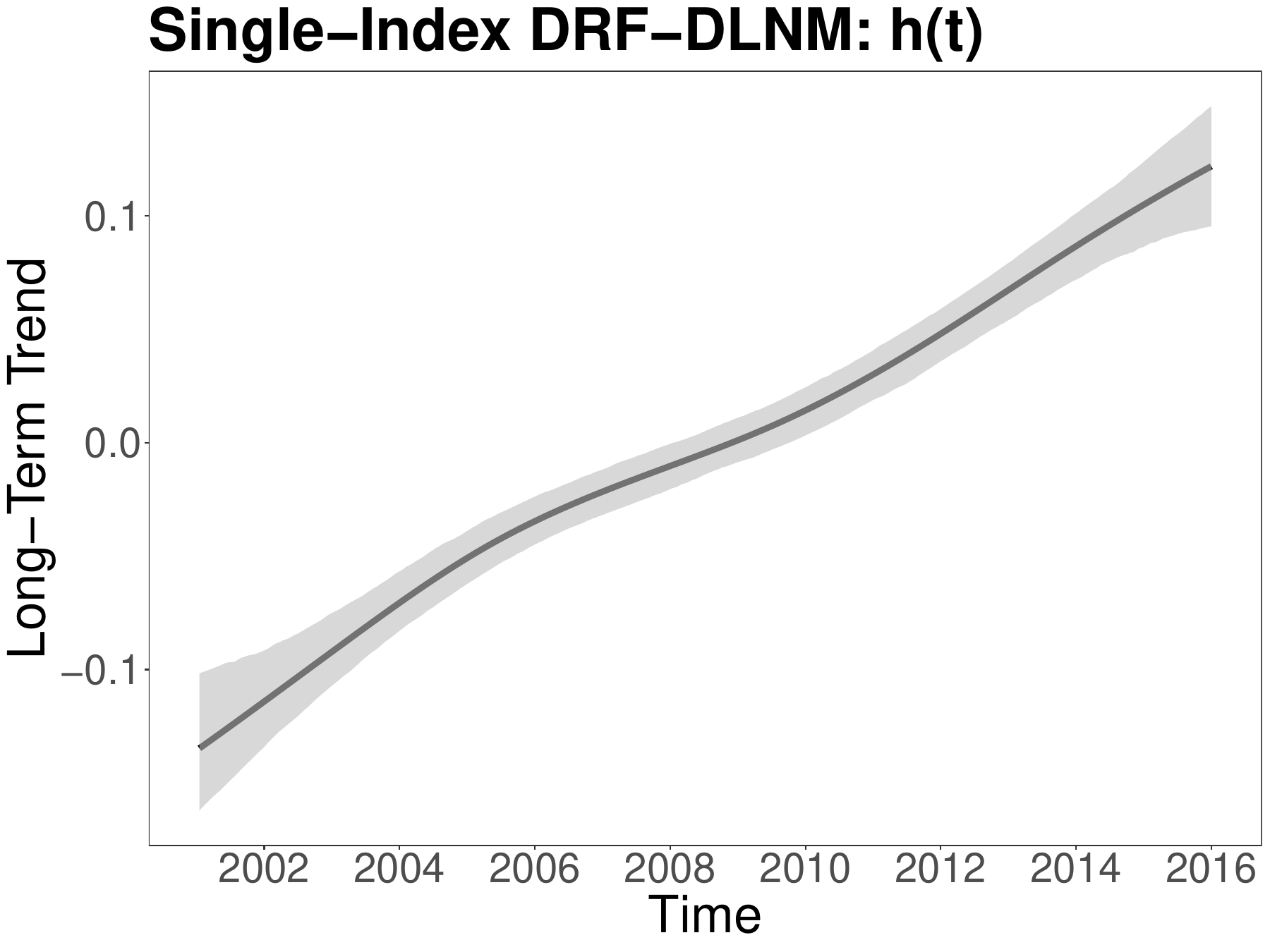}
  \end{subfigure}
  \begin{subfigure}[t]{0.32\textwidth}
    \includegraphics[width=\linewidth]{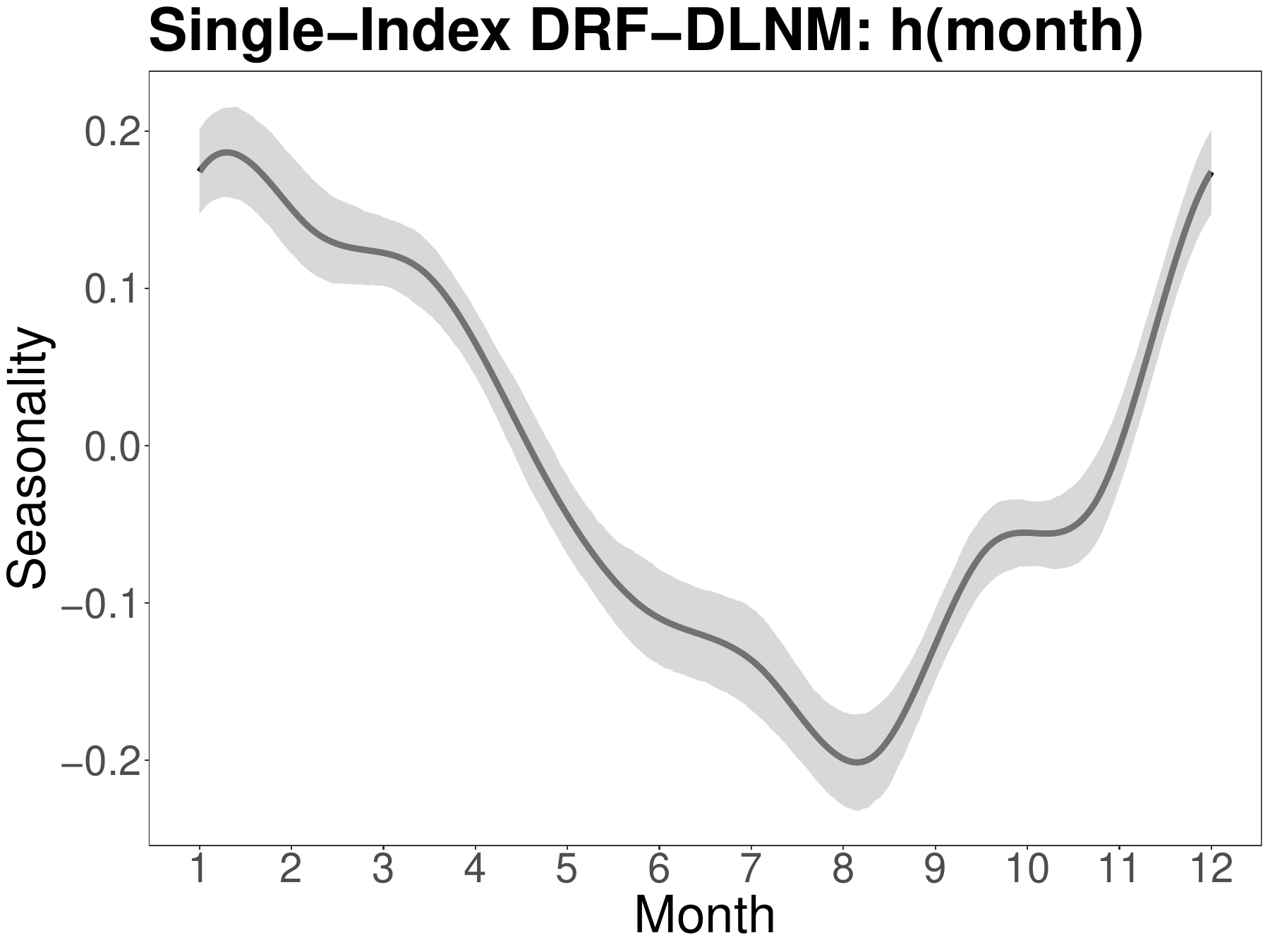}
  \end{subfigure}
  \begin{subfigure}[t]{0.32\textwidth}
    \includegraphics[width=\linewidth]{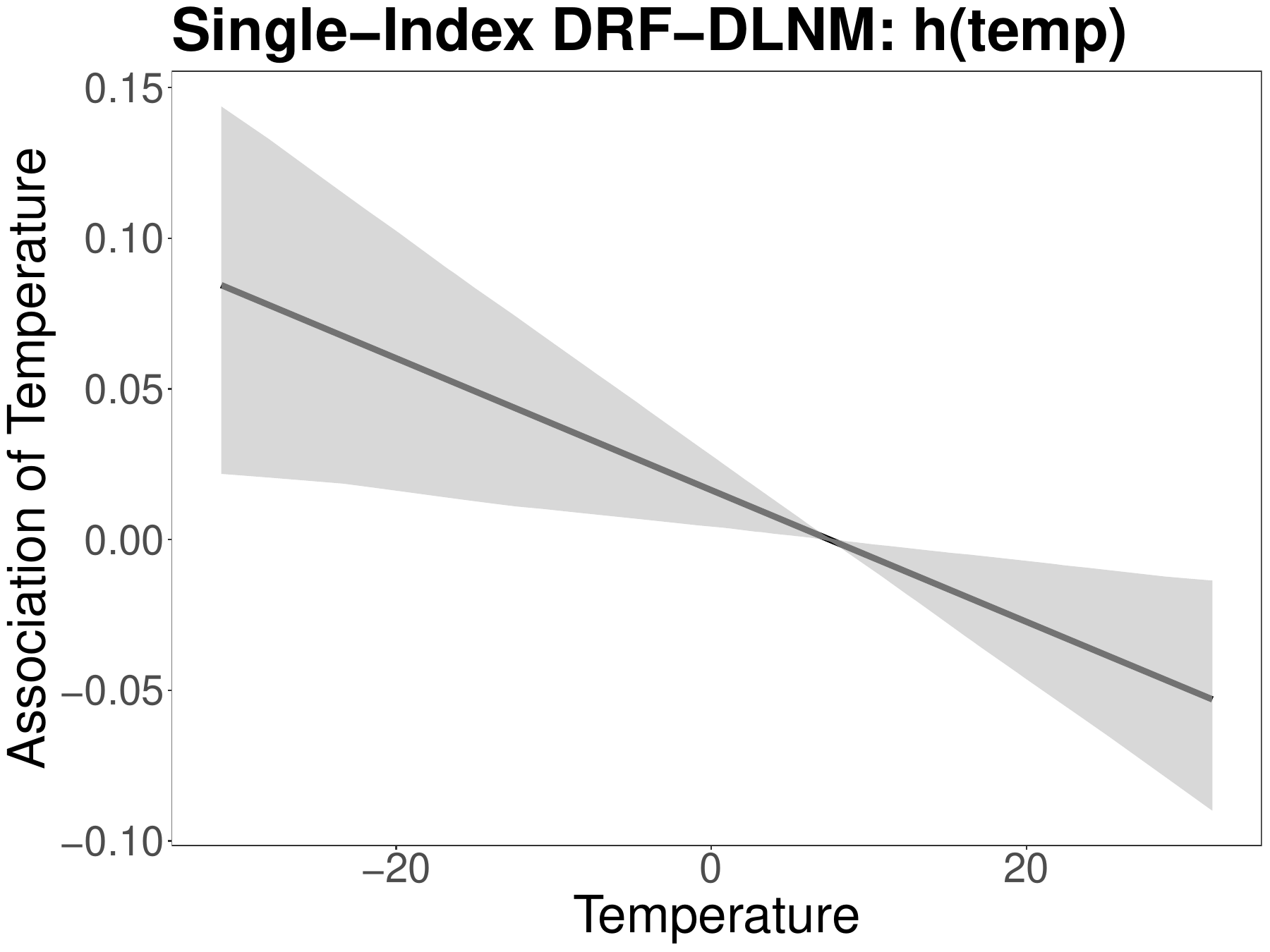}
  \end{subfigure}
  
  \medskip

  \begin{subfigure}[t]{0.32\textwidth}
    \includegraphics[width=\linewidth]{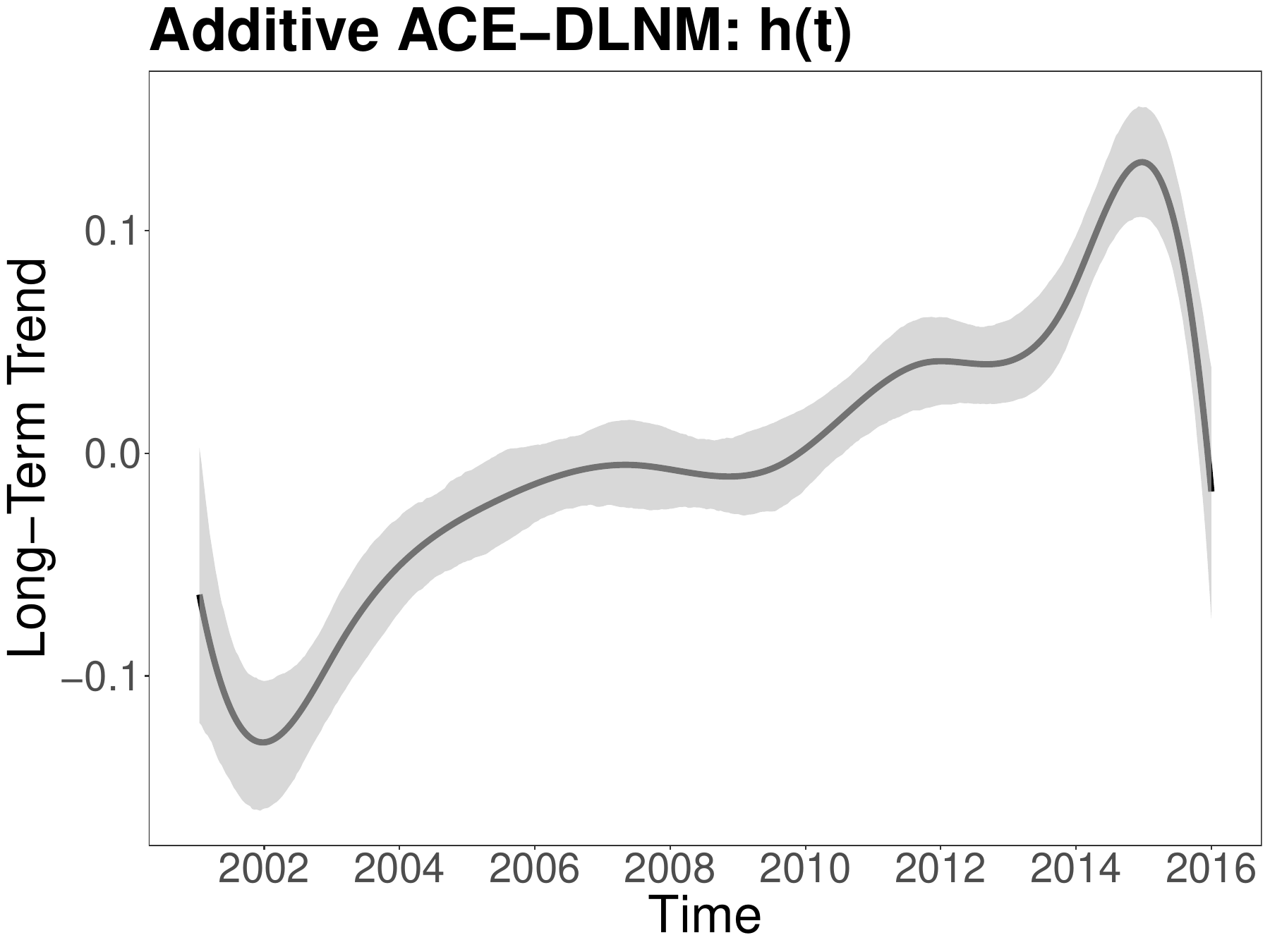}
  \end{subfigure}
  \begin{subfigure}[t]{0.32\textwidth}
    \includegraphics[width=\linewidth]{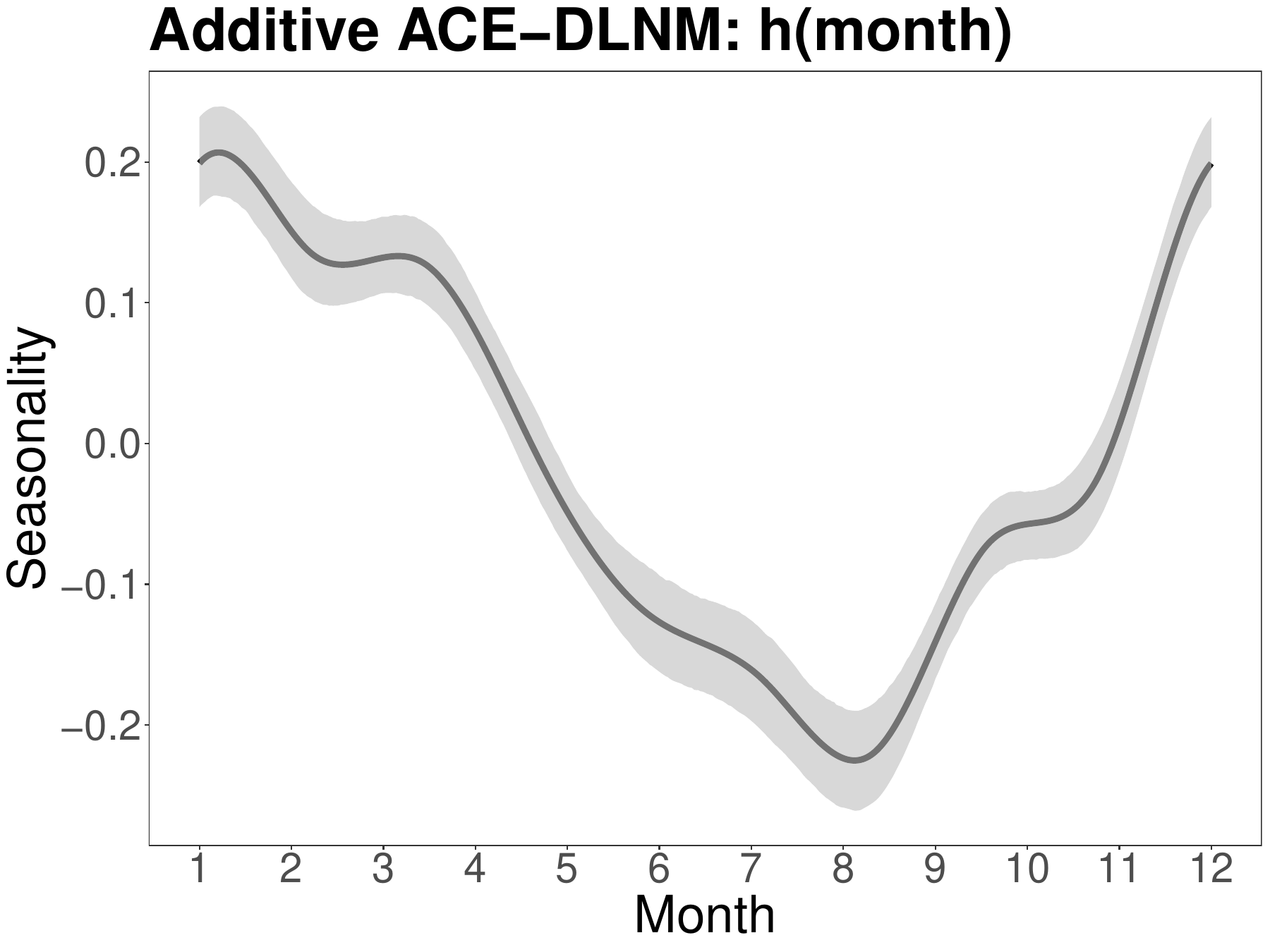}
  \end{subfigure}
  \begin{subfigure}[t]{0.32\textwidth}
    \includegraphics[width=\linewidth]{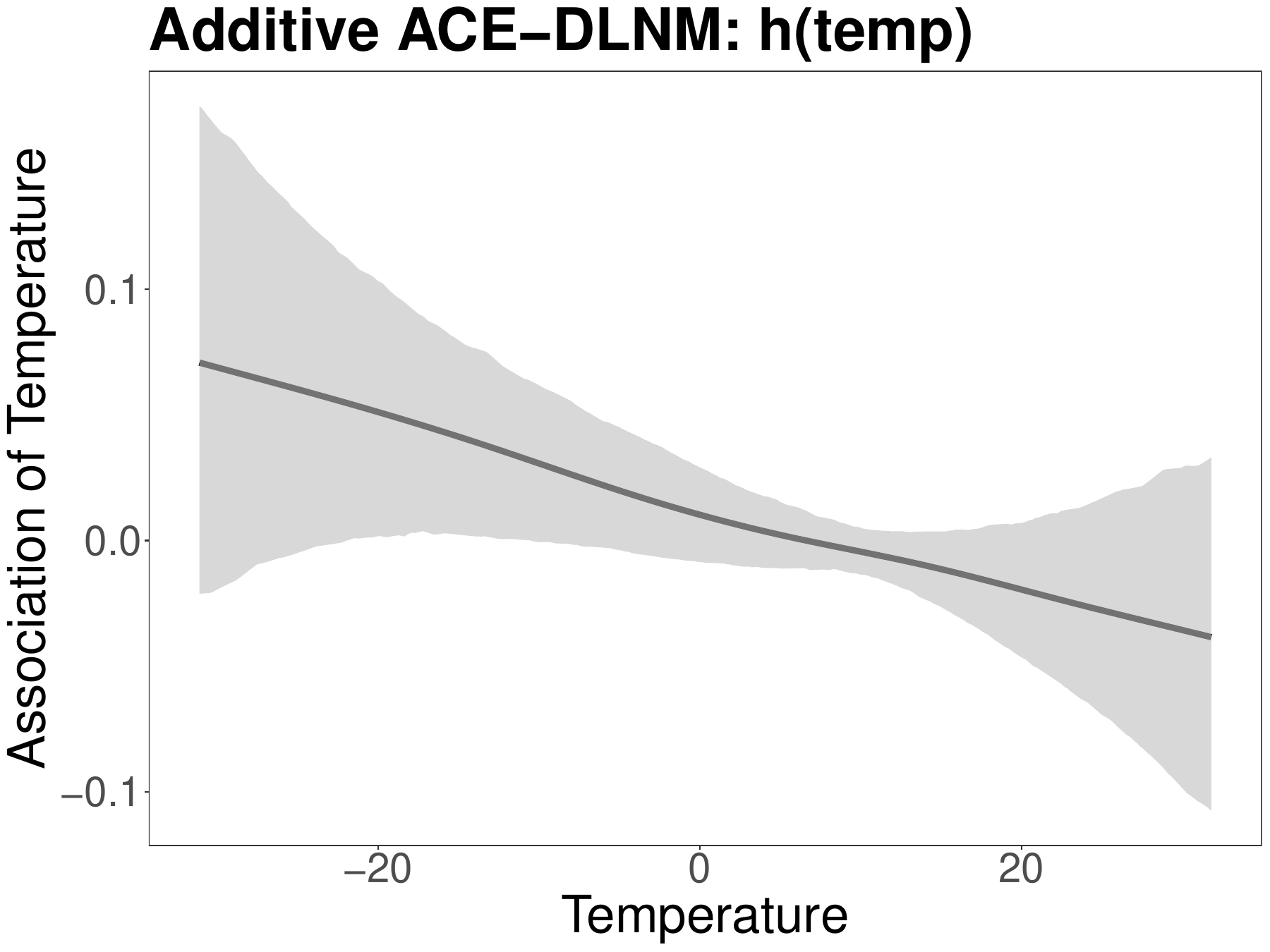}
  \end{subfigure}

  \medskip

  \begin{subfigure}[t]{0.32\textwidth}
    \includegraphics[width=\linewidth]{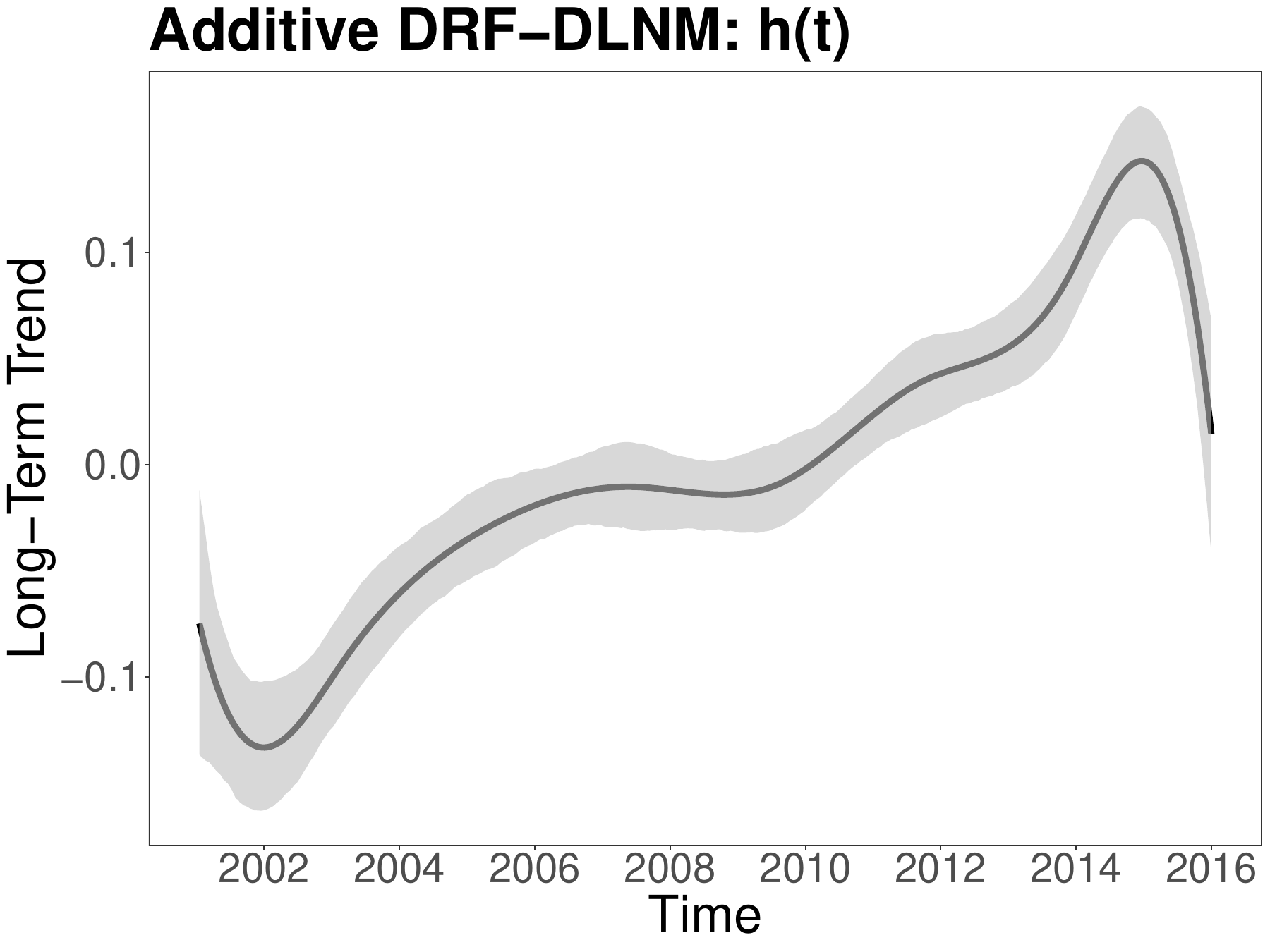}
  \end{subfigure}
  \begin{subfigure}[t]{0.32\textwidth}
    \includegraphics[width=\linewidth]{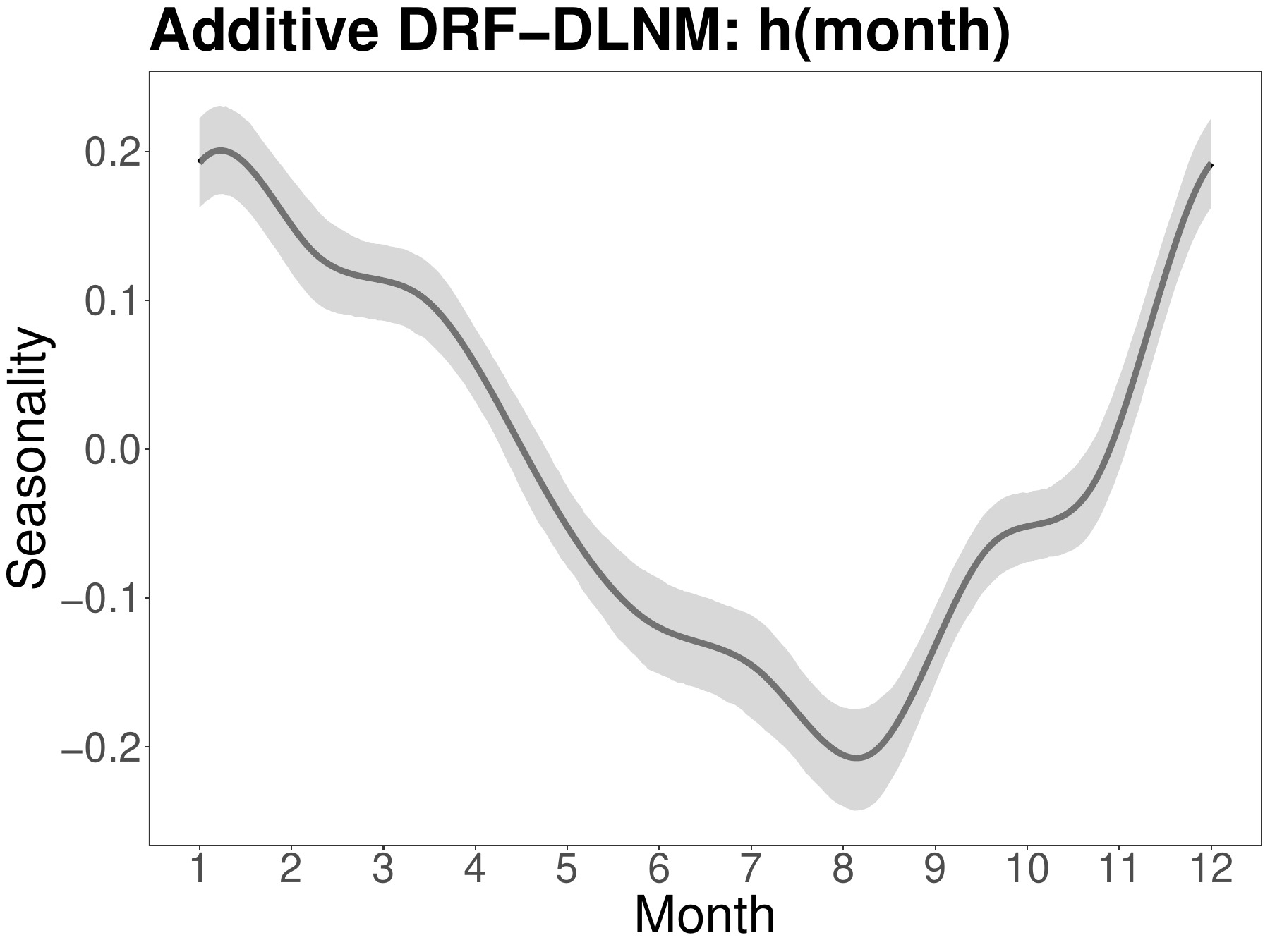}
  \end{subfigure}
  \begin{subfigure}[t]{0.32\textwidth}
    \includegraphics[width=\linewidth]{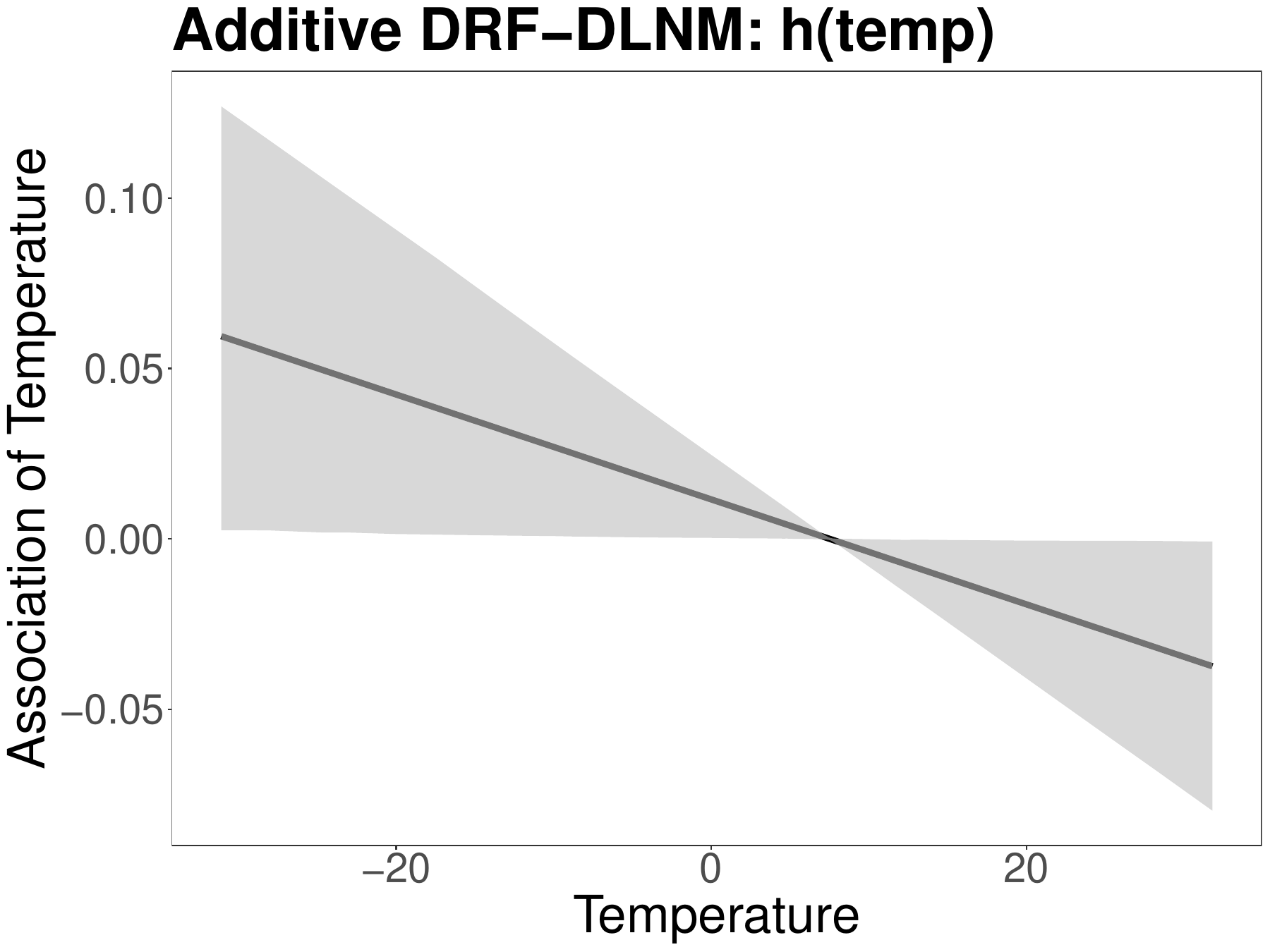}
  \end{subfigure}

  \caption{Estimated covariate functions from the single-index ACE-DLNM (first row),  the single-index DRf-DLNM (second row), the additive ACE-DLNM (third row) and the additive DRF-DLNM (fourth row), for respiratory mortality. The covariate functions include the long-term trend $h_1(t)$ (first column), the seasonality $h_2(\text{Month})$ and the temperature association $h_3(\text{Temp})$.}
\end{figure}

\subsubsection{Circulatory Mortality}
\begin{figure}[H]
  \centering
  \begin{subfigure}[t]{0.32\textwidth}
    \includegraphics[width=\linewidth]{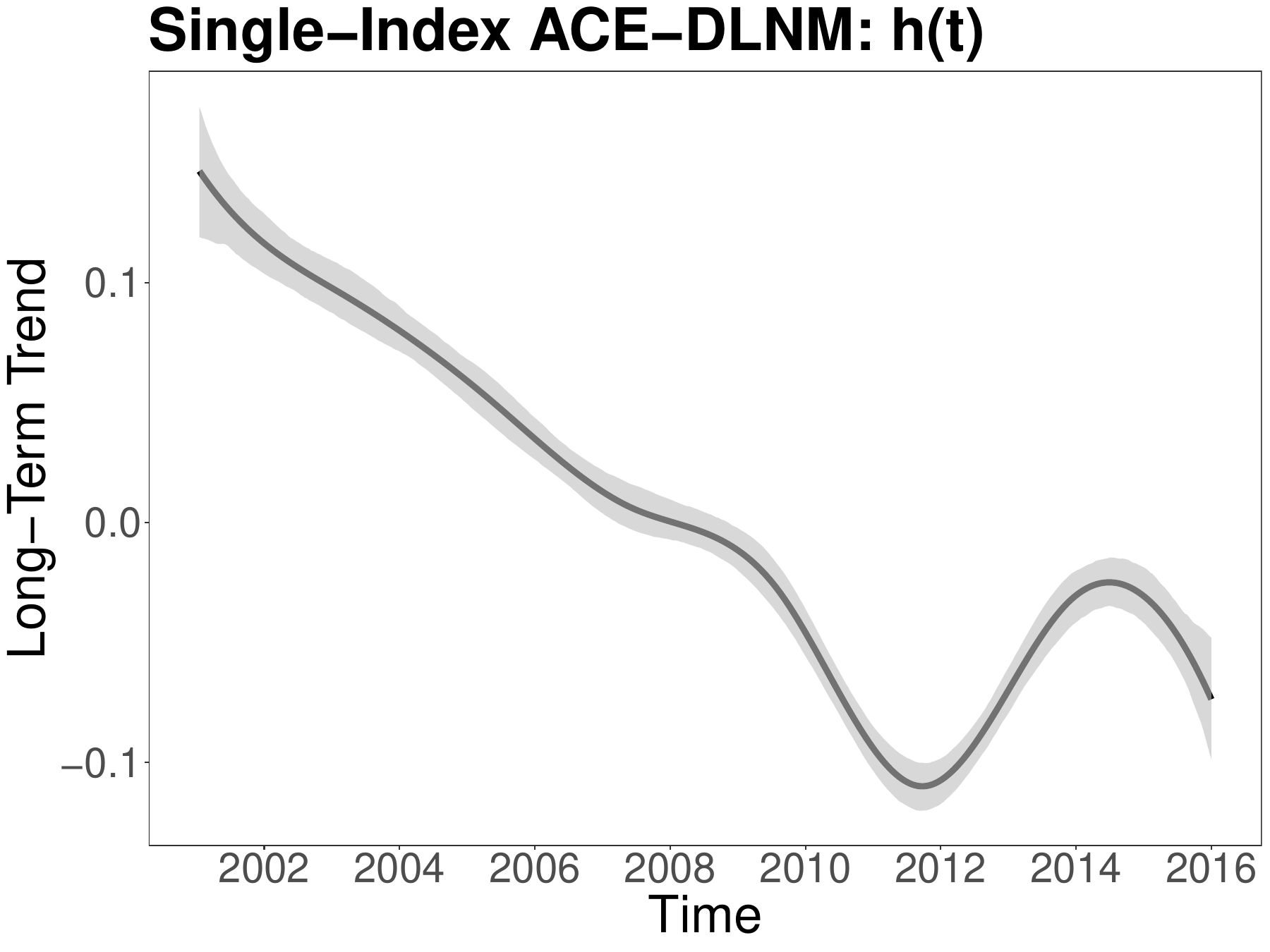}
  \end{subfigure}
  \begin{subfigure}[t]{0.32\textwidth}
    \includegraphics[width=\linewidth]{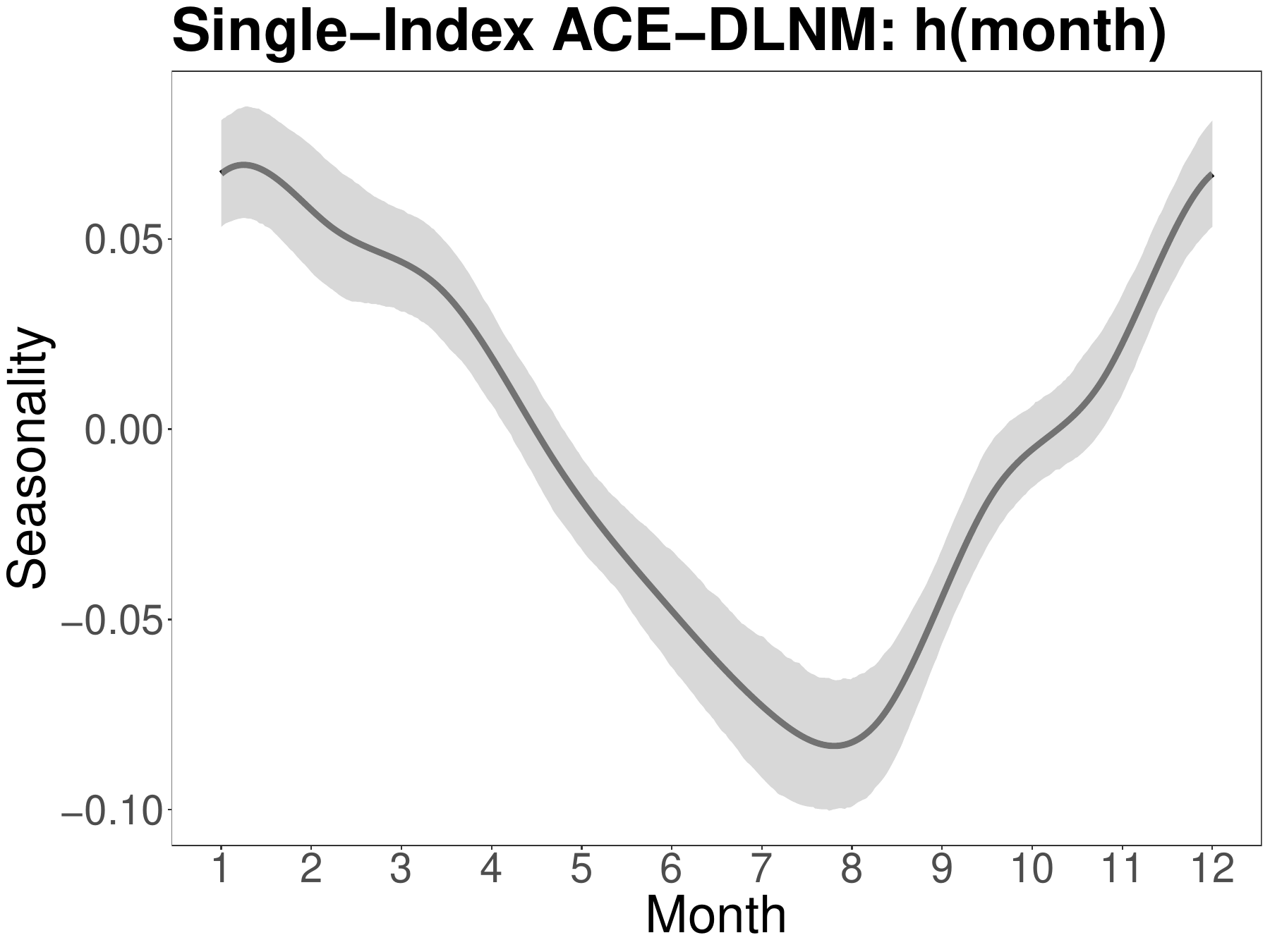}
  \end{subfigure}
  \begin{subfigure}[t]{0.32\textwidth}
    \includegraphics[width=\linewidth]{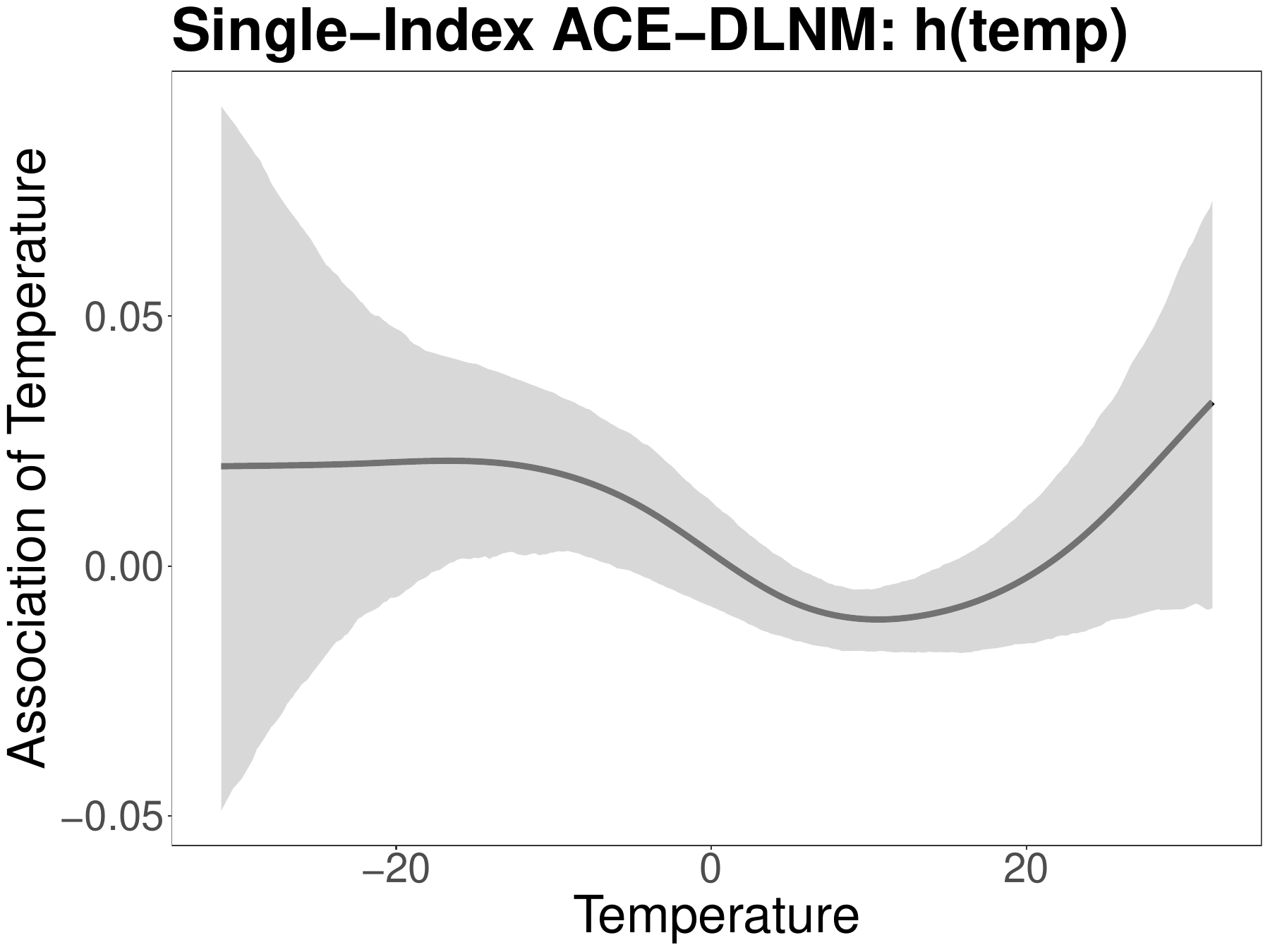}
  \end{subfigure}

  \medskip

  \begin{subfigure}[t]{0.32\textwidth}
    \includegraphics[width=\linewidth]{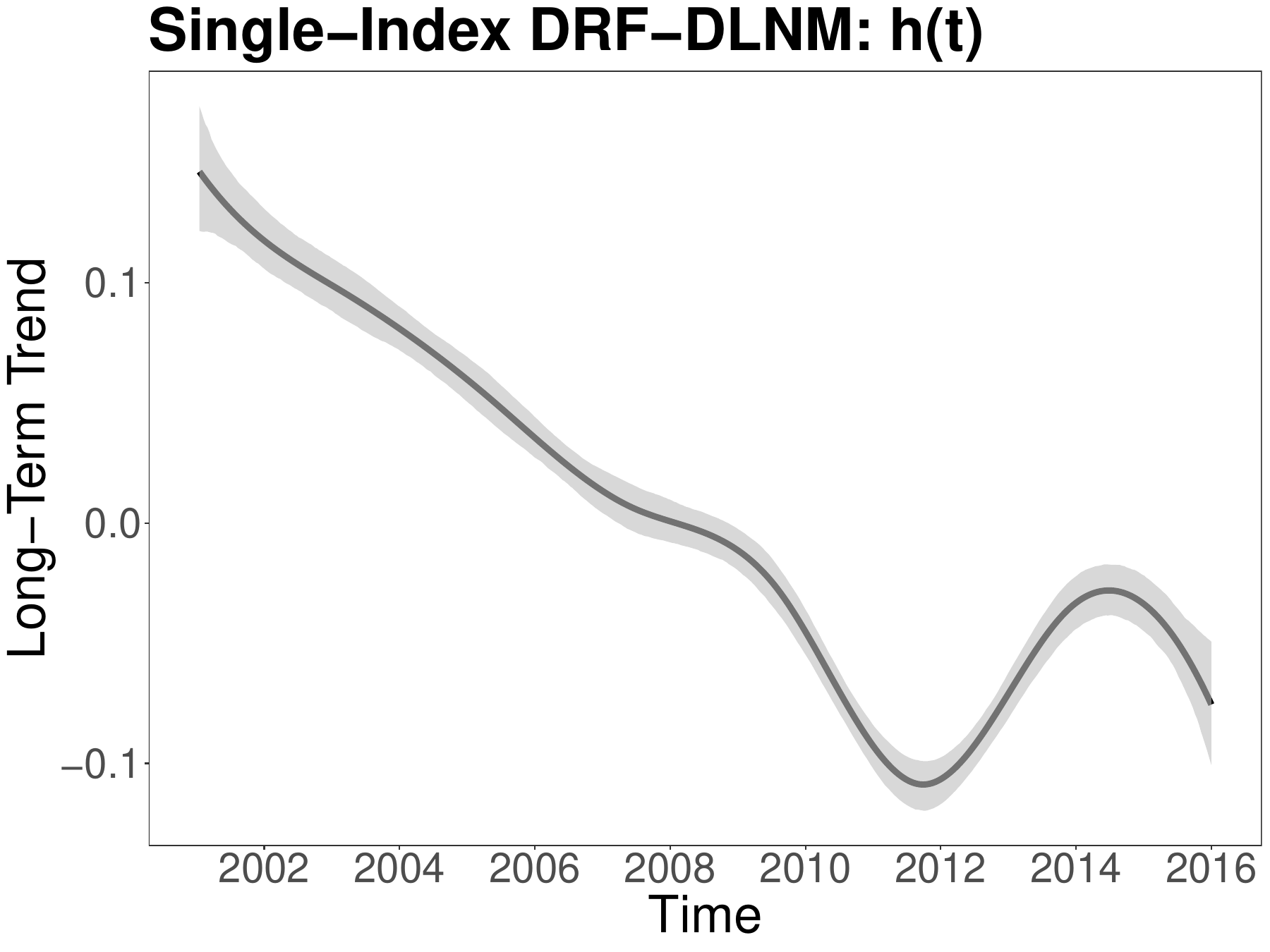}
  \end{subfigure}
  \begin{subfigure}[t]{0.32\textwidth}
    \includegraphics[width=\linewidth]{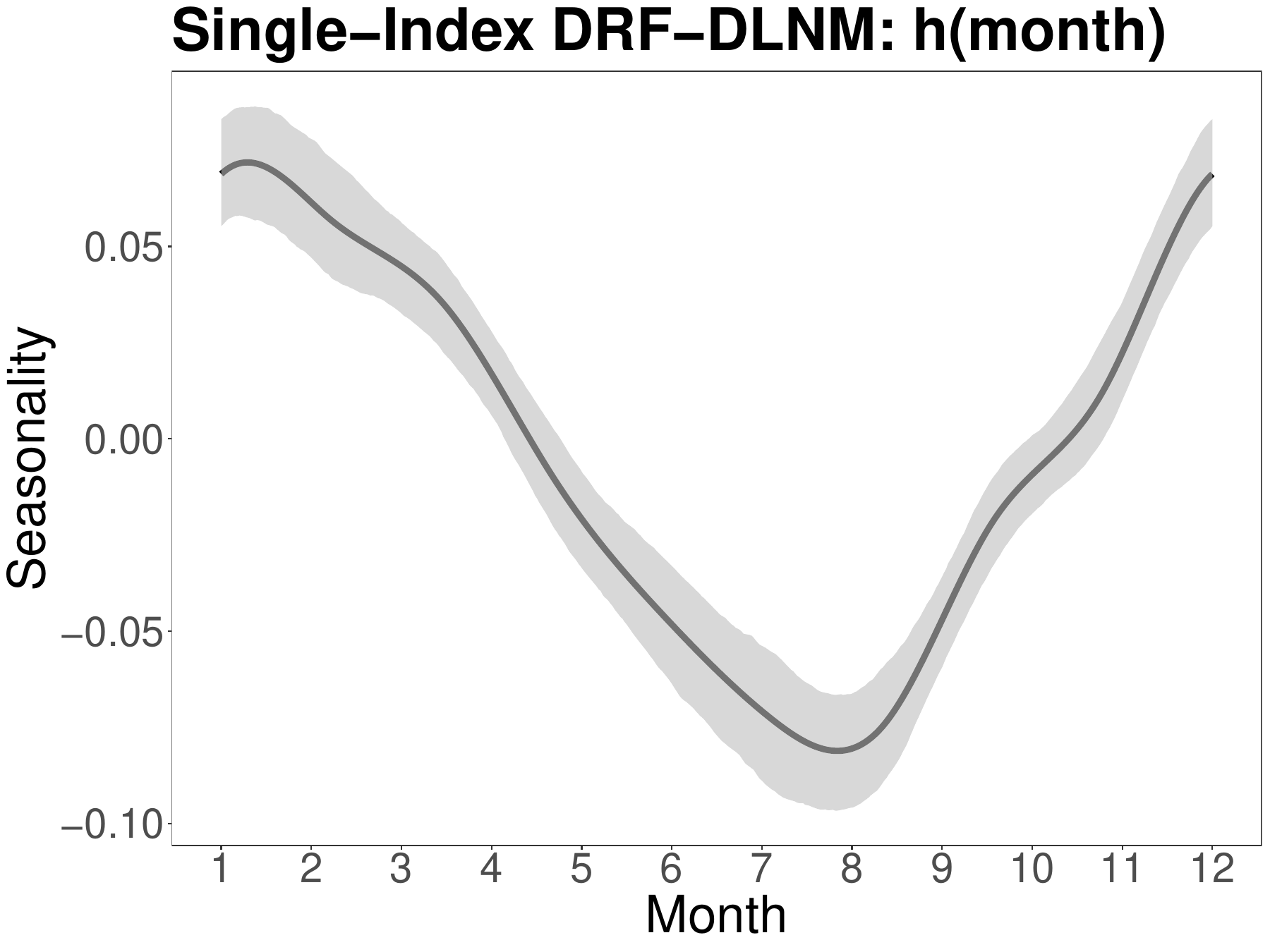}
  \end{subfigure}
  \begin{subfigure}[t]{0.32\textwidth}
    \includegraphics[width=\linewidth]{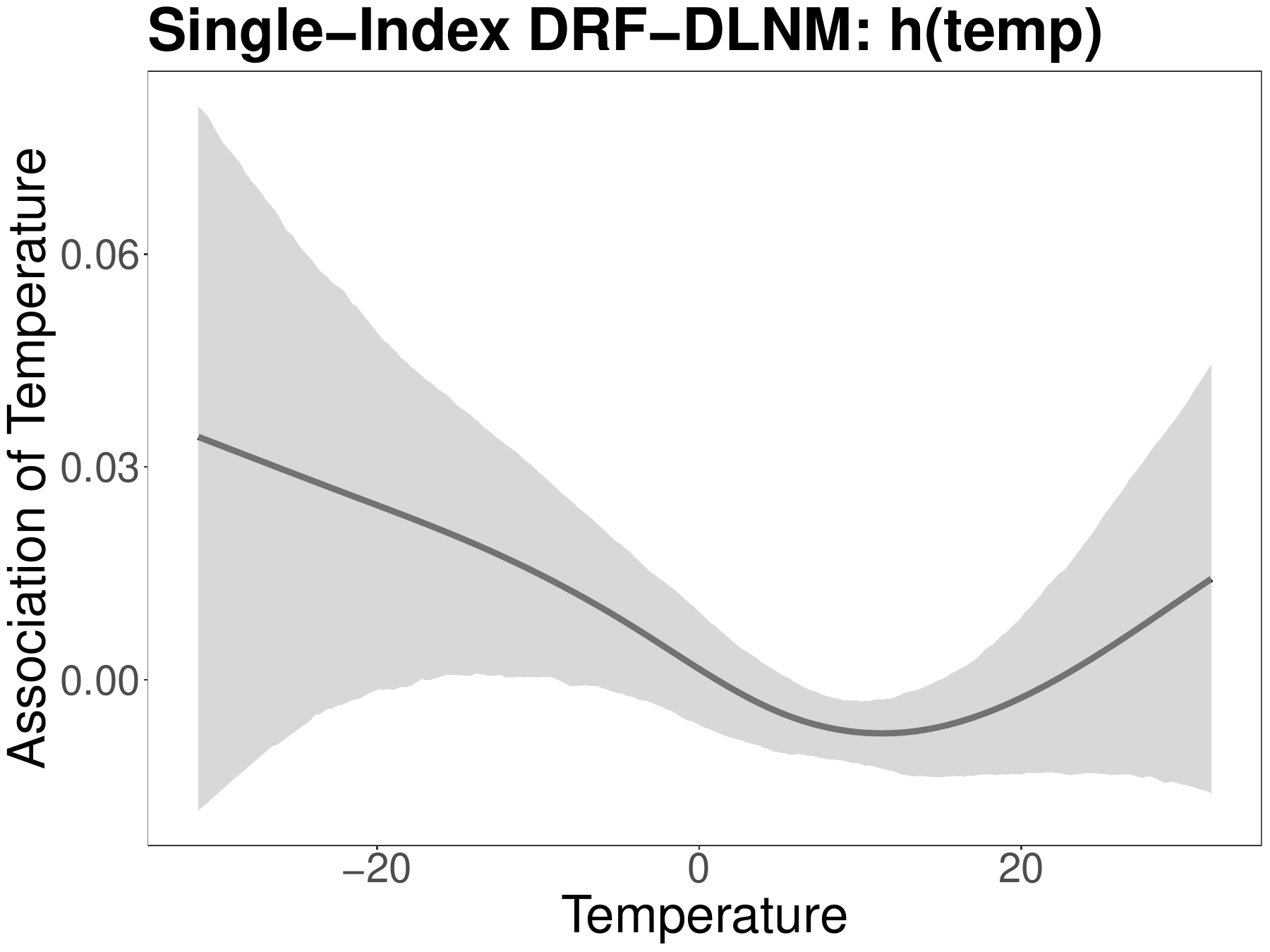}
  \end{subfigure}
  
  \medskip

  \begin{subfigure}[t]{0.32\textwidth}
    \includegraphics[width=\linewidth]{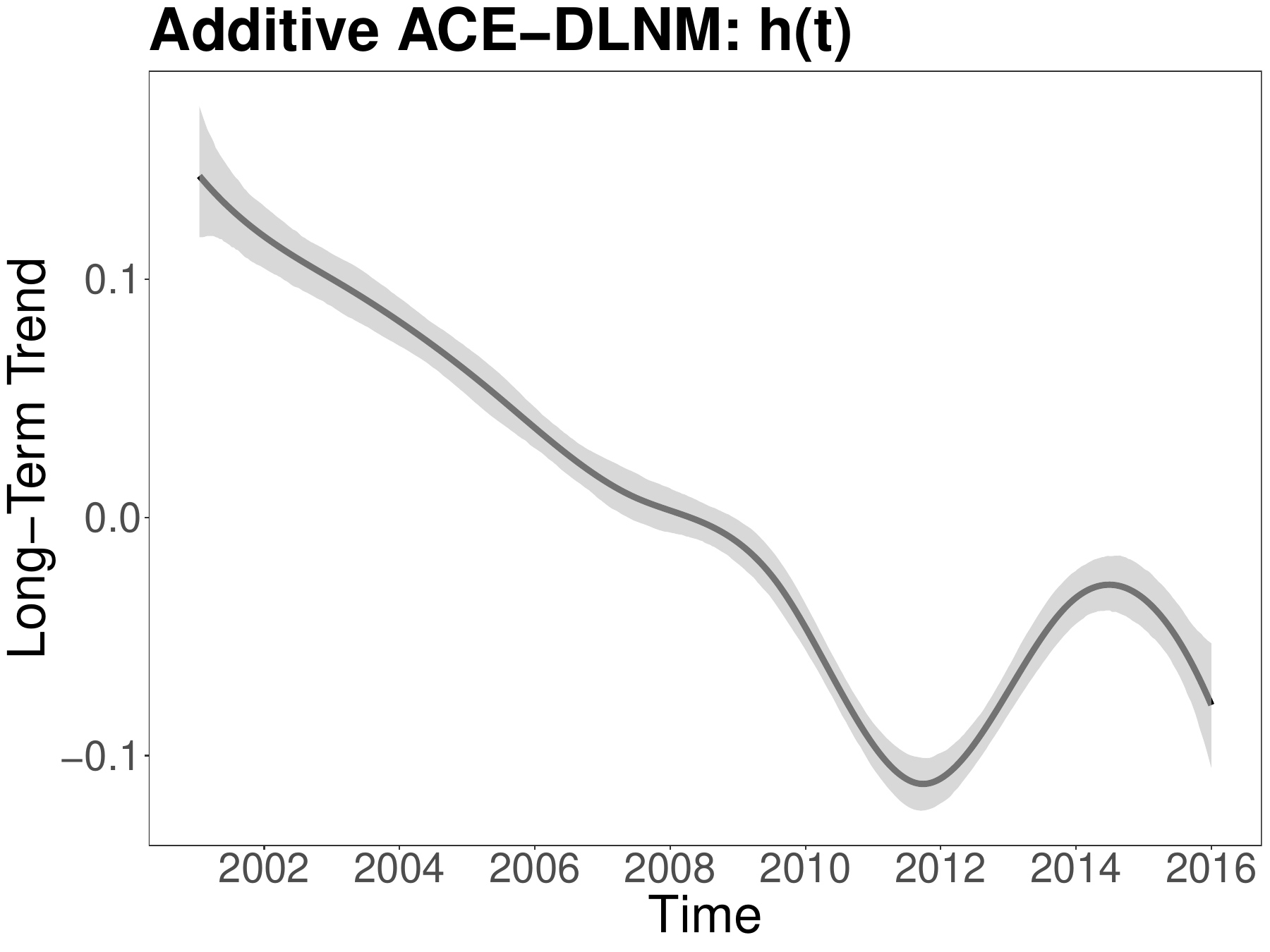}
  \end{subfigure}
  \begin{subfigure}[t]{0.32\textwidth}
    \includegraphics[width=\linewidth]{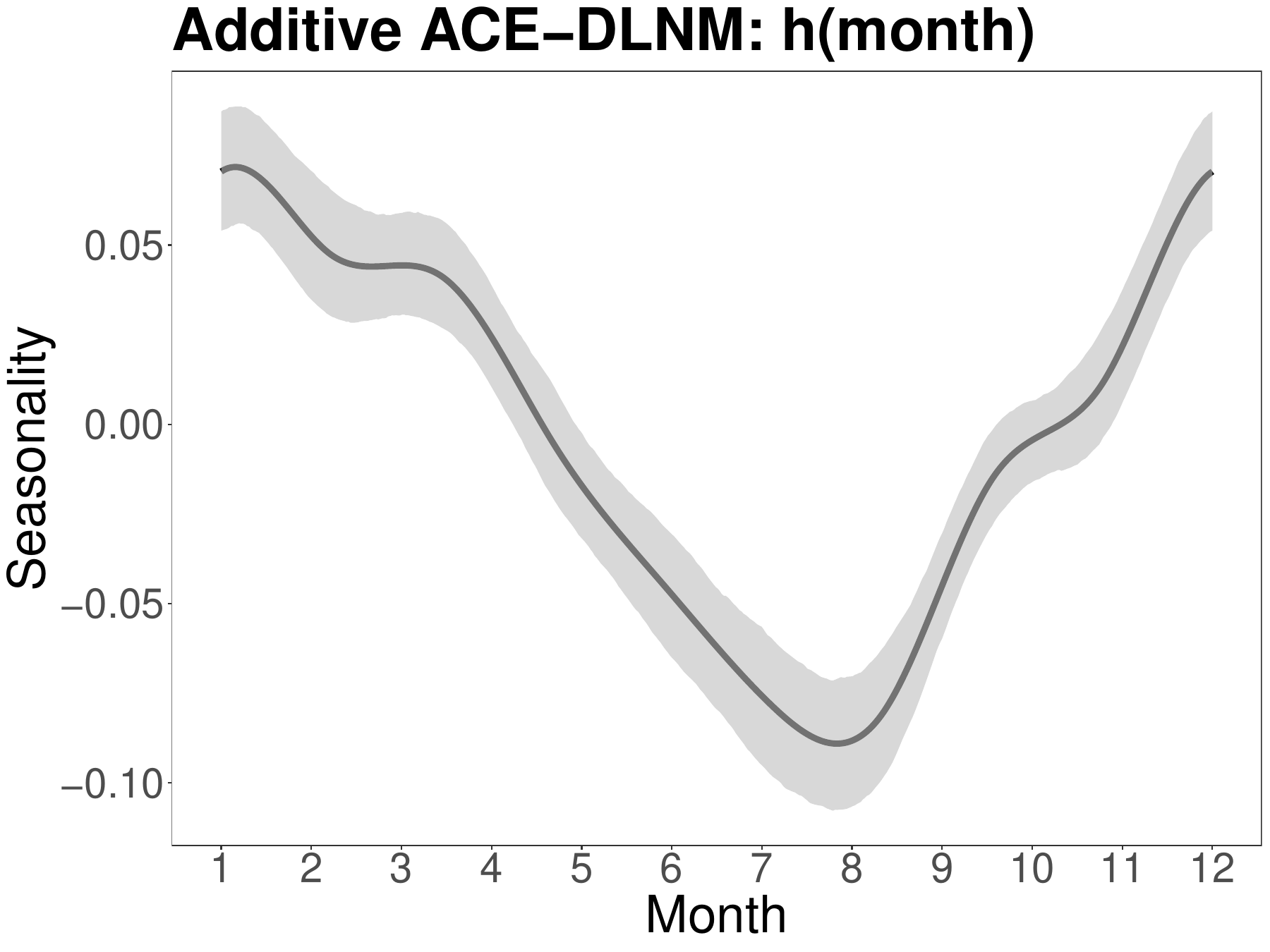}
  \end{subfigure}
  \begin{subfigure}[t]{0.32\textwidth}
    \includegraphics[width=\linewidth]{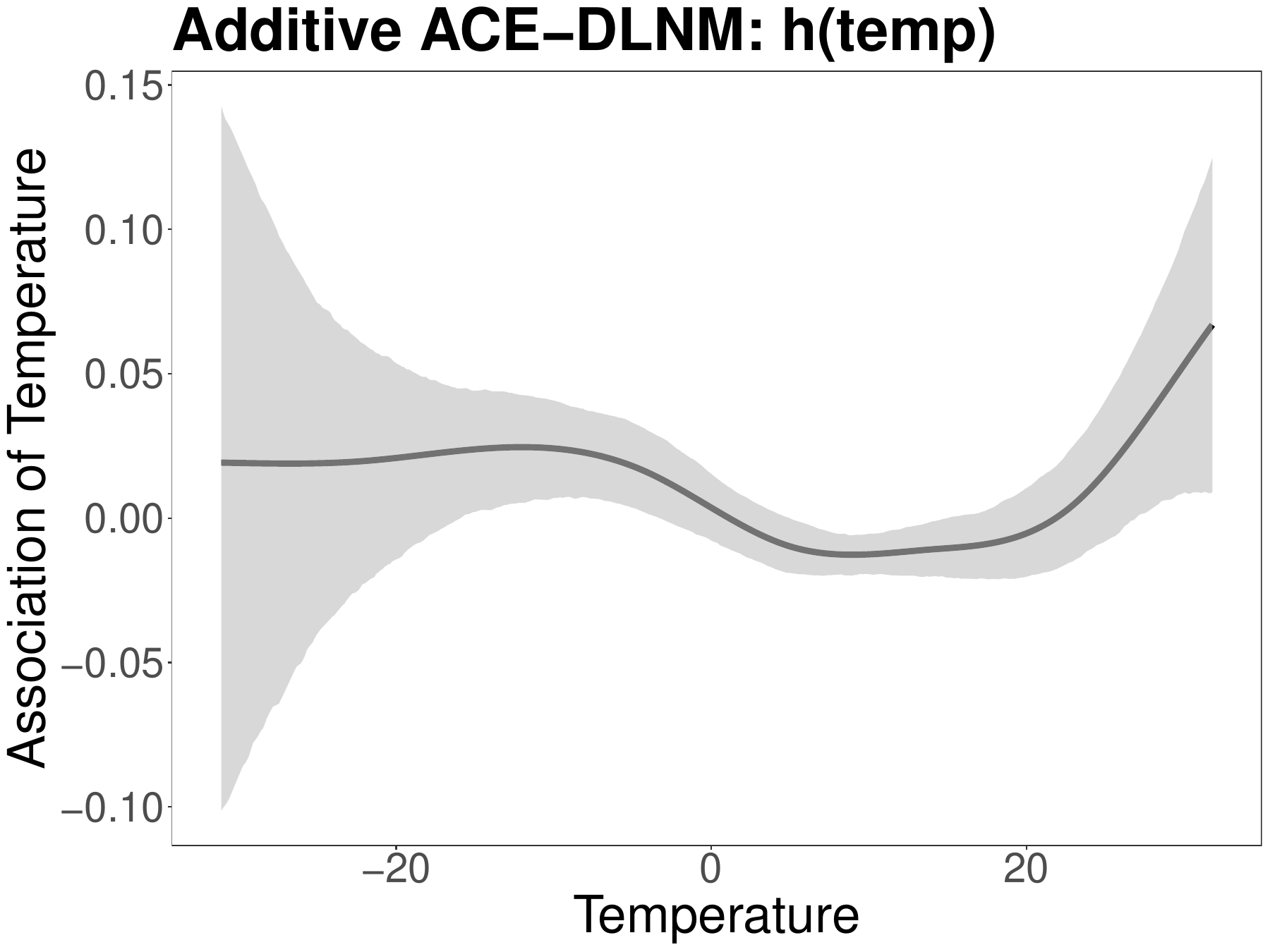}
  \end{subfigure}

  \medskip

  \begin{subfigure}[t]{0.32\textwidth}
    \includegraphics[width=\linewidth]{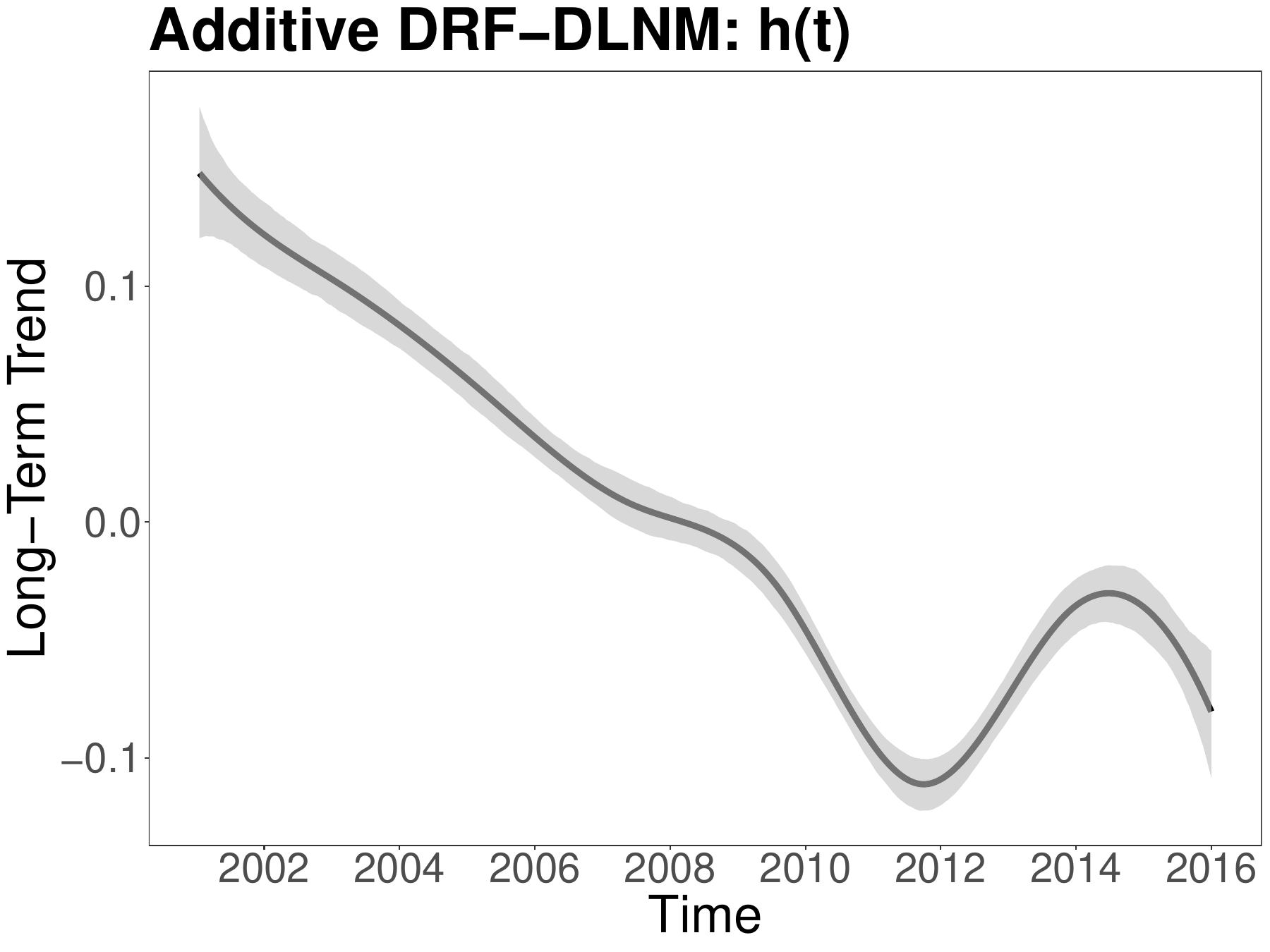}
  \end{subfigure}
  \begin{subfigure}[t]{0.32\textwidth}
    \includegraphics[width=\linewidth]{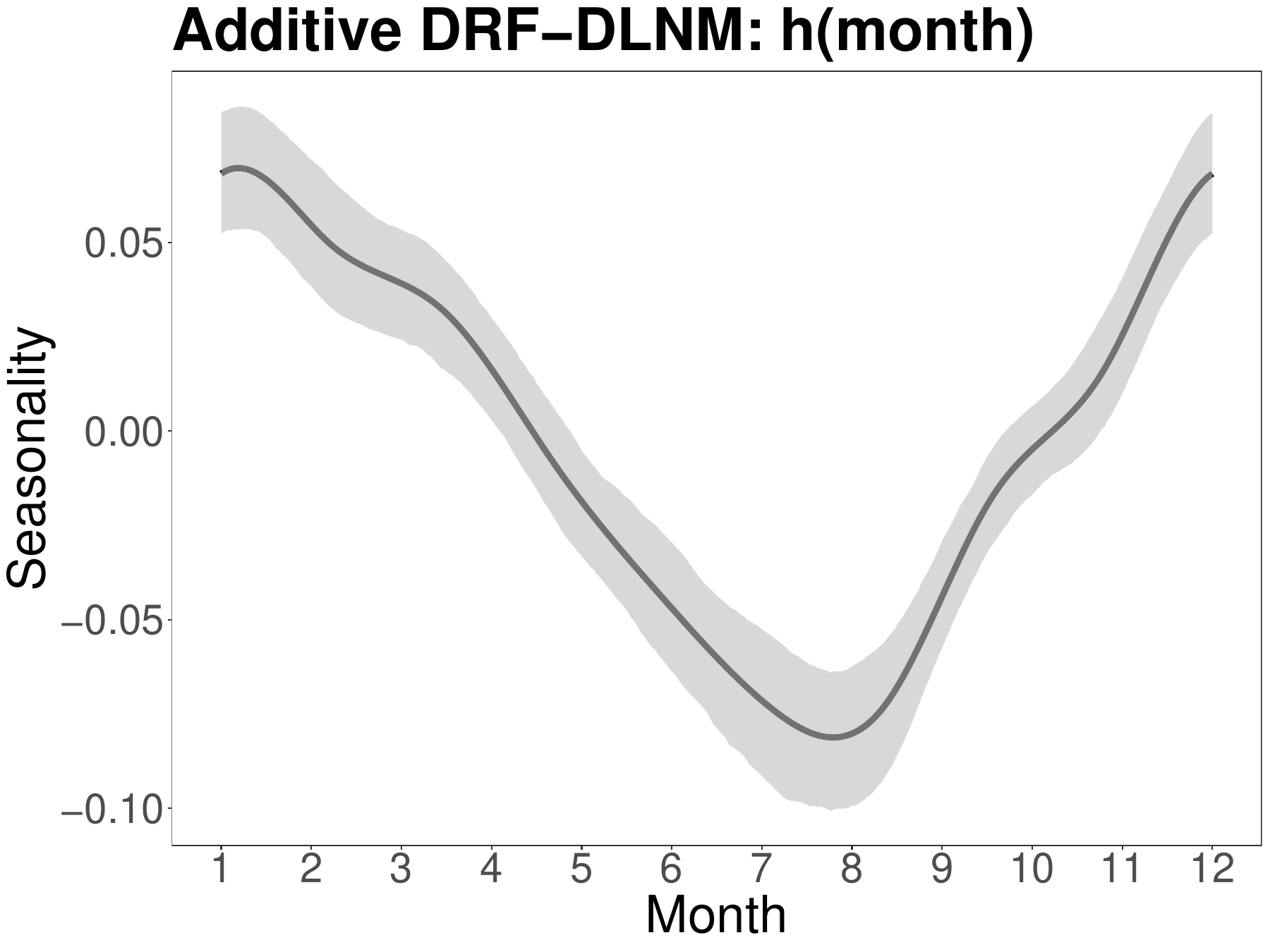}
  \end{subfigure}
  \begin{subfigure}[t]{0.32\textwidth}
    \includegraphics[width=\linewidth]{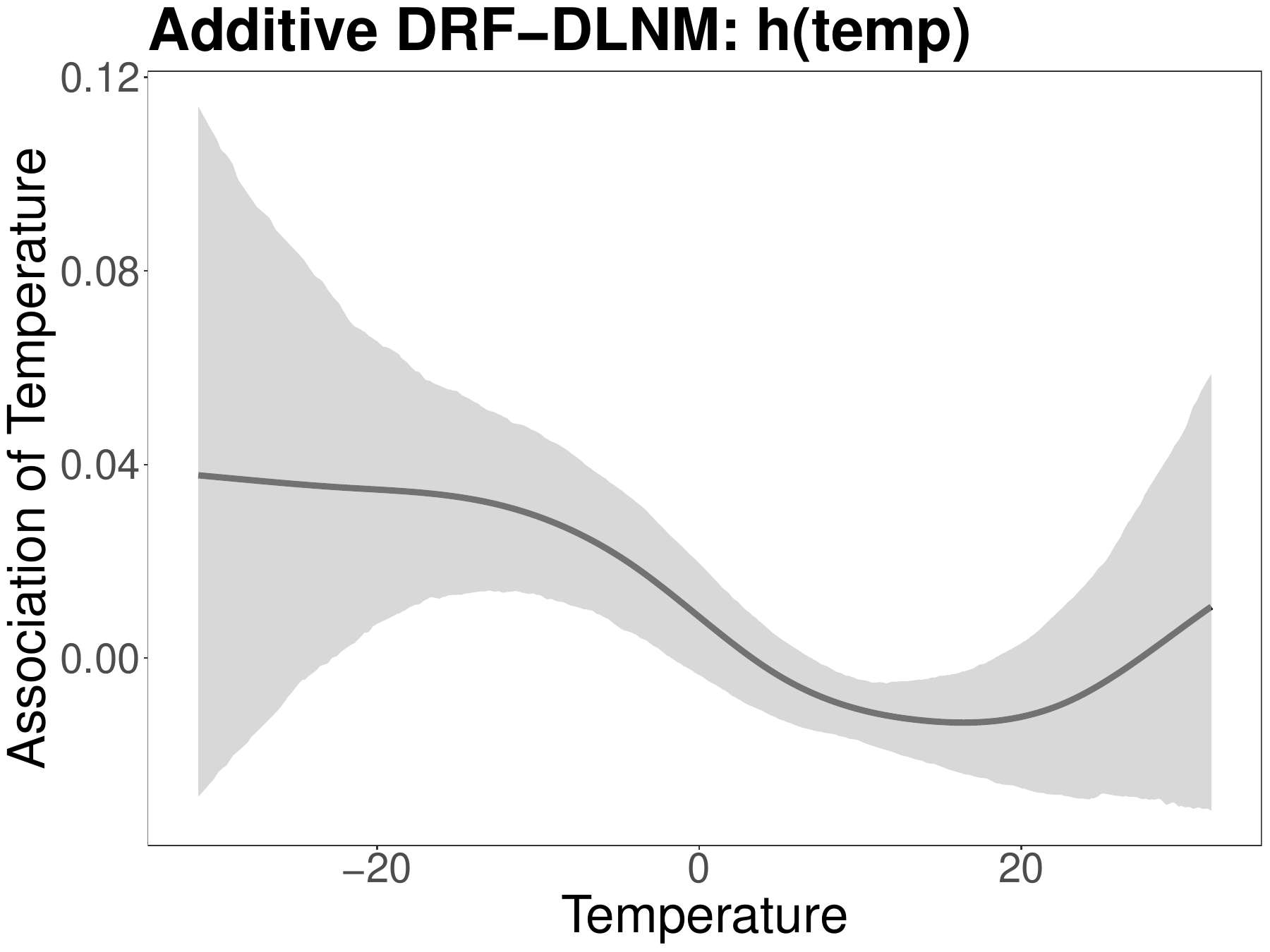}
  \end{subfigure}

  \caption{Estimated covariate functions from the single-index ACE-DLNM (first row),  the single-index DRf-DLNM (second row), the additive ACE-DLNM (third row) and the additive DRF-DLNM (fourth row), for circulatory mortality. The covariate functions include the long-term trend $h_1(t)$ (first column), the seasonality $h_2(\text{Month})$ and the temperature association $h_3(\text{Temp})$.}
\end{figure}

\subsubsection{All-Cause Mortality}
\begin{figure}[H]
  \centering
  \begin{subfigure}[t]{0.32\textwidth}
    \includegraphics[width=\linewidth]{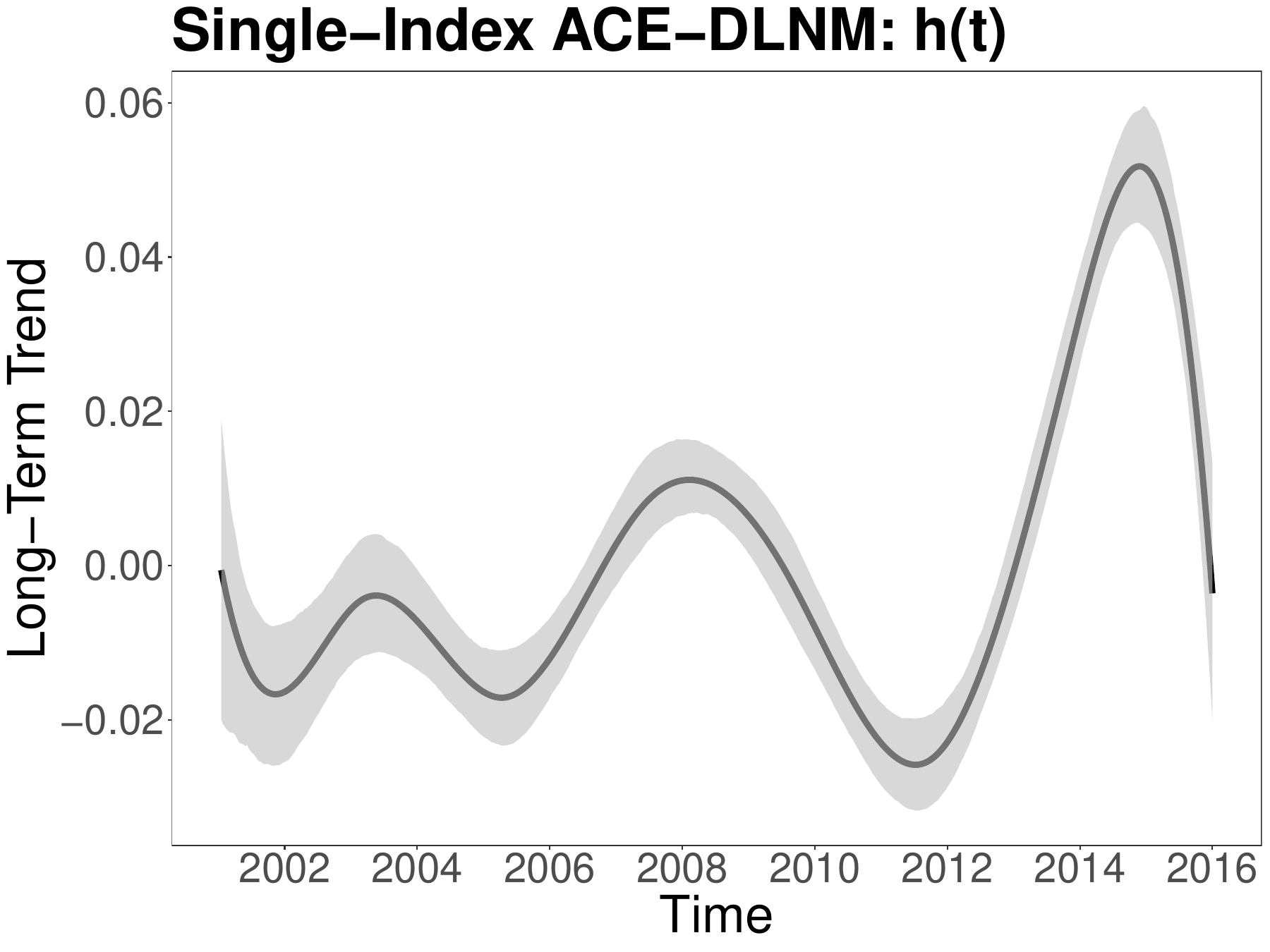}
  \end{subfigure}
  \begin{subfigure}[t]{0.32\textwidth}
    \includegraphics[width=\linewidth]{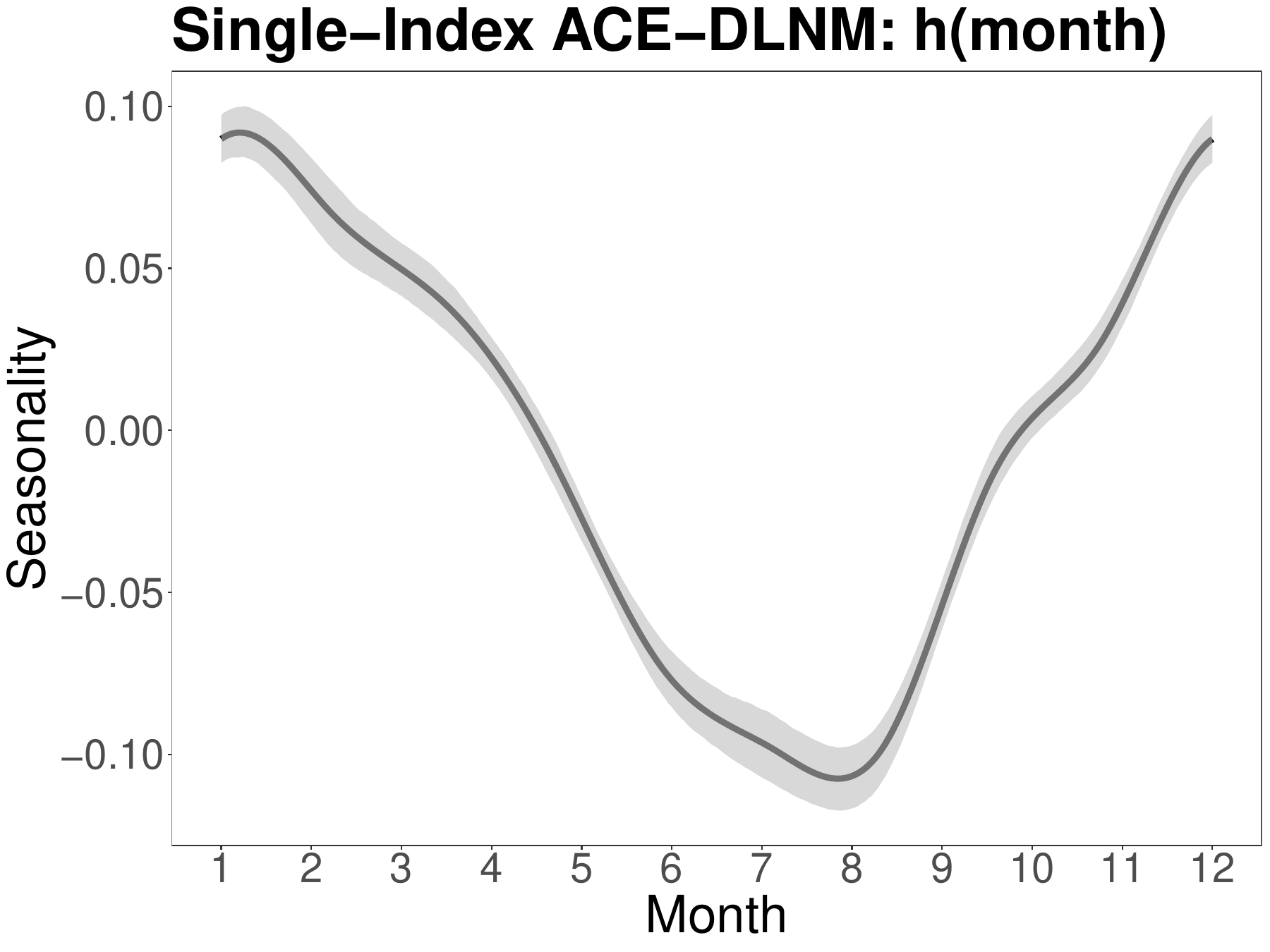}
  \end{subfigure}
  \begin{subfigure}[t]{0.32\textwidth}
    \includegraphics[width=\linewidth]{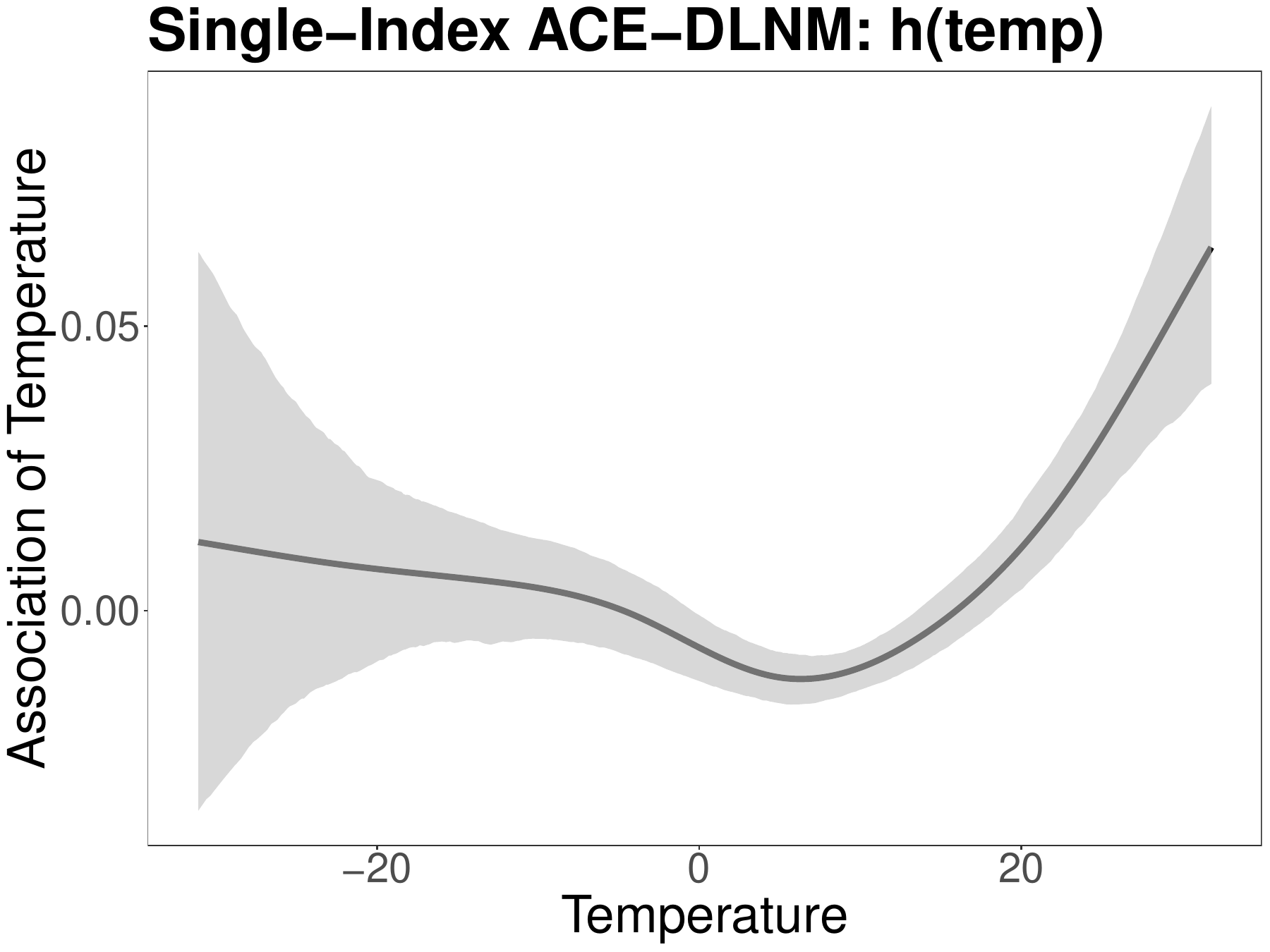}
  \end{subfigure}

  \medskip

  \begin{subfigure}[t]{0.32\textwidth}
    \includegraphics[width=\linewidth]{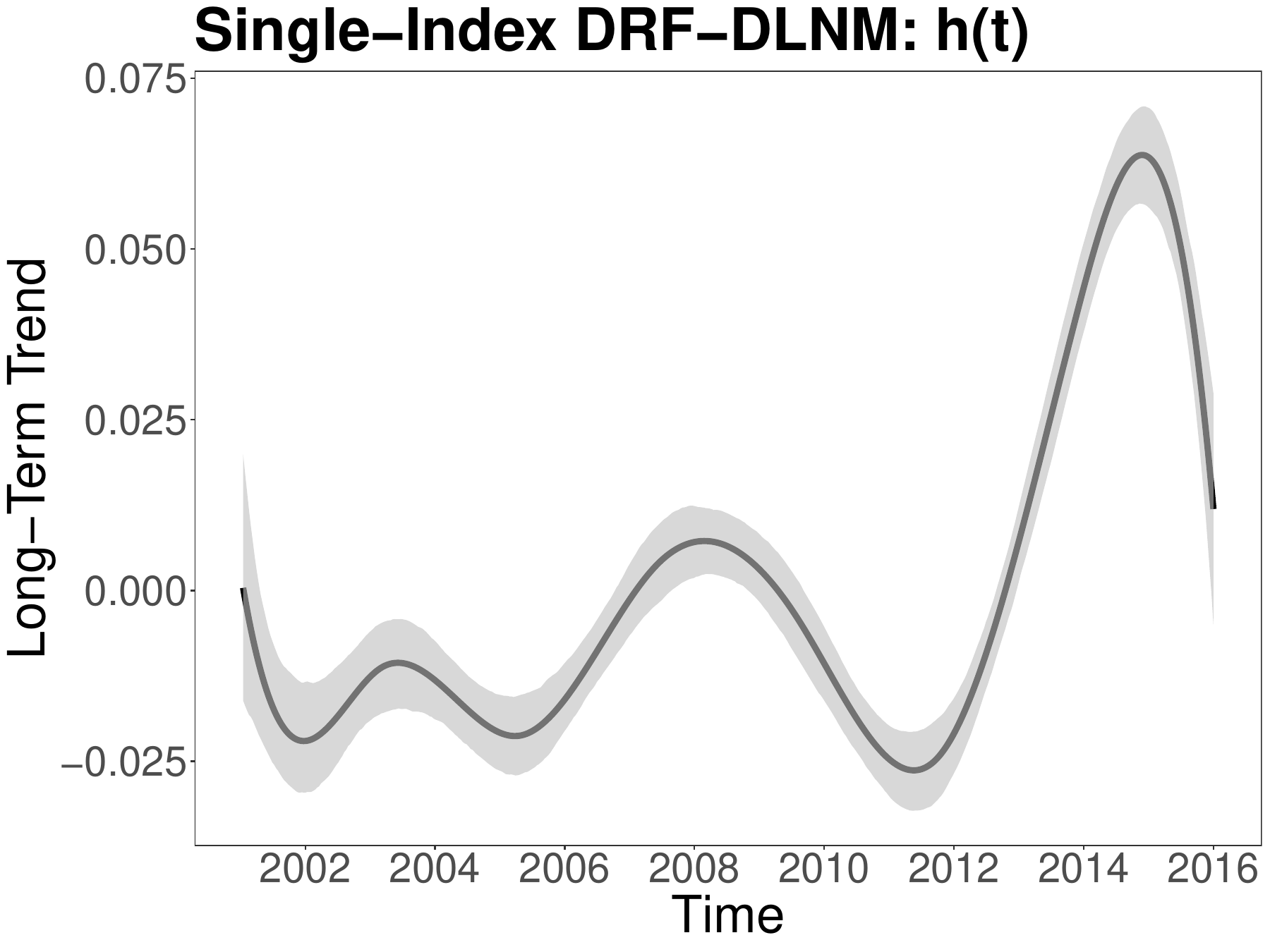}
  \end{subfigure}
  \begin{subfigure}[t]{0.32\textwidth}
    \includegraphics[width=\linewidth]{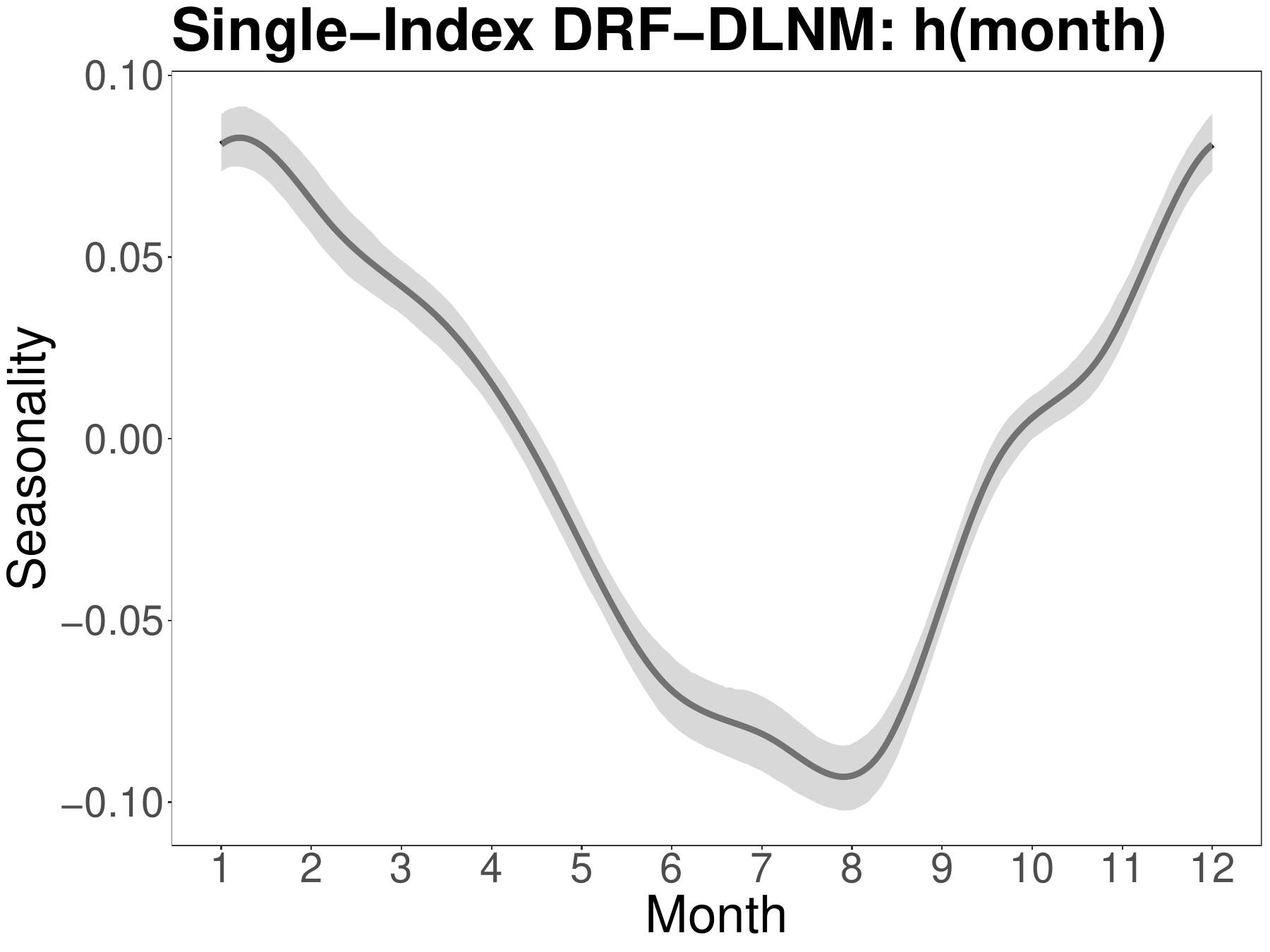}
  \end{subfigure}
  \begin{subfigure}[t]{0.32\textwidth}
    \includegraphics[width=\linewidth]{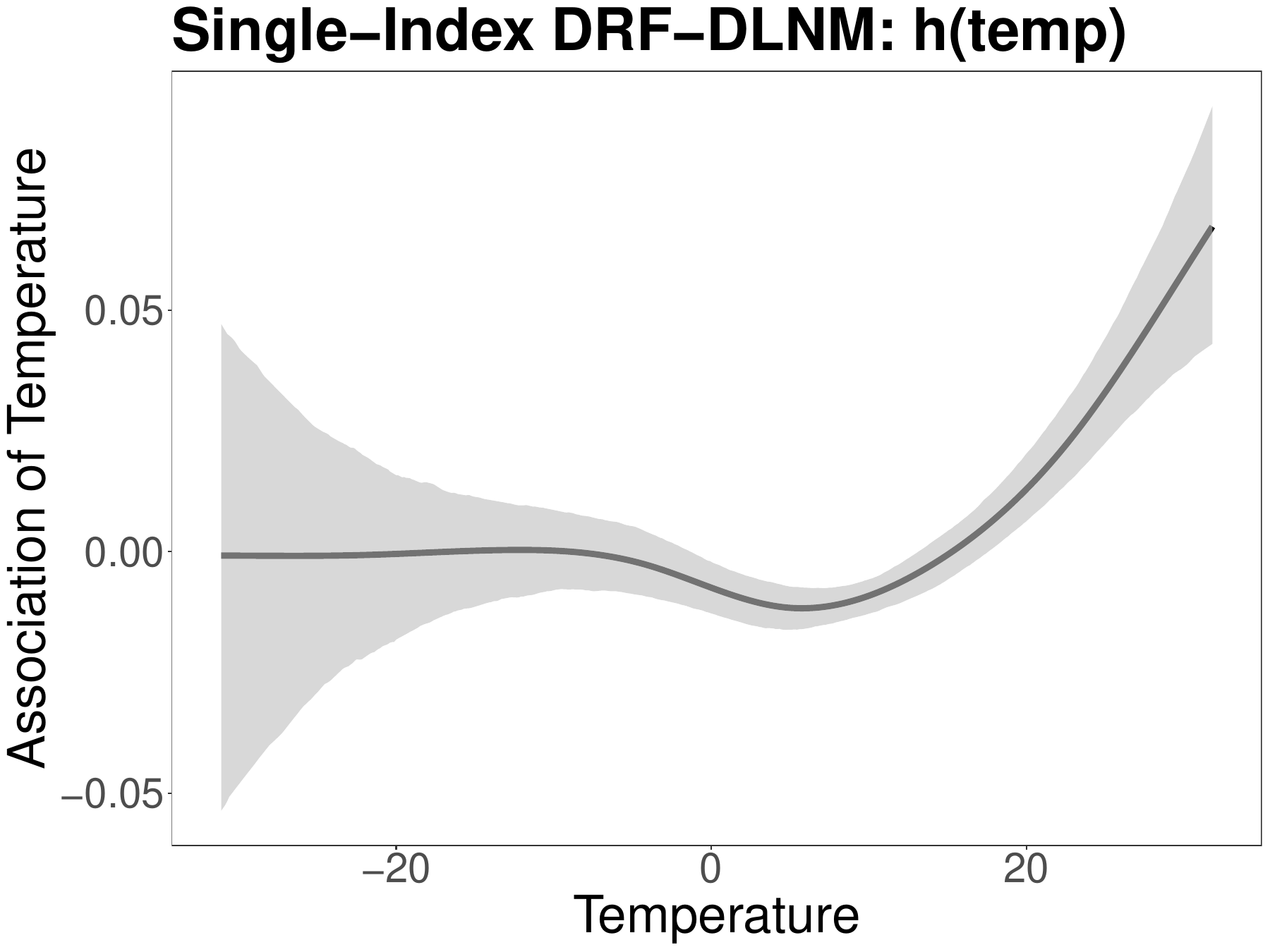}
  \end{subfigure}
  
  \medskip

  \begin{subfigure}[t]{0.32\textwidth}
    \includegraphics[width=\linewidth]{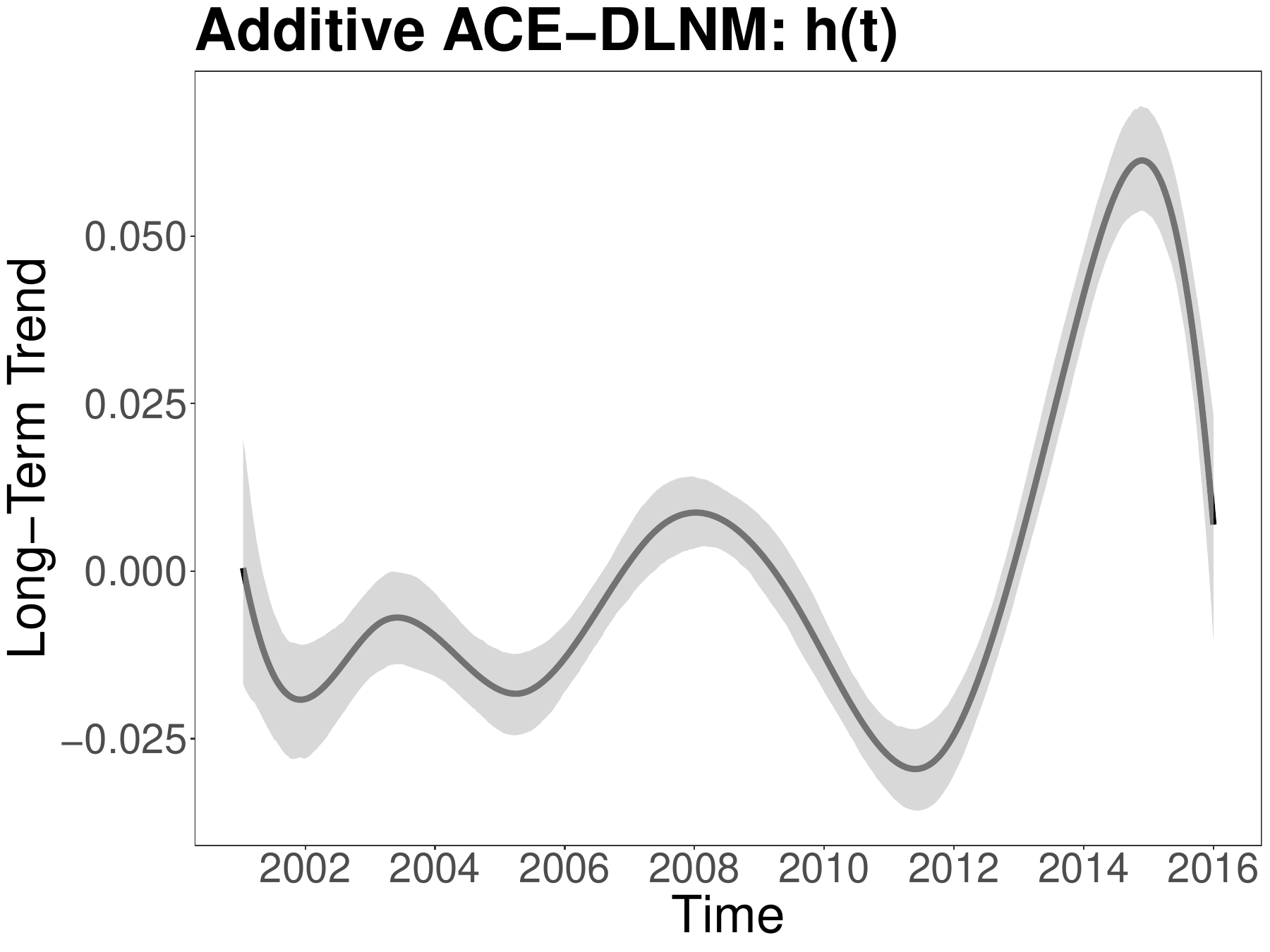}
  \end{subfigure}
  \begin{subfigure}[t]{0.32\textwidth}
    \includegraphics[width=\linewidth]{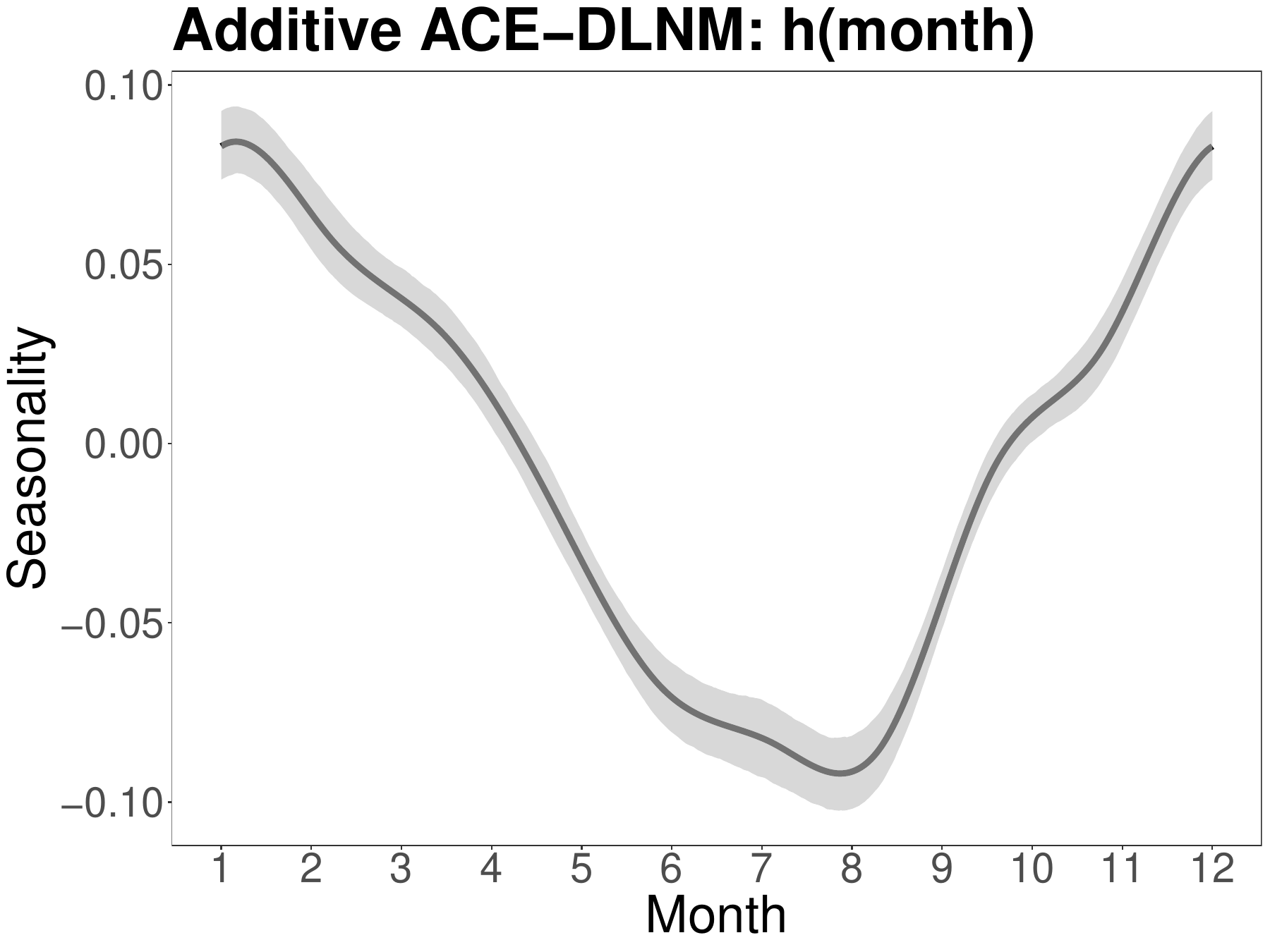}
  \end{subfigure}
  \begin{subfigure}[t]{0.32\textwidth}
    \includegraphics[width=\linewidth]{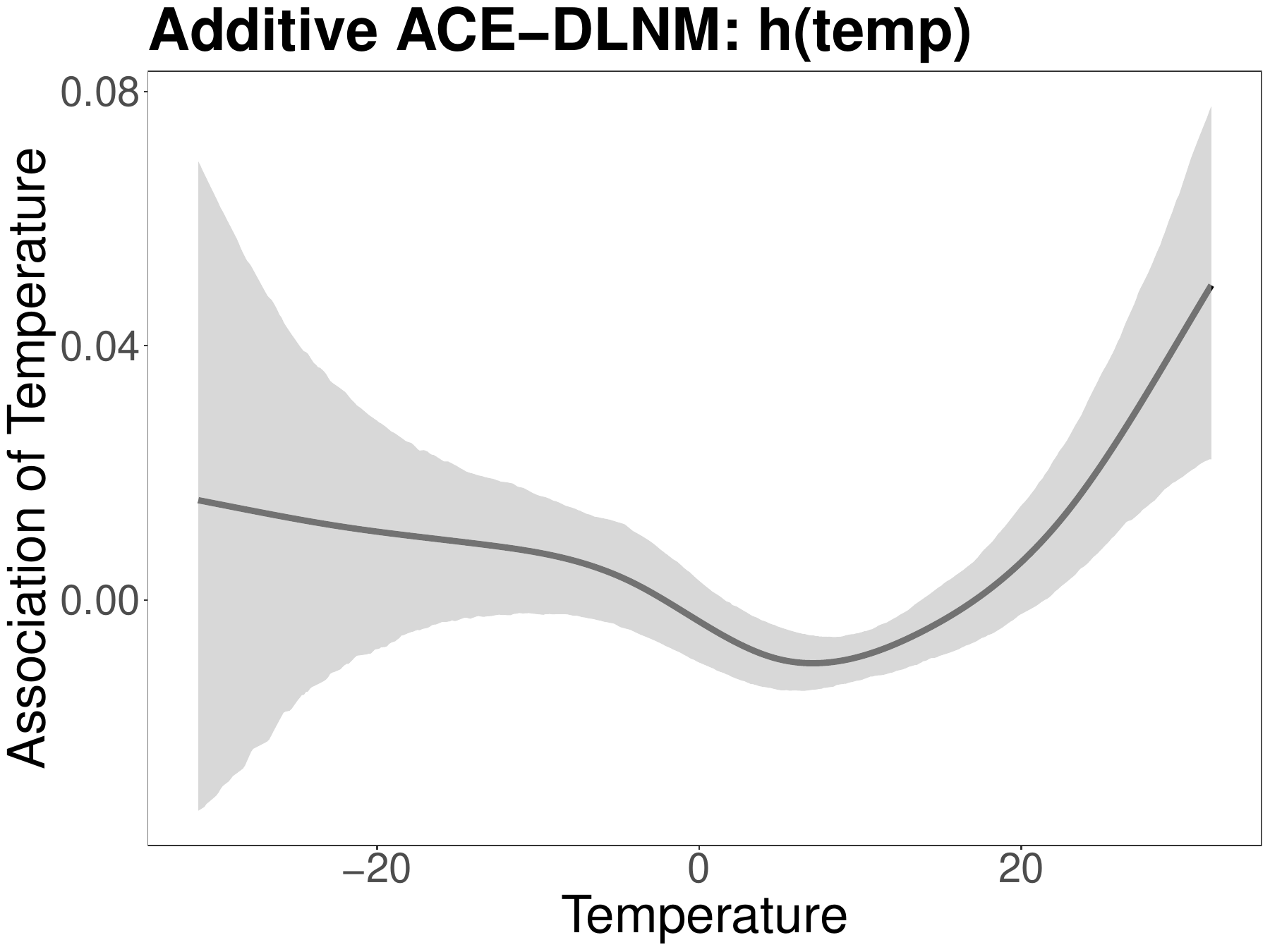}
  \end{subfigure}

  \medskip

  \begin{subfigure}[t]{0.32\textwidth}
    \includegraphics[width=\linewidth]{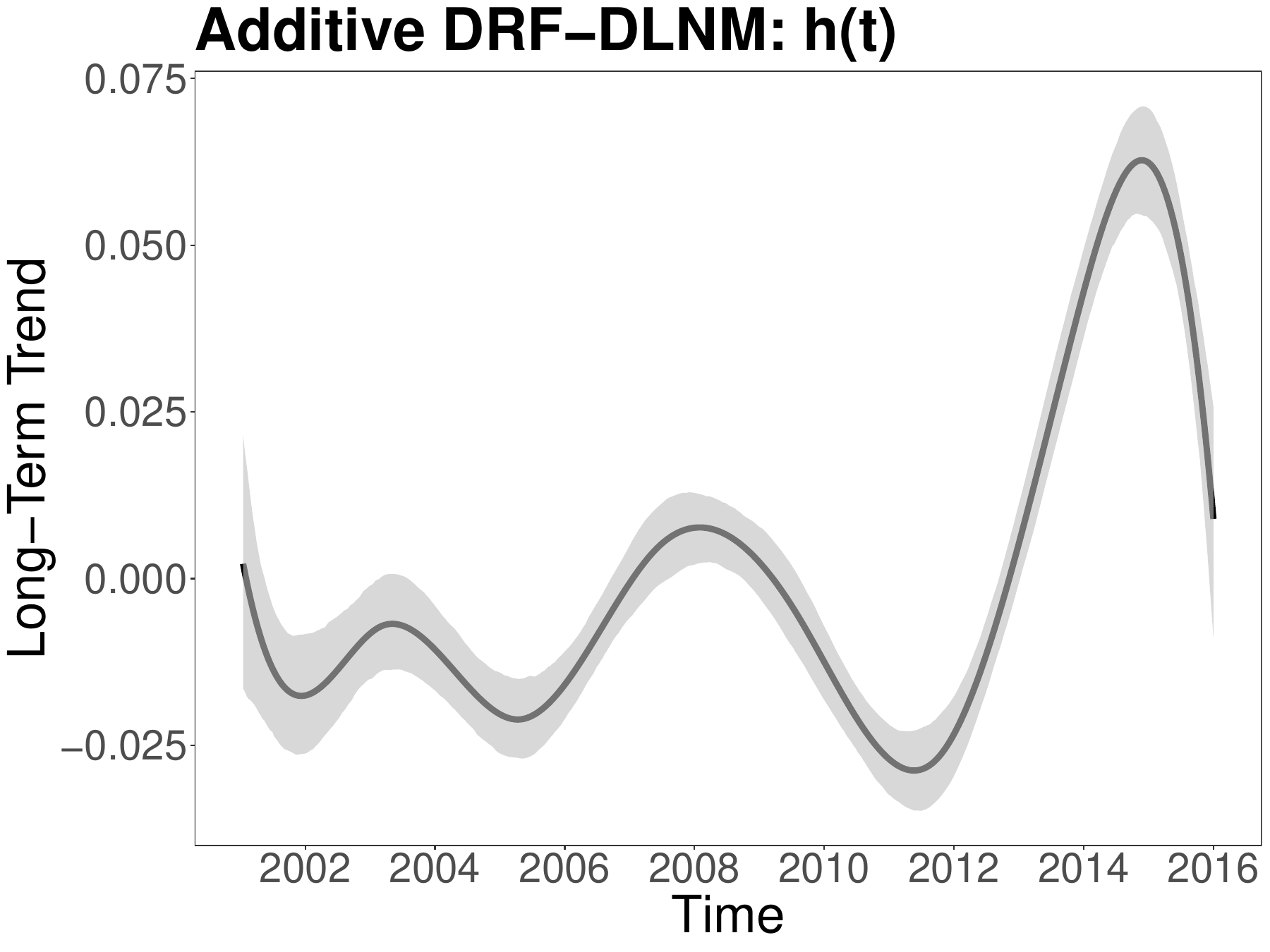}
  \end{subfigure}
  \begin{subfigure}[t]{0.32\textwidth}
    \includegraphics[width=\linewidth]{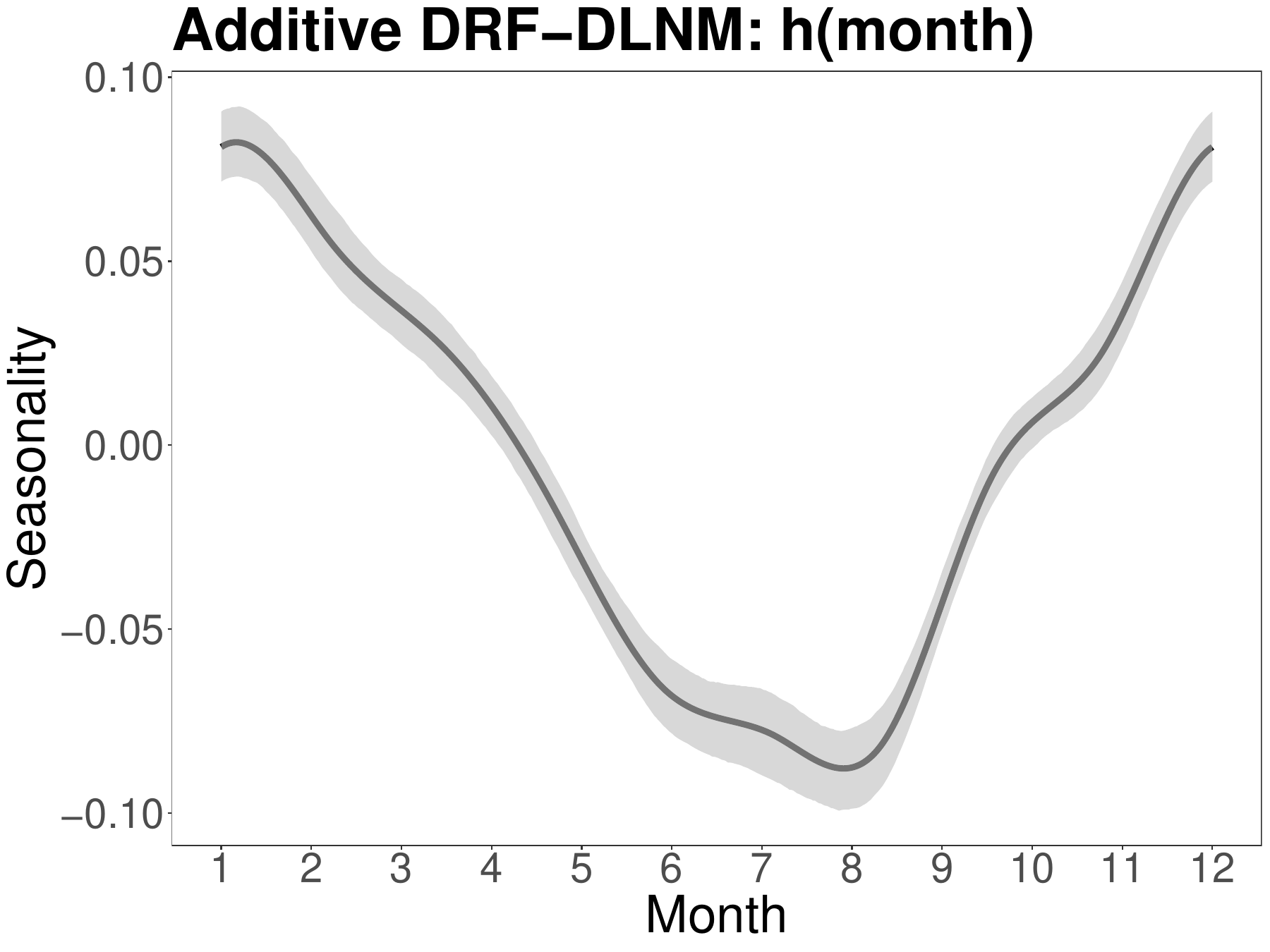}
  \end{subfigure}
  \begin{subfigure}[t]{0.32\textwidth}
    \includegraphics[width=\linewidth]{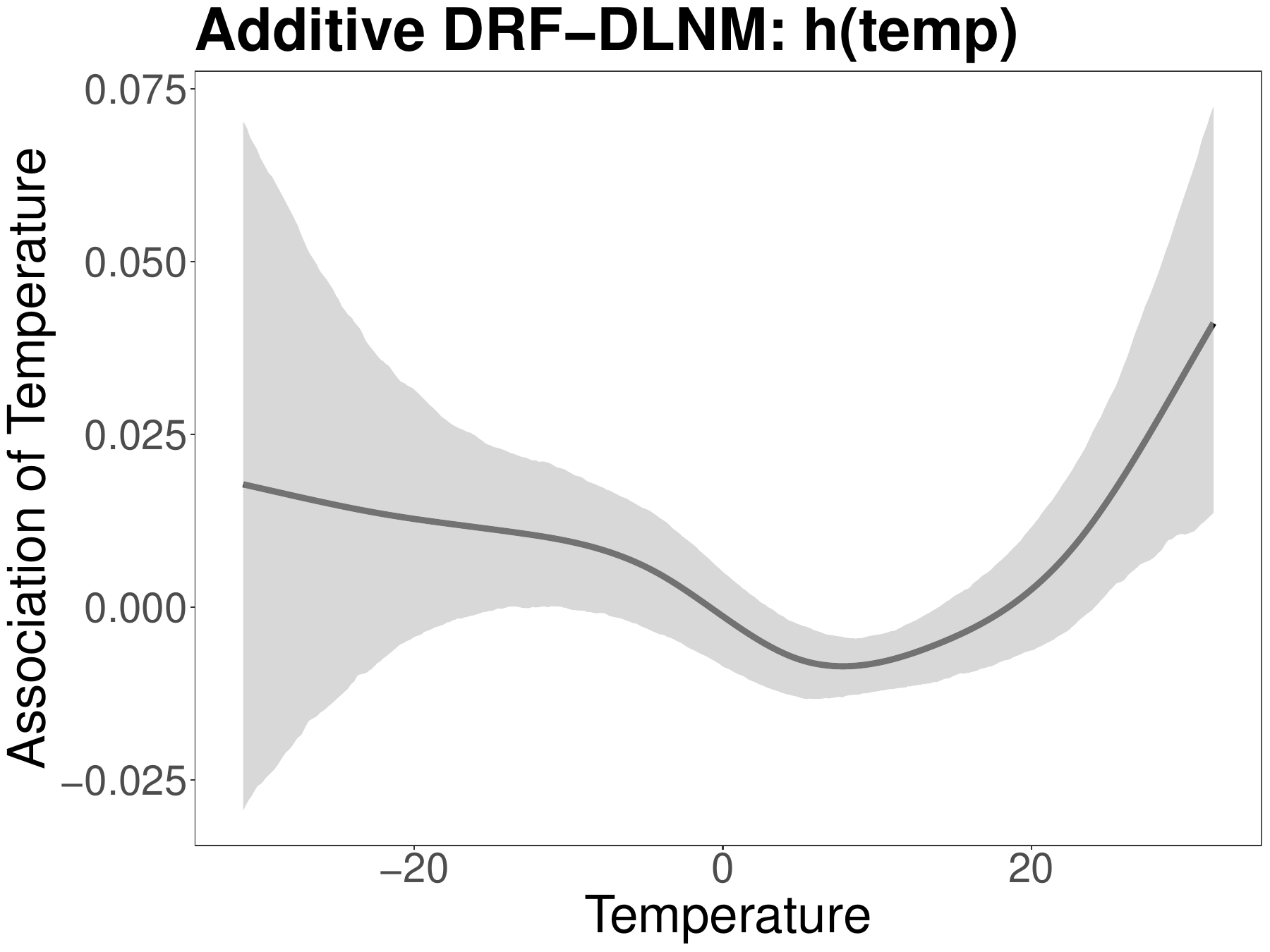}
  \end{subfigure}

  \caption{Estimated covariate functions from the single-index ACE-DLNM (first row),  the single-index DRf-DLNM (second row), the additive ACE-DLNM (third row) and the additive DRF-DLNM (fourth row), for all-cause mortality. The covariate functions include the long-term trend $h_1(t)$ (first column), the seasonality $h_2(\text{Month})$ and the temperature association $h_3(\text{Temp})$.}
\end{figure}

\clearpage
\subsection{Diagnostics}
We report the randomized quantile residual \citep{dunn1996randomized} plots and the QQ-plot across all CDs. 
We also examine the randomized quantile residual plot, QQ-plot and autocorrelation function plot for each CD, and present the results for Waterloo and Toronto as examples. The diagnostics suggest adequate model fit. 
The diagnostics for all-cause mortality in Toronto are the least favorable among all outcomes and CDs, but the autocorrelations remain small in magnitude and do not suggest meaningful temporal dependence. 

\vspace{-0.5cm}

\subsubsection{Respiratory Mortality}
\vspace{-0.5cm}
\begin{figure}[H]
      \centering
      \begin{subfigure}[t]{0.65\textwidth}
        \includegraphics[width=\linewidth]{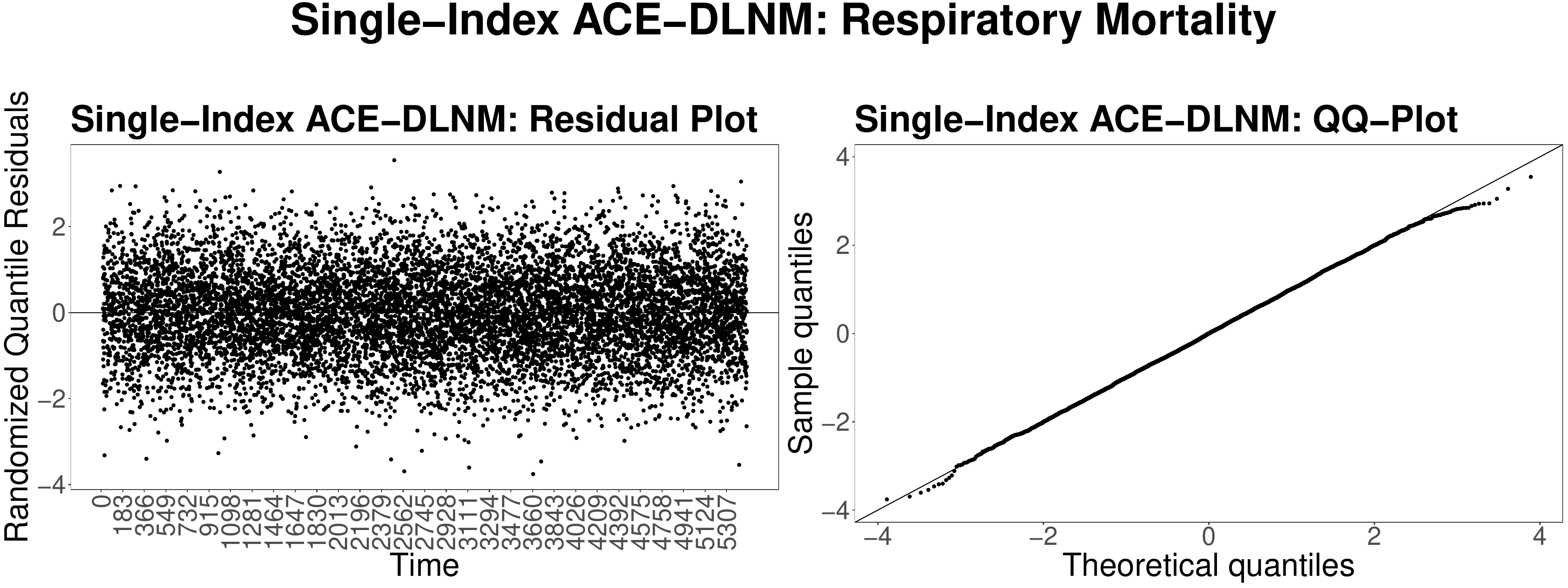}
      \end{subfigure}
    
      \medskip
    
      \begin{subfigure}[t]{0.65\textwidth}
        \includegraphics[width=\linewidth]{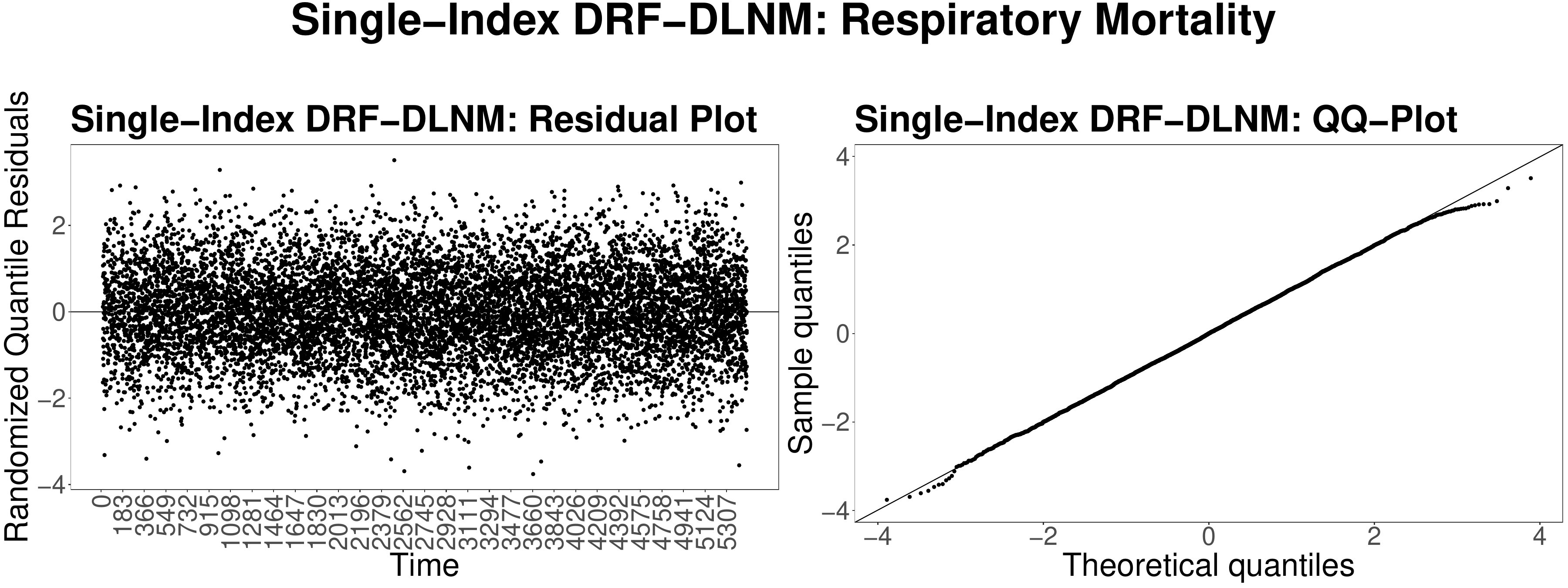}
      \end{subfigure}
    
      \medskip
    
      \begin{subfigure}[t]{0.65\textwidth}
        \includegraphics[width=\linewidth]{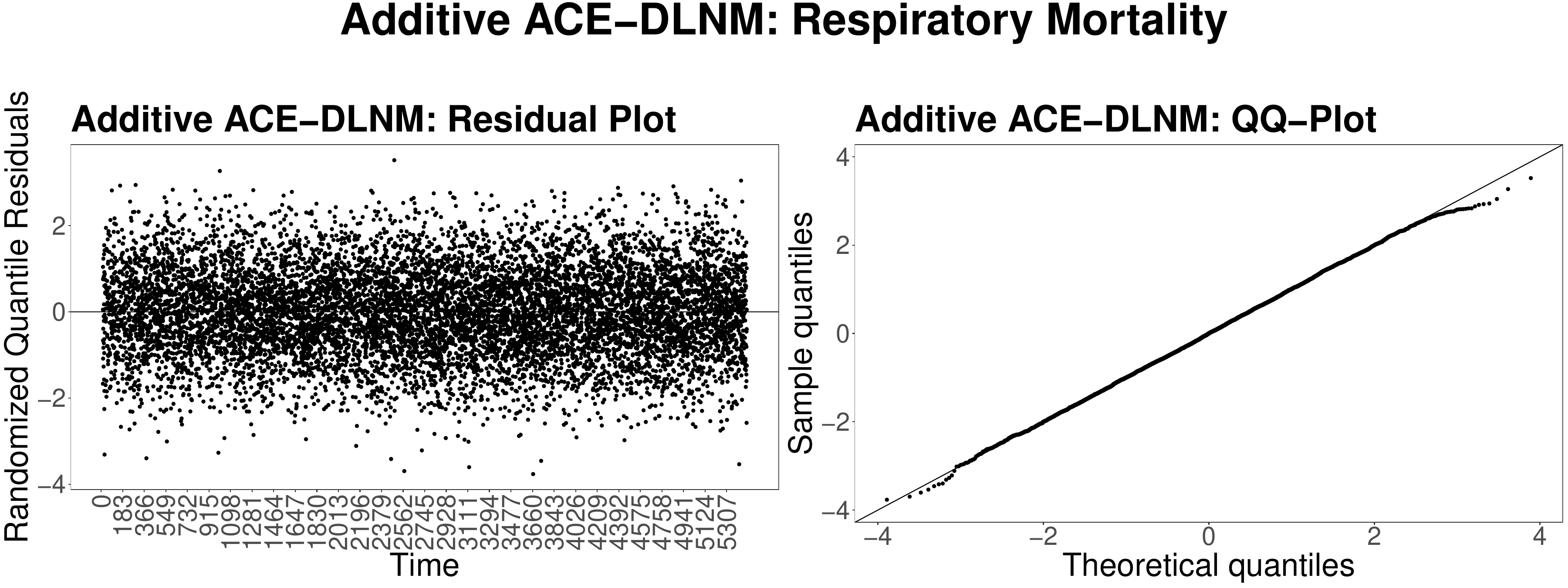}
      \end{subfigure}
    
      \medskip
    
      \begin{subfigure}[t]{0.65\textwidth}
        \includegraphics[width=\linewidth]{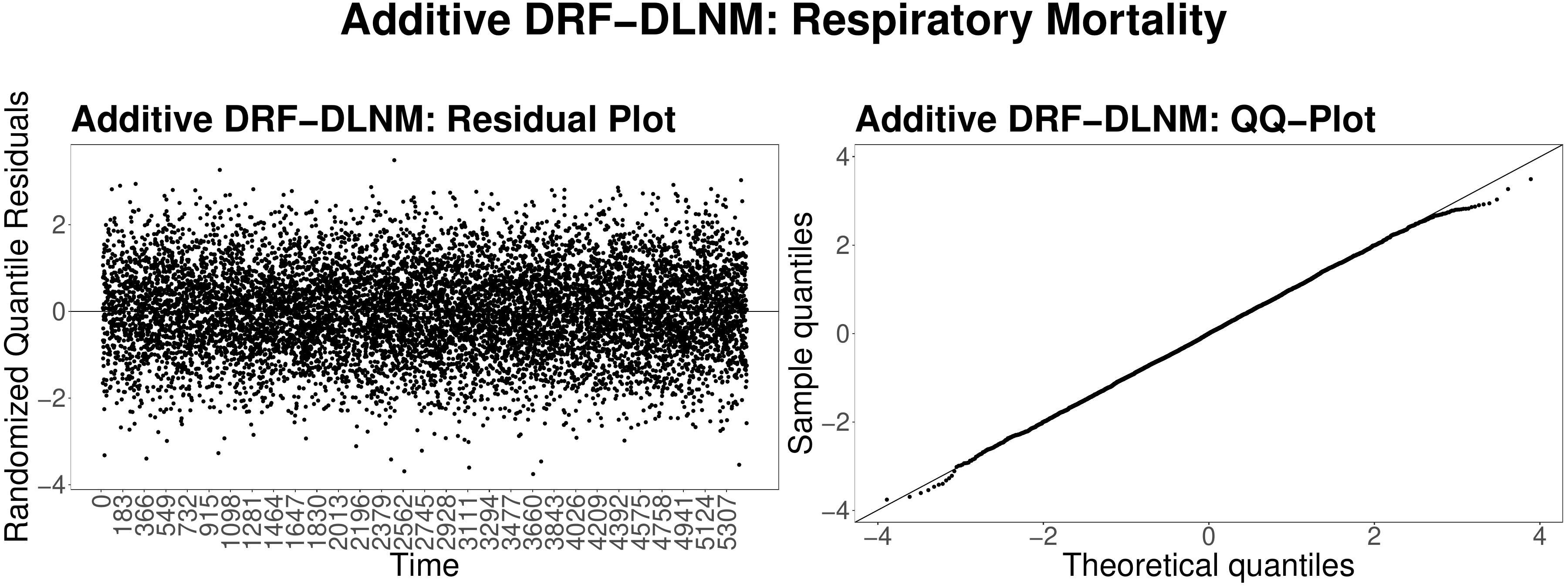}
      \end{subfigure}
      
      \caption{Randomized quantile residuals for respiratory mortality across all CDs. }
\end{figure}

\begin{figure}[H]
      \centering
      \begin{subfigure}[t]{\textwidth}
        \includegraphics[width=\linewidth]{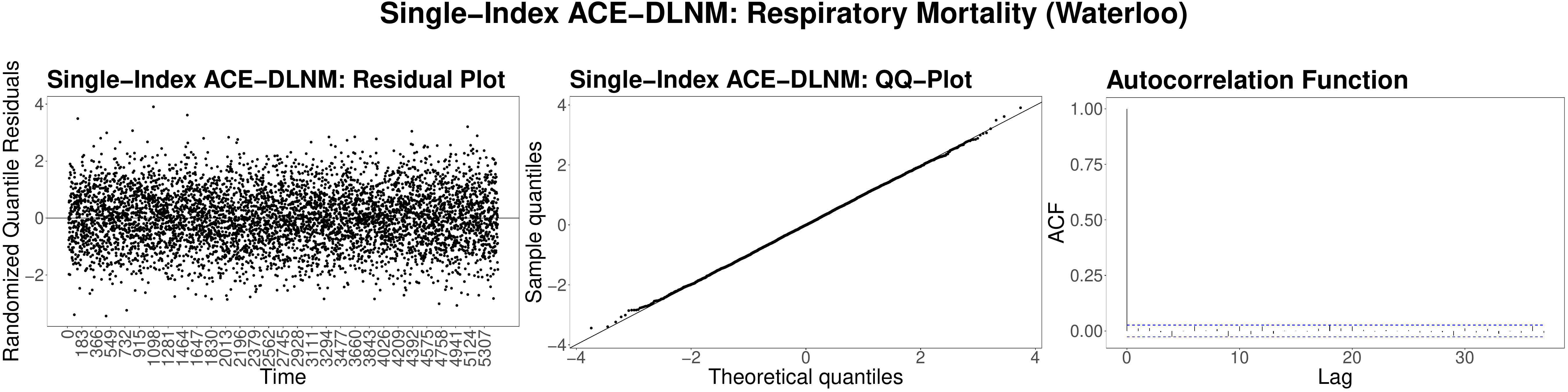}
      \end{subfigure}
    
      \medskip
    
      \begin{subfigure}[t]{\textwidth}
        \includegraphics[width=\linewidth]{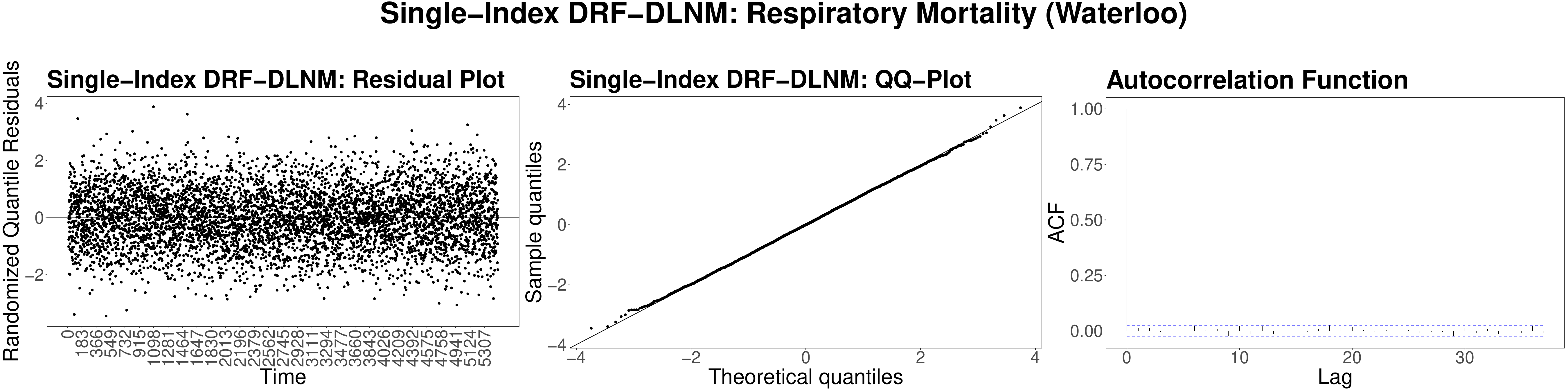}
      \end{subfigure}
    
      \medskip
    
      \begin{subfigure}[t]{\textwidth}
        \includegraphics[width=\linewidth]{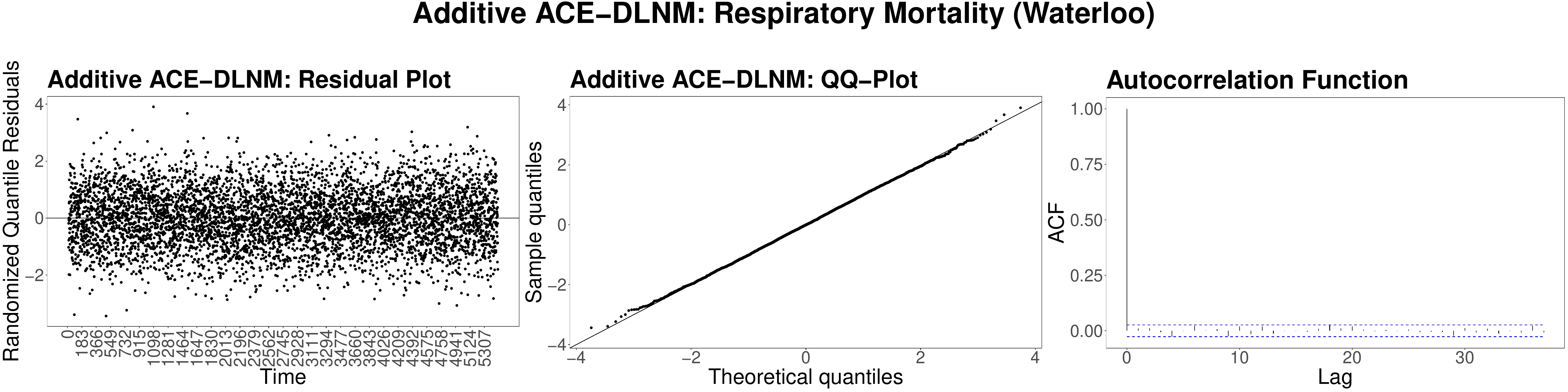}
      \end{subfigure}
    
      \medskip
    
      \begin{subfigure}[t]{\textwidth}
        \includegraphics[width=\linewidth]{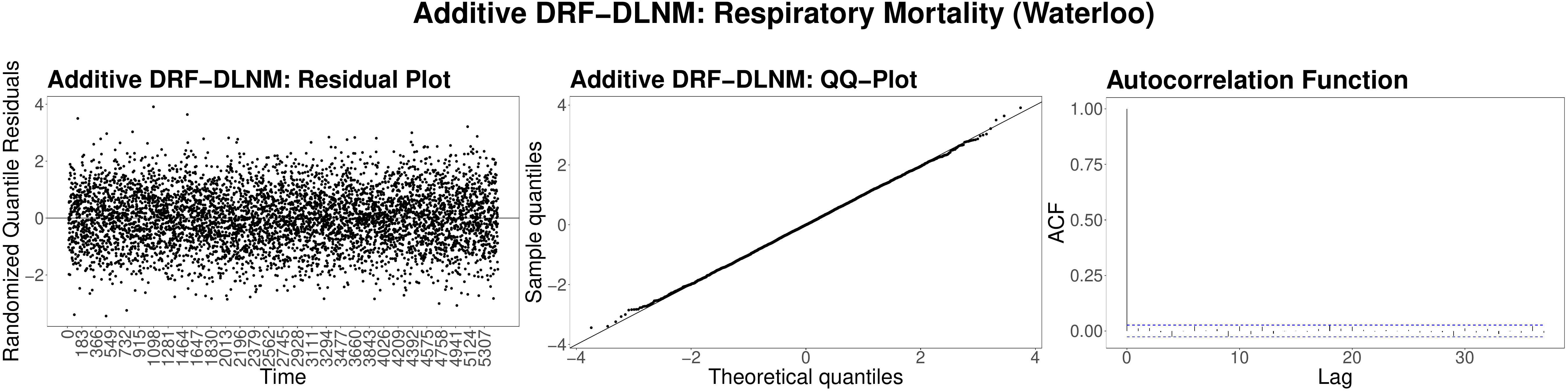}
      \end{subfigure}
      
      \caption{Randomized quantile residuals for respiratory mortality for Waterloo. }
\end{figure}

\begin{figure}[H]
      \centering
      \begin{subfigure}[t]{\textwidth}
        \includegraphics[width=\linewidth]{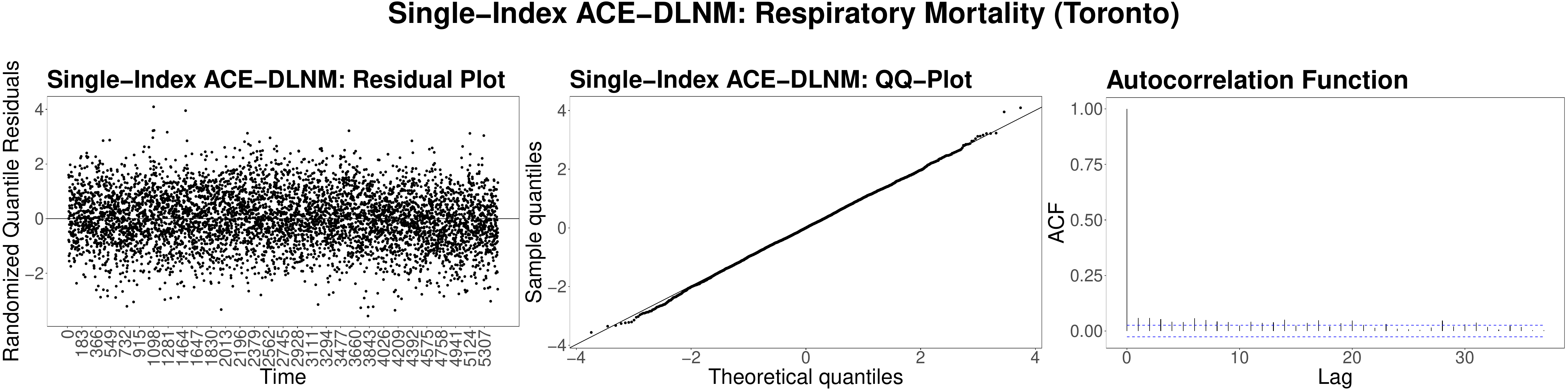}
      \end{subfigure}
    
      \medskip
    
      \begin{subfigure}[t]{\textwidth}
        \includegraphics[width=\linewidth]{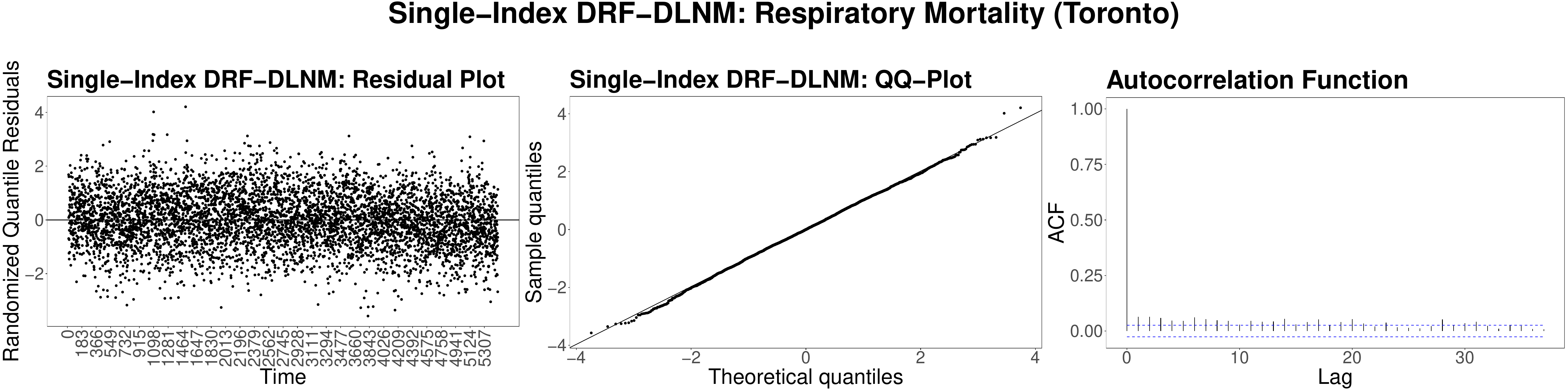}
      \end{subfigure}
    
      \medskip
    
      \begin{subfigure}[t]{\textwidth}
        \includegraphics[width=\linewidth]{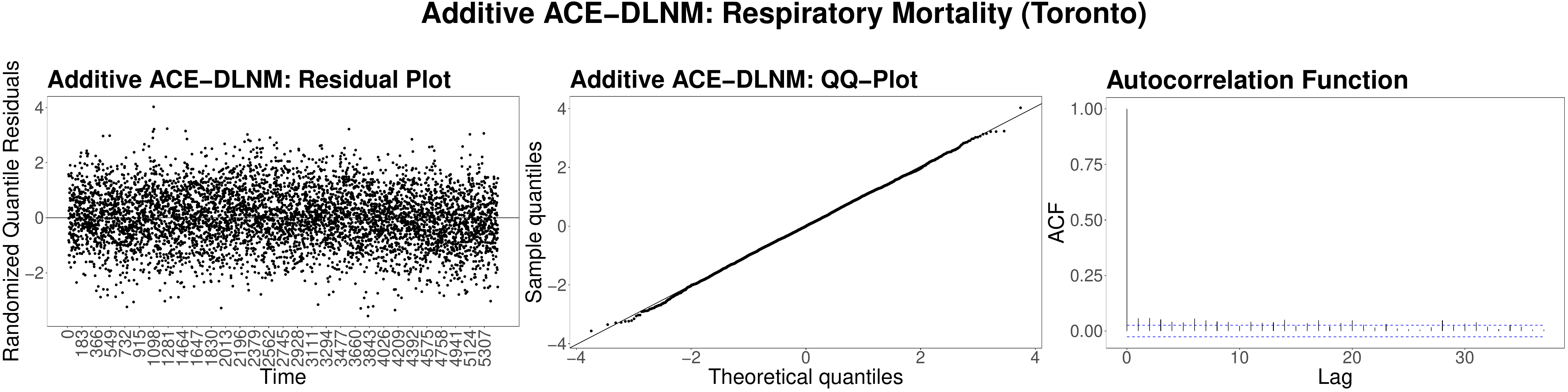}
      \end{subfigure}
    
      \medskip
    
      \begin{subfigure}[t]{\textwidth}
        \includegraphics[width=\linewidth]{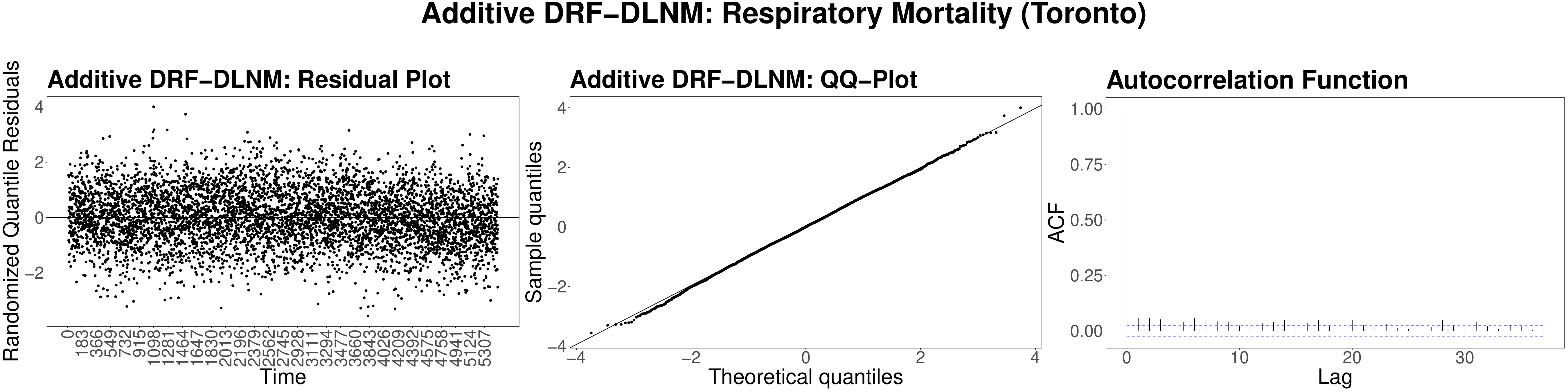}
      \end{subfigure}
      
      \caption{Randomized quantile residuals for respiratory mortality for Toronto. }
\end{figure}

\subsubsection{Circulatory Mortality}
\begin{figure}[H]
      \centering
      \begin{subfigure}[t]{0.65\textwidth}
        \includegraphics[width=\linewidth]{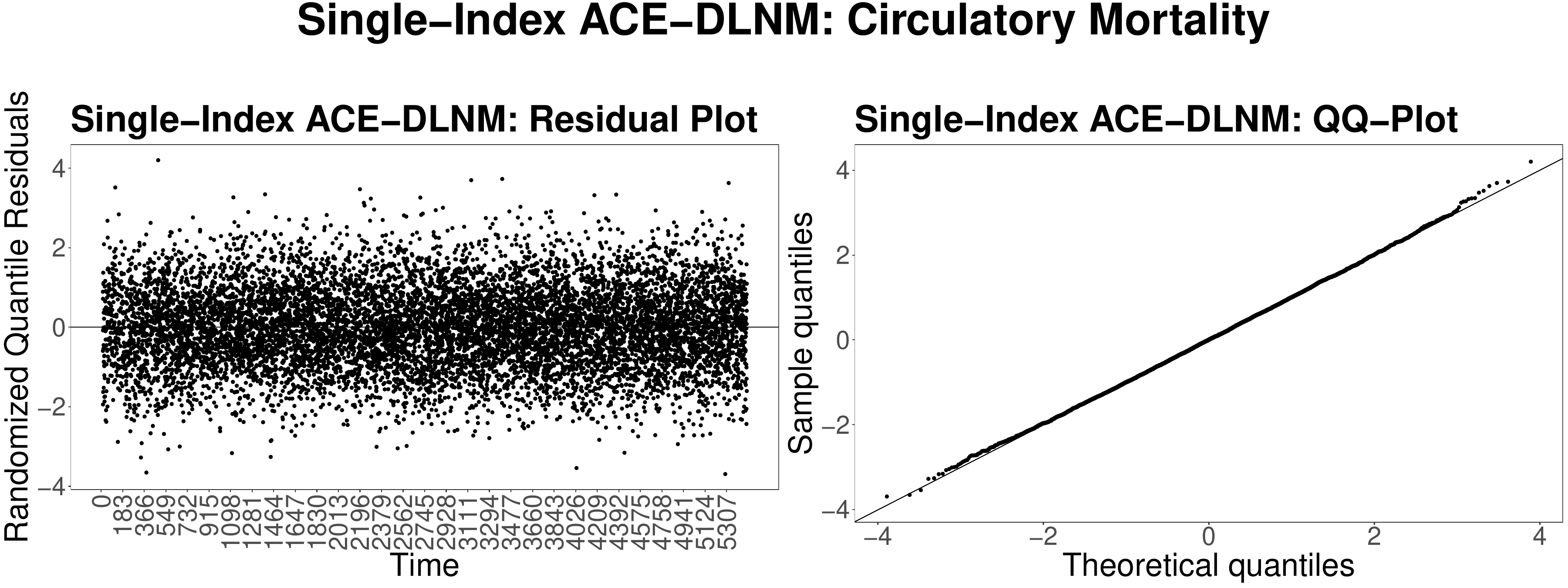}
      \end{subfigure}
    
      \medskip
    
      \begin{subfigure}[t]{0.65\textwidth}
        \includegraphics[width=\linewidth]{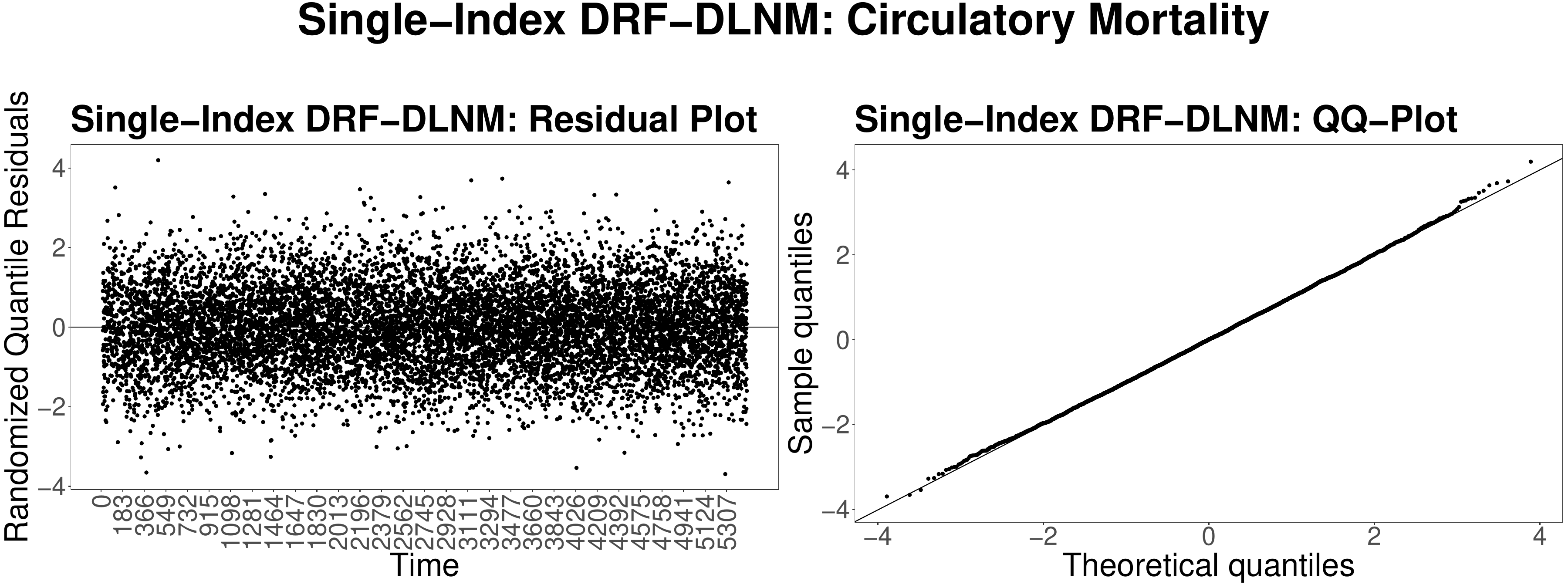}
      \end{subfigure}
    
      \medskip
    
      \begin{subfigure}[t]{0.65\textwidth}
        \includegraphics[width=\linewidth]{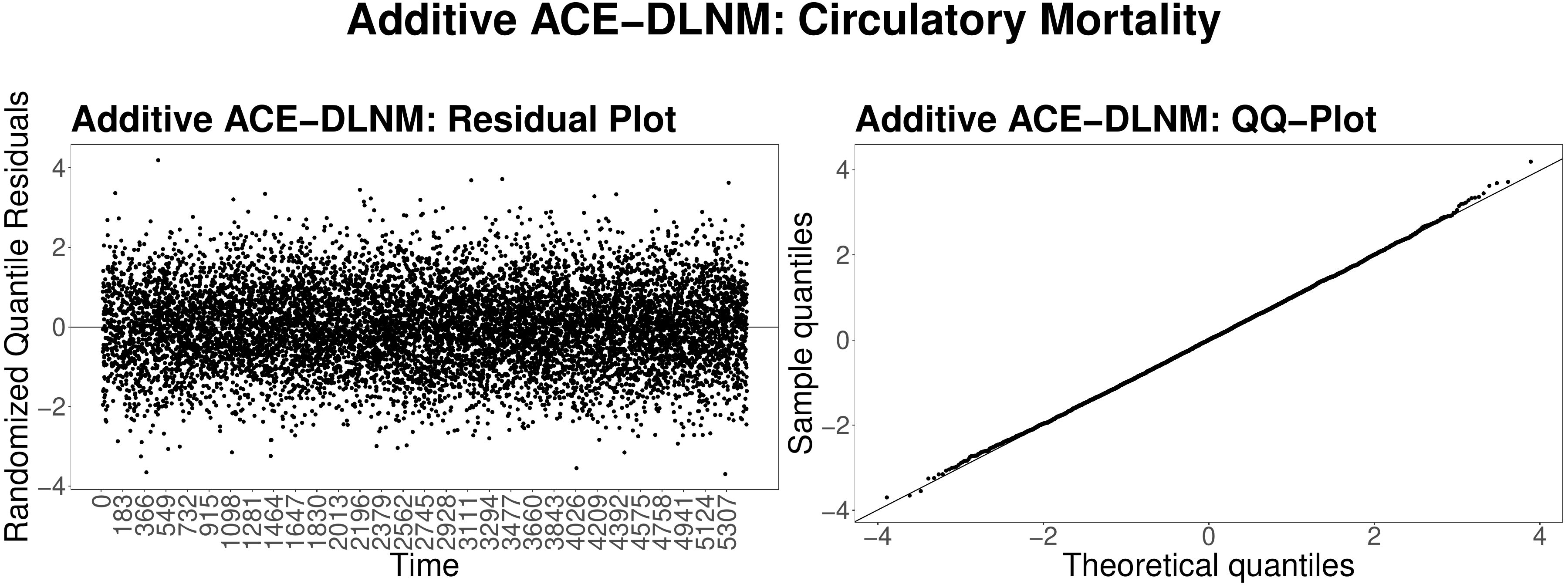}
      \end{subfigure}
    
      \medskip
    
      \begin{subfigure}[t]{0.65\textwidth}
        \includegraphics[width=\linewidth]{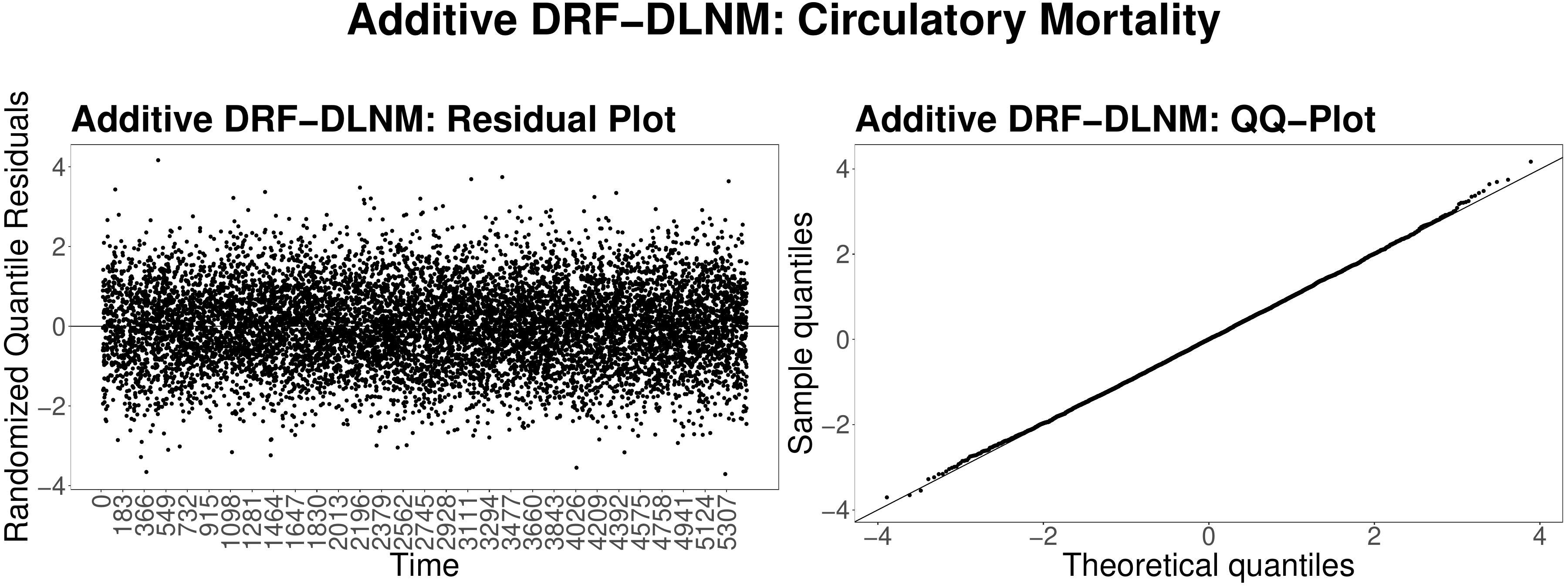}
      \end{subfigure}
      
      \caption{Randomized quantile residuals for circulatory mortality across all CDs. }
\end{figure}

\begin{figure}[H]
      \centering
      \begin{subfigure}[t]{\textwidth}
        \includegraphics[width=\linewidth]{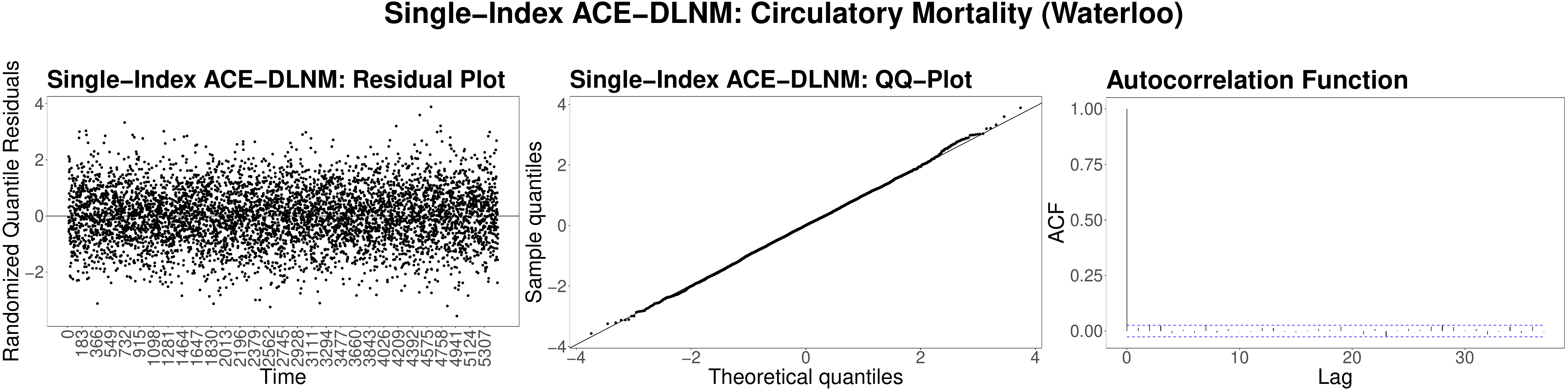}
      \end{subfigure}
    
      \medskip
    
      \begin{subfigure}[t]{\textwidth}
        \includegraphics[width=\linewidth]{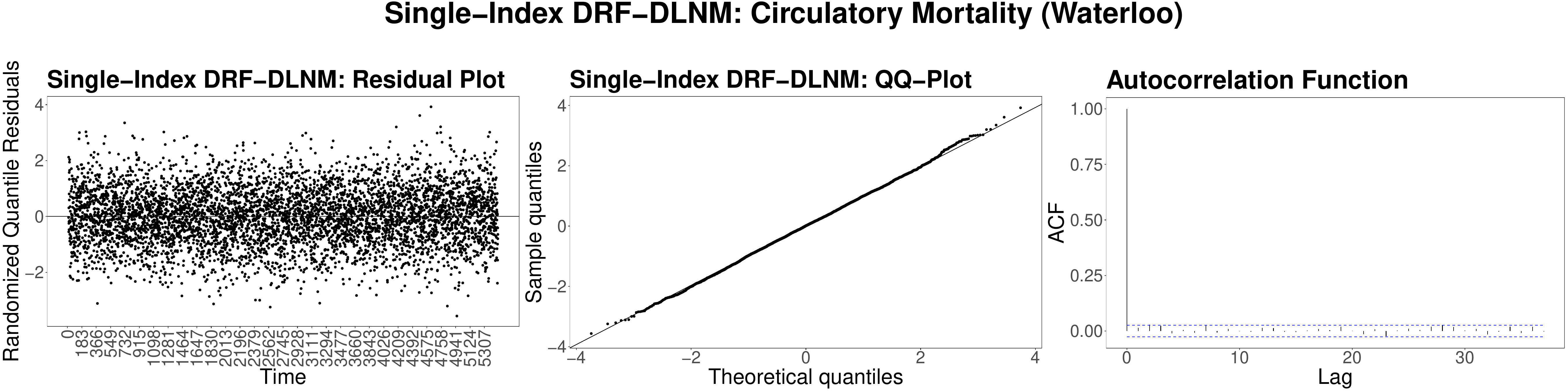}
      \end{subfigure}
    
      \medskip
    
      \begin{subfigure}[t]{\textwidth}
        \includegraphics[width=\linewidth]{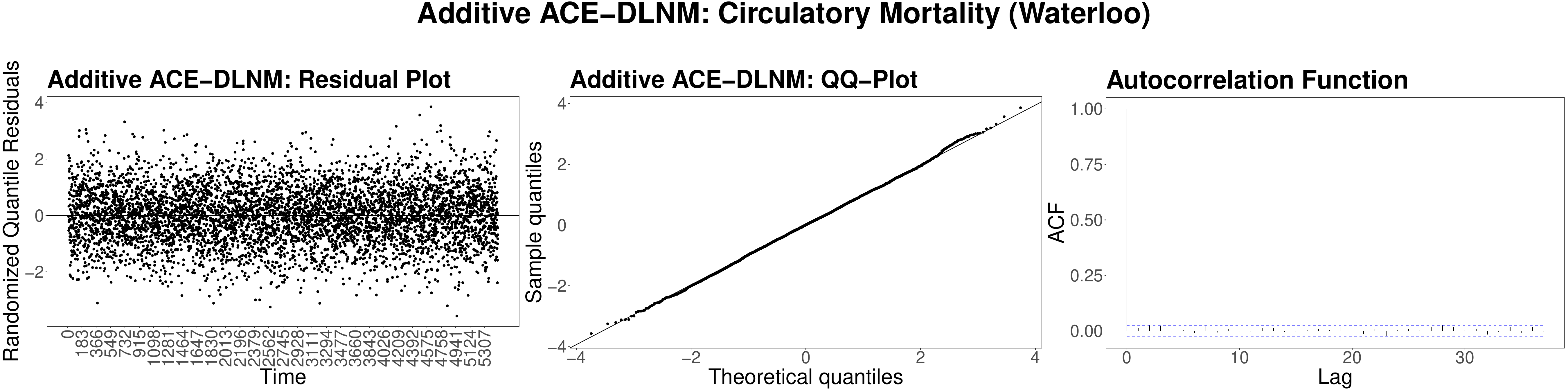}
      \end{subfigure}
    
      \medskip
    
      \begin{subfigure}[t]{\textwidth}
        \includegraphics[width=\linewidth]{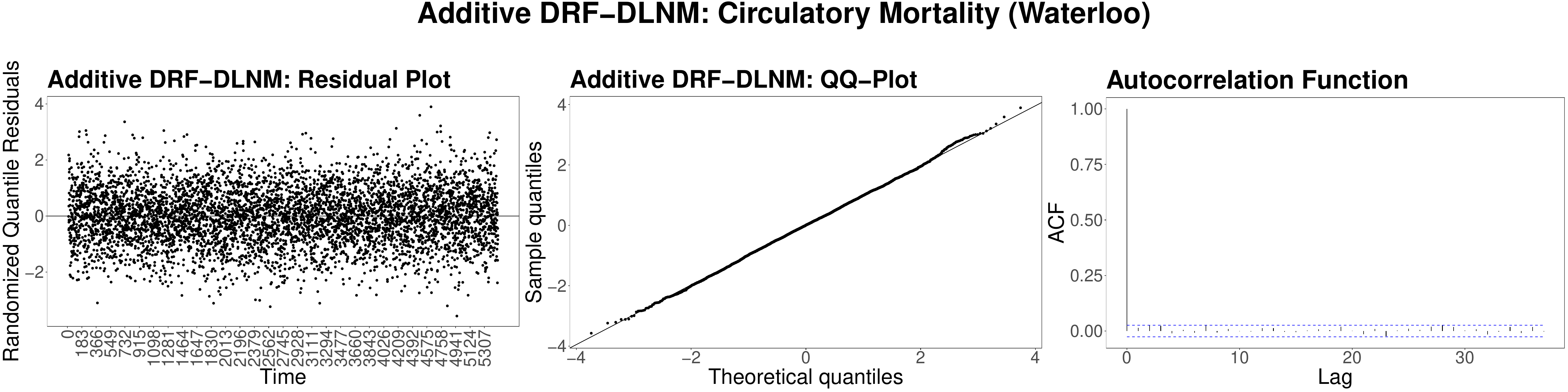}
      \end{subfigure}
      
      \caption{Randomized quantile residuals for circulatory mortality for Waterloo. }
\end{figure}

\begin{figure}[H]
      \centering
      \begin{subfigure}[t]{\textwidth}
        \includegraphics[width=\linewidth]{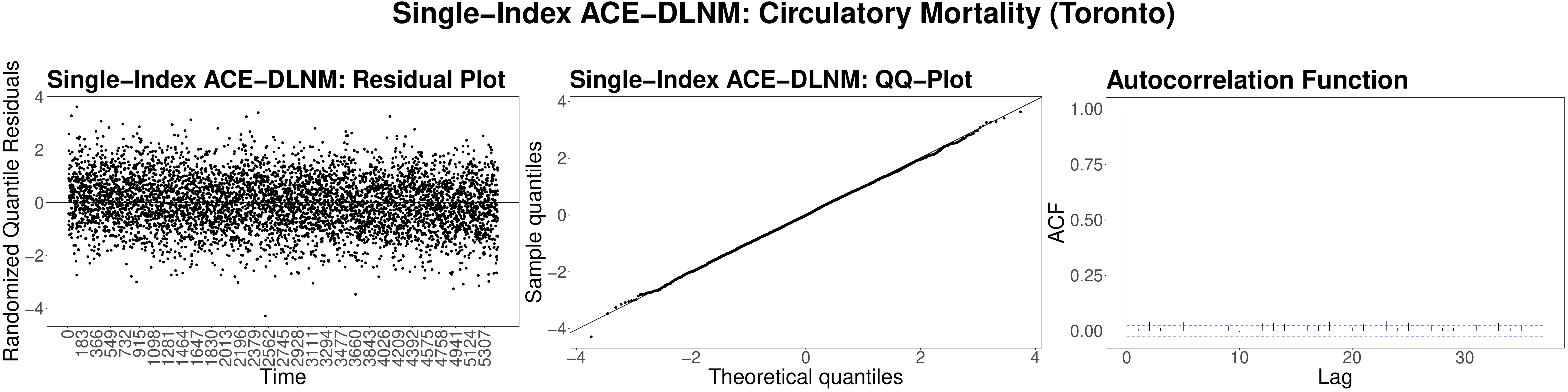}
      \end{subfigure}
    
      \medskip
    
      \begin{subfigure}[t]{\textwidth}
        \includegraphics[width=\linewidth]{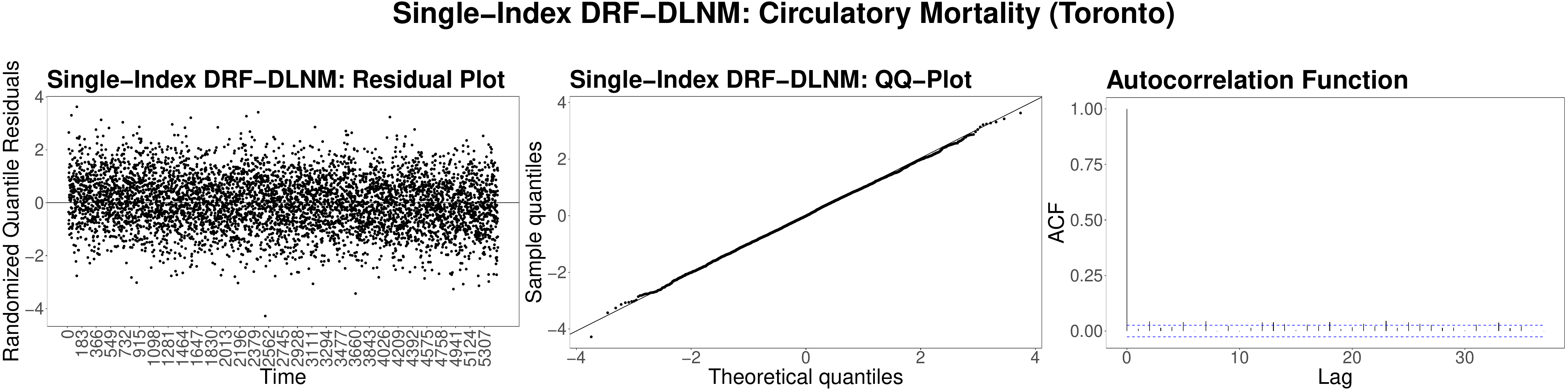}
      \end{subfigure}
    
      \medskip
    
      \begin{subfigure}[t]{\textwidth}
        \includegraphics[width=\linewidth]{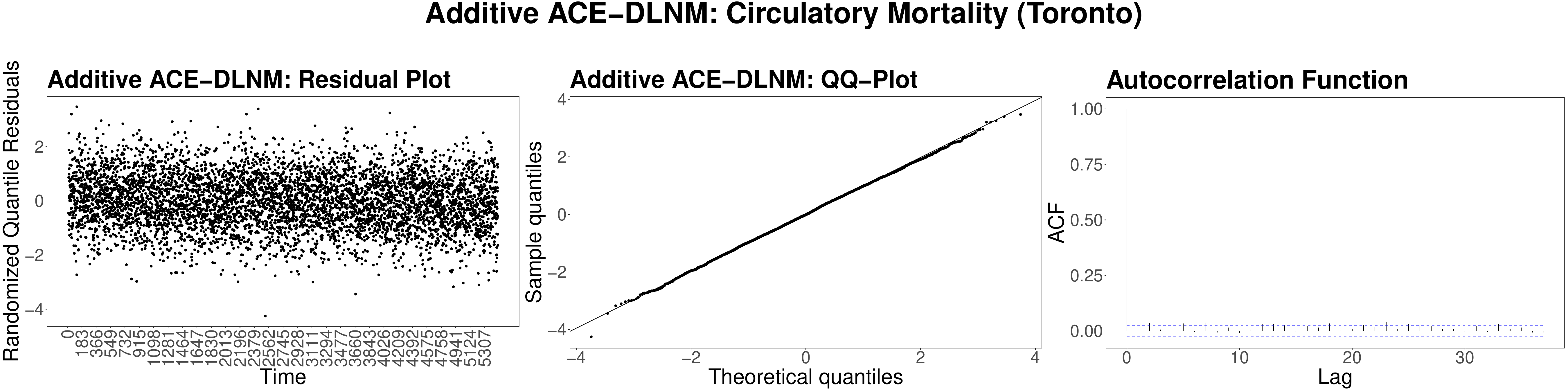}
      \end{subfigure}
    
      \medskip
    
      \begin{subfigure}[t]{\textwidth}
        \includegraphics[width=\linewidth]{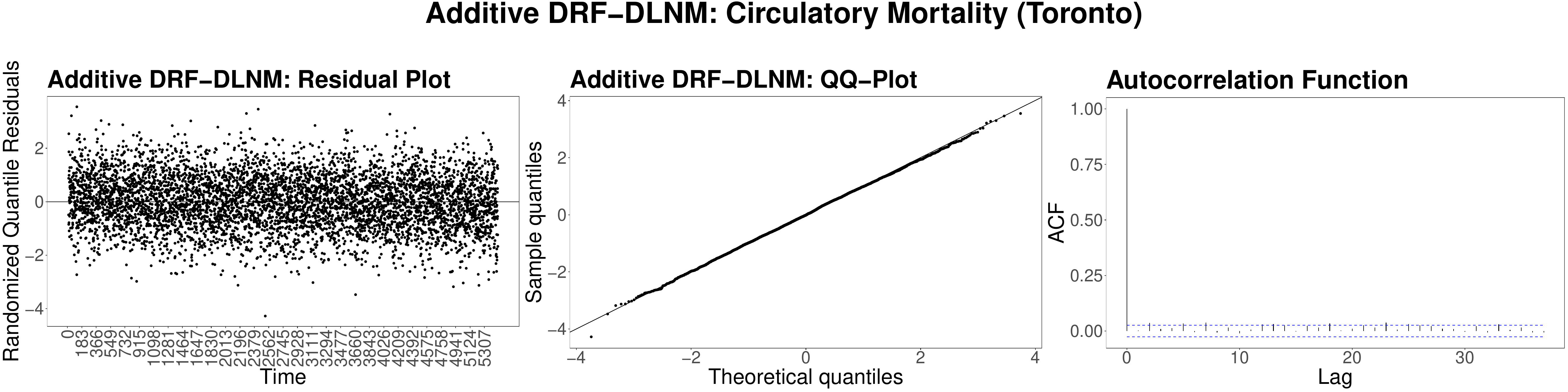}
      \end{subfigure}
      
      \caption{Randomized quantile residuals for circulatory mortality for Toronto. }
\end{figure}

\subsubsection{All-Cause Mortality}
\begin{figure}[H]
      \centering
      \begin{subfigure}[t]{0.65\textwidth}
        \includegraphics[width=\linewidth]{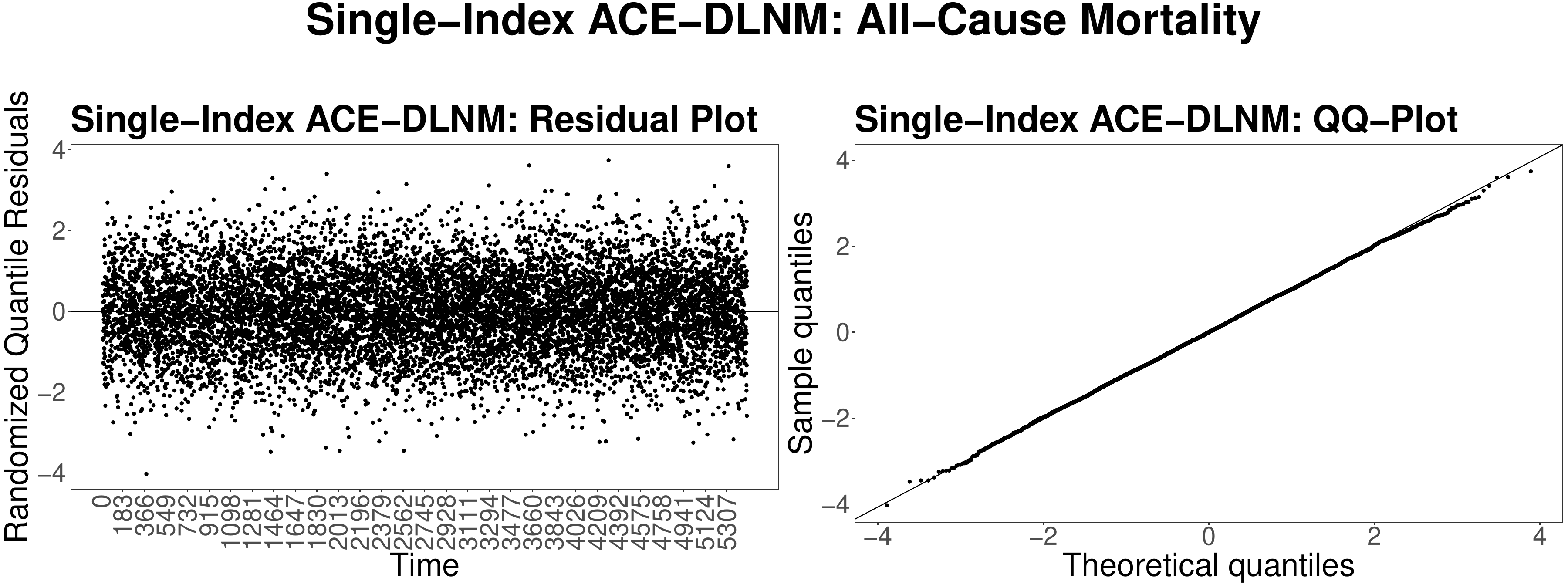}
      \end{subfigure}
    
      \medskip
    
      \begin{subfigure}[t]{0.65\textwidth}
        \includegraphics[width=\linewidth]{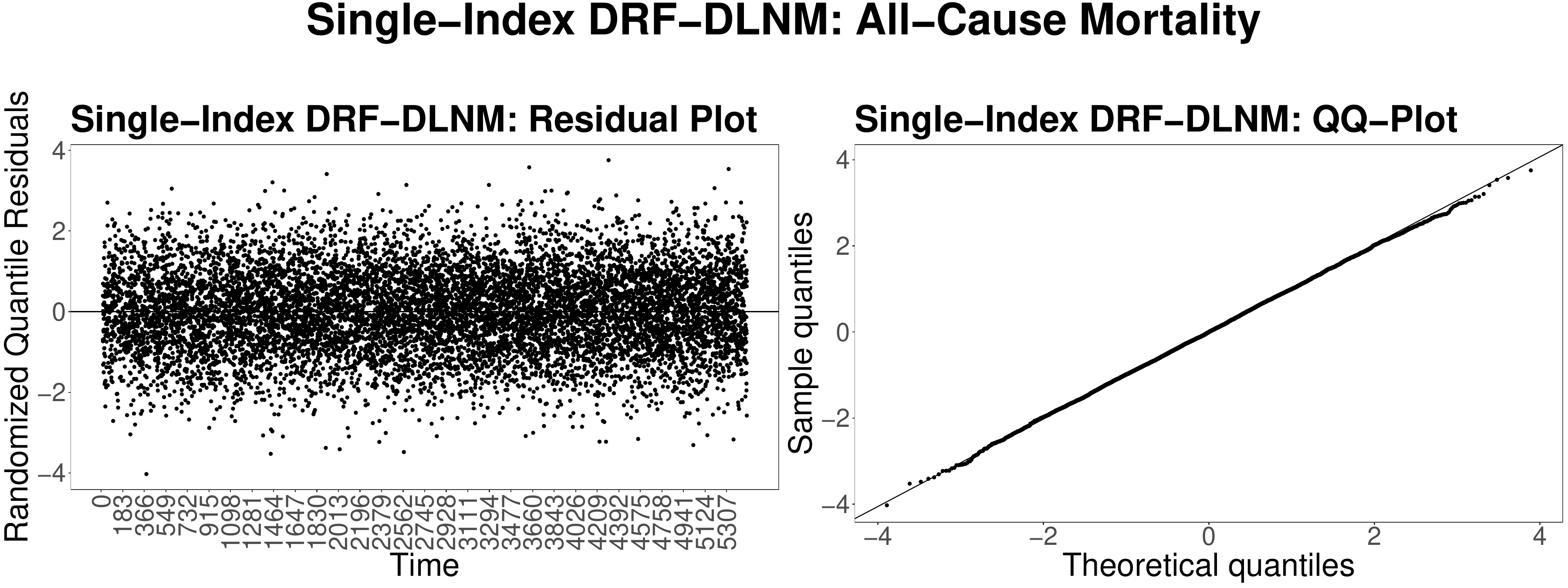}
      \end{subfigure}
    
      \medskip
    
      \begin{subfigure}[t]{0.65\textwidth}
        \includegraphics[width=\linewidth]{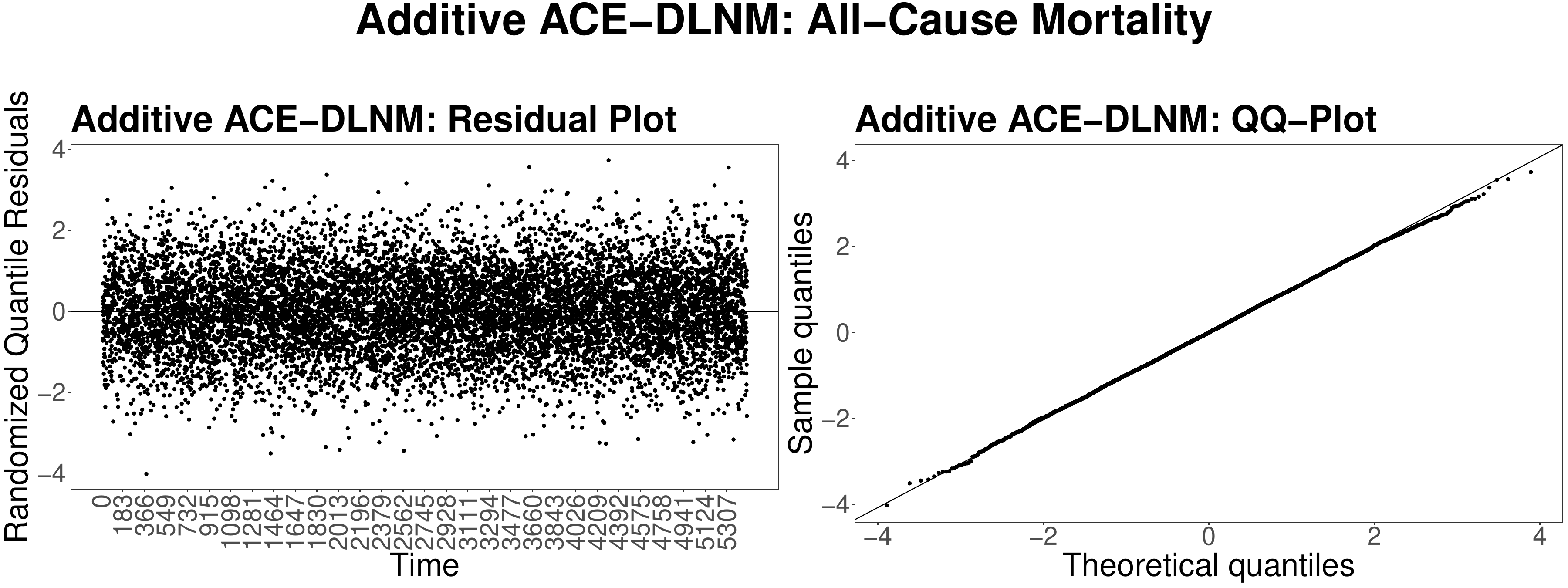}
      \end{subfigure}
    
      \medskip
    
      \begin{subfigure}[t]{0.65\textwidth}
        \includegraphics[width=\linewidth]{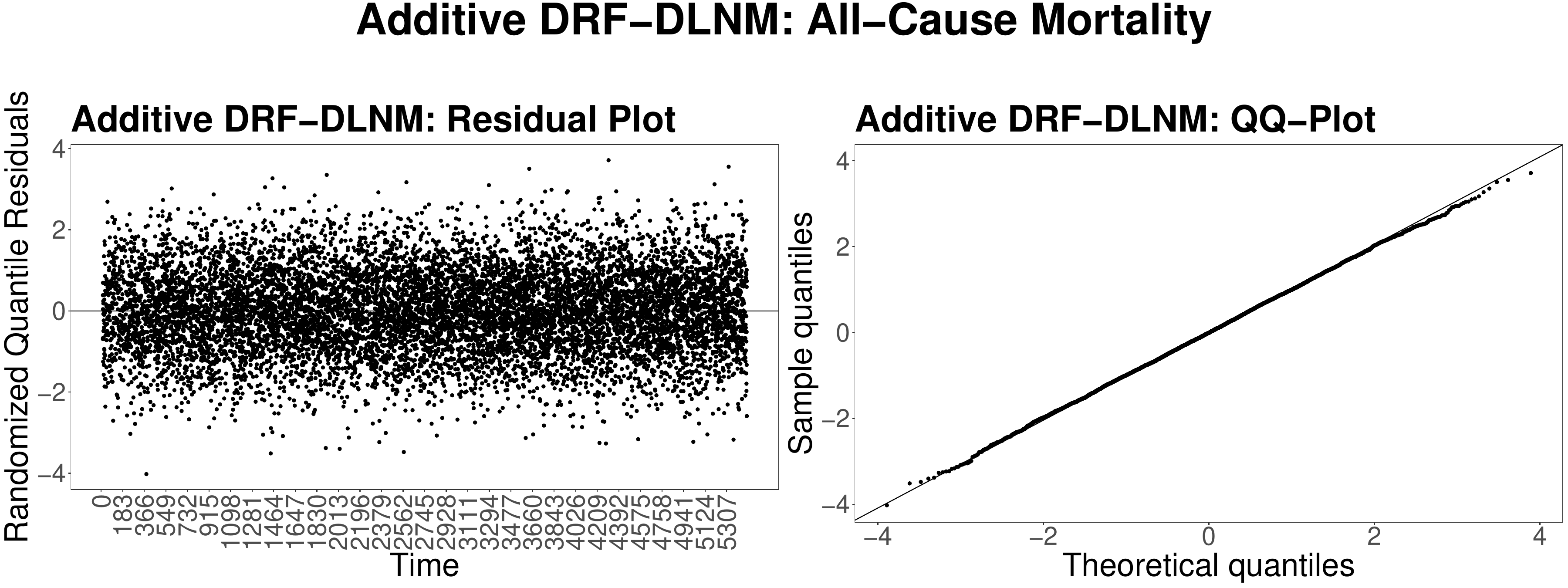}
      \end{subfigure}
      
      \caption{Randomized quantile residuals for all-cause mortality across all CDs. }
\end{figure}

\begin{figure}[H]
      \centering
      \begin{subfigure}[t]{\textwidth}
        \includegraphics[width=\linewidth]{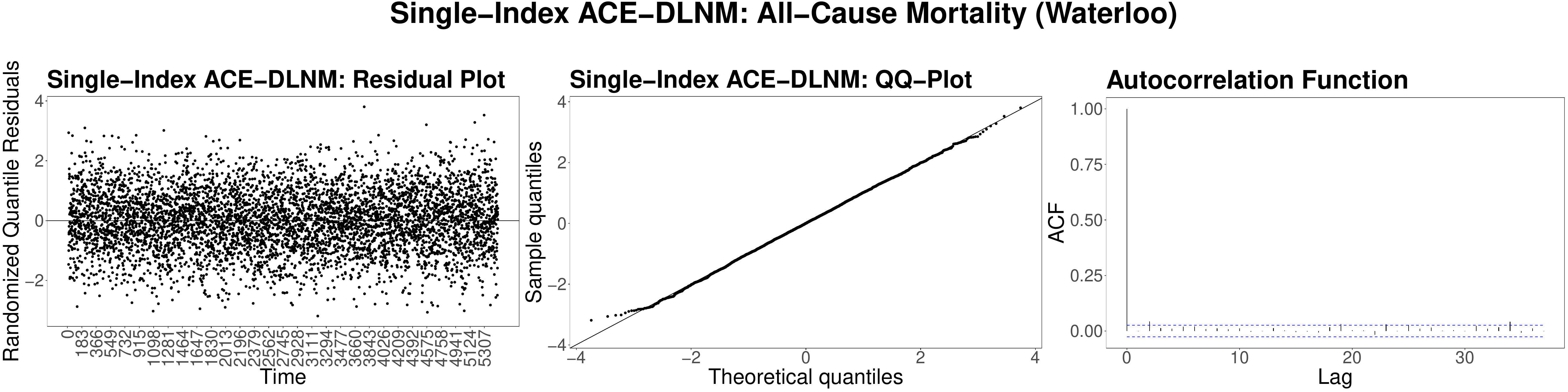}
      \end{subfigure}
    
      \medskip
    
      \begin{subfigure}[t]{\textwidth}
        \includegraphics[width=\linewidth]{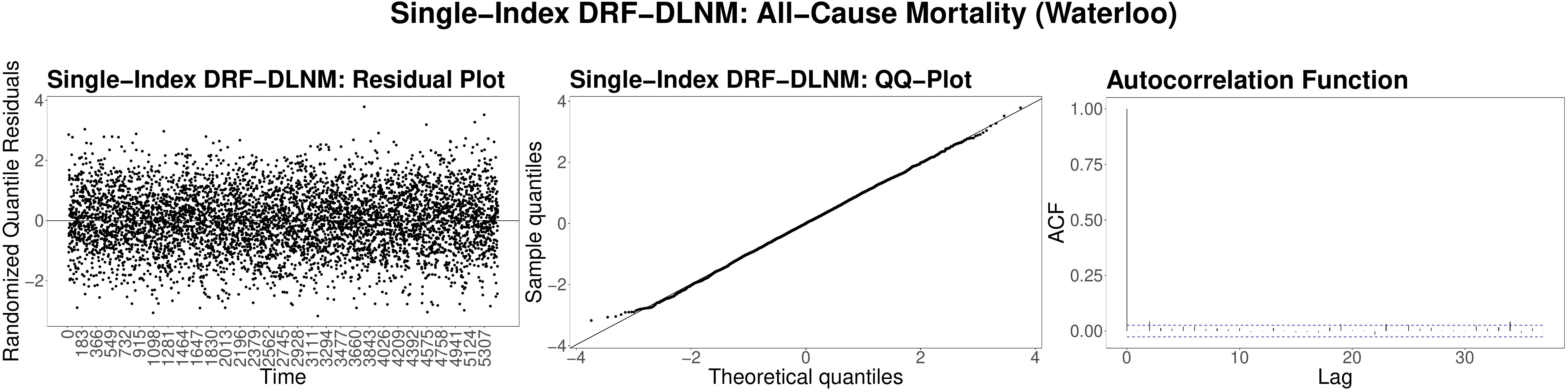}
      \end{subfigure}
    
      \medskip
    
      \begin{subfigure}[t]{\textwidth}
        \includegraphics[width=\linewidth]{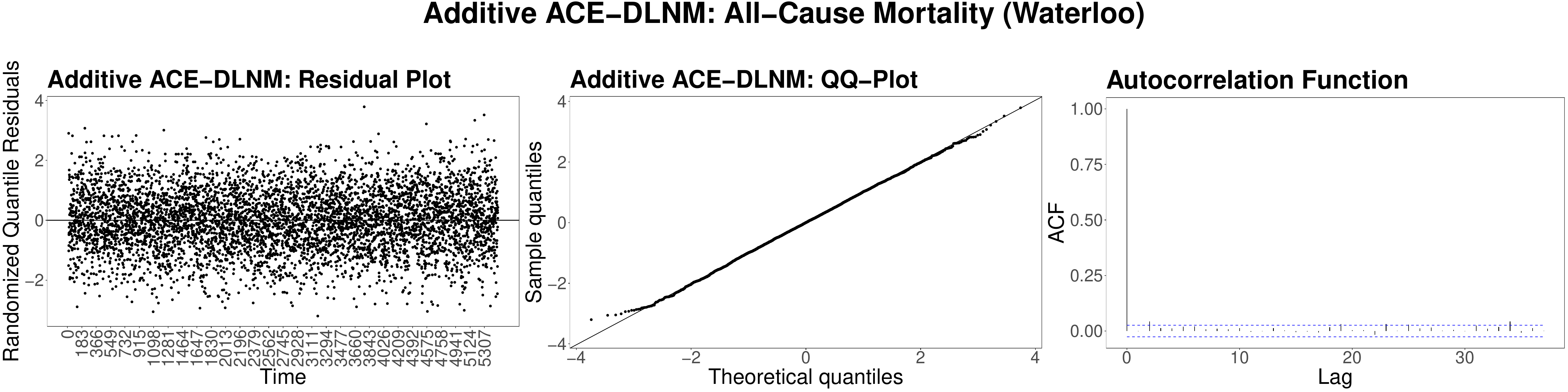}
      \end{subfigure}
    
      \medskip
    
      \begin{subfigure}[t]{\textwidth}
        \includegraphics[width=\linewidth]{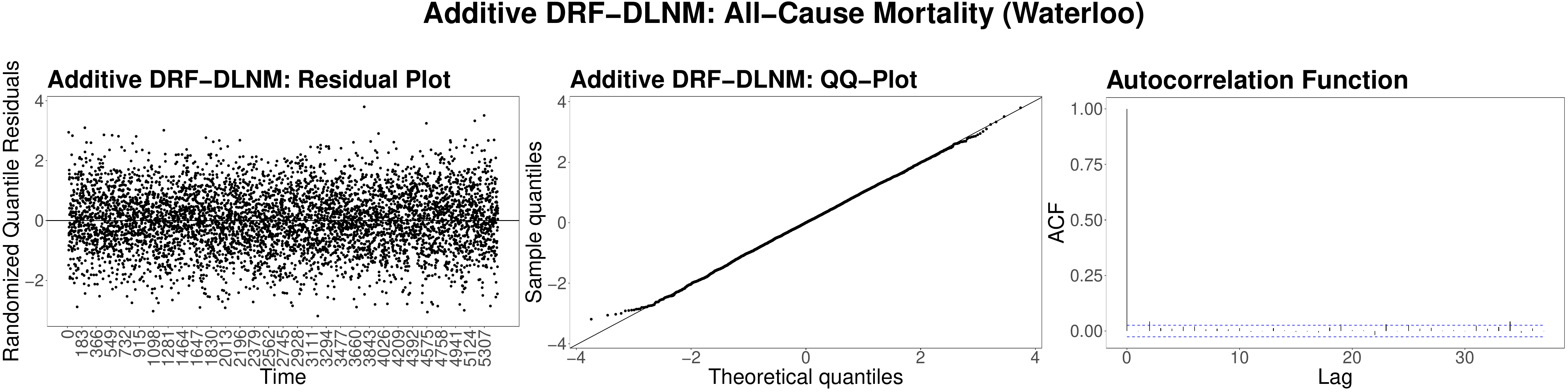}
      \end{subfigure}
      
      \caption{Randomized quantile residuals for all-cause mortality for Waterloo. }
\end{figure}

\begin{figure}[H]
      \centering
      \begin{subfigure}[t]{\textwidth}
        \includegraphics[width=\linewidth]{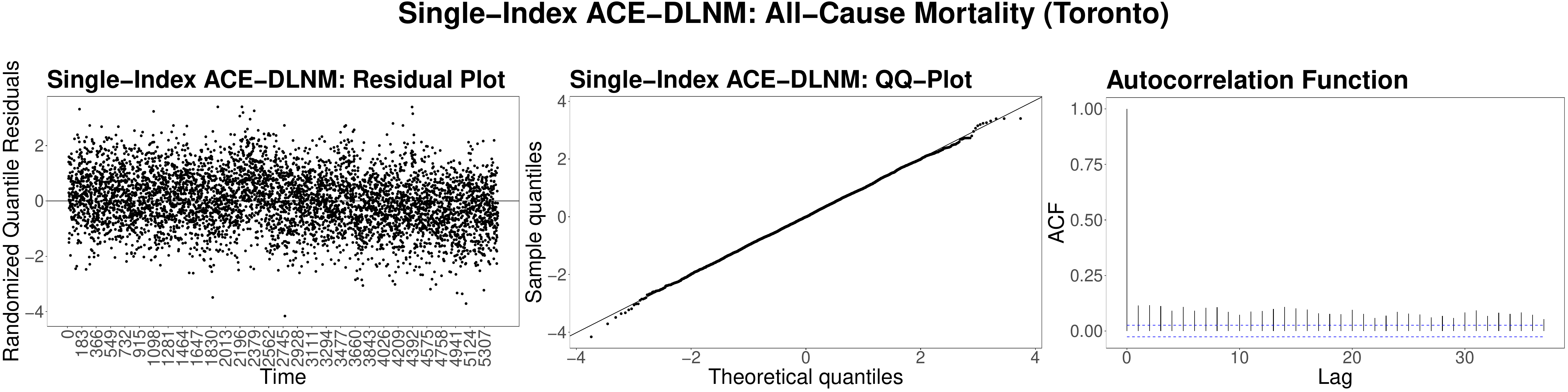}
      \end{subfigure}
    
      \medskip
    
      \begin{subfigure}[t]{\textwidth}
        \includegraphics[width=\linewidth]{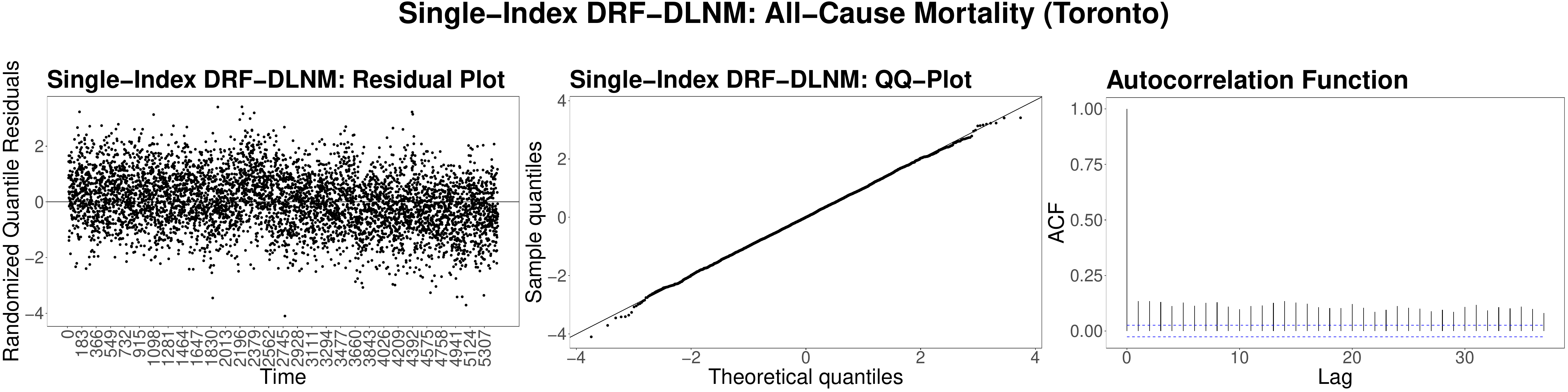}
      \end{subfigure}
    
      \medskip
    
      \begin{subfigure}[t]{\textwidth}
        \includegraphics[width=\linewidth]{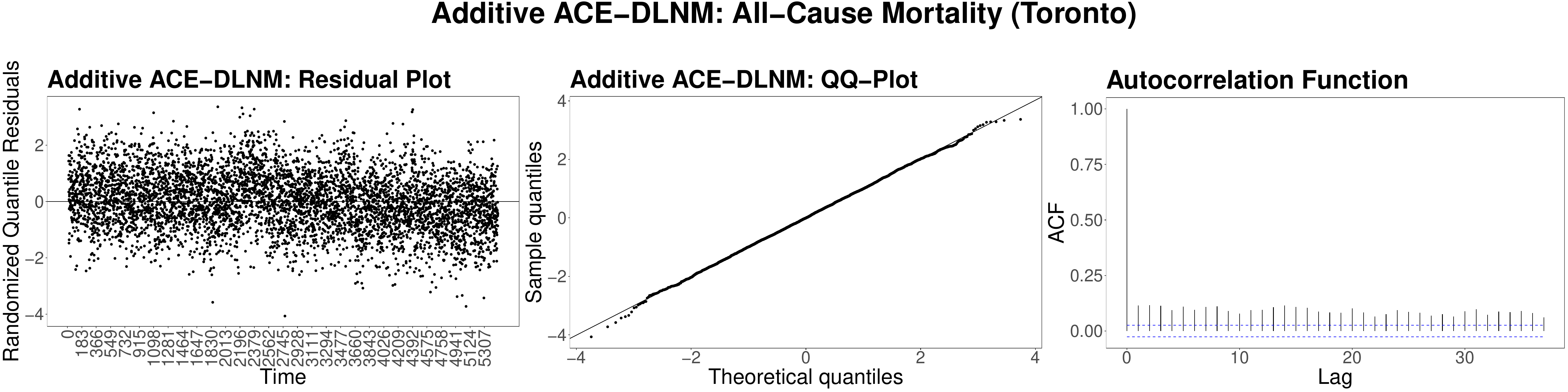}
      \end{subfigure}
    
      \medskip
    
      \begin{subfigure}[t]{\textwidth}
        \includegraphics[width=\linewidth]{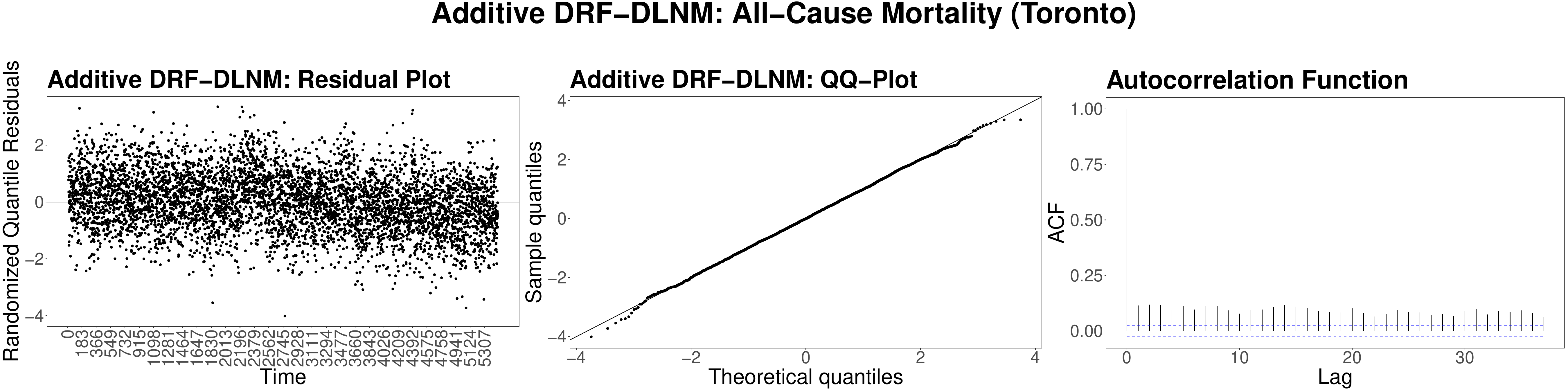}
      \end{subfigure}
      
      \caption{Randomized quantile residuals for all-cause mortality for Toronto. }
\end{figure}

\clearpage

\subsection{Rate Ratio}
\begin{figure}[H]
    \centering
    \includegraphics[width=\linewidth]{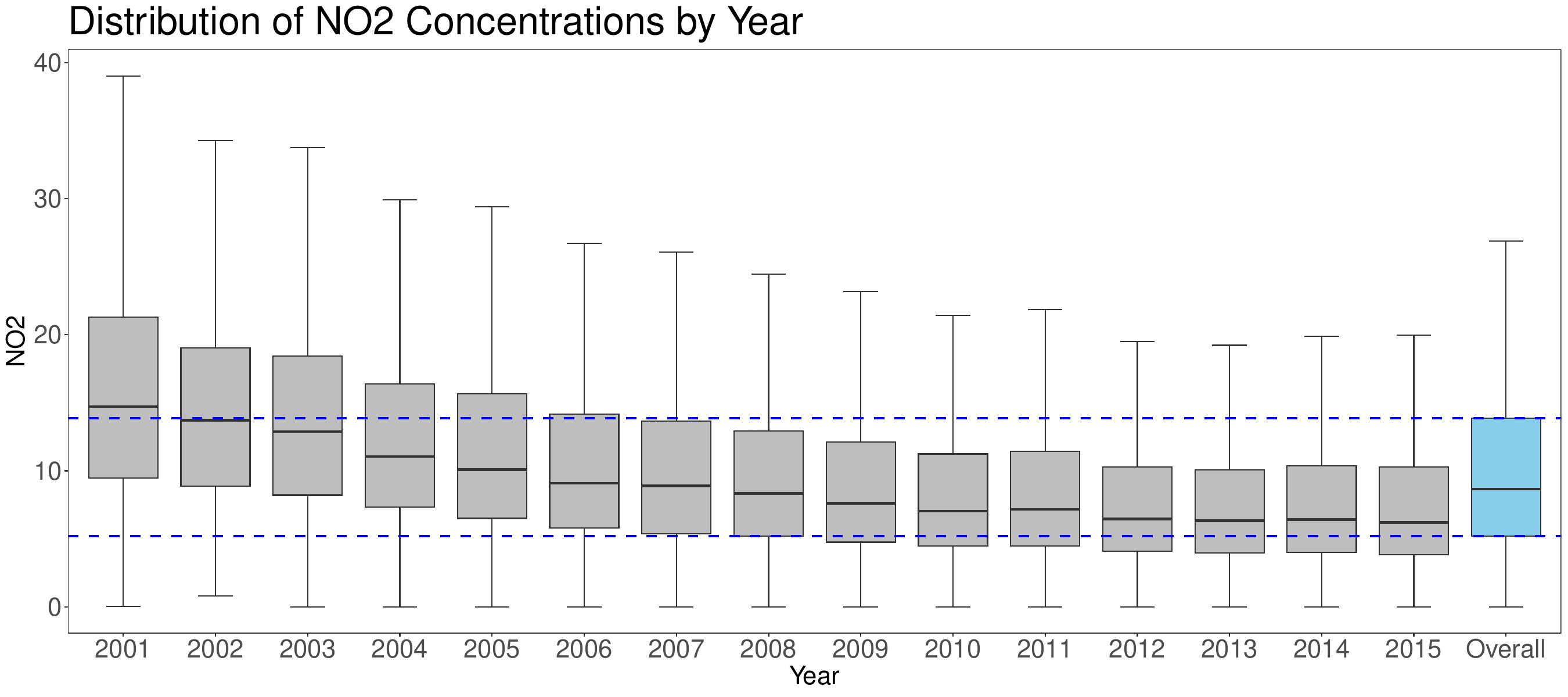}
    \caption{Boxplots of NO$_2$ concentrations by year. The boxplot for the overall distribution across all years is shown in blue. }
    \label{fig:box-NO2}
\end{figure}

\begin{figure}[H]
    \centering
    \includegraphics[width=0.9\linewidth]{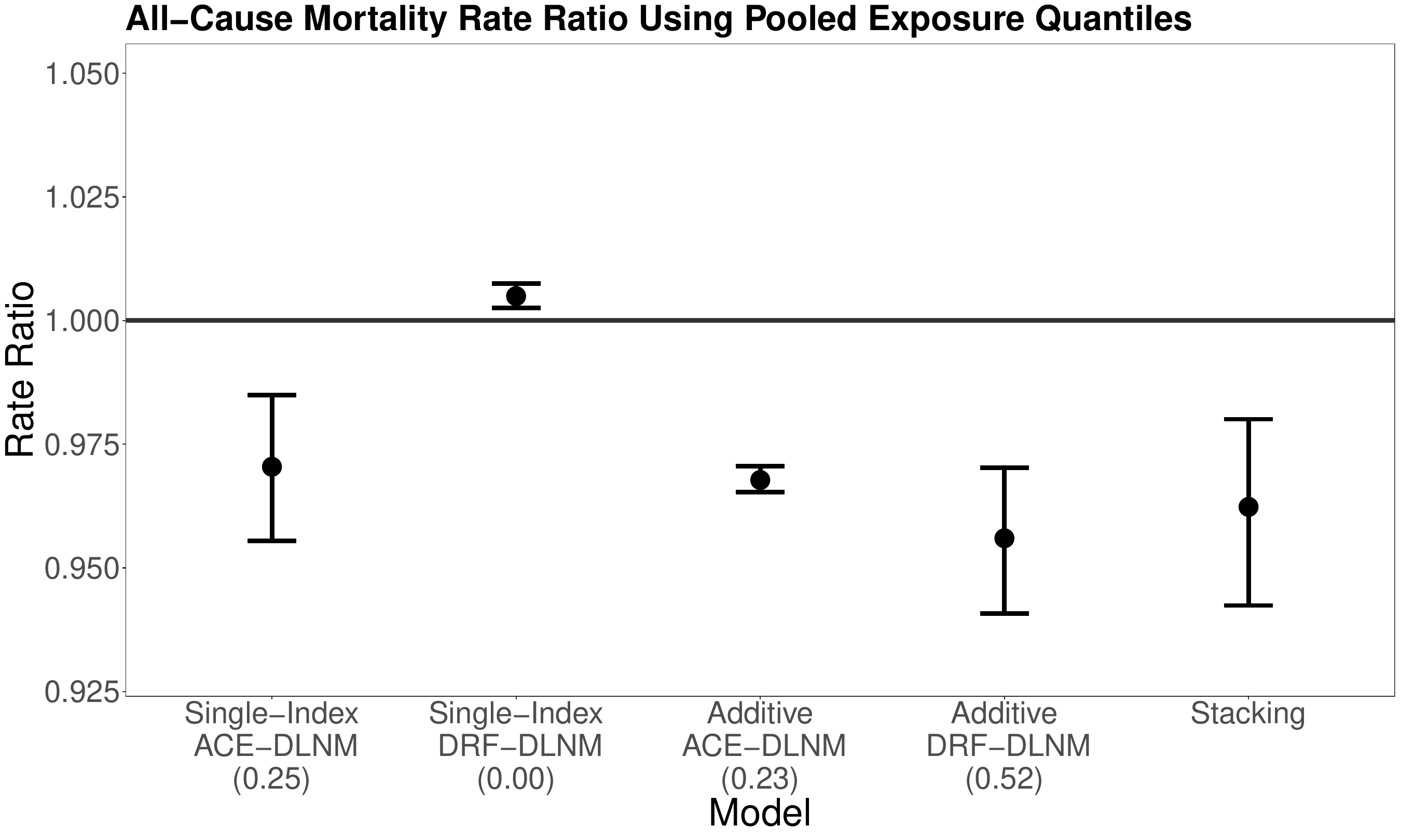}
    \caption{Rate ratio of all-cause mortality comparing scenarios where all three air pollutants are at their 75\% quantiles versus their 25\% quantiles. The quantiles are calculated based on the pooled distribution of concentration across all CDs and years.
    }
    \label{fig:RRall}
\end{figure}

\begin{figure}[H]
    \centering
    \includegraphics[width=0.9\linewidth]{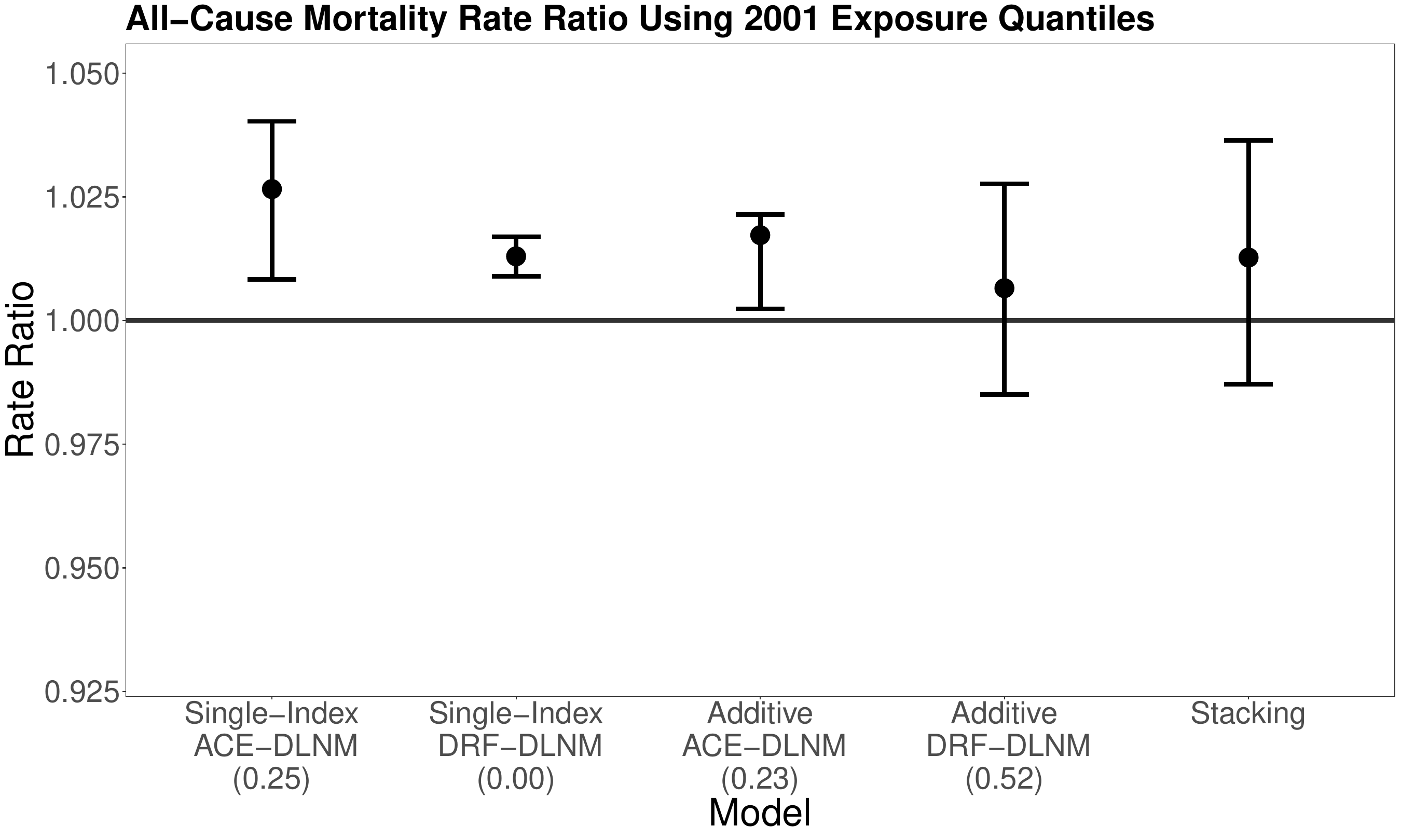}
    \caption{Rate ratio of all-cause mortality comparing scenarios where all three air pollutants are at their 75\% quantiles versus their 25\% quantiles. The quantiles are calculated based on the pooled distribution of concentration across all CDs in 2001.
    }
    \label{fig:RR2001}
\end{figure}

A commonly used estimand is the rate ratio corresponding to a contrast between two fixed exposure levels, such as the 75\% and 25\% percentiles of all observed concentrations, representing relatively high and low levels (e.g. \citealt{krall2013short}). 
In our data analysis, however, we observe strong temporal trends in exposure levels. 
Figure \ref{fig:box-NO2} shows that NO$_2$ levels decline from 2001 to 2015, and the 75\% quantile from the pooled distribution across all years is below the median of NO$_2$ levels in 2001. It is difficult to define two fixed exposure levels that consistently represent relatively high and low levels over time. 
We observe that the estimates are sensitive to the choice of quantiles: quantiles based on pooled distributions lead to significant negative associations between air pollution and all-cause mortality (Figure \ref{fig:RRall}), while quantiles based on exposure levels in 2001 yield positive point estimates with wide confidence intervals (Figure \ref{fig:RR2001}). 
This motivates us to consider an estimand that acknowledges the scale and temporal trends of the observed pollution levels. Therefore, we use the relative mortality reduction in Section 7.

\subsection{CD-Specific Associations}
\label{ss:CD-specific}
\begin{figure}[H]
    \centering
    \includegraphics[width=\linewidth]{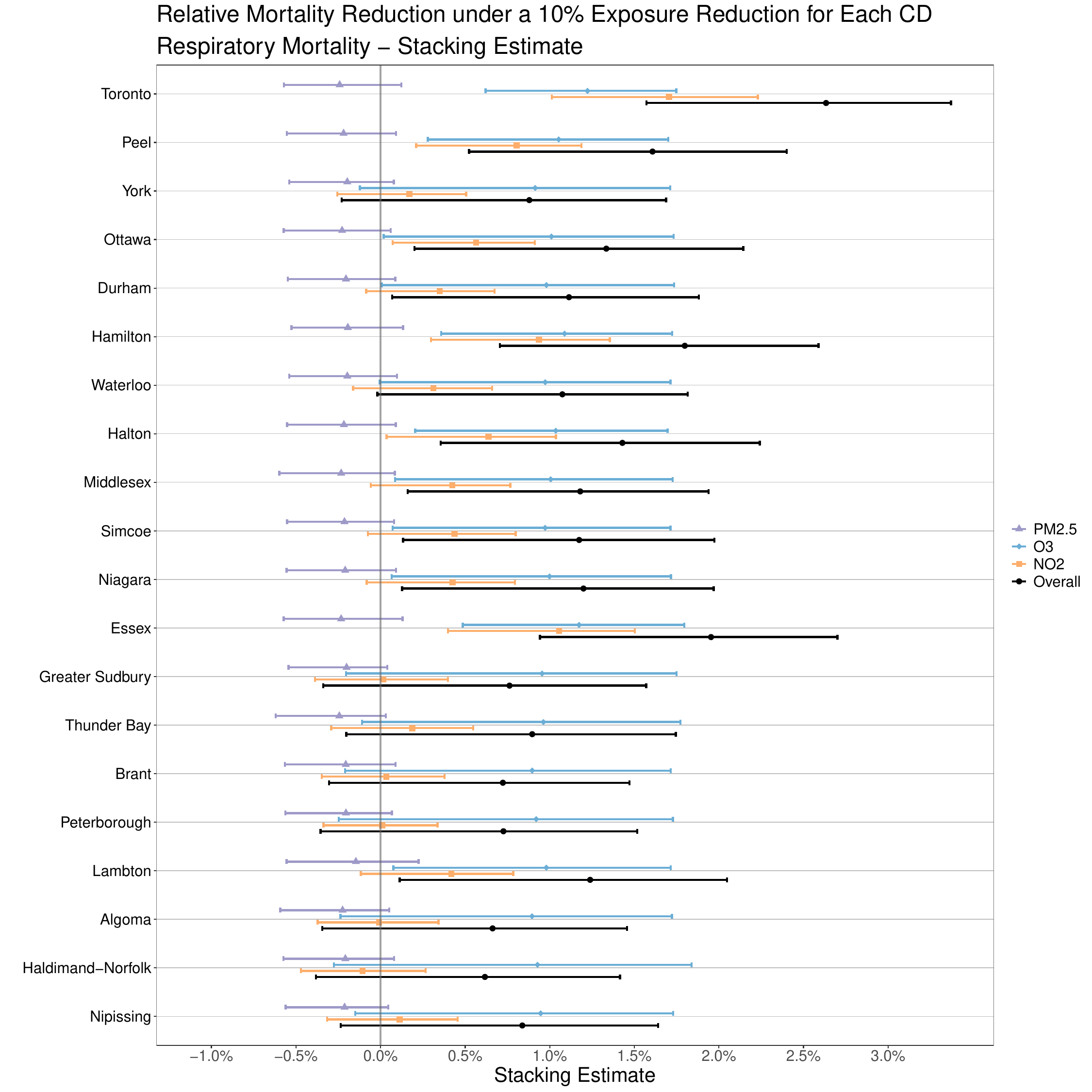}
    \caption{Relative mortality reduction for respiratory mortality under a 10\% exposure reduction relative to the observed exposure levels for each CD. The point estimates and 95\% confidence intervals are obtained using model stacking. 
    The overall effect corresponds to a 10\% reduction in all three pollutants, while the individual effects correspond to a 10\% reduction in one pollutant while holding the others at their observed levels. CD are ordered by population size. }
\end{figure}

\begin{figure}[H]
    \centering
    \includegraphics[width=\linewidth]{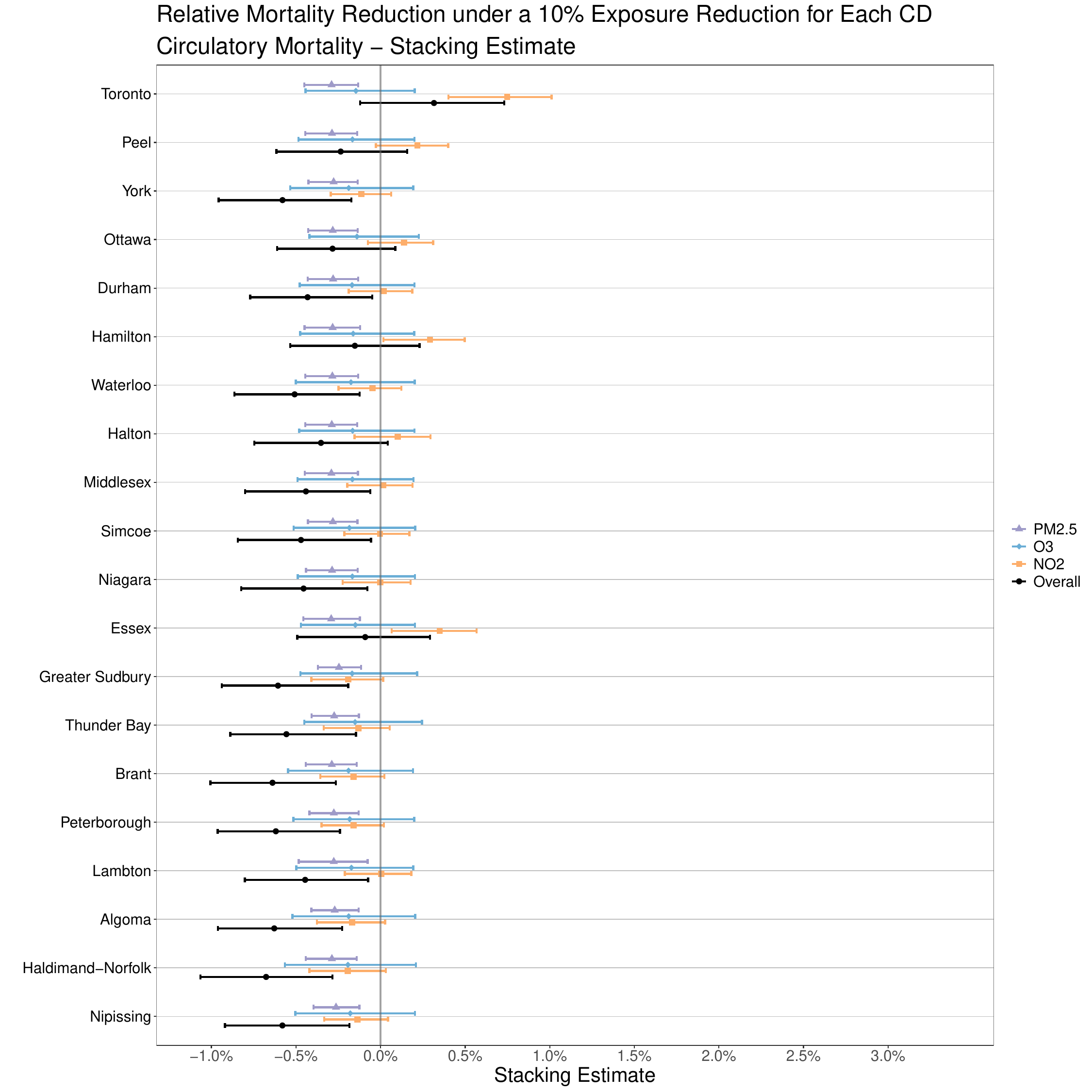}
    \caption{Relative mortality reduction for circulatory mortality under a 10\% exposure reduction relative to the observed exposure levels for each CD. The point estimates and 95\% confidence intervals are obtained using model stacking. 
    The overall effect corresponds to a 10\% reduction in all three pollutants, while the individual effects correspond to a 10\% reduction in one pollutant while holding the others at their observed levels. CD are ordered by population size. }
\end{figure}

\begin{figure}[H]
    \centering
    \includegraphics[width=\linewidth]{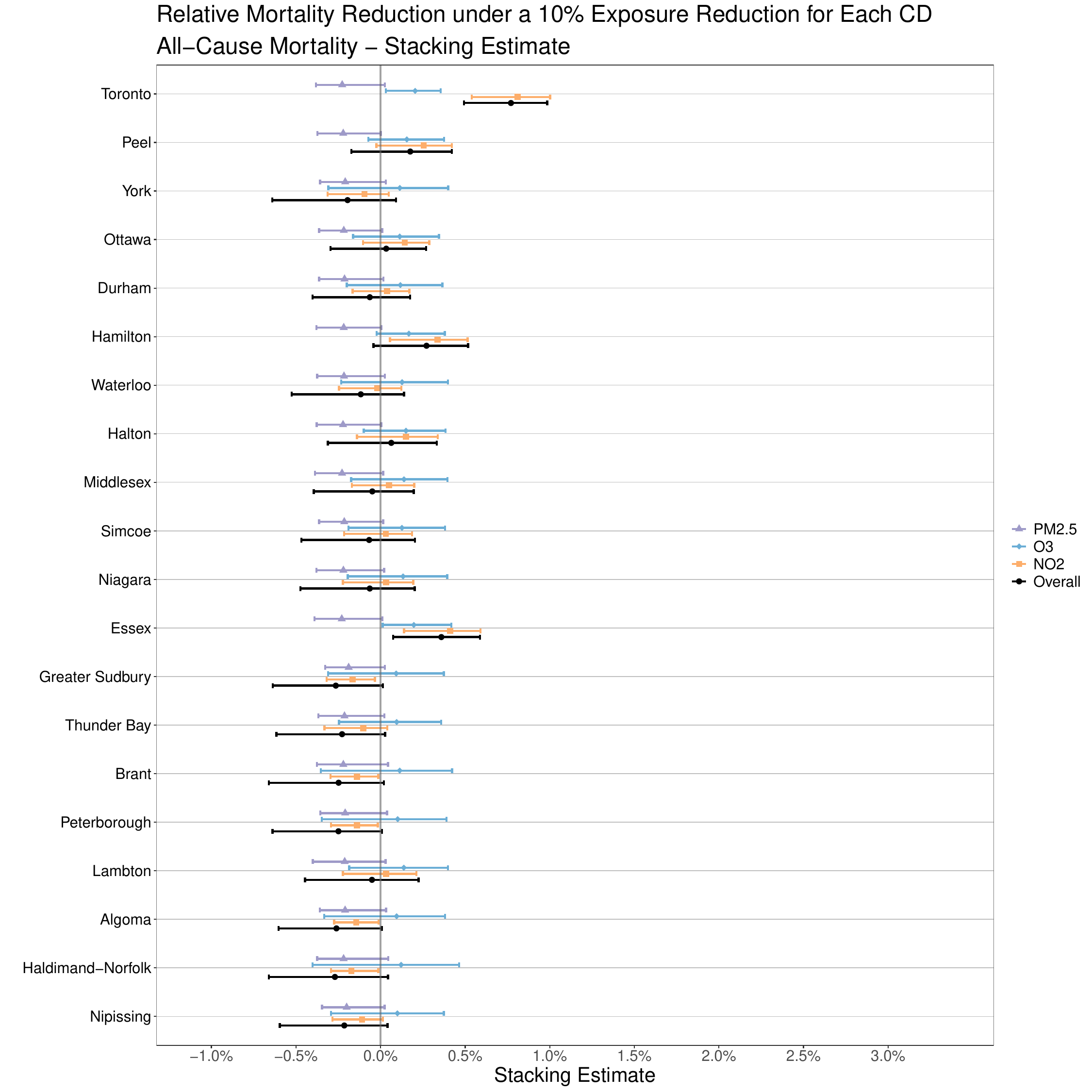}
    \caption{Relative mortality reduction for all-cause mortality under a 10\% exposure reduction relative to the observed exposure levels for each CD. The point estimates and 95\% confidence intervals are obtained using model stacking. 
    The overall effect corresponds to a 10\% reduction in all three pollutants, while the individual effects correspond to a 10\% reduction in one pollutant while holding the others at their observed levels. CD are ordered by population size. }
\end{figure}

\subsection{Year-Specific Association}

\begin{figure}[H]
    \centering
    \includegraphics[width=1.1\linewidth]{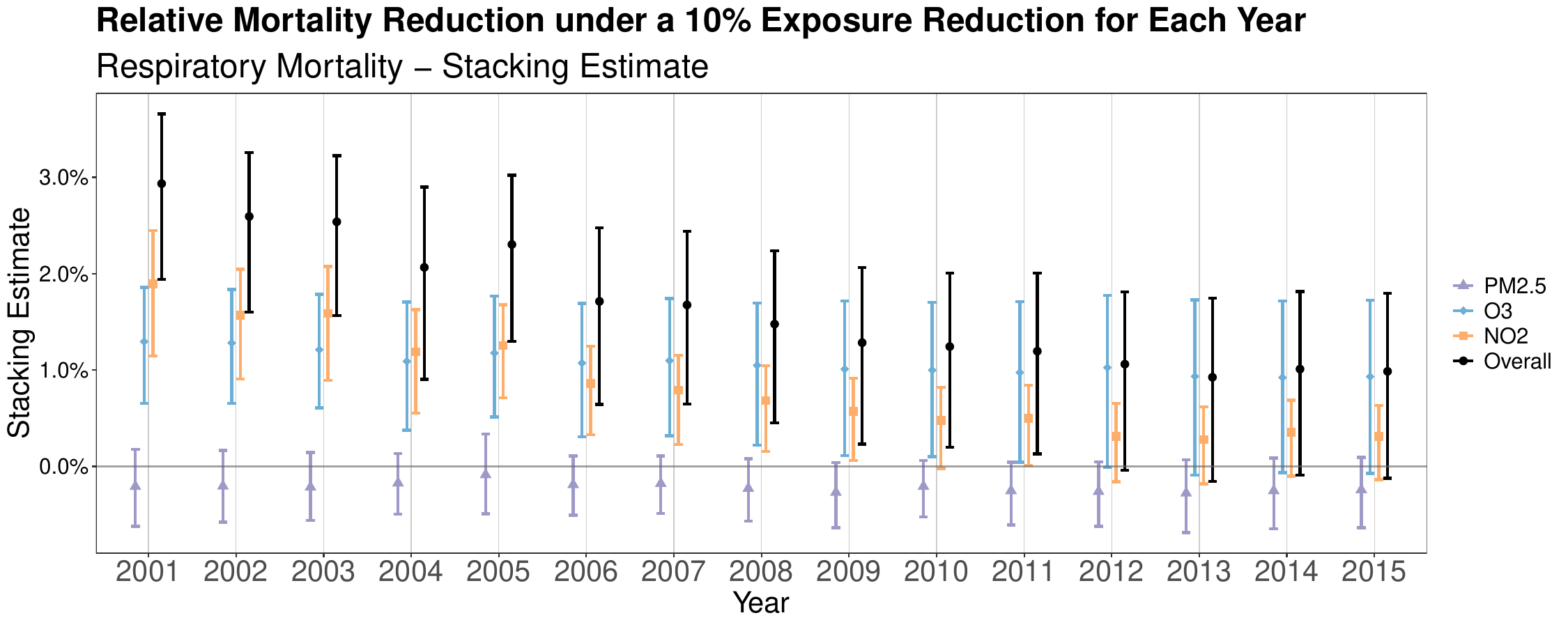}
    \caption{Year-specific relative respiratory mortality reduction under a 10\% exposure reduction. Point estimates and 95\% confidence intervals from model stacking are shown. The overall effect corresponds to a 10\% reduction in all pollutants, and the individual effects correspond to a 10\% reduction in one pollutant holding the others at observed levels. }
\end{figure}

\begin{figure}[H]
    \centering
    \includegraphics[width=1.1\linewidth]{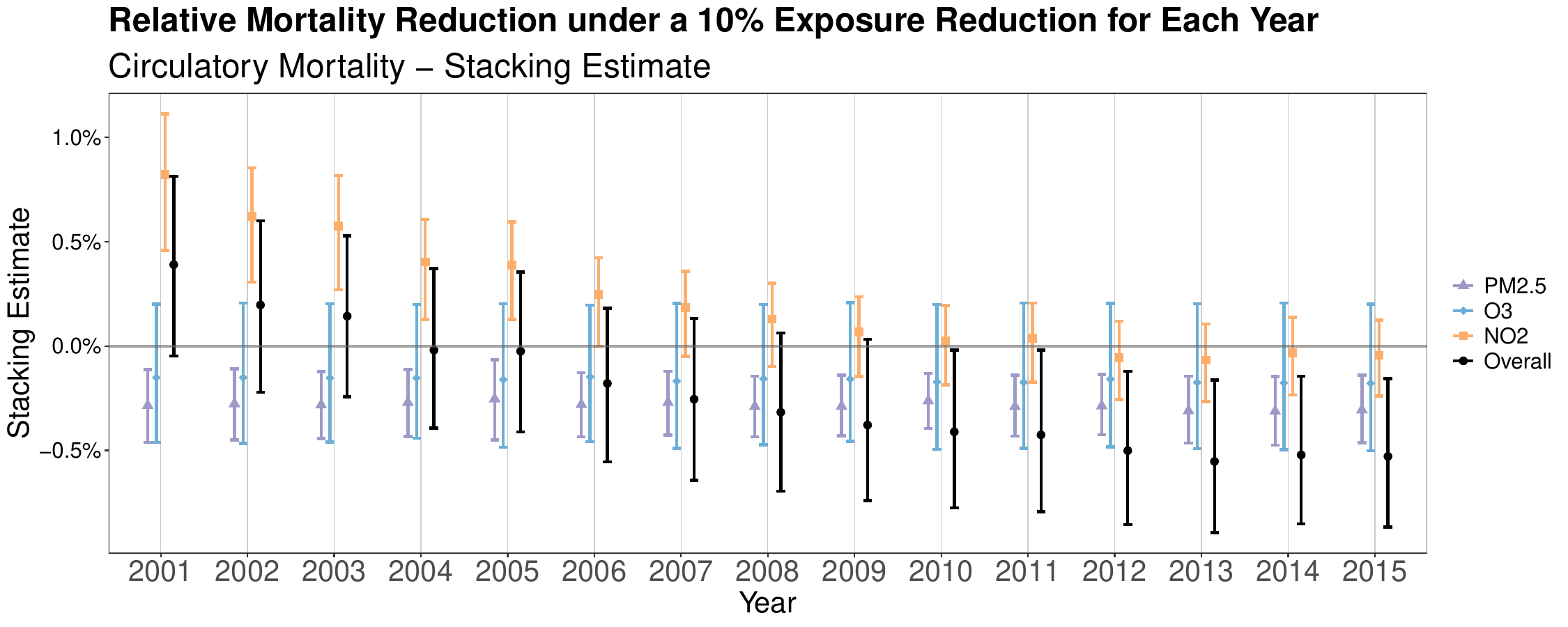}
    \caption{Year-specific relative circulatory mortality reduction under a 10\% exposure reduction. Point estimates and 95\% confidence intervals from model stacking are shown. The overall effect corresponds to a 10\% reduction in all pollutants, and the individual effects correspond to a 10\% reduction in one pollutant holding the others at observed levels. }
\end{figure}

\clearpage

\subsection{Sensitivity Analysis}

\begin{figure}[H]
    \centering
    \includegraphics[width=0.89\linewidth]{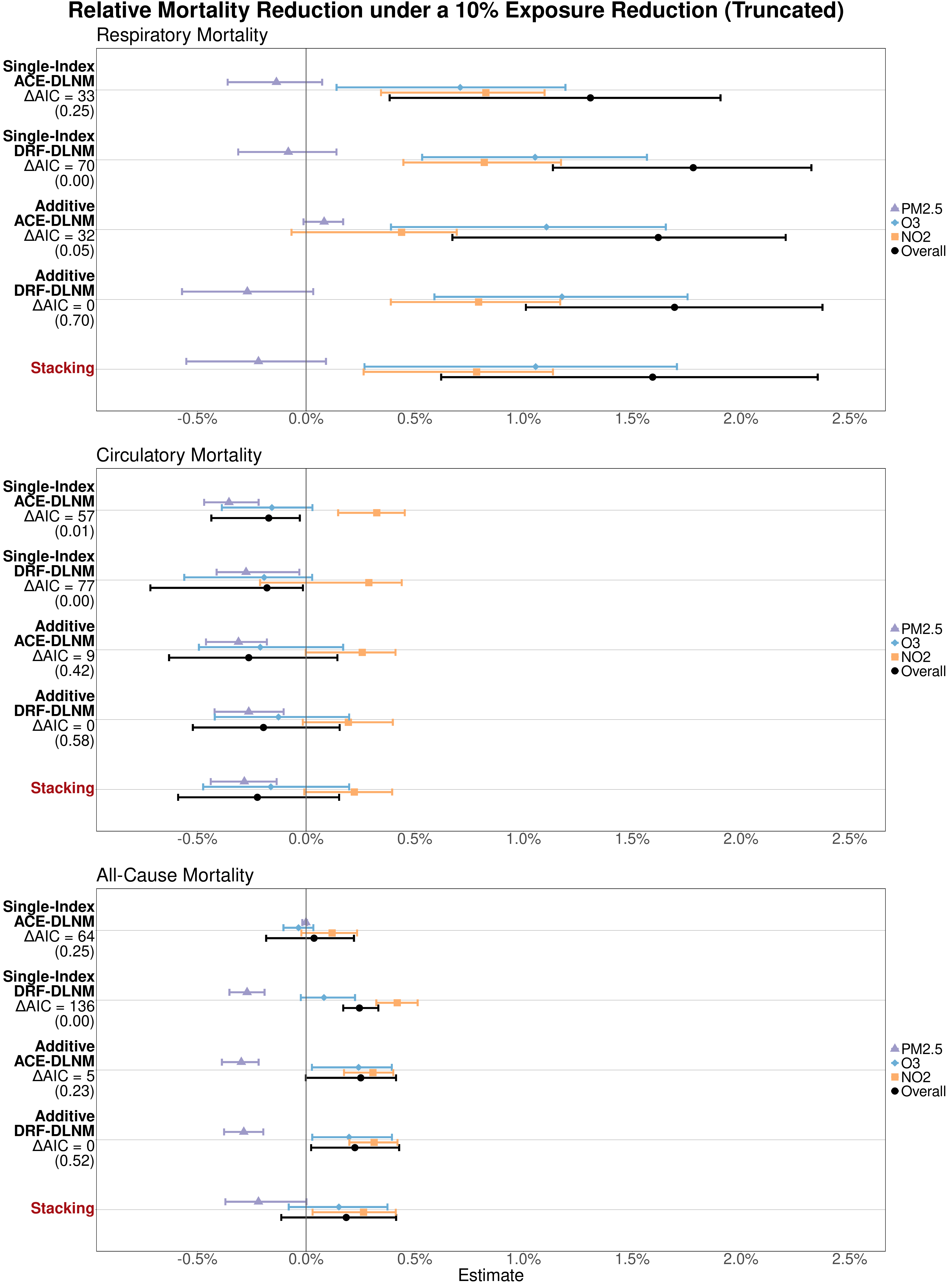}
    \caption{Sensitivity analysis results: relative mortality reduction under a 10\% exposure reduction while reduced levels are truncated at the CD-specific annual minimum. The results are identical to the primary analysis in Section 7.1.}
\end{figure}

\end{document}